\documentclass[aps,prx,twocolumn,amsmath,amssymb,floatfix,superscriptaddress]{revtex4-2}
\usepackage[latin9,utf8]{inputenc}
\usepackage{stmaryrd}
\usepackage{graphicx}
\usepackage{subfigure}
\usepackage{multirow}
\usepackage{mathtools}
\usepackage{xcolor}
\usepackage{textcomp}
\usepackage{amssymb}
\usepackage{gensymb}
\usepackage{calligra}
\usepackage{braket}
\usepackage{float}
\usepackage{bbold}

\pdfoutput=1

\usepackage{xcolor,mathrsfs,dsfont}
\definecolor{darkblue}{RGB}{0,0,150}
\definecolor{nightblue}{RGB}{0,0,100}
\definecolor{evergreen}{RGB}{0,120,0}
    
\usepackage{graphicx,mathtools,bm}
\graphicspath{{./Figures/}}
\usepackage[colorlinks = true,
            linkcolor = blue,
            urlcolor  = blue,
            citecolor = blue,
            anchorcolor = blue]{hyperref}

\let \oldbm \bm
\renewcommand{\vec}[1]{\oldbm{#1}}

\usepackage[english]{babel}
\usepackage[babel,kerning=true,spacing=true]{microtype}
\usepackage[utf8]{inputenc}

\newcommand{\dd}{\mathrm{d}}
\newcommand{\inv}{\ensuremath{^{-1}}}
\DeclareMathOperator{\sgn}{sgn}

\def\up{\uparrow}

\def\bk{{\vec k}}

\def\bq{{\vec q}}

\def\bR{{\vec R}}

\def\bm{{\vec m}}

\def\br{{\vec r}}
\def\bq{{\vec q}}

\def\bsigma{{\boldsymbol \sigma}}
\def\sgn{\mathop{\mathrm{sgn}}}
\def\tr{\mathop{\mathrm{tr}}}
\def\im{\mathop{\mathrm{im}}}
\def\re{\mathop{\mathrm{re}}}
\def\rk{\mathop{\mathrm{rk}}}
\def\Z{\mathds{Z}}
\def\T{\mathcal{T}}

\def\P{\mathbb{P}}
\def\id{\mathbb{1}}
\def\S{\mathbb{S}}

\def\V{\mathcal{V}}
\def\G{\mathcal{G}}

\def\H{\mathcal{H}}

\def\inv{^{-1}}

\def\half{{1\over2}}

\newcommand{\proj}[1]{\left| #1 \right\rangle \left\langle #1 \right|}
\newcommand{\projj}[2]{\left| #1 \right\rangle \left\langle #2 \right|}
\newcommand{\ev}[1]{\left\langle #1 \right\rangle}

\makeatother

\newcommand{\SIref}[1]{#1}
\newcommand{\SIeqref}[1]{\textup{(#1)}}
\begin{document}
\title{Ring states in topological materials}
\author{Raquel Queiroz}
\email{raquel.queiroz@columbia.edu}
\affiliation{Department of Physics, Columbia University, New York, NY 10027, USA}
\affiliation{Center for Computational Quantum Physics, Flatiron Institute, New York, New York 10010, USA}
\author{Roni Ilan}
\affiliation{Raymond and Beverly Sackler School of Physics and Astronomy, Tel Aviv University, Tel Aviv, Israel}
\author{Zhida Song}
\affiliation{Department of Physics, Princeton University, Princeton, NJ 08544, USA}
\author{B. Andrei Bernevig}
\affiliation{Department of Physics, Princeton University, Princeton, NJ 08544, USA}
\affiliation{Donostia International Physics Center, P. Manuel de Lardizabal 4, 20018 Donostia-San Sebastian, Spain}
\affiliation{IKERBASQUE, Basque Foundation for Science, Bilbao, Spain}
\author{Ady Stern}
\affiliation{Department of Condensed Matter Physics,
Weizmann Institute of Science,
Rehovot, Israel}
\date{}

\begin{abstract}
Ingap states in semiconductors are commonly characterized by hydrogenic models within the effective mass approximation. However, when impurities are strong, this description fails. The bound states are non-perturbative with energy pinned to a zero of the impurity-projected Green's function. We show that in the presence of Wannier obstructions, such zeros are irremovable. That is, no symmetric, gap-preserving deformation of the bulk can return the ingap state to the band continuum. We call these states \emph{ring states}: their wavefunctions show a node at the impurity site, and their mixed band character is set by minimizing the level repulsion exerted by the bands above and below the gap. In Chern insulators, the band mixing endows occupied ring states with local circulating currents, remnant of a small edge mode surrounding the vacancies. Clusters of vacancies bind ring states that hybridize to form dispersive chiral edge modes, providing a real-space view of bulk--boundary correspondence.
\end{abstract}
\maketitle

In semiconductors, ingap states bound to impurities are commonly captured by hydrogenic models within the effective-mass approximation~\cite{Kohn.Kohn.1957}, where the bound state is built from a single band edge and its energy depends on the impurity strength as a single phenomenological parameter. When the impurity is strong, this single-band picture fails~\cite{Slater.Koster.1954,Callaway.Callaway.1967,Pantelides.Bernholc.1978,Schlueter.Baraff.1978,Dow.Hjalmarson.1979}: the bound state mixes all momentum eigenstates of the host, its energy is pinned by a zero of the impurity-projected Green's function, and its wavefunction is built from the orbital projections of the bands above and below the gap. The bound state then becomes a property of the band structure as a whole, rather than of any individual band. We ask under what conditions such a state is irremovable, and what protects it.

The answer, as we show, is determined by the topology of the surrounding bands. Topological matter is defined by the existence of Wannier obstructions, which occur when Bloch eigenstates, each with a distinct crystal momentum, cannot be recombined into an orthogonal set of exponentially localized states at specific positions in a unit cell, aligning with the symmetric representation of the site symmetry group~\cite{Bernevig.Bradlyn.2017,Marzari.Brouder.2007,Vanderbilt.Soluyanov.2011, Dai.Yu.2011,Read.Read.2017,Vanderbilt.Marzari.2012, Vishwanath.Po.2018}. In the presence of topological obstructions, symmetric Wannier functions display a polynomial rather than exponential decay around their center. While this nuanced property has not been directly measured, current efforts are directed toward finding the imposed topological boundary modes~\cite{Mele.Kane.2005,Hasan.Hsieh.2008,Zhang.Zhang.2009,Yazdani.Roushan.2009,Kane.Hasan.2010}, and bulk electromagnetic responses which are enforced by these obstructions~\cite{Zhang.Koenig.2007,Burch.Ma.2021,Beidenkopf.Bernevig.2022}. For the impurity problem, the obstruction has a direct spectral consequence: if the orbital removed by the impurity belongs to a band representation that is split across a band gap due to a topological obstruction of the bands, then the orbital spectral weight is necessarily shared between bands above and below the gap. This is reflected in \emph{necessary ingap zeros in the impurity-projected Green's function} (IGF), which become the pinning energies of impurity bound states in the strong-impurity limit.

\begin{figure}[tb]
\centering
\includegraphics[width=\columnwidth]{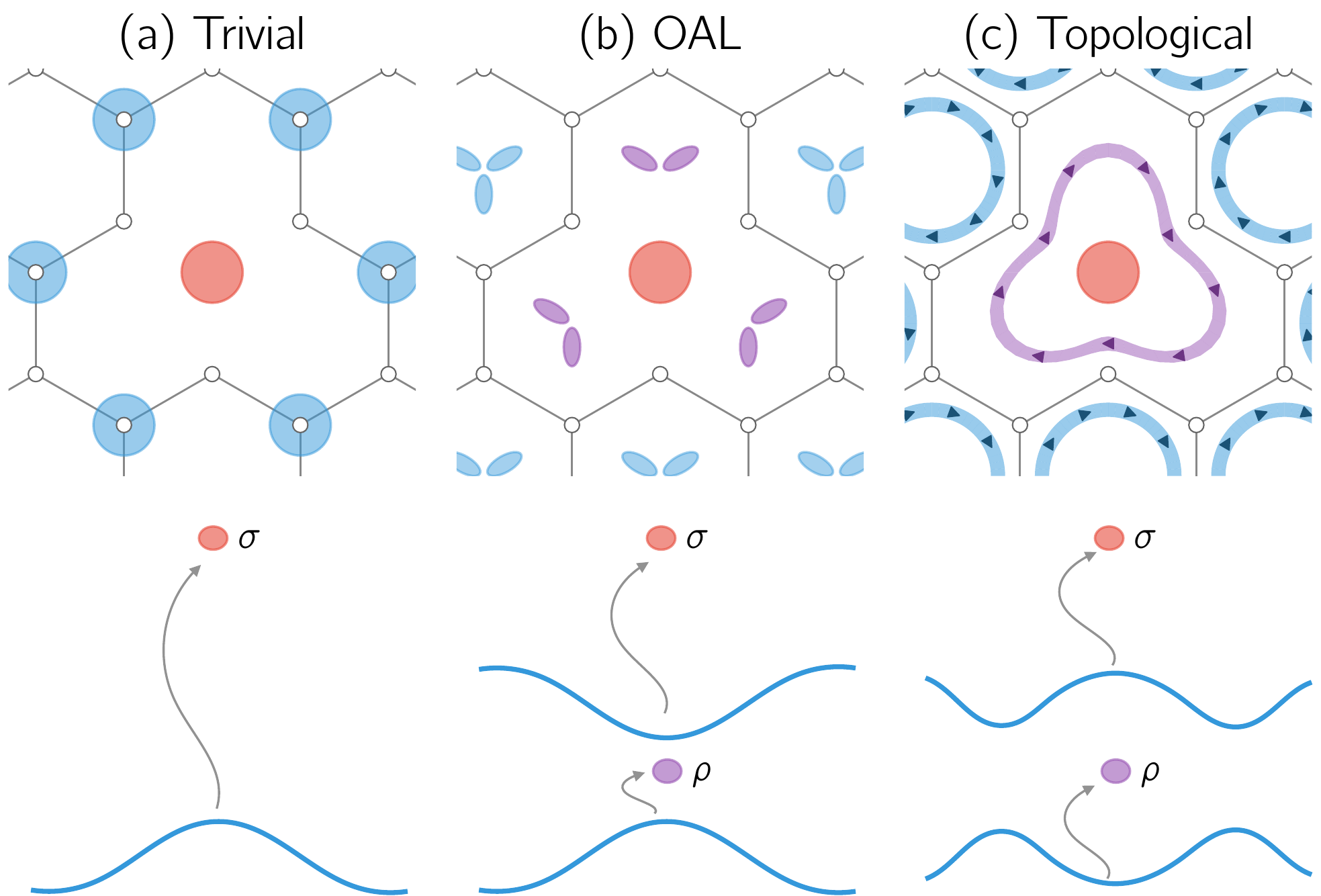}
\caption{Schematic view of vacancy-induced impurity states on a honeycomb lattice. The vacancy removes all orbitals on the central site (red). (a) Trivial insulator: surrounding bulk orbitals (blue) are only weakly affected. (b) Obstructed atomic limit: dangling bonds (purple) appear on symmetry-related sites around the lattice defect. (c) Chern insulator: a ring state (purple) binds at the defect and carries a circulating current around the vacancy. Bottom row: schematic spectrum in each phase, showing the "peeled off" states from a local perturbation $\V=v\proj{\sigma}$, which in the strong-impurity limit $v\to\infty$ creates a bound state $\ket\sigma$ at energy $\sim v$. However, in the obstructed and topological phases, a ring state $\ket\rho$ remains pinned within the band gap.}
\label{fig:schematic}
\end{figure}

We call the resulting strong-impurity states \emph{ring states}. Their universal feature is that they remain at finite energy in the vacancy limit while sharing the symmetry and being orthogonal to the removed local orbitals. In the continuum Chern insulator explored in detail in this work the ring state forms a density node at the defect with a circulating-current texture which appears from the off-diagonal contributions to the IGF. Our setting is therefore a \emph{trivial local defect} of finite rank in a topological or obstructed host. This is distinct from modes bound to topological defects and domain walls, including Jackiw--Rebbi and Majorana modes. Such defects are extensive-rank perturbations whose protected modes are governed by the defect topological index rather than the zeros of a finite-dimensional impurity Green's function~\cite{Rebbi.Jackiw.1976,Vishwanath.Ran.2009,Kane.Teo.2010,Zaanen.Juricic.2012,Ryu.Chiu.2016}. The topologically protected zeros of the IGF discussed in the present work can protect impurity states broadly, including dangling-bond states in obstructed atomic limits such as silicon, diamond, and transition-metal dichalcogenides~\cite{Cardona.Yu.2010,Queiroz.Holbrook.2026,Ugeda.Calugaru.2026,Queiroz.Kang.2025}, impurity resonances in Dirac materials~\cite{Neto.Pereira.2006,Balatsky.Wehling.2014,Gomez-Rodriguez.Ugeda.2010,Balatsky.Huang.2013}; localized states in topological insulators~\cite{Balents.Slager.2015} and Landau level gaps~\cite{Yazdani.Feldman.2016}, and Yu--Shiba--Rusinov states in superconductors~\cite{Balatsky.Zhu.2006}.
In this paper we focus on \emph{topologically protected} ring states which are \emph{enforced} by bulk band topology: unlike ring states in trivial semiconductors such as GaAs, which, although extremely robust, can in principle be removed by symmetric, gap-preserving deformations of the bulk. Earlier work established the existence of such states from the logarithmic band-edge divergence of the full Green's function in two dimensions~\cite{Balents.Slager.2015}; here we show that their topological protection requires only a topological obstruction of the surrounding bands, with no reference to spatial dimension or the shape of the band edge.
With a full understanding of their structure and robustness, ring states may be used as local probes of band topology~\cite{Balents.Slager.2015,Rachel.Diop.2020,Balatsky.Biswas.2010}, as seen in Bi$_2$Se$_3$~\cite{Kapitulnik.Alpichshev.2012} and Co$_3$Sn$_2$S$_2$~\cite{Hasan.Xing.2020,Hasan.Yin.2020,Gao.Chen.2024,Liu.Liu.2022,Beidenkopf.Nag.2024}. Because their existence and finite pinning energies are set by bulk topology rather than local impurity details, ring states are stable strong-impurity states in solid-state systems. Defects also play important roles in optical response, catalysis, and semiconductor characterization~\cite{Javey.Amani.2015, Xie.Xie.2013,Krasheninnikov.Komsa.2015, Lang.Lang.1974}. Ring states also illuminate bulk-boundary correspondence: in the Chern vacancy-cluster setting analyzed in Sec.~\ref{sec:beyondrank1}, clustered vacancies hybridize into dispersive chiral modes, suggesting a real-space picture of edge states as dense arrays of overlapping ring states (Note~\SIref{S11}).

This article is organized as follows: First, we introduce topological obstructions in Sec.~\ref{sec:obstruction}, and the formalism to find impurity states in Sec.~\ref{sec:impurityGF}. We discuss a simple case of two flat bands with an impurity of varying overlap between bands to illustrate the appearance of the ring states and how they can be interpreted as ``antibonding states'' between band projected states in Sec.~\ref{sec:flatbands}. We show in Sec.~\ref{sec:levelrepulsion} that their pinning energy minimizes the level repulsion exerted by the band states above and below the gap, and associate this energy with the zeros of the impurity projected Green's function. We generalize our arguments to dispersive bands in Sec.~\ref{sec:dispersive} and show the spectral and wavefunction evolution of ring states with impurity strength in Sec.~\ref{sec:evolution}. We finally discuss the concrete example of ring states in a Chern insulator in Sec.~\ref{sec:chern} and their relation with boundary states in Sec.~\ref{sec:beyondrank1}.

\section*{RESULTS}

\section{Ring states}\label{sec:rings}
\subsection{Topological obstructions}\label{sec:obstruction}
We consider a general tight-binding Hamiltonian, constructed from a set of basis states $\ket{\phi^\sigma_i}$ at lattice sites $\bq_i$ transforming under an irreducible representation $\sigma$ of the local site symmetry group
of $\bq_i$\cite{Aroyo.Elcoro.2017,Wondratschek.Aroyo.2006},
\begin{align}\H=\sum_{ij,\sigma\tau}h_{ij}^{\sigma\tau}\projj{\phi_i^\sigma}{\phi_j^\tau}=\sum_{\bk\alpha}\varepsilon_{\alpha \bk} \proj{\psi^\alpha_\bk}.\label{ham}\end{align}
The Hamiltonian is diagonalized by Bloch eigenstates with energies $\varepsilon_{\alpha \bk}$, carrying a band index $\alpha$ and crystal momentum $\bk$. For a local orbital at the impurity site $\bq_0$, we write the short-form $\ket{\sigma}\equiv\ket{\phi^\sigma_0}$.

We are interested in what a topological obstruction in a band $\alpha$ can say about the protection of states induced by impurities in gaps between bands. In short, the impurity projection of the resolvent Green's function $\G(\omega)=(\omega-\H)\inv$ enforces an ingap zero to which a strong-impurity pole is pinned. To see this, we start by considering the band projector $\P^\alpha\equiv\sum_\bk\proj{\psi^\alpha_\bk}$. A topological obstruction implies that any set of symmetric Wannier states spanning $\P^\alpha$ cannot be exponentially localized~\cite{Marzari.Brouder.2007,Vanderbilt.Soluyanov.2011,Read.Read.2017}. As further discussed in Note~\SIref{S6}, this means the band does not transform under a band representation~\cite{Zak.Zak.1980,Zak.Zak.1981,Bernevig.Bradlyn.2017}. Consequently, the band subspace cannot be adiabatically connected to any atomic subspace $\P^\sigma\equiv\sum_i\proj{\phi^\sigma_i}$, and there is no unitary transformation $U^\dag\P^\alpha U=\P^\sigma$ which takes the band subspace into the atomic subspace. In other words, the overlap between the two spaces is necessarily singular.

To make the Wannier obstruction apparent, consider using the atomic orbitals $\ket{{\phi^\sigma_i}}$ as trial states. Projecting into the band we get the non-orthogonal orbitals $\ket{\Upsilon_i}\equiv\P^\alpha\ket{\phi^\sigma_i}$, and the overlap matrix
$S^\alpha_{ij}=\bra{\Upsilon^\alpha_i}\Upsilon^\alpha_j\rangle$~\cite{Vanderbilt.Thonhauser.2006} has at least one zero eigenvalue, guaranteeing the states cannot be orthogonalized~\cite{Vanderbilt.Thonhauser.2006,Vanderbilt.Soluyanov.2012yvd}. Conversely, because an isolated band is invariant under lattice translations, having one local orbital $\ket{\sigma}\equiv\ket{\phi^\sigma_0}$ entirely in that band would also put all translated orbitals $\ket{\phi_i^\sigma}$ in the same band. The band would then realize the atomic basis generated by $\ket{\sigma}$. A Wannier obstruction forbids this, so the orbital \emph{must have nonzero spectral weight} $s_\alpha\equiv\bra{\sigma}\P^\alpha\ket{\sigma}$ across bands $\alpha$, below and above the topological (or obstructed) gap, and equivalently the projected states $\Upsilon^\alpha_0$ have \emph{necessarily} nonzero norm $\|\Upsilon_0^{\alpha}\|^2=s_\alpha$ across multiple bands.

This realization motivates us to consider a perturbation that affects a single local orbital,
\begin{align}\V=v\proj{\sigma},\end{align}
which effectively removes it from the system at large $v$. Since $\ket{\sigma}$ has nonzero weight in multiple bands, the perturbation affects them all. In fact, as we prove below, and is schematically shown in Fig.~\ref{fig:schematic}, this rank-1 perturbation removes \emph{multiple} states from the bands which implies, by conservation of the size of the Hilbert space, the creation of bound states within spectral gaps. Those leftover ingap states are what we call ring states.

\subsection{Impurity bound state from the Green's function}\label{sec:impurityGF}
 
Bound states of the perturbed Hamiltonian at energies $E$ outside the band continuum, $(\H+\V)\ket{\varphi_E}=E\ket{\varphi_E}$, can be found using the bare resolvent Green's function
\begin{align}\G(\omega)=\sum_{\alpha \bk}\frac{\proj{\psi^\alpha_{\bk}}}{\omega-\varepsilon_{\alpha \bk}},\label{eq:gf}\end{align}
where $\ket{\psi^\alpha_\bk}$ are band eigenstates with energies $\varepsilon_{\alpha\bk}$. As further expanded in Note~\SIref{S1}, outside the band continuum, bound states satisfy
\begin{align}
\ket{\varphi_E}=\G(E)\V\ket{\varphi_E},\label{eq:boundstatesimp}
\end{align}
corresponding to poles of the $T$-matrix\cite{Slater.Koster.1954,Callaway.Callaway.1967} \begin{align}
    \T(\omega)=\V[1-\G(\omega)\V]\inv.
\end{align}

For an impurity affecting a single orbital $\ket{\sigma}$, the only nonzero matrix element of $\T(\omega)$ that may contain a bound state pole is
\begin{align}t^\sigma(\omega)=\bra{\sigma}\T(\omega)\ket{\sigma}=\frac{v}{1-vg^\sigma(\omega)},\label{eq:tmatrix}\end{align}
where $g^\sigma(\omega)\equiv\bra{\sigma}\G(\omega)\ket{\sigma}$ is the \emph{impurity projected Green's function}. Bound states occur at energies where $g^\sigma(E)=1/v$. In the strong impurity limit $v\to\pm\infty$, these are pinned to energies where $g^\sigma(E)=0$.

Writing the retarded IGF as \begin{align}
    g^\sigma(\omega+i0^+)=\mu^\sigma(\omega)-i\pi\nu^\sigma(\omega),
\end{align} where $\nu^\sigma$ is the projected density of states (positive in the band continuum) and $\mu^\sigma$ is the real part, zeros of $g^\sigma$ require both $\nu^\sigma=0$ and $\mu^\sigma=0$, which can only occur in spectral gaps. At energies far from the spectrum, $g^\sigma\sim1/\omega$, so there is always exactly one bound state at large energy $E\sim v$. This follows from Weyl's inequalities~\cite{Tao.Tao.2012}: the number of eigenstates of $\H+\V$ outside the spectral width of $\H$ is at most the rank of $\V$, here chosen to be one. Any additional zeros of $g^\sigma$ must therefore lie in gaps between bands. Note~\SIref{S3} gives the corresponding state-counting result, separating this high-energy impurity branch from the finite-energy zeros.

\subsection{Exact bound state solution for two flat bands}
\label{sec:flatbands}
Let us first consider the simple yet enlightening case of two non-degenerate flat bands, $\alpha=1,2$. Without loss of generality, they are located at energies $\varepsilon_1=0$ and $\varepsilon_2=\Delta$. The tight-binding Hilbert space in
\eqref{ham} contains two orthogonal basis orbitals $\sigma=1,2$ per unit cell. We define the band projected states $\ket{\Upsilon^1}\equiv\P^1\ket\sigma$ of norm $s_1$ and $\ket{\Upsilon^2}\equiv\P^2\ket\sigma$ of norm $s_2$, satisfying $s_1+s_2=1$. The difference $\delta s=s_1-s_2$ characterizes the relative overlap of the impurity $\V$ with both bands. 

Introducing the bare flat band Green's function 
\begin{align}\G(\omega)=\sum_\alpha\frac{\P^\alpha}{\omega-\varepsilon_\alpha}.\label{eq:gfflat}\end{align}
in the bound state equality \eqref{eq:boundstatesimp}, we find the general form for a bound state in the flat band limit
\begin{align}\ket{\varphi_E}=
\sum_\alpha\frac{v\lambda_E}{E-\varepsilon_\alpha}\ket{\Upsilon^{\alpha}},\label{eq:boundstategen}\end{align}
with $\lambda_E\equiv\bra{\sigma}{\varphi_E}\rangle$ a scalar reflecting the overlap between the bound state $\ket{\varphi_E}$ and the affected orbital, or bare impurity eigenstate, $\ket{\sigma}$.
For large $v$, we consider the only two possible bound states. One, with a large overlap with $\ket{\sigma}$ which we call $\ket{\sigma_E}$, 
with $\lambda_E\sim 1$. The energy of this bound state is $E\sim v$ and exists independently of the details of the Hamiltonian $\H$.
A second solution $\ket{\rho_E}$ exists exclusively in multi-band Hamiltonians when $v\lambda_E$ remains finite as $v\to\infty$ by acquiring a vanishingly small overlap $\lambda_E\to0$ with the bare impurity eigenstates $\ket{\sigma}$. In other words, to guarantee its finite energy $E$, the ingap state avoids the impurity by the orthogonality of its wavefunction. Since $\ket{\rho_E}$ transforms under the same symmetry representation as the impurity eigenstate $\ket{\sigma}$, this orthogonality can only be assured by spatially avoiding the bare impurity eigenstate $\ket{\sigma}$, forming a node at the impurity site. That is, the state $\ket{\rho_E}$ forms a \emph{ring} around the impurity.

\begin{figure}[tb]
\centering
\includegraphics[width=\columnwidth]{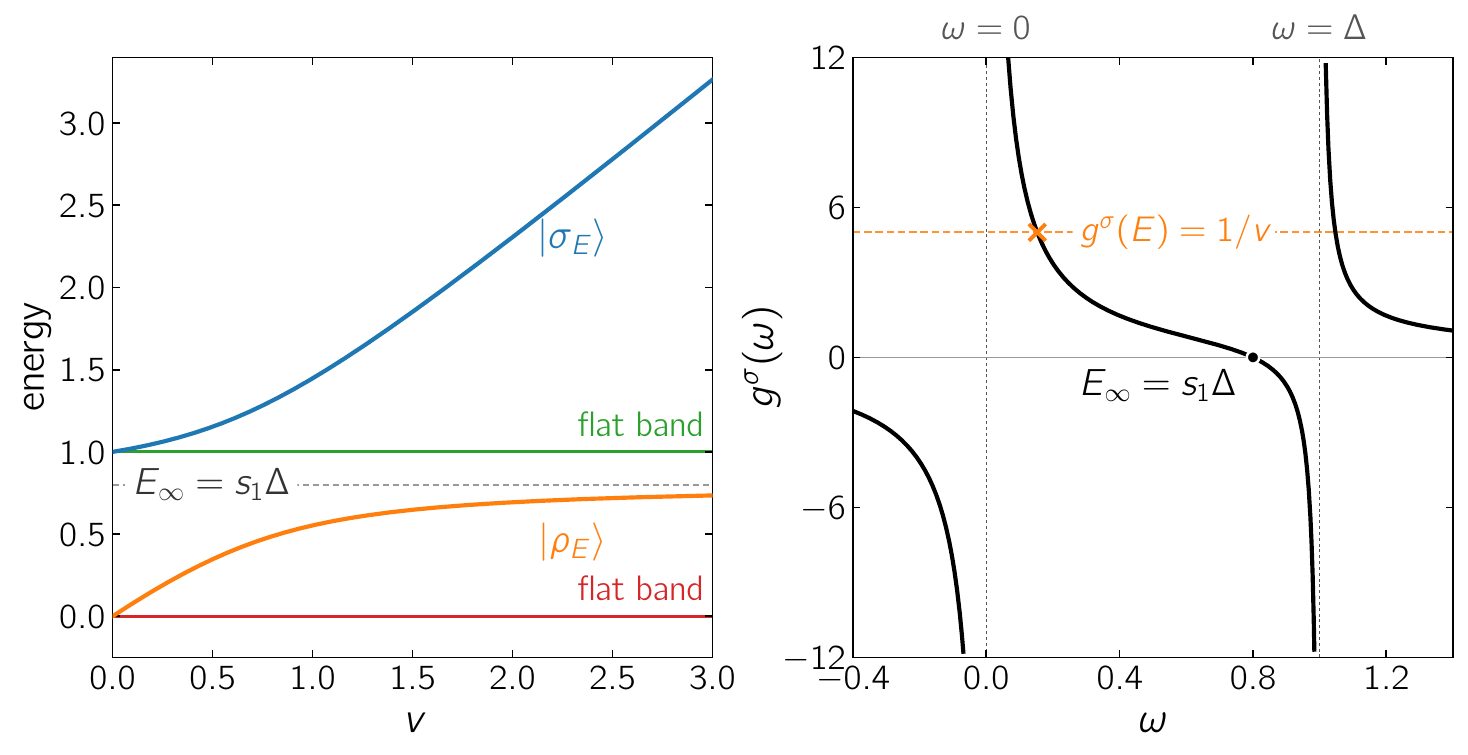}
\caption{Evolution of the bound state energies for two flat bands separated by a gap $\Delta=1$, caused by an impurity that overlaps with both bands, with a relative overlap of $\delta s=0.6$. At $v\to\infty$, the pinning energy $E_\infty$ of the ingap bound state $\ket\rho$ reflects how much the impurity state $\ket{\sigma}$, overlaps with the bands above and below the gap, therefore reflecting the orbital content of the unperturbed bands. The right panel shows the impurity-projected Green's function $g^\sigma(\omega)$, \eqref{eq:gsflat}: positive residues on both flat-band poles at $\omega=0$ and $\omega=\Delta$ force a single zero in the gap at $\omega=E_\infty=s_1\Delta$, the strong-impurity pinning energy of the ring state. The dashed line marks $g^\sigma(E)=1/v$, and its intersection (\(\times\)) with $g^\sigma$ gives the bound-state energy at finite $v$.}
\label{fig:evolution}
\end{figure}

Now we show that the energy of the state $\ket{\rho_E}$  in the large $v$ limit is  determined
by the unperturbed Hamiltonian $\H$ and the affected state up to order $O(1/v$), becoming largely independent of the magnitude of the impurity $v$ in the limit $v\to\infty$. 
Projecting Eq.~\eqref{eq:boundstategen} onto $\ket\sigma$ gives the bound-state condition 
\begin{align}
g^\sigma(\omega)=\sum_\alpha\frac{s_\alpha}{\omega-\varepsilon_\alpha}=\frac{s_1}{\omega}+\frac{s_2}{\omega-\Delta}=\frac{1}{v}.\label{eq:gsflat}
\end{align}
It has a positive-residue pole at each band energy with $s_\alpha\neq 0$, and decreases monotonically between consecutive poles, so it must change sign at some ingap energy $\varepsilon_1<E_\infty<\varepsilon_2$ (Fig.~\ref{fig:evolution}). Solving the bound-state condition explicitly yields two solutions (Note~\SIref{S2}),
\begin{equation}
E_\pm=\half({\Delta+v\pm\sqrt{\Delta^2 -2v\delta s\Delta+v^2}}),\label{eq:flatbandsolsmain}
\end{equation}
for $\delta s=s_1-s_2$, plotted in Fig.~\ref{fig:evolution}. We consider without loss of generality $v>0$ and $s_1>s_2$. In this case, $E_+\sim v+\Delta(1-\delta s)/2$ and $E_-\sim \Delta(1+\delta s)/2$, to first order in $1/v$. 
As long as $|\delta s|\neq1$, there are \emph{two} solutions that deviate from the unperturbed band energies. At infinite $v$, the $\ket\rho$ state remains in the gap, at a distance $\Delta s_1$ from the bottom band and $\Delta s_2$ from the top band. At the limit $\delta s=1$, where $\ket\sigma$ lies entirely in one band, Eq.~\eqref{eq:flatbandsolsmain} reduces to $E_+=v$ and $E_-=\Delta$: only the impurity state $\ket{\sigma_E}$ survives, and the ring state collapses onto the upper band edge.

As we explicitly demonstrate in Note~\SIref{S2}, the $\ket{\rho_E}$ state is normalizable provided both projected states $\ket{\Upsilon^\alpha}$ can be normalized. The limiting form of the ingap state is independent of $v$,  \begin{align}\ket{\rho_E}\to\ket{\rho}=\sqrt{s_2/s_1}\ket{\Upsilon^1}-\sqrt{s_1/s_2}\ket{\Upsilon^2}.\label{eq:limring}\end{align} This asymptotic form can be interpreted as an ``antibonding state" between the two projected states $\ket{\Upsilon^\alpha}$ removed from the bands above and below the gap by the impurity $\V$, orthogonal to the eigenstate of the bare impurity $\V$ or ``bonding state" $\ket{\sigma}=\ket{\Upsilon^1}+\ket{\Upsilon^2}$. Both states are normalized. The overlap between the ring state and the original unperturbed bands is opposed to the overlap of the impurity state. That is $\bra{\rho}\P^1\ket{\rho}=s_2$ and $\bra{\rho}\P^2\ket{\rho}=s_1$.

At this stage we have not invoked band topology. \emph{Existence} of an ingap state in the flat band limit requires $0<s_\alpha<1$ on both bands. \emph{Topological protection} of the ring state requires the strictly stronger condition that neither band $\alpha$ transforms under the band representation induced by $\ket\sigma$ at $\bq_0$ which, as discussed in Sec.\ref{sec:obstruction} forbids $s_\alpha=1$; If there is no obstruction, the overlap on one band can be adiabatically taken to 1, and the ring state removed. If the impurity sits on an atomic site, as is common for vacancies, this implies that all obstructed atomic, fragile~\cite{Vishwanath.Po.2018,Bernevig.Bradlyn.2019nh,Yang.Ahn.2019pt,Bouhon.Slager.2020np}, and stable topological phases stabilize ring states in the flat band limit; If the impurity sits away from an atomic site, only stable topological phases are guaranteed to host ring states.

\subsection{Level repulsion interpretation}\label{sec:levelrepulsion}

In Sec.~\ref{sec:flatbands} we constructed $\ket\rho$ explicitly in the flat-band limit and found it pinned at an energy $E_\infty$ inside the gap, set by the orbital weights $s_1,s_2$ of $\ket\sigma$ on the two bands. We now interpret this pinning physically. The bands above and below the gap each exert a level repulsion on the ingap state: the lower band pushes it up, the upper band pushes it down. The ring state settles at the energy $E_\infty$ where these two opposing forces balance. This balance is encoded in the impurity-projected Green's function, as we now show.

From strong-coupling perturbation theory (Note~\SIref{S4}), the bound-state energy $E(v)$ evolves with the impurity strength according to the Newton equation
\begin{align}
\partial_v^2 E= -2\lambda_E^2\mu^\sigma(E),\label{eq:forceeq}
\end{align}
where $\mu^\sigma(\omega)=\re g^\sigma(\omega+i0^+)$ is the real part of the retarded impurity-projected Green's function and $\lambda_E\equiv\langle\sigma|\varphi_E\rangle$. The minus sign reflects differentiation with respect to $v$ rather than the strong-coupling variable $1/v$; the leading nonlinear correction to Eq.~\eqref{eq:forceeq} is of order $\lambda_E^4$ and is derived in Note~\SIref{S4}. Equation~\eqref{eq:forceeq} admits a direct mechanical reading: $\mu^\sigma(E)$ is the effective \emph{force} exerted on a state of energy $E$ by all other eigenstates of $\H$ through their overlap with $\ket\sigma$, and zeros of $\mu^\sigma$ are fixed points of this flow.

The stability of a fixed point is determined by the slope of $\mu^\sigma$. A zero with $\partial_\omega\mu^\sigma(E_\infty)<0$ is an attractor: along the physical spectral branch, the energy approaches this zero as the impurity strength increases in magnitude, while the impurity overlap vanishes. A zero with positive slope is a repeller. Because $\mu^\sigma(\omega)$ decreases monotonically outside the spectrum of $\H$, ingap zeros are always attractors, while repellers can only appear inside band continua. We denote these two types of fixed points as \emph{spectral attractors} and \emph{spectral repellers} respectively. Hellmann--Feynman gives $\partial_v E=\lambda_E^2$, so the bound state moves only as fast as it touches the impurity. Finite $E$ as $v\to\infty$ then forces $\lambda_E\to 0$: the state pulls off the impurity site, and the flow stops at $E_\infty$.

The location of $E_\infty$ inside the gap is therefore a direct readout of the orbital content of $\ket\sigma$. In Fig.~\ref{fig:evolution}, $E_\infty=0.8\Delta$ tells us that $\ket\sigma$ is $80\%$ supported in the lower band and $20\%$ in the upper band; correspondingly $\ket\rho$ carries the opposite content and lives mostly on the upper band. A local negative potential would produce the same ring state, with only the regular impurity state $\ket{\sigma_E}$ flipping sign. Beyond the flat-band limit this picture extends in two directions, addressed in the next two sections: a completeness argument shows that an obstructed band representation forbids absorbing $\ket\rho$ into the perturbed bands (Sec.~\ref{sec:proofrings}), and the Hilbert transform of $\nu^\sigma$ supplies the same fixed-point protection for dispersive bands (Sec.~\ref{sec:dispersive}).

\subsection{Necessity of ring states in topological gaps}\label{sec:proofrings}

We now show that, in the vacancy limit, the finite-energy state constructed above cannot be absorbed into the perturbed extended bands when the original bands are obstructed relative to the atomic representation defined by $\ket{\sigma}$. Let us assume the ring state $\ket{\rho}$ can be absorbed into a perturbed band $\tilde\P^\alpha$ formed by the extended scattering eigenstates in the presence of $\V$. 
Completeness ensures that
\begin{align}\sum_\alpha\tilde\P^\alpha+\proj{\sigma}+\proj{\rho}=\sum_\alpha\P^\alpha=\id.\label{eq:completeness}\end{align}
which implies the overlap $\bra{\sigma}\tilde \P^\alpha\ket{\sigma}=0$. Therefore, the combined subspace $\tilde\P^\alpha+\proj{\rho}$ has no overlap with $\ket{\sigma}$ for any perturbed band $\alpha$. In contrast, the overlap $\bra{\sigma} \P^\alpha\ket{\sigma}>0$ strictly, guaranteed by the topological obstruction. Therefore, there is no lattice-symmetric unitary transformation $U$ such that $U^\dag(\tilde\P^\alpha+\proj{\rho})U=\P^\alpha$, or equivalently $\tilde\P^\alpha+\proj{\rho}\simeq\P^\alpha$.

Take a two-band model $\alpha=1,2$. Suppose it were possible to absorb $\ket{\rho}$ into one of the perturbed bands via a lattice-symmetric unitary, $\tilde\P^1+\proj{\rho}\simeq\P^1$. Completeness then requires $\tilde\P^2+\proj{\sigma}\simeq\P^2$ under the same transformation. By translation invariance, the same transformation enforces the $\sigma$ orbital to be translated to $\ket{\phi^\sigma_i}$ at every site $i$, making band $2$ equivalent to the atomic limit defined by $\ket{\sigma}$. This contradicts the topological obstruction; hence $\tilde\P^1+\proj{\rho}\not\simeq\P^1$ for any lattice-symmetric unitary transformation. The argument requires only that the bands are separable and closed under the lattice symmetries, not that the bands are flat.

\subsection{Phase shifts and the Friedel sum rule}\label{sec:dispersive}

We extend the analysis to dispersive bands $\varepsilon_{\alpha\bk}$, where the impurity-projected Green's function takes the Lehmann form
\begin{align}g^\sigma(\omega)=\frac{1}{N}\sum_{\alpha,\bk}\frac{s^\sigma_{\alpha \bk}}{\omega-\varepsilon_{\alpha \bk}},\end{align}
with weights $s^\sigma_{\alpha\bk}\equiv|\langle\phi^\sigma_\bk|\psi_{\alpha\bk}\rangle|^2$ defined from the band Bloch states $\ket{\psi_{\alpha\bk}}$ of $\H$ and the local-orbital Bloch sums $\ket{\phi^\sigma_\bk}=N^{-1/2}\sum_i e^{i\bk\cdot\bq_i}\ket{\phi^\sigma_i}$ built from the basis orbitals of Sec.~\ref{sec:obstruction}; the band-integrated weights are $s_\alpha=N^{-1}\sum_\bk s^\sigma_{\alpha\bk}$.

For finite $N$, whenever $\ket\sigma$ has orbital weight on at least one band on each side of the gap ($s_\alpha>0$ on such a band on each side), $g^\sigma$ has positive-weight poles below and above the gap, and Cauchy interlacing between the topmost positive-weight pole below the gap and the bottommost positive-weight pole above forces a real zero of $g^\sigma$ inside the gap (Note~\SIref{S7.1}). This zero exists in trivial and topological phases alike; the question is whether it survives the thermodynamic limit.

In the thermodynamic limit the discrete pole structure of $g^\sigma$ collapses; writing the retarded Green's function as $g^\sigma(\omega+i0^+)=\mu^\sigma(\omega)-i\pi\nu^\sigma(\omega)$, only the continuous functions $\mu^\sigma$ and $\nu^\sigma$ survive. To determine whether the ingap zero of $\mu^\sigma$ (the spectral attractor) survives in the thermodynamic limit, we track the sign changes of $\mu^\sigma$ across the bands using the scattering phase shift~\cite{Friedel.Friedel.1958,Mahan.Mahan.2000,Economou.Economou}
\begin{align}\eta^\sigma(\omega)\equiv\arg\!\big[1-v\,g^\sigma(\omega+i0^+)\big],\end{align}
with $\eta^\sigma\to\arg g^\sigma+\pi~(\mathrm{mod}~2\pi)$ as $v\to+\infty$ (Note~\SIref{S1}). Inside any gap, $\nu^\sigma=0$ forces $g^\sigma=\mu^\sigma$ real, so $\arg g^\sigma$ takes values $0$ or $\pi$ according to the sign of $\mu^\sigma$; a sign change of $\mu^\sigma$ between the two edges of a band corresponds to a $\pi$ jump in $\eta^\sigma$ across the band. Physically, this jump counts states removed from the band by the impurity, which can be computed by the impurity-induced change in the total density of states given by the derivative of the phase shift~\cite{Hewson.Hewson.1993},
\begin{align}\delta\nu^\sigma(\omega)=-\frac{1}{\pi}\partial_\omega\eta^\sigma(\omega),\end{align}
and its integral across band $\alpha$ counts the states pulled from the band continuum by the impurity,
\begin{align}N^{\alpha}=\int_{\varepsilon^\alpha_{\rm min}}^{\varepsilon^\alpha_{\rm max}} \frac{1}{\pi}\partial_\omega\eta^\sigma(\omega)\,d\omega,\label{eq:friedel}\end{align}
equal to $1$ at rank-one $\V$ when the change in the continuously followed phase $\eta^\sigma$ across the band is $\pi$ (one state removed) and $0$ when it vanishes (none). At general rank, the counts obey the generalized Friedel sum rule
\begin{align}\sum_{\alpha}N^{\alpha}=N_{\rm ring}+\rk\V,\label{eq:friedelsum}\end{align}
with $\rk\V$ counting high-energy bound states at energies $\sim v$ assumed to be larger than the bandwidth. This Friedel sum rule expresses conservation of the total number of states, since in our construction the potential only shifts the energy of states rather than changing the dimension of the Hilbert space. The ring states are guaranteed to exist when the total number of states removed from the continuum by $\V$ is larger than the rank of $\V$. Focusing again on a rank 1 impurity, \eqref{eq:friedel} implies that there are multiple $\pi$ phase shifts across bands, and consequently a zero of $\mu^\sigma$ in between. This zero is a spectral attractor in the sense of Sec.~\ref{sec:levelrepulsion}, and pins the ring state energy at finite energy as $v\to\pm\infty$.

\begin{figure}[tb]
\centering
\includegraphics[width=\columnwidth]{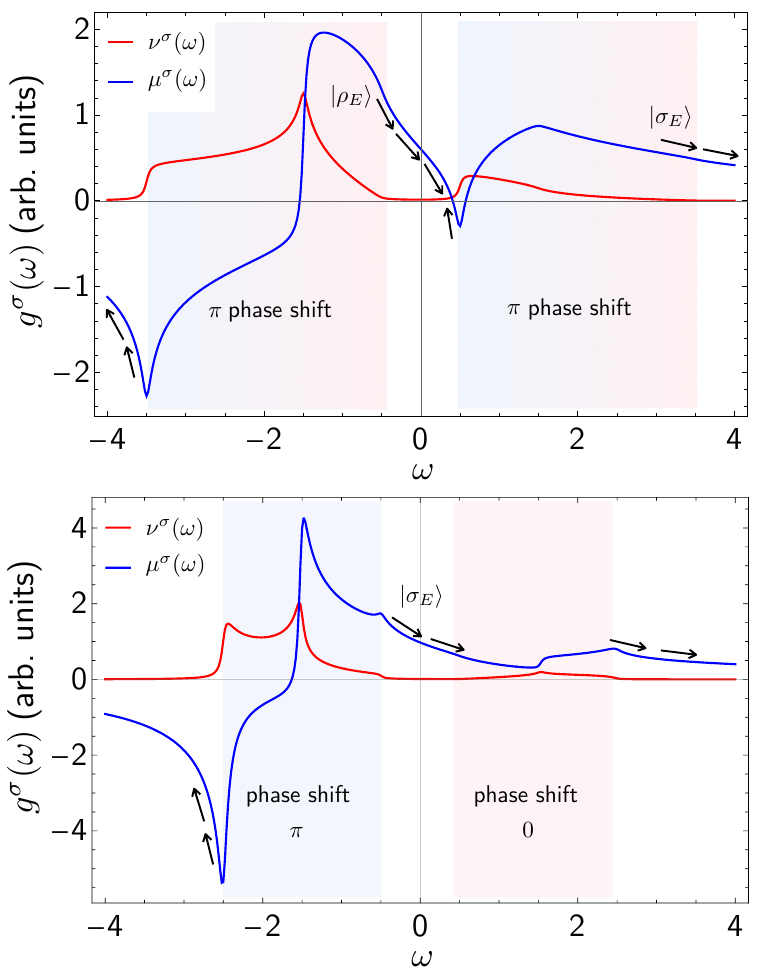}
\caption{Impurity projected Green's function $g^\sigma(\omega)=\mu^\sigma(\omega)-i\pi\nu^\sigma(\omega)$ for an orbital vacancy in a Chern insulator in the topological (top panel) and trivial (bottom panel) phases. The topological obstruction is reflected in an irremovable $\pi$ phase shift across \emph{both} topological bands, and a topologically robust zero of $\mu^\sigma(\omega)$ in the band gap. This zero is a finite energy spectral attractor that binds a state in the strong impurity limit. The energy of the ring state $\ket{\rho_E}$ flows towards a fixed energy $E_\infty$ that minimizes the level repulsion from the valence and conduction bands, and its wavefunction to a state $\ket{\rho}$ orthogonal to $\ket{\sigma}$. On the other hand, the trivial state $\ket{\sigma_E}$ is taken to infinite energy in the same limit, adiabatically connected to the impurity eigenstate $\ket{\sigma}$. We consider $\upsilon=1$ and $b=1$, and $m=-0.5$ for the topological and $m=+0.5$ in the trivial case. The projection shown is the $p$ orbital in the conventions of Eq.~\eqref{eq:cherncont} ($\sigma_z\ket{p}=-\ket{p}$), whose trivial-phase weight resides in the valence band, matching the band-representation narrative of the text.}
\label{fig:phaseshift}
\end{figure}

Let us now illustrate how the phase shifts in the Green's function appear from band inversions. 
Let us consider a two-dimensional insulator formed by two local orbitals $\sigma=s,p$ forming two bands, one transforming under an $s$ band representation (valence) and a $p$ band representation (conduction). By inverting the band character between the top of the valence band and the bottom of the conduction band, the insulator enters a Chern insulating phase. Note that here we describe the transition qualitatively and leave the quantitative analysis to Sec.~\ref{sec:chern}. The impurity projected Green's function is shown in Fig.~\ref{fig:phaseshift}. Focusing on one orbital projection, say $\sigma$ is an $s$ orbital, 
we expect, in the trivial case, both the top and bottom band edges of the valence band to be of $s$ type, leading to a step function in $\nu^s$ at both ends of the valence band. Correspondingly, its Hilbert transform $\mu^s$ will show a divergence at both band edges with opposite signs, leading to a $\pi$ phase shift across the band. On the other hand, the conduction band has a $p$ character on the band edges, and $\mu^\sigma$ is finite and positive below and above the top band, and therefore there is a zero phase shift across the conduction band. This is compatible with an adiabatic deformation of both bands into the atomic limit, in which $\nu^s=0$ in the conduction band.
After the band inversion, the band edge shifts across the gap from one band to another (top panel), and the divergence in $\mu^\sigma$ appears at the bottom of the conduction band rather than the top of the valence band. This imposes a $\pi$ phase shift across both the valence \emph{and} conduction bands, unlike in the trivial phase. 
The topological phase transition is characterized by creating two additional zeros of $\mu^\sigma$, an attractor in the gap, and a repeller in the conduction band. This means that a strong $s$-like impurity will now remove a state from the conduction band, as we expected from the topological case.

\subsection{Topological protection of the phase shift}\label{sec:protection}

We now ask whether the ingap zero of $\mu^\sigma$ is topologically protected, that is, whether it survives symmetric, gap-preserving deformations of the band dispersion. Such deformations act on the projected density of states $\nu^\sigma$, to which $\mu^\sigma$ is related by a Hilbert transform. If the bands are topological, $\nu^\sigma$ cannot be made to vanish identically on either gap-adjacent band, since this would imply $\sigma$ is compactly localized in a single unit cell in the subspace of the other band, which is a contradiction. In a trivial band the orbital overlap is not protected, $\nu^\sigma$ can be deformed to zero, and the ingap zero of $\mu^\sigma$ dissolves into the continuum.

It then follows that the attractor in the gap and the repeller in the adjacent band cannot annihilate each other without closing the bulk gap. Annihilation would require two zeros of $\mu^\sigma$ with opposite slopes to merge at the band edge, i.e., $\mu^\sigma=\partial_\omega\mu^\sigma=0$ simultaneously. Assuming a band-edge profile $\nu^\sigma\sim\omega^\gamma$ up to a cutoff energy $W$, the slope through the Hilbert transform is $\partial_\omega\mu^\sigma\sim 1/\omega$ for $\gamma=0$, diverges logarithmically as $\partial_\omega\mu^\sigma\sim-\log(1-W/\omega)$ for $\gamma=1$, and approaches $\partial_\omega\mu^\sigma\sim-W^{\gamma-1}/(\gamma-1)$ for $\gamma>1$ (Note~\SIref{S7.2}); a divergent edge density ($\gamma<0$) only sharpens the bound. In all cases and in every spatial dimension the slope of $\mu^\sigma$ approaching the band edge is strictly negative, the annihilation is forbidden, and the ingap zero is a \emph{topologically protected ring state}.

Stepping back, only two routes can remove a topologically protected ring state from the gap (Note~\SIref{S7.2}): (i) the bulk gap closes, allowing the ingap attractor and the adjacent-band repeller to merge in a saddle-node bifurcation at $\mu^\sigma=\partial_\omega\mu^\sigma=0$, a bulk topological phase transition; (ii) the orbital weight $s^\sigma_\alpha$ drains to zero on a gap-adjacent band, so $\nu^\sigma$ on that band vanishes identically and the ingap zero dissolves into the continuum: the \emph{atomic limit} in which $\sigma$ decouples from that band. The Wannier obstruction blocks (ii) at any finite parameter: the atomic limit is precisely what the obstruction forbids. The strongest deformation of the dispersion compatible with the obstruction is the \emph{flat-band limit}, in which the bands collapse to delta-function poles while $s^\sigma_\alpha$ remains bounded away from zero on both sides of the gap, and Sec.~\ref{sec:flatbands} already showed that the ingap zero survives there. The band-edge-slope reasoning here is the general dispersive-band counterpart of the abstract Wannier-obstruction necessity argument of Sec.~\ref{sec:proofrings}: no adiabatic deformation of the perturbed bands absorbs the ring state back into either gap-adjacent band, recovering the same conclusion through the spectral structure of $\mu^\sigma$ and extending it to any band-edge profile, any spatial dimension, and the regime in which the protected state appears as a resonance overlapping the band continuum. The protection is the symmetry-enforced obstruction, not the gap itself: breaking the protecting symmetry lifts the obstruction, and the ingap zero can then exit through a band edge with the gap wide open.

Notice that when $\nu^\sigma$ grows slowly with energy, which occurs for example when the minimum of the band is not at the band inversion point, as in Dirac semimetals or topological semimetals with indirect gaps~\cite{Beidenkopf.Nag.2024}, the attractor zero of $\mu^\sigma$ can overlap the band continuum where $\nu^\sigma\neq 0$ and no longer sit strictly inside the gap. The ring state then hybridizes with delocalized band states. In the narrow-resonance regime, where the projected density of states and the slope of $\mu^\sigma$ vary slowly over the resonance width, its width is $\Gamma^\sigma\simeq2\pi\nu^\sigma/|\partial_\omega\mu^\sigma|$, evaluated at the resonance energy. The same Hilbert-transform constraint that forbids $\mu^\sigma=\partial_\omega\mu^\sigma=0$ at a band edge also forbids it at the resonance energy, so $\Gamma^\sigma$ remains finite and the lifetime is bounded away from zero. Note~\SIref{S7.3} shows that even in this overlap regime the ring state is topologically protected and cannot be adiabatically absorbed back into the perturbed band.

\subsection{Spectral evolution of ring states} \label{sec:evolution}
\begin{figure}[tb]
\centering
\includegraphics[width=\columnwidth]{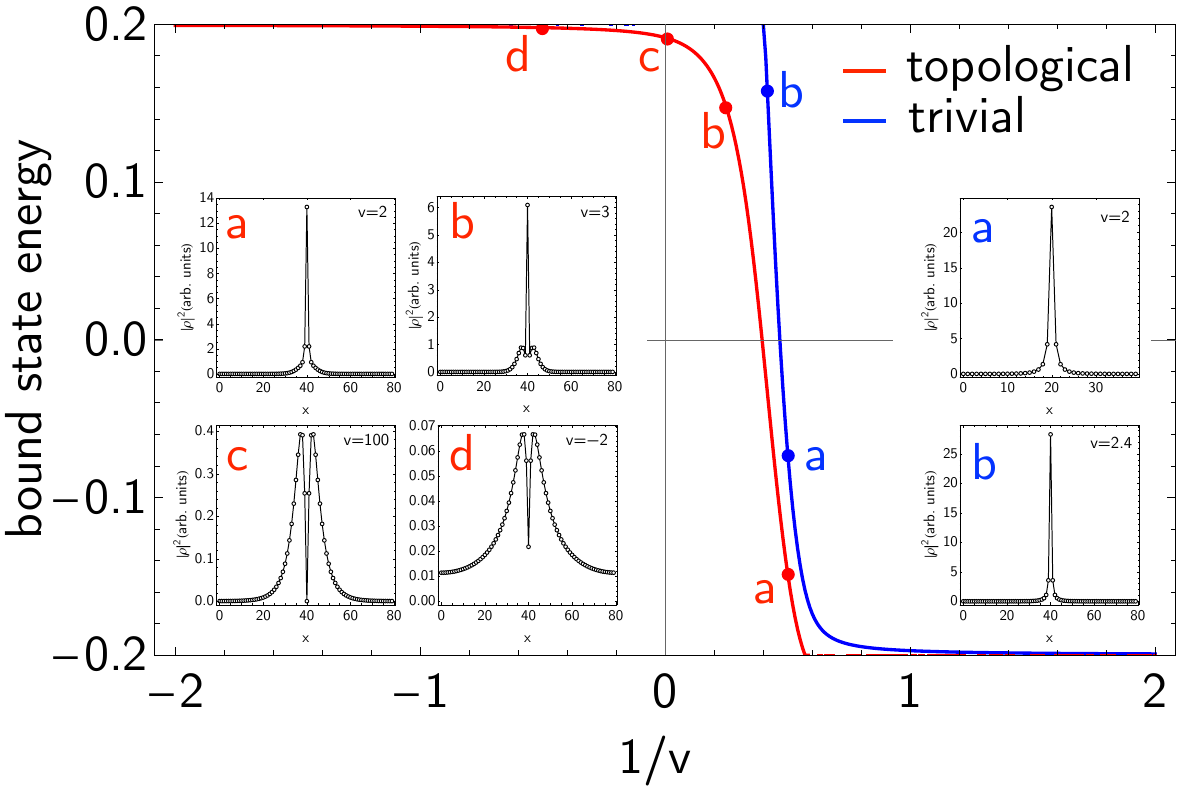}
\caption{\textbf{Spectral evolution of impurity states in topological and trivial phases.} Ingap spectrum of $\H+\V$ as a function of $1/v$ for the topological ($m=-0.2$) and trivial ($m=+0.2$) phases of a Chern insulator on a square lattice with $\upsilon=1$ and $b=1.5$, which approximates well the continuum limit \eqref{eq:cherncont} (lattice form in Note~\SIref{S5}, Eq.~\SIeqref{66}). Since the horizontal axis is $1/v$, increasing impurity strength corresponds to moving toward the strong-coupling point $1/v=0$. In the trivial case, the impurity state becomes sharper with increasing $v$, crossing through the bulk gap and upper bands. In the topological case the state remains in the gap by suppressing its weight at the impurity's core. The same impurity $\V$ leads to drastically distinct bound states depending on the global geometry of the hosting system's bands. The formation of ring states is expected for \emph{any} symmetric local perturbation in topological bands, while in trivial bands there is always a choice of $\V$ for which a ring state is not formed. The channel shown is the $p$ projection, with topological pinning energy $E^p_\infty\approx+0.19$; the $s$ channel is its mirror $E\to-E$, with $E^s_\infty\approx-0.2$ (Fig.~\ref{fig:ring}).}
\label{fig:flow}
\end{figure}

We now describe the evolution of the bound state energy as a function of impurity strength $v$, shown in Fig.~\ref{fig:flow}. The ring state $\ket{\rho_E}$ converges to a state $\ket{\rho}$ orthogonal to the impurity eigenstate $\ket{\sigma}$, while a regular impurity state $\ket{\sigma_E}$ remains adiabatically connected to $\ket{\sigma}$. From \eqref{eq:tmatrix}, the energy of any bound state is determined by $\mu^\sigma(E)=1/v$. When $v\rightarrow\pm\infty$, a state at energy of order $v$ is created, corresponding to the asymptotic behavior $\mu^\sigma\sim 1/\omega$ far outside the spectrum of $\H$. The ring state, in contrast, sits at an energy where $\mu^\sigma(\omega)$ crosses zero linearly.

It is illuminating to plot the energy evolution as a function of $v^{-1}$ rather than $v$. This parametrization, introduced in Ref.~\cite{Balatsky.Huang.2013}, identifies $v=+\infty$ with $v=-\infty$ to form a compact loop. Along this loop, the ring state is transferred between bands and across the gap, effectively being ``pumped'' between partner bands whose combined subspace contains $\ket{\sigma}$.

In the topological phase with a $\pi$ phase shift across both bands, when $v^{-1}\rightarrow +\infty$ the impurity is weak and repulsive, and the ingap state appears just above the valence band. At $v^{-1}\rightarrow-\infty$ the impurity is weak and attractive, and the ingap state sits just below the conduction band. Since $\mu^\sigma$ is continuous in the gap, the energy evolves continuously as a function of $v^{-1}$, even when $v^{-1}$ changes sign. The ingap state crosses the topological gap smoothly, moving from the valence to the conduction band as $v^{-1}$ evolves from $+\infty$ to $-\infty$, see Fig.~\ref{fig:flow}. At the strong-coupling point $v^{-1}=0$, the state stays at finite energy in the gap by developing a node at the impurity site.

In a trivial phase with a $\pi$ phase shift only in the valence band, the state removed from that band becomes increasingly localized as $v$ grows, eventually passing through the conduction band and exiting the gap at high energy. In this case, a weak attractive potential does not produce an ingap state.

Section~\ref{sec:chern} works out the ring state explicitly in the Chern continuum.

\section{Ring state of a Chern insulator} \label{sec:chern}

We now illustrate the ring state in the solvable case of an $s$-$p$ band inversion in a spinless Chern insulator~\cite{Zhang.Qi.2006}. Close to a topological phase transition, the band is well described by a continuum isotropic two-band model $\H=\sum_\bk c^\dag_\bk h(\bk)c_\bk$ where $c^\dag_{\bk\sigma}$ creates a basis state $\ket{\phi^\sigma_\bk}$ of orbital character $\sigma=s,p$ acted on by $\sigma_z$ with eigenvalue $+1$ ($s$) and $-1$ ($p$), and the Hamiltonian in Fourier space is
\begin{align}
    h(\bk)=(m+b k^2)\sigma_z-\upsilon k_x\sigma_x-\upsilon k_y\sigma_y\label{eq:cherncont}.
\end{align}
The $-\upsilon(k_x\sigma_x+k_y\sigma_y)$ term is an \emph{orbital} mixing between $s$ and $p$, not spin-orbit coupling. We define $m_E^2=m^2-E^2$ and introduce two momentum scales $k_1=m_E/\upsilon$ and $k_2=\upsilon/b$, which determine the long-distance and short-distance behavior of $h(\bk)$ (the exact pole scales $\kappa_{1,2}$ are derived in Note~\SIref{S8}, Eq.~\SIeqref{105}). For $m<0$ and $b,\upsilon>0$, $h(\bk)$ is in a topological phase.

\begin{figure}[tb]
\centering
\includegraphics[width=\columnwidth]{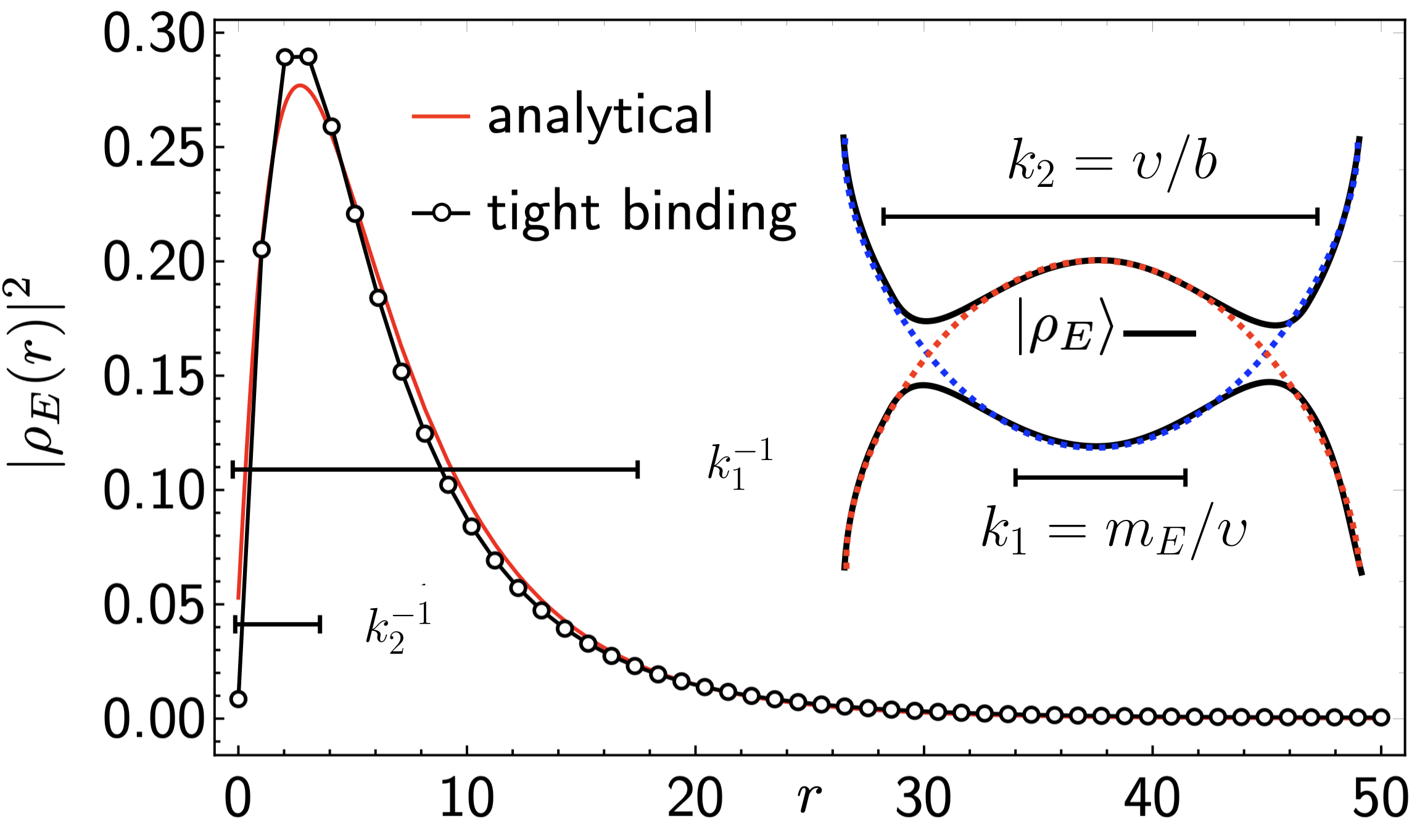}
\caption{Radial profile of an ingap ring state caused by a local potential in a Chern insulator \eqref{eq:cherncont}. The ingap state reflects the band inversion in momentum space since the orbital content at short and large distances is inverted. At the core, where the orbital character matches the impurity, the wavefunction is suppressed. It is followed by a maximum and an exponential decay at large distances. The two length scales annotated on the plot are the leading-order momentum scales $k_1=m_E/\upsilon$ (long distance) and $k_2=\upsilon/b$ (short distance); the exact pole scales $\kappa_{1,2}$ (Note~\SIref{S8}) refine these by logarithmic $\Lambda$-corrections. Inset: scheme of the band structure showing the inversion and the energy/length scales that define the ring state. Solution for the ring state at $E_\infty \approx -0.199$ with model parameters $m=-0.2$, $\upsilon=1$, $b=1.5$, $\Lambda=20$.}
\label{fig:ring}
\end{figure}

Consider an impurity of $s$ character. For $m<0$, $b>0$, the valence band has $s$ character near $k=0$, coinciding with the impurity, and crosses over to $p$ character at $k\gtrsim k_2$; the conduction band does the opposite. A localized solution from Bloch states of the inverted bands therefore behaves differently at different length scales.

Using $\br=re^{i\varphi}$, the orbital-diagonal part of the real-space Green's function $G(\omega;\br)\equiv\bra{\br}\G(\omega)\ket{0}$ is $\varphi$-independent and reads
\begin{align}&G(\omega;\br)=\!\!\sum_{\alpha=1,2}\!\!(-1)^\alpha\frac{\omega\sigma_0+\!(m-b \kappa_\alpha^2)\sigma_z}{2\pi b^2(\kappa_2^2-\kappa_1^2)}K_0(\kappa_\alpha r)\label{eq:chernGF}\end{align}
with $K_0$ a modified Bessel function of the second kind. The off-diagonal part of the Green's function, derived in Note~\SIref{S8.3}, takes the compact form
\begin{align}
F(\omega;\br) = \frac{i\,f(\omega;r)}{2}\left[e^{i\varphi}\sigma_- + e^{-i\varphi}\sigma_+\right],
\label{eq:Fcompact}
\end{align}
with $\sigma_\pm\equiv\sigma_x\pm i\sigma_y$ and $f(\omega;r)=\frac{\upsilon[\kappa_1 K_1(\kappa_1 r)-\kappa_2 K_1(\kappa_2 r)]}{2\pi b^2(\kappa_2^2-\kappa_1^2)}$, where $K_1$ is the modified Bessel function of the second kind of order one and $\kappa_{1,2}$ are evaluated at energy $\omega$ (see supplementary information). Acting on $\ket{s}$ gives $F\ket{s}=F_{ps}\ket{p}=ie^{i\varphi}f(\omega;r)\ket{p}$, so the ring state $\langle\br|\rho_E\rangle=[G(E;\br)+F(E;\br)]\ket{s}$ takes the explicit form
\begin{align}
\ket{\rho_E(\br)} = g(E;r)\ket{s} + i e^{i\varphi} f(E;r)\ket{p},\label{eq:ringfull}
\end{align}
with $g(E;r)=G_{ss}(E;r)$.

The energy is set by the standard $T$-matrix pole condition $\re g^s(E)=1/v$, which in the strong-impurity limit $v\to\infty$ pins $E_\infty$ at the gap zero $g^s(E_\infty)=0$. The detailed solution in Note~\SIref{S8} also separates this finite-energy ring branch from the weak-impurity band-edge state and from the high-energy $\sigma$ branch, whose normalized wavefunction approaches the local impurity orbital as $v\to\infty$. The radial profile of the ring state $|\rho(E_\infty;r)|^2$ is shown in Fig.~\ref{fig:ring}.

\begin{figure}[tb]
\centering
\includegraphics[width=\columnwidth]{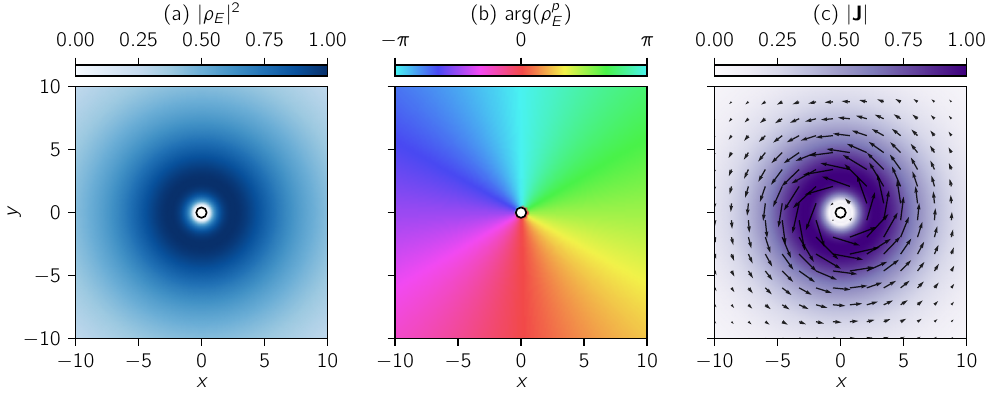}
\caption{\textbf{Probability density, phase winding, and circulating current of a ring state.} (a) Probability density $|\rho_E(\br)|^2$ of the ring state in the continuum Chern insulator model \eqref{eq:cherncont}, showing the characteristic ring structure around the vacancy (open circle). (b) Phase of the $p$-component $\arg(\rho_E^p)$, exhibiting a $2\pi$ winding around the origin. (c) Local current pattern (arrows: direction, color: magnitude). The persistent circulation arises from the orbital $\bk\cdot\bsigma$ coupling, which mixes the $s$ and $p$ components and drives an azimuthal current through their interference. Parameters: $m=-0.2$, $\upsilon=1$, $b=1.5$, $\Lambda=20$, $E_\infty\approx -0.199$.}
\label{fig:ringcurrent}
\end{figure}

When the ring state is occupied, the two components together produce a local azimuthal current, shown in Fig.~\ref{fig:ringcurrent}, built from phase-gradient and hybridization contributions that combine to circulate around the vacancy (Note~\SIref{S9}). In the continuum model these contributions give
\begin{align}
j_\varphi(r)=-\frac{2b|f|^2}{r}-2\upsilon\,g f,
\label{eq:jphimain}
\end{align}

where the first term comes from the $p$-component phase gradient and the second from $s$-$p$ hybridization. The full wavefunction is single-valued and we do not claim a gauge-independent Chern number for the impurity state; rather, the circulating current is a local remnant of the chiral edge mode of a topological domain, surviving even when the domain shrinks to a single vacancy. In Fig.~\ref{fig:ringcurrent}, $\mathbf J$ is the single-particle current density contributed by the occupied ring state in the minimal continuum model. In this sense the circulation reflects the topological nature, or twist, of the bulk orbital texture in the minimal model, rather than defining an independent invariant of the impurity wavefunction. Note~\SIref{S9} also checks this interpretation in a quadratic $C=2$ continuum model, where the off-diagonal component has double winding, and in Haldane-model numerics, where the coherent ring pattern and persistent current distinguish the topological vacancy response from a trivial one.

The ring state structure reflects the competition between two length scales in the Hamiltonian~\eqref{eq:cherncont}: the short distance $k_2^{-1}$, below which the curvature term $bk^2$ dictates the orbital character, and the long distance $k_1^{-1}$, above which the mass term $m$ takes over. These scales determine which Bessel function appears in the diagonal $G(\omega;r)$ (Eq.~\eqref{eq:chernGF}) and the off-diagonal $F(\omega;r)$ (Eq.~\eqref{eq:Fcompact}) terms of the Green's function, resulting in the non-monotonic behavior of $|\rho_E(r)|^2$. The band inversion is particularly apparent in the intermediate region $k_2^{-1}<r<k_1^{-1}$, where the ring state shows a logarithmic decay due to the admixture of orbital character in the bands in the momentum window $k_1<k<k_2$, see Fig.~\ref{fig:ring}. Ring states are correspondingly large, $r_2\gg a$, when their energy approaches the band edge and the mass gap is small, and have a pronounced core $r_1>a$ when the band inversion takes only a small portion of the Brillouin zone. If the band inversion is large, $k_1\sim\pi/a$, then the core is narrow $r_1\sim2a$. 
The $k^2$ term regularizes the large-momentum orbital texture, allowing a well-defined topological invariant. A lattice or an explicit momentum cutoff is needed for a finite bandwidth. Without it, $\mu^\sigma$ becomes monotonic in $\omega$ with its only zero pushed to $\omega=\pm\infty$, so the would-be ring state escapes to infinite energy and the number of states in the gap is not conserved. A ring state is guaranteed only for a properly regularized theory in which the topological obstruction is well-defined.

\section{Beyond rank 1 perturbations}\label{sec:beyondrank1}

The power of the present analysis is that it describes on equal footing topologically stable ingap states in any band structure, and therefore can be straightforwardly generalized to perturbations $\V$ affecting tight-binding Hamiltonians, from topological insulators to crystalline topological phases. The natural first generalization of our discussion is to consider perturbations of higher rank than one, which we now briefly discuss.

By higher rank we still mean a local impurity problem: $\V$ acts on a finite set of local orbitals, or on a finite collection of vacancy sites, while the asymptotic bulk Hamiltonian remains the same topological host. This is different from a domain wall or topological defect, where a mass, order parameter, or crystalline texture changes across space. If such a domain wall is written as a perturbation of either uniform bulk Hamiltonian, its rank is extensive, scaling with the size of one spatial region or of the interface, and the protected modes are counted by an interface or defect index theorem rather than by the finite-dimensional impurity Green's-function matrix used below. Thus a macroscopic vacancy cluster and a smooth mass domain wall can realize the same boundary mode in the large-size limit, but they approach it from opposite microscopic descriptions: the domain wall starts from an extended interface between two bulk Hamiltonians, whereas the vacancy construction starts from finite-rank trivial defects whose individual ring states hybridize into the boundary band.

Let us consider, for example, four-band models of topological insulators such as the Bernevig-Hughes-Zhang~\cite{Zhang.Bernevig.2006q49} or Kane-Mele models~\cite{Mele.Kane.2005}, the latter defined and characterized in Note~\SIref{S5}. Considering a rank-two perturbation $\V$ that preserves time-reversal symmetry, we have $\rk\V=2$, corresponding to a local potential that affects a Kramers pair of a given orbital quantum number. When both the Hamiltonian and the perturbation conserve the chosen spin component, we can carry out our analysis for each spin sector and find a Kramers pair of degenerate ring states at some pinning energy $E_\infty$ within the gap determined by the clean Hamiltonian. That is, the (ingap) zero eigenvalues of $\G(\omega)$ projected into the two-dimensional eigenspace of $\V$. A fragile topological phase~\cite{Vishwanath.Po.2018,Bernevig.Bradlyn.2019nh,Yang.Ahn.2019pt,Bouhon.Slager.2020np} protected for example by the combination of time-reversal and an in-plane inversion $C_2\T$ is expected to bind ring states at a local $C_2\T$ symmetric perturbation at a lattice site. 
The simplest example of a symmetric perturbation, irrespective of the system under study, is a lattice vacancy, the absence of all orbitals at one lattice site $\bq_0$, and therefore, ring states bound to vacancies can be a good indicator of topological obstructions. For a local perturbation that acts on several impurity eigenstates $\ket{\phi^\sigma_0}$, $\ket{\phi^\tau_0}$ at the same site, which we write as $\ket{\sigma}\equiv\ket{\phi^\sigma_0}$ and $\ket{\tau}\equiv\ket{\phi^\tau_0}$, the rank-one condition $g^\sigma(E_\infty)=0$ is replaced by the poles of the $T$-matrix $\T=\V[1-\G\V]^{-1}$, i.e.\ the secular equation
\begin{align}
\det\big[\,1 - g(E_\infty)\,v\,\big]=0,\label{eq:Tpole}
\end{align}
where $g(\omega)$ and $v$ are matrices in the impurity-orbital basis, with elements $g^{\sigma\tau}(\omega)\equiv\bra{\sigma}(\omega-\H)^{-1}\ket{\tau}$ and $v^{\sigma\tau}=\bra{\sigma}\V\ket{\tau}$. A lattice vacancy removes all orbitals at the site, sending every eigenvalue of $v$ to infinity; then $v^{-1}\to0$ and the condition collapses to $\det g(E_\infty)=0$, so $g$ has a zero eigenvalue in the gap. When $v$ instead mixes strong and weak channels, the poles of $\T$ go beyond this limit: only a strongly coupled channel pins near a zero eigenvalue of $g$, while a weakly coupled one yields an ingap root of $\det[1-g(\omega)v]$ only when its coupling is large enough to pull a solution into the gap, so the channel structure of $v$ selects which ingap branches survive. Because $g(\omega)$ is Hermitian in a spectral gap and $\partial_\omega g$, with elements $-\bra{\sigma}(\omega-\H)^{-2}\ket{\tau}$, is negative semi-definite, its eigenvalues move monotonically with $\omega$. In the vacancy limit the number of ring-state branches is then counted by how many eigenvalues of $g$ change sign between the two gap edges, reducing to the scalar zero of $g^\sigma$ when only one impurity eigenstate is affected. Summing these sign-changing matrix branches together with the high-energy impurity branches gives the matrix version of the Friedel counting in Eq.~\eqref{eq:friedelsum}.

\begin{figure}[tb]
\centering
\includegraphics[width=\columnwidth]{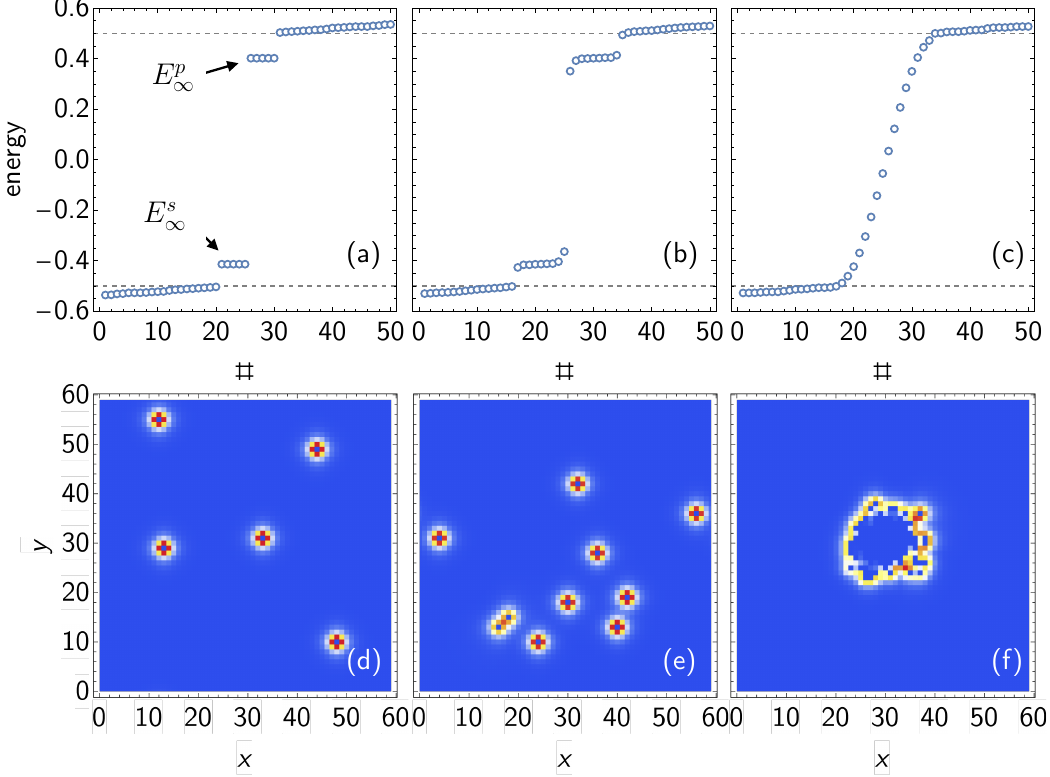}
\caption{\textbf{Hybridization of vacancy-bound ring states into chiral boundary modes.} Ingap spectrum (a-c) and local ingap density of states (d-f) of random vacancies in a Chern insulator, defined in \eqref{eq:cherncont} with $m=-0.5$. The band edge is marked by a dashed gray line. (a,d) Dilute vacancies with minimal overlap between them, their energy is located at $E_\infty^\sigma$ determined by the zeros of $\mu^\sigma$. At each impurity site there are two ring states $\ket{\rho^s}$ and $\ket{\rho^p}$ located at the energy that minimizes the level repulsion of bands below and above the gap. Since in our model the lowest band is predominantly $p$, level repulsion places the $s$ state closest to the top of the valence band ($E^s_\infty\approx-0.36$) and the $p$ state closest to the bottom of the conduction band ($E^p_\infty\approx+0.36$). (b,e) Increasing the density of impurities, some start overlapping, leading to delocalization around the small vacancy cluster and splitting in energy, deviating from $E_\infty^\sigma$. (c,f) Choosing the vacancies to be Gaussian distributed around the center of the sample, leads to a large vacant cluster. The ring states delocalize along the edge of the cluster filling the topological gap and gaining a linear dispersion seen by the equal energy spacing between states around zero energy. These states are naturally interpreted as finite-size precursors of the chiral boundary mode around the vacancy cluster.}
\label{fig:impurityhyb}
\end{figure}

Expanding our argument to potentials that affect multiple sites, $\V=\sum_iv\proj{\bq_i}$, more ingap states can be created, however their exact number is not topologically protected. A limit worth considering is combining multiple vacancies to create an edge. Let us consider that momentum along the edge is a good quantum number, say $k_y$. 
Therefore, we can work in the Fourier basis along the $y$ direction and decompose the edge perturbation as $\V=\sum_{k\sigma}v\ket{\phi^\sigma_{x_0,k_y}}\bra{\phi^\sigma_{x_0,k_y}}$ with $\ket{\phi^\sigma_{x,k_y}}=\sum_ye^{iyk_y}\ket{\phi^\sigma_{x,y}}$. Since also $\G(\omega)$ is diagonal in momentum space, the $T$-matrix is diagonalized by $k_y$. Ingap states are expected at the condition $\det(\G(E)\V-1)=0$, as it is discussed in Ref.~\cite{Balents.Slager.2015}.
 Ingap zeros of $\G(\omega;k_y)$ will exist only for a subset of momenta in the edge Brillouin zone up to the momenta at which the boundary states merge into bulk states. Note the region in $k_y$ for which edge bound states are found can be changed adiabatically, which leads to a change in the number of ingap states. The energy of the edge mode at $v\to\infty$ will follow the dispersion of the zeros in $\G$. 
Here, we wish to highlight that ring states, due to their spatial extent, hybridize with each other to form clusters of dispersing delocalized modes surrounding macroscopically large defects. To show this point, we present in Fig.~\ref{fig:impurityhyb} a numerical simulation of random vacancies in a Chern insulator with varying density. If the vacancies are far apart, as shown in panels (a) and (d), they bind ring states at their pinning energies $E^\sigma_\infty$ determined by the zeros of the local Green's function. When ring states come close to each other, they hybridize and split in energy, panel (b) and (e); finally evolving into the dispersive chiral modes surrounding large trivial puddles in the system, simulated in our case by choosing random vacancies with a Gaussian distribution centered at a particular position in the sample, panel (c) and (f). Note that in the last case, the number of ingap states does not match the number of impurities, or rank of $\V$, implying that when impurity states hybridize, a subset of them can be reabsorbed into the band continuum and become fully extended. The ingap states become delocalized around the impurity and gain a dispersion, as it can be concluded from the equal energy spacing of the energy states that cross the topological gap.

\section*{DISCUSSION}

We have shown that a trivial local defect, a vacancy or finite-rank impurity, necessarily binds a protected ingap state when the host bands carry a Wannier obstruction relative to the impurity site. The impurity potential $\V$ at position $\bq_0$, transforming under a representation $\sigma$ of the impurity site symmetry group, defines a target atomic limit in the space group of the clean system. When the bands above and below the gap are topologically distinct from this atomic limit, the impurity-projected Green’s function $\mu^\sigma(\omega)$ acquires an irremovable zero inside the gap, and the strong-impurity bound state is pinned to it. The ring state wavefunction develops a node at the impurity site and acquires its orbital character from the bands on both sides of the gap, with the pinning energy $E_\infty$ set by minimizing the level repulsion between them. The impurity itself plays no special role beyond selecting a symmetry channel: removing an orbital that participates in the quantum interference underlying band topology leaves behind a remnant that cannot be erased without changing the topology itself.

The pinning of deep impurity states to zeros of $\mu^\sigma$ is not unique to topological insulators with gapless boundary modes. Color centers in silicon and diamond, which are obstructed atomic limits, are properly topologically protected ring states whose dangling-bond zeros of $\mu^\sigma$ are irremovable~\cite{Dow.Hjalmarson.1979,Pantelides.Bernholc.1978,Cardona.Yu.2010}. In trivial semiconductors such as GaAs, deep states share the same pinning phenomenology but can in principle be removed by symmetric, gap-preserving deformations of the bulk, even if in practice they are extremely robust. In all cases, once the perturbation is strong enough, ring states are insensitive to the impurity strength, the local electrostatic environment, and the detailed shape of the defect potential. The pinning energy is fixed by the bulk band structure and the symmetry channel of the impurity orbital alone; the local details become irrelevant. Close to a band inversion, the spatial extent of the ring is set by the ratio of velocity to gap, $\xi \sim \upsilon/\Delta$: for topological insulator surfaces with $\upsilon \sim 2$--$5$~eV\AA\ and gaps $\Delta \sim 10$--$100$~meV, this gives $\xi \sim 2$--$50$~nm. When the attractor zero of $\mu^\sigma$ overlaps the band continuum, as on the gapless surface of Bi$_2$Se$_3$, the ring state hybridizes with extended surface states and becomes a finite-lifetime resonance with power-law rather than exponential decay~\cite{Kapitulnik.Alpichshev.2012}; the topological protection persists in this regime (Note~\SIref{S7.3}). In Chern insulators, the occupied ring state carries a circulating current around the defect and a winding of the relative phase between its orbital components, both absent in the trivial case; these are local remnants of the chiral edge mode surviving when the topological domain shrinks to a single vacancy.

As shown in Sec.~\ref{sec:beyondrank1}, when vacancies are close enough to overlap, their ring states hybridize and the number of ingap states is not conserved: some are reabsorbed into the band continuum while others delocalize along the cluster boundary and acquire a linear dispersion, evolving into finite-size precursors of the chiral edge mode. This provides a local mechanism for bulk-boundary correspondence, assembling the boundary mode from overlapping single-vacancy ring states rather than imposing it through a global topological index (Note~\SIref{S11}).

The spectral-attractor mechanism applies whenever the bands carry a Wannier obstruction relative to the impurity site, regardless of the specific topological invariant involved. This includes Chern insulators, time-reversal invariant topological insulators, obstructed atomic limits, and fragile topological phases; the only requirement is that the orbital spectral weight of the impurity representation is split across the gap by the obstruction. The scalar pinning condition $g^\sigma(E_\infty)=0$ generalizes to a matrix zero eigenvalue of $g^{\sigma\tau}(\omega)$ for perturbations affecting multiple orbitals or Kramers pairs, and the zero-counting argument holds in arbitrary spatial dimension (Note~\SIref{S10}). The mechanism rests on a fundamental incompatibility between the eigenspace of the impurity and the eigenspace of the bands: the local orbital cannot be reconstructed from the Bloch states of any single band when a Wannier obstruction is present, and this incompatibility is encoded as a zero of the projected Green’s function. The same structure appears in interacting systems, where the many-body Green’s function $\G(\omega)$ develops zeros that signal the incompatibility between single-particle excitations and the correlated ground state~\cite{Dzyaloshinskii.2003,Gurarie.2011,Fabrizio.Blason.2023,Sangiovanni.Wagner.2023}. Whether the obstruction is between a local orbital and a topological band, or between a single-particle sector and a many-body Hilbert space, the zero of the projected Green’s function is the spectral signature of their incompatibility.

\section*{METHODS}

The analytical derivations and numerical methods are described in the main text and Supplemental Information.

\section*{RESOURCE AVAILABILITY}

\subsection*{Lead contact}

Requests for further information and resources should be directed to and will be fulfilled by the lead contact, Raquel Queiroz (raquel.queiroz@columbia.edu).

\subsection*{Materials availability}

This study did not generate new materials.

\subsection*{Data and code availability}

\begin{itemize}
    \item All data reported in this paper will be shared by the lead contact upon request.
    \item All original code will be shared by the lead contact upon request.
    \item Any additional information required to reanalyze the results reported in this paper is available from the lead contact upon request.
\end{itemize}

\section*{ACKNOWLEDGMENTS}
 R.Q. acknowledges fruitful discussions with H. Beidenkopf, J. Pixley, F. Schindler, J. Cano and N. Verma. R.Q. and A.S. thank the Yukawa Institute for Theoretical Physics for hospitality. R.Q. is supported by the National Science Foundation under Award No. DMR-2340394.  R.I. is supported by the Israeli Science Foundation under grant No. 1790/18 and U.S.-Israel Binational Science Foundation (BSF) Grant No. 2018226.  B.A.B. is supported by the European Research Council (ERC) under the European Union’s Horizon 2020 research and innovation program (grant agreement No. 101020833), the ONR Grant No. N00014-20-1-2303, the Schmidt Fund for Innovative Research, Simons Investigator Grant No. 404513, the Gordon and Betty Moore Foundation through the EPiQS Initiative, Grant GBMF11070 and Grant No. GBMF8685 towards the Princeton theory program. Further support was provided by the NSF-MRSEC Grant No. DMR-2011750, BSF Israel US foundation Grant No. 2018226, the Princeton Global Network Funds.
A.S. was supported by grants from the ERC under the European Union's Horizon 2020 research and innovation programme (Grant Agreements LEGOTOP No. 788715), the DFG (CRC/Transregio 183, EI 519/71), the ISF and the ISF Quantum Science and Technology (2478/24).
The Flatiron Institute is a division of the Simons Foundation.

\section*{AUTHOR CONTRIBUTIONS}

All authors contributed to the conceptual development of the work, the analysis and interpretation of the results, and the preparation and revision of the manuscript.

\section*{DECLARATION OF INTERESTS}

The authors declare no competing interests.

\bibliography{RingStates}

\begin{thebibliography}{78}%
\makeatletter
\providecommand \@ifxundefined [1]{%
 \@ifx{#1\undefined}
}%
\providecommand \@ifnum [1]{%
 \ifnum #1\expandafter \@firstoftwo
 \else \expandafter \@secondoftwo
 \fi
}%
\providecommand \@ifx [1]{%
 \ifx #1\expandafter \@firstoftwo
 \else \expandafter \@secondoftwo
 \fi
}%
\providecommand \natexlab [1]{#1}%
\providecommand \enquote  [1]{``#1''}%
\providecommand \bibnamefont  [1]{#1}%
\providecommand \bibfnamefont [1]{#1}%
\providecommand \citenamefont [1]{#1}%
\providecommand \href@noop [0]{\@secondoftwo}%
\providecommand \href [0]{\begingroup \@sanitize@url \@href}%
\providecommand \@href[1]{\@@startlink{#1}\@@href}%
\providecommand \@@href[1]{\endgroup#1\@@endlink}%
\providecommand \@sanitize@url [0]{\catcode `\\12\catcode `\$12\catcode
  `\&12\catcode `\#12\catcode `\^12\catcode `\_12\catcode `\%12\relax}%
\providecommand \@@startlink[1]{}%
\providecommand \@@endlink[0]{}%
\providecommand \url  [0]{\begingroup\@sanitize@url \@url }%
\providecommand \@url [1]{\endgroup\@href {#1}{\urlprefix }}%
\providecommand \urlprefix  [0]{URL }%
\providecommand \Eprint [0]{\href }%
\providecommand \doibase [0]{https://doi.org/}%
\providecommand \selectlanguage [0]{\@gobble}%
\providecommand \bibinfo  [0]{\@secondoftwo}%
\providecommand \bibfield  [0]{\@secondoftwo}%
\providecommand \translation [1]{[#1]}%
\providecommand \BibitemOpen [0]{}%
\providecommand \bibitemStop [0]{}%
\providecommand \bibitemNoStop [0]{.\EOS\space}%
\providecommand \EOS [0]{\spacefactor3000\relax}%
\providecommand \BibitemShut  [1]{\csname bibitem#1\endcsname}%
\let\auto@bib@innerbib\@empty
\bibitem [{\citenamefont {Kohn}(1957)}]{Kohn.Kohn.1957}%
  \BibitemOpen
  \bibfield  {author} {\bibinfo {author} {\bibfnamefont {W.}~\bibnamefont
  {Kohn}},\ }\bibfield  {title} {\bibinfo {title} {{Shallow Impurity States in
  Silicon and Germanium}},\ }\href
  {https://doi.org/10.1016/s0081-1947(08)60104-6} {\bibfield  {journal}
  {\bibinfo  {journal} {Solid State Physics}\ }\textbf {\bibinfo {volume}
  {5}},\ \bibinfo {pages} {257} (\bibinfo {year} {1957})}\BibitemShut {NoStop}%
\bibitem [{\citenamefont {Koster}\ and\ \citenamefont
  {Slater}(1954)}]{Slater.Koster.1954}%
  \BibitemOpen
  \bibfield  {author} {\bibinfo {author} {\bibfnamefont {G.~F.}\ \bibnamefont
  {Koster}}\ and\ \bibinfo {author} {\bibfnamefont {J.~C.}\ \bibnamefont
  {Slater}},\ }\bibfield  {title} {\bibinfo {title} {{Wave Functions for
  Impurity Levels}},\ }\href {https://doi.org/10.1103/physrev.95.1167}
  {\bibfield  {journal} {\bibinfo  {journal} {Physical Review}\ }\textbf
  {\bibinfo {volume} {95}},\ \bibinfo {pages} {1167} (\bibinfo {year}
  {1954})}\BibitemShut {NoStop}%
\bibitem [{\citenamefont {Callaway}(1967)}]{Callaway.Callaway.1967}%
  \BibitemOpen
  \bibfield  {author} {\bibinfo {author} {\bibfnamefont {J.}~\bibnamefont
  {Callaway}},\ }\bibfield  {title} {\bibinfo {title} {{t Matrix and Phase
  Shifts in Solid-State Scattering Theory}},\ }\href
  {https://doi.org/10.1103/physrev.154.515} {\bibfield  {journal} {\bibinfo
  {journal} {Physical Review}\ }\textbf {\bibinfo {volume} {154}},\ \bibinfo
  {pages} {515} (\bibinfo {year} {1967})}\BibitemShut {NoStop}%
\bibitem [{\citenamefont {Bernholc}\ and\ \citenamefont
  {Pantelides}(1978)}]{Pantelides.Bernholc.1978}%
  \BibitemOpen
  \bibfield  {author} {\bibinfo {author} {\bibfnamefont {J.}~\bibnamefont
  {Bernholc}}\ and\ \bibinfo {author} {\bibfnamefont {S.~T.}\ \bibnamefont
  {Pantelides}},\ }\bibfield  {title} {\bibinfo {title} {{Scattering-theoretic
  method for defects in semiconductors. I. Tight-binding description of
  vacancies in Si, Ge, and GaAs}},\ }\href
  {https://doi.org/10.1103/physrevb.18.1780} {\bibfield  {journal} {\bibinfo
  {journal} {Physical Review B}\ }\textbf {\bibinfo {volume} {18}},\ \bibinfo
  {pages} {1780} (\bibinfo {year} {1978})}\BibitemShut {NoStop}%
\bibitem [{\citenamefont {Baraff}\ and\ \citenamefont
  {Schlüter}(1978)}]{Schlueter.Baraff.1978}%
  \BibitemOpen
  \bibfield  {author} {\bibinfo {author} {\bibfnamefont {G.~A.}\ \bibnamefont
  {Baraff}}\ and\ \bibinfo {author} {\bibfnamefont {M.}~\bibnamefont
  {Schlüter}},\ }\bibfield  {title} {\bibinfo {title} {{Self-Consistent
  Green's-Function Calculation of the Ideal Si Vacancy}},\ }\href
  {https://doi.org/10.1103/physrevlett.41.892} {\bibfield  {journal} {\bibinfo
  {journal} {Physical Review Letters}\ }\textbf {\bibinfo {volume} {41}},\
  \bibinfo {pages} {892} (\bibinfo {year} {1978})}\BibitemShut {NoStop}%
\bibitem [{\citenamefont {Hjalmarson}\ \emph {et~al.}(1979)\citenamefont
  {Hjalmarson}, \citenamefont {Vogl}, \citenamefont {Wolford},\ and\
  \citenamefont {Dow}}]{Dow.Hjalmarson.1979}%
  \BibitemOpen
  \bibfield  {author} {\bibinfo {author} {\bibfnamefont {H.~P.}\ \bibnamefont
  {Hjalmarson}}, \bibinfo {author} {\bibfnamefont {P.}~\bibnamefont {Vogl}},
  \bibinfo {author} {\bibfnamefont {D.~J.}\ \bibnamefont {Wolford}},\ and\
  \bibinfo {author} {\bibfnamefont {J.~D.}\ \bibnamefont {Dow}},\ }\bibfield
  {title} {\bibinfo {title} {{Theory of Substitutional Deep Traps in Covalent
  Semiconductors}},\ }\href {https://doi.org/10.1103/physrevlett.44.810}
  {\bibfield  {journal} {\bibinfo  {journal} {Physical Review Letters}\
  }\textbf {\bibinfo {volume} {44}},\ \bibinfo {pages} {810} (\bibinfo {year}
  {1979})}\BibitemShut {NoStop}%
\bibitem [{\citenamefont {Bradlyn}\ \emph {et~al.}(2017)\citenamefont
  {Bradlyn}, \citenamefont {Elcoro}, \citenamefont {Cano}, \citenamefont
  {Vergniory}, \citenamefont {Wang}, \citenamefont {Felser}, \citenamefont
  {Aroyo},\ and\ \citenamefont {Bernevig}}]{Bernevig.Bradlyn.2017}%
  \BibitemOpen
  \bibfield  {author} {\bibinfo {author} {\bibfnamefont {B.}~\bibnamefont
  {Bradlyn}}, \bibinfo {author} {\bibfnamefont {L.}~\bibnamefont {Elcoro}},
  \bibinfo {author} {\bibfnamefont {J.}~\bibnamefont {Cano}}, \bibinfo {author}
  {\bibfnamefont {M.}~\bibnamefont {Vergniory}}, \bibinfo {author}
  {\bibfnamefont {Z.}~\bibnamefont {Wang}}, \bibinfo {author} {\bibfnamefont
  {C.}~\bibnamefont {Felser}}, \bibinfo {author} {\bibfnamefont
  {M.}~\bibnamefont {Aroyo}},\ and\ \bibinfo {author} {\bibfnamefont {B.~A.}\
  \bibnamefont {Bernevig}},\ }\bibfield  {title} {\bibinfo {title}
  {{Topological quantum chemistry}},\ }\href
  {https://doi.org/10.1038/nature23268} {\bibfield  {journal} {\bibinfo
  {journal} {Nature}\ }\textbf {\bibinfo {volume} {547}},\ \bibinfo {pages}
  {298} (\bibinfo {year} {2017})},\ \Eprint {https://arxiv.org/abs/1703.02050}
  {1703.02050} \BibitemShut {NoStop}%
\bibitem [{\citenamefont {Brouder}\ \emph {et~al.}(2007)\citenamefont
  {Brouder}, \citenamefont {Panati}, \citenamefont {Calandra}, \citenamefont
  {Mourougane},\ and\ \citenamefont {Marzari}}]{Marzari.Brouder.2007}%
  \BibitemOpen
  \bibfield  {author} {\bibinfo {author} {\bibfnamefont {C.}~\bibnamefont
  {Brouder}}, \bibinfo {author} {\bibfnamefont {G.}~\bibnamefont {Panati}},
  \bibinfo {author} {\bibfnamefont {M.}~\bibnamefont {Calandra}}, \bibinfo
  {author} {\bibfnamefont {C.}~\bibnamefont {Mourougane}},\ and\ \bibinfo
  {author} {\bibfnamefont {N.}~\bibnamefont {Marzari}},\ }\bibfield  {title}
  {\bibinfo {title} {{Exponential Localization of Wannier Functions in
  Insulators}},\ }\href {https://doi.org/10.1103/physrevlett.98.046402}
  {\bibfield  {journal} {\bibinfo  {journal} {Physical Review Letters}\
  }\textbf {\bibinfo {volume} {98}},\ \bibinfo {pages} {046402} (\bibinfo
  {year} {2007})},\ \Eprint {https://arxiv.org/abs/cond-mat/0606726}
  {cond-mat/0606726} \BibitemShut {NoStop}%
\bibitem [{\citenamefont {Soluyanov}\ and\ \citenamefont
  {Vanderbilt}(2011)}]{Vanderbilt.Soluyanov.2011}%
  \BibitemOpen
  \bibfield  {author} {\bibinfo {author} {\bibfnamefont {A.~A.}\ \bibnamefont
  {Soluyanov}}\ and\ \bibinfo {author} {\bibfnamefont {D.}~\bibnamefont
  {Vanderbilt}},\ }\bibfield  {title} {\bibinfo {title} {{Wannier
  representation of Z2 topological insulators}},\ }\href
  {https://doi.org/10.1103/physrevb.83.035108} {\bibfield  {journal} {\bibinfo
  {journal} {Physical Review B}\ }\textbf {\bibinfo {volume} {83}},\ \bibinfo
  {pages} {035108} (\bibinfo {year} {2011})},\ \Eprint
  {https://arxiv.org/abs/1009.1415} {1009.1415} \BibitemShut {NoStop}%
\bibitem [{\citenamefont {Yu}\ \emph {et~al.}(2011)\citenamefont {Yu},
  \citenamefont {Qi}, \citenamefont {Bernevig}, \citenamefont {Fang},\ and\
  \citenamefont {Dai}}]{Dai.Yu.2011}%
  \BibitemOpen
  \bibfield  {author} {\bibinfo {author} {\bibfnamefont {R.}~\bibnamefont
  {Yu}}, \bibinfo {author} {\bibfnamefont {X.~L.}\ \bibnamefont {Qi}}, \bibinfo
  {author} {\bibfnamefont {A.}~\bibnamefont {Bernevig}}, \bibinfo {author}
  {\bibfnamefont {Z.}~\bibnamefont {Fang}},\ and\ \bibinfo {author}
  {\bibfnamefont {X.}~\bibnamefont {Dai}},\ }\bibfield  {title} {\bibinfo
  {title} {{Equivalent expression of Z2 topological invariant for band
  insulators using the non-Abelian Berry connection}},\ }\href
  {https://doi.org/10.1103/physrevb.84.075119} {\bibfield  {journal} {\bibinfo
  {journal} {Physical Review B}\ }\textbf {\bibinfo {volume} {84}},\ \bibinfo
  {pages} {075119} (\bibinfo {year} {2011})},\ \Eprint
  {https://arxiv.org/abs/1101.2011} {1101.2011} \BibitemShut {NoStop}%
\bibitem [{\citenamefont {Read}(2017)}]{Read.Read.2017}%
  \BibitemOpen
  \bibfield  {author} {\bibinfo {author} {\bibfnamefont {N.}~\bibnamefont
  {Read}},\ }\bibfield  {title} {\bibinfo {title} {{Compactly supported Wannier
  functions and algebraic K-theory}},\ }\href
  {https://doi.org/10.1103/physrevb.95.115309} {\bibfield  {journal} {\bibinfo
  {journal} {Physical Review B}\ }\textbf {\bibinfo {volume} {95}},\ \bibinfo
  {pages} {115309} (\bibinfo {year} {2017})},\ \Eprint
  {https://arxiv.org/abs/1608.04696} {1608.04696} \BibitemShut {NoStop}%
\bibitem [{\citenamefont {Marzari}\ \emph {et~al.}(2012)\citenamefont
  {Marzari}, \citenamefont {Mostofi}, \citenamefont {Yates}, \citenamefont
  {Souza},\ and\ \citenamefont {Vanderbilt}}]{Vanderbilt.Marzari.2012}%
  \BibitemOpen
  \bibfield  {author} {\bibinfo {author} {\bibfnamefont {N.}~\bibnamefont
  {Marzari}}, \bibinfo {author} {\bibfnamefont {A.~A.}\ \bibnamefont
  {Mostofi}}, \bibinfo {author} {\bibfnamefont {J.~R.}\ \bibnamefont {Yates}},
  \bibinfo {author} {\bibfnamefont {I.}~\bibnamefont {Souza}},\ and\ \bibinfo
  {author} {\bibfnamefont {D.}~\bibnamefont {Vanderbilt}},\ }\bibfield  {title}
  {\bibinfo {title} {{Maximally localized Wannier functions: Theory and
  applications}},\ }\href {https://doi.org/10.1103/revmodphys.84.1419}
  {\bibfield  {journal} {\bibinfo  {journal} {Reviews of Modern Physics}\
  }\textbf {\bibinfo {volume} {84}},\ \bibinfo {pages} {1419} (\bibinfo {year}
  {2012})},\ \Eprint {https://arxiv.org/abs/1112.5411} {1112.5411} \BibitemShut
  {NoStop}%
\bibitem [{\citenamefont {Po}\ \emph {et~al.}(2018)\citenamefont {Po},
  \citenamefont {Watanabe},\ and\ \citenamefont
  {Vishwanath}}]{Vishwanath.Po.2018}%
  \BibitemOpen
  \bibfield  {author} {\bibinfo {author} {\bibfnamefont {H.~C.}\ \bibnamefont
  {Po}}, \bibinfo {author} {\bibfnamefont {H.}~\bibnamefont {Watanabe}},\ and\
  \bibinfo {author} {\bibfnamefont {A.}~\bibnamefont {Vishwanath}},\ }\bibfield
   {title} {\bibinfo {title} {{Fragile Topology and Wannier Obstructions}},\
  }\href {https://doi.org/10.1103/physrevlett.121.126402} {\bibfield  {journal}
  {\bibinfo  {journal} {Physical Review Letters}\ }\textbf {\bibinfo {volume}
  {121}},\ \bibinfo {pages} {126402} (\bibinfo {year} {2018})},\ \Eprint
  {https://arxiv.org/abs/1709.06551} {1709.06551} \BibitemShut {NoStop}%
\bibitem [{\citenamefont {Kane}\ and\ \citenamefont
  {Mele}(2005)}]{Mele.Kane.2005}%
  \BibitemOpen
  \bibfield  {author} {\bibinfo {author} {\bibfnamefont {C.~L.}\ \bibnamefont
  {Kane}}\ and\ \bibinfo {author} {\bibfnamefont {E.~J.}\ \bibnamefont
  {Mele}},\ }\bibfield  {title} {\bibinfo {title} {{Quantum Spin Hall Effect in
  Graphene}},\ }\href {https://doi.org/10.1103/physrevlett.95.226801}
  {\bibfield  {journal} {\bibinfo  {journal} {Physical Review Letters}\
  }\textbf {\bibinfo {volume} {95}},\ \bibinfo {pages} {226801} (\bibinfo
  {year} {2005})},\ \Eprint {https://arxiv.org/abs/cond-mat/0411737}
  {cond-mat/0411737} \BibitemShut {NoStop}%
\bibitem [{\citenamefont {Hsieh}\ \emph {et~al.}(2008)\citenamefont {Hsieh},
  \citenamefont {Qian}, \citenamefont {Wray}, \citenamefont {Xia},
  \citenamefont {Hor}, \citenamefont {Cava},\ and\ \citenamefont
  {Hasan}}]{Hasan.Hsieh.2008}%
  \BibitemOpen
  \bibfield  {author} {\bibinfo {author} {\bibfnamefont {D.}~\bibnamefont
  {Hsieh}}, \bibinfo {author} {\bibfnamefont {D.}~\bibnamefont {Qian}},
  \bibinfo {author} {\bibfnamefont {L.}~\bibnamefont {Wray}}, \bibinfo {author}
  {\bibfnamefont {Y.}~\bibnamefont {Xia}}, \bibinfo {author} {\bibfnamefont
  {Y.~S.}\ \bibnamefont {Hor}}, \bibinfo {author} {\bibfnamefont {R.~J.}\
  \bibnamefont {Cava}},\ and\ \bibinfo {author} {\bibfnamefont {M.~Z.}\
  \bibnamefont {Hasan}},\ }\bibfield  {title} {\bibinfo {title} {{A topological
  Dirac insulator in a quantum spin Hall phase}},\ }\href
  {https://doi.org/10.1038/nature06843} {\bibfield  {journal} {\bibinfo
  {journal} {Nature}\ }\textbf {\bibinfo {volume} {452}},\ \bibinfo {pages}
  {970} (\bibinfo {year} {2008})},\ \Eprint {https://arxiv.org/abs/0902.1356}
  {0902.1356} \BibitemShut {NoStop}%
\bibitem [{\citenamefont {Zhang}\ \emph {et~al.}(2009)\citenamefont {Zhang},
  \citenamefont {Liu}, \citenamefont {Qi}, \citenamefont {Dai}, \citenamefont
  {Fang},\ and\ \citenamefont {Zhang}}]{Zhang.Zhang.2009}%
  \BibitemOpen
  \bibfield  {author} {\bibinfo {author} {\bibfnamefont {H.}~\bibnamefont
  {Zhang}}, \bibinfo {author} {\bibfnamefont {C.-X.}\ \bibnamefont {Liu}},
  \bibinfo {author} {\bibfnamefont {X.-L.}\ \bibnamefont {Qi}}, \bibinfo
  {author} {\bibfnamefont {X.}~\bibnamefont {Dai}}, \bibinfo {author}
  {\bibfnamefont {Z.}~\bibnamefont {Fang}},\ and\ \bibinfo {author}
  {\bibfnamefont {S.-C.}\ \bibnamefont {Zhang}},\ }\bibfield  {title} {\bibinfo
  {title} {{Topological insulators in Bi2Se3, Bi2Te3 and Sb2Te3 with a single
  Dirac cone on the surface}},\ }\href {https://doi.org/10.1038/nphys1270}
  {\bibfield  {journal} {\bibinfo  {journal} {Nature Physics}\ }\textbf
  {\bibinfo {volume} {5}},\ \bibinfo {pages} {438} (\bibinfo {year}
  {2009})}\BibitemShut {NoStop}%
\bibitem [{\citenamefont {Roushan}\ \emph {et~al.}(2009)\citenamefont
  {Roushan}, \citenamefont {Seo}, \citenamefont {Parker}, \citenamefont {Hor},
  \citenamefont {Hsieh}, \citenamefont {Qian}, \citenamefont {Richardella},
  \citenamefont {Hasan}, \citenamefont {Cava},\ and\ \citenamefont
  {Yazdani}}]{Yazdani.Roushan.2009}%
  \BibitemOpen
  \bibfield  {author} {\bibinfo {author} {\bibfnamefont {P.}~\bibnamefont
  {Roushan}}, \bibinfo {author} {\bibfnamefont {J.}~\bibnamefont {Seo}},
  \bibinfo {author} {\bibfnamefont {C.~V.}\ \bibnamefont {Parker}}, \bibinfo
  {author} {\bibfnamefont {Y.~S.}\ \bibnamefont {Hor}}, \bibinfo {author}
  {\bibfnamefont {D.}~\bibnamefont {Hsieh}}, \bibinfo {author} {\bibfnamefont
  {D.}~\bibnamefont {Qian}}, \bibinfo {author} {\bibfnamefont {A.}~\bibnamefont
  {Richardella}}, \bibinfo {author} {\bibfnamefont {M.~Z.}\ \bibnamefont
  {Hasan}}, \bibinfo {author} {\bibfnamefont {R.~J.}\ \bibnamefont {Cava}},\
  and\ \bibinfo {author} {\bibfnamefont {A.}~\bibnamefont {Yazdani}},\
  }\bibfield  {title} {\bibinfo {title} {{Topological surface states protected
  from backscattering by chiral spin texture}},\ }\href
  {https://doi.org/10.1038/nature08308} {\bibfield  {journal} {\bibinfo
  {journal} {Nature}\ }\textbf {\bibinfo {volume} {460}},\ \bibinfo {pages}
  {1106} (\bibinfo {year} {2009})},\ \Eprint {https://arxiv.org/abs/0908.1247}
  {0908.1247} \BibitemShut {NoStop}%
\bibitem [{\citenamefont {Hasan}\ and\ \citenamefont
  {Kane}(2010)}]{Kane.Hasan.2010}%
  \BibitemOpen
  \bibfield  {author} {\bibinfo {author} {\bibfnamefont {M.~Z.}\ \bibnamefont
  {Hasan}}\ and\ \bibinfo {author} {\bibfnamefont {C.~L.}\ \bibnamefont
  {Kane}},\ }\bibfield  {title} {\bibinfo {title} {{Colloquium: Topological
  insulators}},\ }\href {https://doi.org/10.1103/revmodphys.82.3045} {\bibfield
   {journal} {\bibinfo  {journal} {Reviews of Modern Physics}\ }\textbf
  {\bibinfo {volume} {82}},\ \bibinfo {pages} {3045} (\bibinfo {year}
  {2010})},\ \Eprint {https://arxiv.org/abs/1002.3895} {1002.3895} \BibitemShut
  {NoStop}%
\bibitem [{\citenamefont {K{\"o}nig}\ \emph {et~al.}(2007)\citenamefont
  {K{\"o}nig}, \citenamefont {Wiedmann}, \citenamefont {Br{\"u}ne},
  \citenamefont {Roth}, \citenamefont {Buhmann}, \citenamefont {Molenkamp},
  \citenamefont {Qi},\ and\ \citenamefont {Zhang}}]{Zhang.Koenig.2007}%
  \BibitemOpen
  \bibfield  {author} {\bibinfo {author} {\bibfnamefont {M.}~\bibnamefont
  {K{\"o}nig}}, \bibinfo {author} {\bibfnamefont {S.}~\bibnamefont {Wiedmann}},
  \bibinfo {author} {\bibfnamefont {C.}~\bibnamefont {Br{\"u}ne}}, \bibinfo
  {author} {\bibfnamefont {A.}~\bibnamefont {Roth}}, \bibinfo {author}
  {\bibfnamefont {H.}~\bibnamefont {Buhmann}}, \bibinfo {author} {\bibfnamefont
  {L.~W.}\ \bibnamefont {Molenkamp}}, \bibinfo {author} {\bibfnamefont {X.-L.}\
  \bibnamefont {Qi}},\ and\ \bibinfo {author} {\bibfnamefont {S.-C.}\
  \bibnamefont {Zhang}},\ }\bibfield  {title} {\bibinfo {title} {{Quantum Spin
  Hall Insulator State in HgTe Quantum Wells}},\ }\href
  {https://doi.org/10.1126/science.1148047} {\bibfield  {journal} {\bibinfo
  {journal} {Science}\ }\textbf {\bibinfo {volume} {318}},\ \bibinfo {pages}
  {766} (\bibinfo {year} {2007})},\ \Eprint {https://arxiv.org/abs/0710.0582}
  {0710.0582} \BibitemShut {NoStop}%
\bibitem [{\citenamefont {Ma}\ \emph {et~al.}(2021)\citenamefont {Ma},
  \citenamefont {Grushin},\ and\ \citenamefont {Burch}}]{Burch.Ma.2021}%
  \BibitemOpen
  \bibfield  {author} {\bibinfo {author} {\bibfnamefont {Q.}~\bibnamefont
  {Ma}}, \bibinfo {author} {\bibfnamefont {A.~G.}\ \bibnamefont {Grushin}},\
  and\ \bibinfo {author} {\bibfnamefont {K.~S.}\ \bibnamefont {Burch}},\
  }\bibfield  {title} {\bibinfo {title} {{Topology and geometry under the
  nonlinear electromagnetic spotlight}},\ }\href
  {https://doi.org/10.1038/s41563-021-00992-7} {\bibfield  {journal} {\bibinfo
  {journal} {Nature Materials}\ }\textbf {\bibinfo {volume} {20}},\ \bibinfo
  {pages} {1601} (\bibinfo {year} {2021})},\ \Eprint
  {https://arxiv.org/abs/2103.03269} {2103.03269} \BibitemShut {NoStop}%
\bibitem [{\citenamefont {Bernevig}\ \emph {et~al.}(2022)\citenamefont
  {Bernevig}, \citenamefont {Felser},\ and\ \citenamefont
  {Beidenkopf}}]{Beidenkopf.Bernevig.2022}%
  \BibitemOpen
  \bibfield  {author} {\bibinfo {author} {\bibfnamefont {B.~A.}\ \bibnamefont
  {Bernevig}}, \bibinfo {author} {\bibfnamefont {C.}~\bibnamefont {Felser}},\
  and\ \bibinfo {author} {\bibfnamefont {H.}~\bibnamefont {Beidenkopf}},\
  }\bibfield  {title} {\bibinfo {title} {{Progress and prospects in magnetic
  topological materials}},\ }\href {https://doi.org/10.1038/s41586-021-04105-x}
  {\bibfield  {journal} {\bibinfo  {journal} {Nature}\ }\textbf {\bibinfo
  {volume} {603}},\ \bibinfo {pages} {41} (\bibinfo {year} {2022})},\ \Eprint
  {https://arxiv.org/abs/2203.02890} {2203.02890} \BibitemShut {NoStop}%
\bibitem [{\citenamefont {Jackiw}\ and\ \citenamefont
  {Rebbi}(1976)}]{Rebbi.Jackiw.1976}%
  \BibitemOpen
  \bibfield  {author} {\bibinfo {author} {\bibfnamefont {R.}~\bibnamefont
  {Jackiw}}\ and\ \bibinfo {author} {\bibfnamefont {C.}~\bibnamefont {Rebbi}},\
  }\bibfield  {title} {\bibinfo {title} {{Solitons with fermion number 1/2}},\
  }\href {https://doi.org/10.1103/PhysRevD.13.3398} {\bibfield  {journal}
  {\bibinfo  {journal} {Physical Review D}\ }\textbf {\bibinfo {volume} {13}},\
  \bibinfo {pages} {3398} (\bibinfo {year} {1976})}\BibitemShut {NoStop}%
\bibitem [{\citenamefont {Ran}\ \emph {et~al.}(2009)\citenamefont {Ran},
  \citenamefont {Zhang},\ and\ \citenamefont
  {Vishwanath}}]{Vishwanath.Ran.2009}%
  \BibitemOpen
  \bibfield  {author} {\bibinfo {author} {\bibfnamefont {Y.}~\bibnamefont
  {Ran}}, \bibinfo {author} {\bibfnamefont {Y.}~\bibnamefont {Zhang}},\ and\
  \bibinfo {author} {\bibfnamefont {A.}~\bibnamefont {Vishwanath}},\ }\bibfield
   {title} {\bibinfo {title} {{One-dimensional topologically protected modes in
  topological insulators with lattice dislocations}},\ }\href
  {https://doi.org/10.1038/nphys1220} {\bibfield  {journal} {\bibinfo
  {journal} {Nature Physics}\ }\textbf {\bibinfo {volume} {5}},\ \bibinfo
  {pages} {298} (\bibinfo {year} {2009})},\ \Eprint
  {https://arxiv.org/abs/0810.5121} {0810.5121} \BibitemShut {NoStop}%
\bibitem [{\citenamefont {Teo}\ and\ \citenamefont
  {Kane}(2010)}]{Kane.Teo.2010}%
  \BibitemOpen
  \bibfield  {author} {\bibinfo {author} {\bibfnamefont {J.~C.~Y.}\
  \bibnamefont {Teo}}\ and\ \bibinfo {author} {\bibfnamefont {C.~L.}\
  \bibnamefont {Kane}},\ }\bibfield  {title} {\bibinfo {title} {{Topological
  defects and gapless modes in insulators and superconductors}},\ }\href
  {https://doi.org/10.1103/physrevb.82.115120} {\bibfield  {journal} {\bibinfo
  {journal} {Physical Review B}\ }\textbf {\bibinfo {volume} {82}},\ \bibinfo
  {pages} {115120} (\bibinfo {year} {2010})},\ \Eprint
  {https://arxiv.org/abs/1006.0690} {1006.0690} \BibitemShut {NoStop}%
\bibitem [{\citenamefont {Juricic}\ \emph {et~al.}(2012)\citenamefont
  {Juricic}, \citenamefont {Mesaros}, \citenamefont {Slager},\ and\
  \citenamefont {Zaanen}}]{Zaanen.Juricic.2012}%
  \BibitemOpen
  \bibfield  {author} {\bibinfo {author} {\bibfnamefont {V.}~\bibnamefont
  {Juricic}}, \bibinfo {author} {\bibfnamefont {A.}~\bibnamefont {Mesaros}},
  \bibinfo {author} {\bibfnamefont {R.-J.}\ \bibnamefont {Slager}},\ and\
  \bibinfo {author} {\bibfnamefont {J.}~\bibnamefont {Zaanen}},\ }\bibfield
  {title} {\bibinfo {title} {{Universal Probes of Two-Dimensional Topological
  Insulators: Dislocation and {$\pi$} Flux}},\ }\href
  {https://doi.org/10.1103/PhysRevLett.108.106403} {\bibfield  {journal}
  {\bibinfo  {journal} {Physical Review Letters}\ }\textbf {\bibinfo {volume}
  {108}},\ \bibinfo {pages} {106403} (\bibinfo {year} {2012})}\BibitemShut
  {NoStop}%
\bibitem [{\citenamefont {Chiu}\ \emph {et~al.}(2016)\citenamefont {Chiu},
  \citenamefont {Teo}, \citenamefont {Schnyder},\ and\ \citenamefont
  {Ryu}}]{Ryu.Chiu.2016}%
  \BibitemOpen
  \bibfield  {author} {\bibinfo {author} {\bibfnamefont {C.-K.}\ \bibnamefont
  {Chiu}}, \bibinfo {author} {\bibfnamefont {J.~C.~Y.}\ \bibnamefont {Teo}},
  \bibinfo {author} {\bibfnamefont {A.~P.}\ \bibnamefont {Schnyder}},\ and\
  \bibinfo {author} {\bibfnamefont {S.}~\bibnamefont {Ryu}},\ }\bibfield
  {title} {\bibinfo {title} {{Classification of topological quantum matter with
  symmetries}},\ }\href {https://doi.org/10.1103/revmodphys.88.035005}
  {\bibfield  {journal} {\bibinfo  {journal} {Reviews of Modern Physics}\
  }\textbf {\bibinfo {volume} {88}},\ \bibinfo {pages} {035005} (\bibinfo
  {year} {2016})},\ \Eprint {https://arxiv.org/abs/1505.03535} {1505.03535}
  \BibitemShut {NoStop}%
\bibitem [{\citenamefont {Yu}\ and\ \citenamefont
  {Cardona}(2010)}]{Cardona.Yu.2010}%
  \BibitemOpen
  \bibfield  {author} {\bibinfo {author} {\bibfnamefont {P.~Y.}\ \bibnamefont
  {Yu}}\ and\ \bibinfo {author} {\bibfnamefont {M.}~\bibnamefont {Cardona}},\
  }\bibfield  {title} {\bibinfo {title} {{Fundamentals of Semiconductors,
  Physics and Materials Properties}},\ }\bibfield  {journal} {\bibinfo
  {journal} {Graduate Texts in Physics}\ }\href
  {https://doi.org/10.1007/978-3-642-00710-1} {10.1007/978-3-642-00710-1}
  (\bibinfo {year} {2010})\BibitemShut {NoStop}%
\bibitem [{\citenamefont {Holbrook}\ \emph {et~al.}(2026)\citenamefont
  {Holbrook}, \citenamefont {Ingham}, \citenamefont {Kaplan}, \citenamefont
  {Holtzman}, \citenamefont {Bierman}, \citenamefont {Hou}, \citenamefont
  {Olsen}, \citenamefont {Nashabeh}, \citenamefont {Li}, \citenamefont {Liu},
  \citenamefont {Zhu}, \citenamefont {Qiu}, \citenamefont {Rhodes},
  \citenamefont {Barmak}, \citenamefont {Hone}, \citenamefont {Queiroz},\ and\
  \citenamefont {Pasupathy}}]{Queiroz.Holbrook.2026}%
  \BibitemOpen
  \bibfield  {author} {\bibinfo {author} {\bibfnamefont {M.}~\bibnamefont
  {Holbrook}}, \bibinfo {author} {\bibfnamefont {J.}~\bibnamefont {Ingham}},
  \bibinfo {author} {\bibfnamefont {D.}~\bibnamefont {Kaplan}}, \bibinfo
  {author} {\bibfnamefont {L.~N.}\ \bibnamefont {Holtzman}}, \bibinfo {author}
  {\bibfnamefont {B.}~\bibnamefont {Bierman}}, \bibinfo {author} {\bibfnamefont
  {B.}~\bibnamefont {Hou}}, \bibinfo {author} {\bibfnamefont {N.}~\bibnamefont
  {Olsen}}, \bibinfo {author} {\bibfnamefont {L.}~\bibnamefont {Nashabeh}},
  \bibinfo {author} {\bibfnamefont {Y.}~\bibnamefont {Li}}, \bibinfo {author}
  {\bibfnamefont {S.}~\bibnamefont {Liu}}, \bibinfo {author} {\bibfnamefont
  {X.}~\bibnamefont {Zhu}}, \bibinfo {author} {\bibfnamefont {D.~Y.}\
  \bibnamefont {Qiu}}, \bibinfo {author} {\bibfnamefont {D.}~\bibnamefont
  {Rhodes}}, \bibinfo {author} {\bibfnamefont {K.}~\bibnamefont {Barmak}},
  \bibinfo {author} {\bibfnamefont {J.~C.}\ \bibnamefont {Hone}}, \bibinfo
  {author} {\bibfnamefont {R.}~\bibnamefont {Queiroz}},\ and\ \bibinfo {author}
  {\bibfnamefont {A.~N.}\ \bibnamefont {Pasupathy}},\ }\bibfield  {title}
  {\bibinfo {title} {{Real-space imaging of the band topology of transition
  metal dichalcogenides}},\ }\href {https://doi.org/10.1038/s41567-026-03197-4}
  {\bibfield  {journal} {\bibinfo  {journal} {Nature Physics}\ }\textbf
  {\bibinfo {volume} {22}},\ \bibinfo {pages} {680} (\bibinfo {year}
  {2026})}\BibitemShut {NoStop}%
\bibitem [{\citenamefont {C{\u a}lug{\u a}ru}\ \emph
  {et~al.}(2026)\citenamefont {C{\u a}lug{\u a}ru}, \citenamefont {Jiang},
  \citenamefont {Guo} \emph {et~al.}}]{Ugeda.Calugaru.2026}%
  \BibitemOpen
  \bibfield  {author} {\bibinfo {author} {\bibfnamefont {D.}~\bibnamefont {C{\u
  a}lug{\u a}ru}}, \bibinfo {author} {\bibfnamefont {Y.}~\bibnamefont {Jiang}},
  \bibinfo {author} {\bibfnamefont {H.}~\bibnamefont {Guo}}, \emph {et~al.},\
  }\bibfield  {title} {\bibinfo {title} {{Observation of an obstructed atomic
  band in a transition metal dichalcogenide}},\ }\href
  {https://doi.org/10.1038/s41567-026-03196-5} {\bibfield  {journal} {\bibinfo
  {journal} {Nature Physics}\ }\textbf {\bibinfo {volume} {22}},\ \bibinfo
  {pages} {686} (\bibinfo {year} {2026})}\BibitemShut {NoStop}%
\bibitem [{\citenamefont {Kang}\ \emph {et~al.}(2025)\citenamefont {Kang},
  \citenamefont {Queiroz},\ and\ \citenamefont {Muechler}}]{Queiroz.Kang.2025}%
  \BibitemOpen
  \bibfield  {author} {\bibinfo {author} {\bibfnamefont {H.}~\bibnamefont
  {Kang}}, \bibinfo {author} {\bibfnamefont {R.}~\bibnamefont {Queiroz}},\ and\
  \bibinfo {author} {\bibfnamefont {L.}~\bibnamefont {Muechler}},\ }\bibfield
  {title} {\bibinfo {title} {{Electronic and optical properties of
  lanthanide-doped {MoS}$_2$: Impact of ionic size and orbital configuration
  mismatch}},\ }\href {https://doi.org/10.1103/kpsg-z8tp} {\bibfield  {journal}
  {\bibinfo  {journal} {Physical Review B}\ }\textbf {\bibinfo {volume}
  {112}},\ \bibinfo {pages} {174107} (\bibinfo {year} {2025})}\BibitemShut
  {NoStop}%
\bibitem [{\citenamefont {Pereira}\ \emph {et~al.}(2006)\citenamefont
  {Pereira}, \citenamefont {Guinea}, \citenamefont {Santos}, \citenamefont
  {Peres},\ and\ \citenamefont {Neto}}]{Neto.Pereira.2006}%
  \BibitemOpen
  \bibfield  {author} {\bibinfo {author} {\bibfnamefont {V.~M.}\ \bibnamefont
  {Pereira}}, \bibinfo {author} {\bibfnamefont {F.}~\bibnamefont {Guinea}},
  \bibinfo {author} {\bibfnamefont {J.~M. B. L.~d.}\ \bibnamefont {Santos}},
  \bibinfo {author} {\bibfnamefont {N.~M.~R.}\ \bibnamefont {Peres}},\ and\
  \bibinfo {author} {\bibfnamefont {A.~H.~C.}\ \bibnamefont {Neto}},\
  }\bibfield  {title} {\bibinfo {title} {{Disorder Induced Localized States in
  Graphene}},\ }\href {https://doi.org/10.1103/physrevlett.96.036801}
  {\bibfield  {journal} {\bibinfo  {journal} {Physical Review Letters}\
  }\textbf {\bibinfo {volume} {96}},\ \bibinfo {pages} {036801} (\bibinfo
  {year} {2006})},\ \Eprint {https://arxiv.org/abs/cond-mat/0508530}
  {cond-mat/0508530} \BibitemShut {NoStop}%
\bibitem [{\citenamefont {Wehling}\ \emph {et~al.}(2014)\citenamefont
  {Wehling}, \citenamefont {Black-Schaffer},\ and\ \citenamefont
  {Balatsky}}]{Balatsky.Wehling.2014}%
  \BibitemOpen
  \bibfield  {author} {\bibinfo {author} {\bibfnamefont {T.}~\bibnamefont
  {Wehling}}, \bibinfo {author} {\bibfnamefont {A.}~\bibnamefont
  {Black-Schaffer}},\ and\ \bibinfo {author} {\bibfnamefont {A.}~\bibnamefont
  {Balatsky}},\ }\bibfield  {title} {\bibinfo {title} {{Dirac materials}},\
  }\href {https://doi.org/10.1080/00018732.2014.927109} {\bibfield  {journal}
  {\bibinfo  {journal} {Advances in Physics}\ }\textbf {\bibinfo {volume}
  {63}},\ \bibinfo {pages} {1} (\bibinfo {year} {2014})},\ \Eprint
  {https://arxiv.org/abs/1405.5774} {1405.5774} \BibitemShut {NoStop}%
\bibitem [{\citenamefont {Ugeda}\ \emph {et~al.}(2010)\citenamefont {Ugeda},
  \citenamefont {Brihuega}, \citenamefont {Guinea},\ and\ \citenamefont
  {Gómez-Rodríguez}}]{Gomez-Rodriguez.Ugeda.2010}%
  \BibitemOpen
  \bibfield  {author} {\bibinfo {author} {\bibfnamefont {M.~M.}\ \bibnamefont
  {Ugeda}}, \bibinfo {author} {\bibfnamefont {I.}~\bibnamefont {Brihuega}},
  \bibinfo {author} {\bibfnamefont {F.}~\bibnamefont {Guinea}},\ and\ \bibinfo
  {author} {\bibfnamefont {J.~M.}\ \bibnamefont {Gómez-Rodríguez}},\
  }\bibfield  {title} {\bibinfo {title} {{Missing Atom as a Source of Carbon
  Magnetism}},\ }\href {https://doi.org/10.1103/physrevlett.104.096804}
  {\bibfield  {journal} {\bibinfo  {journal} {Physical Review Letters}\
  }\textbf {\bibinfo {volume} {104}},\ \bibinfo {pages} {096804} (\bibinfo
  {year} {2010})},\ \Eprint {https://arxiv.org/abs/1001.3081} {1001.3081}
  \BibitemShut {NoStop}%
\bibitem [{\citenamefont {Huang}\ \emph {et~al.}(2013)\citenamefont {Huang},
  \citenamefont {Arovas},\ and\ \citenamefont
  {Balatsky}}]{Balatsky.Huang.2013}%
  \BibitemOpen
  \bibfield  {author} {\bibinfo {author} {\bibfnamefont {Z.}~\bibnamefont
  {Huang}}, \bibinfo {author} {\bibfnamefont {D.~P.}\ \bibnamefont {Arovas}},\
  and\ \bibinfo {author} {\bibfnamefont {A.~V.}\ \bibnamefont {Balatsky}},\
  }\bibfield  {title} {\bibinfo {title} {{Impurity scattering in Weyl
  Semimetals and their stability classification}},\ }\href
  {https://doi.org/10.1088/1367-2630/15/12/123019} {\bibfield  {journal}
  {\bibinfo  {journal} {New Journal of Physics}\ }\textbf {\bibinfo {volume}
  {15}},\ \bibinfo {pages} {123019} (\bibinfo {year} {2013})},\ \Eprint
  {https://arxiv.org/abs/1310.0137} {1310.0137} \BibitemShut {NoStop}%
\bibitem [{\citenamefont {Slager}\ \emph {et~al.}(2015)\citenamefont {Slager},
  \citenamefont {Rademaker}, \citenamefont {Zaanen},\ and\ \citenamefont
  {Balents}}]{Balents.Slager.2015}%
  \BibitemOpen
  \bibfield  {author} {\bibinfo {author} {\bibfnamefont {R.-J.}\ \bibnamefont
  {Slager}}, \bibinfo {author} {\bibfnamefont {L.}~\bibnamefont {Rademaker}},
  \bibinfo {author} {\bibfnamefont {J.}~\bibnamefont {Zaanen}},\ and\ \bibinfo
  {author} {\bibfnamefont {L.}~\bibnamefont {Balents}},\ }\bibfield  {title}
  {\bibinfo {title} {{Impurity-bound states and Green's function zeros as local
  signatures of topology}},\ }\href
  {https://doi.org/10.1103/physrevb.92.085126} {\bibfield  {journal} {\bibinfo
  {journal} {Physical Review B}\ }\textbf {\bibinfo {volume} {92}},\ \bibinfo
  {pages} {085126} (\bibinfo {year} {2015})},\ \Eprint
  {https://arxiv.org/abs/1504.04881} {1504.04881} \BibitemShut {NoStop}%
\bibitem [{\citenamefont {Feldman}\ \emph {et~al.}(2016)\citenamefont
  {Feldman}, \citenamefont {Randeria}, \citenamefont {Gyenis}, \citenamefont
  {Wu}, \citenamefont {Ji}, \citenamefont {Cava}, \citenamefont {MacDonald},\
  and\ \citenamefont {Yazdani}}]{Yazdani.Feldman.2016}%
  \BibitemOpen
  \bibfield  {author} {\bibinfo {author} {\bibfnamefont {B.~E.}\ \bibnamefont
  {Feldman}}, \bibinfo {author} {\bibfnamefont {M.~T.}\ \bibnamefont
  {Randeria}}, \bibinfo {author} {\bibfnamefont {A.}~\bibnamefont {Gyenis}},
  \bibinfo {author} {\bibfnamefont {F.}~\bibnamefont {Wu}}, \bibinfo {author}
  {\bibfnamefont {H.}~\bibnamefont {Ji}}, \bibinfo {author} {\bibfnamefont
  {R.~J.}\ \bibnamefont {Cava}}, \bibinfo {author} {\bibfnamefont {A.~H.}\
  \bibnamefont {MacDonald}},\ and\ \bibinfo {author} {\bibfnamefont
  {A.}~\bibnamefont {Yazdani}},\ }\bibfield  {title} {\bibinfo {title}
  {{Observation of a nematic quantum Hall liquid on the surface of bismuth}},\
  }\href {https://doi.org/10.1126/science.aag1715} {\bibfield  {journal}
  {\bibinfo  {journal} {Science}\ }\textbf {\bibinfo {volume} {354}},\ \bibinfo
  {pages} {316} (\bibinfo {year} {2016})},\ \Eprint
  {https://arxiv.org/abs/1610.07613} {1610.07613} \BibitemShut {NoStop}%
\bibitem [{\citenamefont {Balatsky}\ \emph {et~al.}(2006)\citenamefont
  {Balatsky}, \citenamefont {Vekhter},\ and\ \citenamefont
  {Zhu}}]{Balatsky.Zhu.2006}%
  \BibitemOpen
  \bibfield  {author} {\bibinfo {author} {\bibfnamefont {A.~V.}\ \bibnamefont
  {Balatsky}}, \bibinfo {author} {\bibfnamefont {I.}~\bibnamefont {Vekhter}},\
  and\ \bibinfo {author} {\bibfnamefont {J.-X.}\ \bibnamefont {Zhu}},\
  }\bibfield  {title} {\bibinfo {title} {{Impurity-induced states in
  conventional and unconventional superconductors}},\ }\href
  {https://doi.org/10.1103/RevModPhys.78.373} {\bibfield  {journal} {\bibinfo
  {journal} {Reviews of Modern Physics}\ }\textbf {\bibinfo {volume} {78}},\
  \bibinfo {pages} {373} (\bibinfo {year} {2006})}\BibitemShut {NoStop}%
\bibitem [{\citenamefont {Diop}\ \emph {et~al.}(2020)\citenamefont {Diop},
  \citenamefont {Fritz}, \citenamefont {Vojta},\ and\ \citenamefont
  {Rachel}}]{Rachel.Diop.2020}%
  \BibitemOpen
  \bibfield  {author} {\bibinfo {author} {\bibfnamefont {S.-S.}\ \bibnamefont
  {Diop}}, \bibinfo {author} {\bibfnamefont {L.}~\bibnamefont {Fritz}},
  \bibinfo {author} {\bibfnamefont {M.}~\bibnamefont {Vojta}},\ and\ \bibinfo
  {author} {\bibfnamefont {S.}~\bibnamefont {Rachel}},\ }\bibfield  {title}
  {\bibinfo {title} {{Impurity bound states as detectors of topological band
  structures revisited}},\ }\href {https://doi.org/10.1103/physrevb.101.245132}
  {\bibfield  {journal} {\bibinfo  {journal} {Physical Review B}\ }\textbf
  {\bibinfo {volume} {101}},\ \bibinfo {pages} {245132} (\bibinfo {year}
  {2020})},\ \Eprint {https://arxiv.org/abs/2003.09498} {2003.09498}
  \BibitemShut {NoStop}%
\bibitem [{\citenamefont {Biswas}\ and\ \citenamefont
  {Balatsky}(2010)}]{Balatsky.Biswas.2010}%
  \BibitemOpen
  \bibfield  {author} {\bibinfo {author} {\bibfnamefont {R.~R.}\ \bibnamefont
  {Biswas}}\ and\ \bibinfo {author} {\bibfnamefont {A.~V.}\ \bibnamefont
  {Balatsky}},\ }\bibfield  {title} {\bibinfo {title} {{Impurity-induced states
  on the surface of three-dimensional topological insulators}},\ }\href
  {https://doi.org/10.1103/PhysRevB.81.233405} {\bibfield  {journal} {\bibinfo
  {journal} {Physical Review B}\ }\textbf {\bibinfo {volume} {81}},\ \bibinfo
  {pages} {233405} (\bibinfo {year} {2010})}\BibitemShut {NoStop}%
\bibitem [{\citenamefont {Alpichshev}\ \emph {et~al.}(2012)\citenamefont
  {Alpichshev}, \citenamefont {Biswas}, \citenamefont {Balatsky}, \citenamefont
  {Analytis}, \citenamefont {Chu}, \citenamefont {Fisher},\ and\ \citenamefont
  {Kapitulnik}}]{Kapitulnik.Alpichshev.2012}%
  \BibitemOpen
  \bibfield  {author} {\bibinfo {author} {\bibfnamefont {Z.}~\bibnamefont
  {Alpichshev}}, \bibinfo {author} {\bibfnamefont {R.~R.}\ \bibnamefont
  {Biswas}}, \bibinfo {author} {\bibfnamefont {A.~V.}\ \bibnamefont
  {Balatsky}}, \bibinfo {author} {\bibfnamefont {J.~G.}\ \bibnamefont
  {Analytis}}, \bibinfo {author} {\bibfnamefont {J.-H.}\ \bibnamefont {Chu}},
  \bibinfo {author} {\bibfnamefont {I.~R.}\ \bibnamefont {Fisher}},\ and\
  \bibinfo {author} {\bibfnamefont {A.}~\bibnamefont {Kapitulnik}},\ }\bibfield
   {title} {\bibinfo {title} {{STM imaging of impurity resonances on Bi2Se3}},\
  }\href {https://doi.org/10.1103/PhysRevLett.108.206402} {\bibfield  {journal}
  {\bibinfo  {journal} {Physical Review Letters}\ }\textbf {\bibinfo {volume}
  {108}},\ \bibinfo {pages} {206402} (\bibinfo {year} {2012})}\BibitemShut
  {NoStop}%
\bibitem [{\citenamefont {Xing}\ \emph {et~al.}(2020)\citenamefont {Xing},
  \citenamefont {Shen}, \citenamefont {Chen}, \citenamefont {Huang},
  \citenamefont {Gao}, \citenamefont {Zheng}, \citenamefont {Zhang},
  \citenamefont {Li}, \citenamefont {Hu}, \citenamefont {Qian}, \citenamefont
  {Zhao}, \citenamefont {Jin}, \citenamefont {Shi}, \citenamefont {Yin},
  \citenamefont {Wang}, \citenamefont {Jiang}, \citenamefont {Zhang},
  \citenamefont {Qu}, \citenamefont {Yan}, \citenamefont {Fang}, \citenamefont
  {Zhu}, \citenamefont {Dai}, \citenamefont {Hasan}, \citenamefont {Ma},\ and\
  \citenamefont {Xue}}]{Hasan.Xing.2020}%
  \BibitemOpen
  \bibfield  {author} {\bibinfo {author} {\bibfnamefont {Y.}~\bibnamefont
  {Xing}}, \bibinfo {author} {\bibfnamefont {J.}~\bibnamefont {Shen}}, \bibinfo
  {author} {\bibfnamefont {H.}~\bibnamefont {Chen}}, \bibinfo {author}
  {\bibfnamefont {L.}~\bibnamefont {Huang}}, \bibinfo {author} {\bibfnamefont
  {Y.}~\bibnamefont {Gao}}, \bibinfo {author} {\bibfnamefont {Q.}~\bibnamefont
  {Zheng}}, \bibinfo {author} {\bibfnamefont {Y.-Y.}\ \bibnamefont {Zhang}},
  \bibinfo {author} {\bibfnamefont {G.}~\bibnamefont {Li}}, \bibinfo {author}
  {\bibfnamefont {B.}~\bibnamefont {Hu}}, \bibinfo {author} {\bibfnamefont
  {G.}~\bibnamefont {Qian}}, \bibinfo {author} {\bibfnamefont {L.}~\bibnamefont
  {Zhao}}, \bibinfo {author} {\bibfnamefont {S.}~\bibnamefont {Jin}}, \bibinfo
  {author} {\bibfnamefont {O.}~\bibnamefont {Shi}}, \bibinfo {author}
  {\bibfnamefont {J.-X.}\ \bibnamefont {Yin}}, \bibinfo {author} {\bibfnamefont
  {W.}~\bibnamefont {Wang}}, \bibinfo {author} {\bibfnamefont {P.}~\bibnamefont
  {Jiang}}, \bibinfo {author} {\bibfnamefont {H.}~\bibnamefont {Zhang}},
  \bibinfo {author} {\bibfnamefont {F.}~\bibnamefont {Qu}}, \bibinfo {author}
  {\bibfnamefont {B.}~\bibnamefont {Yan}}, \bibinfo {author} {\bibfnamefont
  {Z.}~\bibnamefont {Fang}}, \bibinfo {author} {\bibfnamefont {J.}~\bibnamefont
  {Zhu}}, \bibinfo {author} {\bibfnamefont {X.}~\bibnamefont {Dai}}, \bibinfo
  {author} {\bibfnamefont {M.~Z.}\ \bibnamefont {Hasan}}, \bibinfo {author}
  {\bibfnamefont {X.-C.}\ \bibnamefont {Ma}},\ and\ \bibinfo {author}
  {\bibfnamefont {Q.-K.}\ \bibnamefont {Xue}},\ }\bibfield  {title} {\bibinfo
  {title} {{Localized spin-orbit polaron in magnetic Weyl semimetal
  Co3Sn2S2}},\ }\href {https://doi.org/10.1038/s41467-020-19440-2} {\bibfield
  {journal} {\bibinfo  {journal} {Nature Communications}\ }\textbf {\bibinfo
  {volume} {11}},\ \bibinfo {pages} {5613} (\bibinfo {year}
  {2020})}\BibitemShut {NoStop}%
\bibitem [{\citenamefont {Yin}\ \emph {et~al.}(2020)\citenamefont {Yin},
  \citenamefont {Shumiya}, \citenamefont {Jiang}, \citenamefont {Zhou},
  \citenamefont {Macam}, \citenamefont {Mohammad~Sura}, \citenamefont {Zhang},
  \citenamefont {Cheng}, \citenamefont {Guguchia}, \citenamefont {Li},
  \citenamefont {Wang}, \citenamefont {Litskevich}, \citenamefont {Belopolski},
  \citenamefont {Yang}, \citenamefont {Cochran}, \citenamefont {Chang},
  \citenamefont {Zhang}, \citenamefont {Huang}, \citenamefont {Chuang},
  \citenamefont {Lin}, \citenamefont {Lei}, \citenamefont {Andersen},
  \citenamefont {Wang}, \citenamefont {Jia},\ and\ \citenamefont
  {Hasan}}]{Hasan.Yin.2020}%
  \BibitemOpen
  \bibfield  {author} {\bibinfo {author} {\bibfnamefont {J.-X.}\ \bibnamefont
  {Yin}}, \bibinfo {author} {\bibfnamefont {N.}~\bibnamefont {Shumiya}},
  \bibinfo {author} {\bibfnamefont {Y.}~\bibnamefont {Jiang}}, \bibinfo
  {author} {\bibfnamefont {H.}~\bibnamefont {Zhou}}, \bibinfo {author}
  {\bibfnamefont {G.}~\bibnamefont {Macam}}, \bibinfo {author} {\bibfnamefont
  {H.~O.}\ \bibnamefont {Mohammad~Sura}}, \bibinfo {author} {\bibfnamefont
  {S.~S.}\ \bibnamefont {Zhang}}, \bibinfo {author} {\bibfnamefont {Z.-J.}\
  \bibnamefont {Cheng}}, \bibinfo {author} {\bibfnamefont {Z.}~\bibnamefont
  {Guguchia}}, \bibinfo {author} {\bibfnamefont {Y.}~\bibnamefont {Li}},
  \bibinfo {author} {\bibfnamefont {Q.}~\bibnamefont {Wang}}, \bibinfo {author}
  {\bibfnamefont {M.}~\bibnamefont {Litskevich}}, \bibinfo {author}
  {\bibfnamefont {I.}~\bibnamefont {Belopolski}}, \bibinfo {author}
  {\bibfnamefont {X.~P.}\ \bibnamefont {Yang}}, \bibinfo {author}
  {\bibfnamefont {T.~A.}\ \bibnamefont {Cochran}}, \bibinfo {author}
  {\bibfnamefont {G.}~\bibnamefont {Chang}}, \bibinfo {author} {\bibfnamefont
  {Q.}~\bibnamefont {Zhang}}, \bibinfo {author} {\bibfnamefont {Z.-Q.}\
  \bibnamefont {Huang}}, \bibinfo {author} {\bibfnamefont {F.-C.}\ \bibnamefont
  {Chuang}}, \bibinfo {author} {\bibfnamefont {H.}~\bibnamefont {Lin}},
  \bibinfo {author} {\bibfnamefont {H.}~\bibnamefont {Lei}}, \bibinfo {author}
  {\bibfnamefont {B.~M.}\ \bibnamefont {Andersen}}, \bibinfo {author}
  {\bibfnamefont {Z.}~\bibnamefont {Wang}}, \bibinfo {author} {\bibfnamefont
  {S.}~\bibnamefont {Jia}},\ and\ \bibinfo {author} {\bibfnamefont {M.~Z.}\
  \bibnamefont {Hasan}},\ }\bibfield  {title} {\bibinfo {title} {{Spin-orbit
  quantum impurity in a topological magnet}},\ }\href
  {https://doi.org/10.1038/s41467-020-18111-6} {\bibfield  {journal} {\bibinfo
  {journal} {Nature Communications}\ }\textbf {\bibinfo {volume} {11}},\
  \bibinfo {pages} {4415} (\bibinfo {year} {2020})}\BibitemShut {NoStop}%
\bibitem [{\citenamefont {Chen}\ \emph {et~al.}(2024)\citenamefont {Chen},
  \citenamefont {Xing}, \citenamefont {Tan}, \citenamefont {Huang},
  \citenamefont {Zheng}, \citenamefont {Huang}, \citenamefont {Han},
  \citenamefont {Hu}, \citenamefont {Ye}, \citenamefont {Li}, \citenamefont
  {Xiao}, \citenamefont {Lei}, \citenamefont {Qiu}, \citenamefont {Liu},
  \citenamefont {Yang}, \citenamefont {Wang}, \citenamefont {Yan},\ and\
  \citenamefont {Gao}}]{Gao.Chen.2024}%
  \BibitemOpen
  \bibfield  {author} {\bibinfo {author} {\bibfnamefont {H.}~\bibnamefont
  {Chen}}, \bibinfo {author} {\bibfnamefont {Y.}~\bibnamefont {Xing}}, \bibinfo
  {author} {\bibfnamefont {H.}~\bibnamefont {Tan}}, \bibinfo {author}
  {\bibfnamefont {L.}~\bibnamefont {Huang}}, \bibinfo {author} {\bibfnamefont
  {Q.}~\bibnamefont {Zheng}}, \bibinfo {author} {\bibfnamefont
  {Z.}~\bibnamefont {Huang}}, \bibinfo {author} {\bibfnamefont
  {X.}~\bibnamefont {Han}}, \bibinfo {author} {\bibfnamefont {B.}~\bibnamefont
  {Hu}}, \bibinfo {author} {\bibfnamefont {Y.}~\bibnamefont {Ye}}, \bibinfo
  {author} {\bibfnamefont {Y.}~\bibnamefont {Li}}, \bibinfo {author}
  {\bibfnamefont {Y.}~\bibnamefont {Xiao}}, \bibinfo {author} {\bibfnamefont
  {H.}~\bibnamefont {Lei}}, \bibinfo {author} {\bibfnamefont {X.}~\bibnamefont
  {Qiu}}, \bibinfo {author} {\bibfnamefont {E.}~\bibnamefont {Liu}}, \bibinfo
  {author} {\bibfnamefont {H.}~\bibnamefont {Yang}}, \bibinfo {author}
  {\bibfnamefont {Z.}~\bibnamefont {Wang}}, \bibinfo {author} {\bibfnamefont
  {B.}~\bibnamefont {Yan}},\ and\ \bibinfo {author} {\bibfnamefont {H.-J.}\
  \bibnamefont {Gao}},\ }\bibfield  {title} {\bibinfo {title} {{Atomically
  precise engineering of spin-orbit polarons in a kagome magnetic Weyl
  semimetal}},\ }\href {https://doi.org/10.1038/s41467-024-46729-3} {\bibfield
  {journal} {\bibinfo  {journal} {Nature Communications}\ }\textbf {\bibinfo
  {volume} {15}},\ \bibinfo {pages} {2301} (\bibinfo {year}
  {2024})}\BibitemShut {NoStop}%
\bibitem [{\citenamefont {Liu}\ \emph {et~al.}(2022)\citenamefont {Liu},
  \citenamefont {Liu}, \citenamefont {Xu}, \citenamefont {de~Visser},
  \citenamefont {Wu}, \citenamefont {Sun}, \citenamefont {Fang}, \citenamefont
  {Yan},\ and\ \citenamefont {Chen}}]{Liu.Liu.2022}%
  \BibitemOpen
  \bibfield  {author} {\bibinfo {author} {\bibfnamefont {D.~F.}\ \bibnamefont
  {Liu}}, \bibinfo {author} {\bibfnamefont {E.~K.}\ \bibnamefont {Liu}},
  \bibinfo {author} {\bibfnamefont {Q.~N.}\ \bibnamefont {Xu}}, \bibinfo
  {author} {\bibfnamefont {J.}~\bibnamefont {de~Visser}}, \bibinfo {author}
  {\bibfnamefont {J.~W.}\ \bibnamefont {Wu}}, \bibinfo {author} {\bibfnamefont
  {H.~L.}\ \bibnamefont {Sun}}, \bibinfo {author} {\bibfnamefont
  {C.}~\bibnamefont {Fang}}, \bibinfo {author} {\bibfnamefont {B.~H.}\
  \bibnamefont {Yan}},\ and\ \bibinfo {author} {\bibfnamefont {Y.~L.}\
  \bibnamefont {Chen}},\ }\bibfield  {title} {\bibinfo {title} {{Direct
  observation of the spin-orbit coupling effect in magnetic Weyl semimetal
  Co3Sn2S2}},\ }\href {https://doi.org/10.1038/s41535-021-00392-9} {\bibfield
  {journal} {\bibinfo  {journal} {npj Quantum Materials}\ }\textbf {\bibinfo
  {volume} {7}},\ \bibinfo {pages} {11} (\bibinfo {year} {2022})}\BibitemShut
  {NoStop}%
\bibitem [{\citenamefont {Nag}\ \emph {et~al.}(2026)\citenamefont {Nag},
  \citenamefont {Morali}, \citenamefont {Batyabal}, \citenamefont {Mazhar},
  \citenamefont {Geier}, \citenamefont {Brouwer}, \citenamefont {Felser},
  \citenamefont {Yan}, \citenamefont {Avraham}, \citenamefont {Queiroz},\ and\
  \citenamefont {Beidenkopf}}]{Beidenkopf.Nag.2024}%
  \BibitemOpen
  \bibfield  {author} {\bibinfo {author} {\bibfnamefont {P.~K.}\ \bibnamefont
  {Nag}}, \bibinfo {author} {\bibfnamefont {N.}~\bibnamefont {Morali}},
  \bibinfo {author} {\bibfnamefont {R.}~\bibnamefont {Batyabal}}, \bibinfo
  {author} {\bibfnamefont {A.}~\bibnamefont {Mazhar}}, \bibinfo {author}
  {\bibfnamefont {M.}~\bibnamefont {Geier}}, \bibinfo {author} {\bibfnamefont
  {P.}~\bibnamefont {Brouwer}}, \bibinfo {author} {\bibfnamefont
  {C.}~\bibnamefont {Felser}}, \bibinfo {author} {\bibfnamefont
  {B.}~\bibnamefont {Yan}}, \bibinfo {author} {\bibfnamefont {N.}~\bibnamefont
  {Avraham}}, \bibinfo {author} {\bibfnamefont {R.}~\bibnamefont {Queiroz}},\
  and\ \bibinfo {author} {\bibfnamefont {H.}~\bibnamefont {Beidenkopf}},\
  }\bibfield  {title} {\bibinfo {title} {{From Ring States to Fermi Arc Modes
  in a Ferromagnetic Weyl Semimetal}}} (\bibinfo {year} {2026}),\ \bibinfo
  {note} {in preparation}\BibitemShut {NoStop}%
\bibitem [{\citenamefont {Amani}\ \emph {et~al.}(2015)\citenamefont {Amani},
  \citenamefont {Lien}, \citenamefont {Kiriya}, \citenamefont {Xiao},
  \citenamefont {Azcatl}, \citenamefont {Noh}, \citenamefont {Madhvapathy},
  \citenamefont {Addou}, \citenamefont {KC}, \citenamefont {Dubey},
  \citenamefont {Cho}, \citenamefont {Wallace}, \citenamefont {Lee},
  \citenamefont {He}, \citenamefont {III}, \citenamefont {Zhang}, \citenamefont
  {Yablonovitch},\ and\ \citenamefont {Javey}}]{Javey.Amani.2015}%
  \BibitemOpen
  \bibfield  {author} {\bibinfo {author} {\bibfnamefont {M.}~\bibnamefont
  {Amani}}, \bibinfo {author} {\bibfnamefont {D.-H.}\ \bibnamefont {Lien}},
  \bibinfo {author} {\bibfnamefont {D.}~\bibnamefont {Kiriya}}, \bibinfo
  {author} {\bibfnamefont {J.}~\bibnamefont {Xiao}}, \bibinfo {author}
  {\bibfnamefont {A.}~\bibnamefont {Azcatl}}, \bibinfo {author} {\bibfnamefont
  {J.}~\bibnamefont {Noh}}, \bibinfo {author} {\bibfnamefont {S.~R.}\
  \bibnamefont {Madhvapathy}}, \bibinfo {author} {\bibfnamefont
  {R.}~\bibnamefont {Addou}}, \bibinfo {author} {\bibfnamefont
  {S.}~\bibnamefont {KC}}, \bibinfo {author} {\bibfnamefont {M.}~\bibnamefont
  {Dubey}}, \bibinfo {author} {\bibfnamefont {K.}~\bibnamefont {Cho}}, \bibinfo
  {author} {\bibfnamefont {R.~M.}\ \bibnamefont {Wallace}}, \bibinfo {author}
  {\bibfnamefont {S.-C.}\ \bibnamefont {Lee}}, \bibinfo {author} {\bibfnamefont
  {J.-H.}\ \bibnamefont {He}}, \bibinfo {author} {\bibfnamefont {J.~W.~A.}\
  \bibnamefont {III}}, \bibinfo {author} {\bibfnamefont {X.}~\bibnamefont
  {Zhang}}, \bibinfo {author} {\bibfnamefont {E.}~\bibnamefont
  {Yablonovitch}},\ and\ \bibinfo {author} {\bibfnamefont {A.}~\bibnamefont
  {Javey}},\ }\bibfield  {title} {\bibinfo {title} {{Near-unity
  photoluminescence quantum yield in MoS2}},\ }\href
  {https://doi.org/10.1126/science.aad2114} {\bibfield  {journal} {\bibinfo
  {journal} {Science}\ }\textbf {\bibinfo {volume} {350}},\ \bibinfo {pages}
  {1065} (\bibinfo {year} {2015})}\BibitemShut {NoStop}%
\bibitem [{\citenamefont {Xie}\ \emph {et~al.}(2013)\citenamefont {Xie},
  \citenamefont {Zhang}, \citenamefont {Li}, \citenamefont {Wang},
  \citenamefont {Sun}, \citenamefont {Zhou}, \citenamefont {Zhou},
  \citenamefont {Lou},\ and\ \citenamefont {Xie}}]{Xie.Xie.2013}%
  \BibitemOpen
  \bibfield  {author} {\bibinfo {author} {\bibfnamefont {J.}~\bibnamefont
  {Xie}}, \bibinfo {author} {\bibfnamefont {H.}~\bibnamefont {Zhang}}, \bibinfo
  {author} {\bibfnamefont {S.}~\bibnamefont {Li}}, \bibinfo {author}
  {\bibfnamefont {R.}~\bibnamefont {Wang}}, \bibinfo {author} {\bibfnamefont
  {X.}~\bibnamefont {Sun}}, \bibinfo {author} {\bibfnamefont {M.}~\bibnamefont
  {Zhou}}, \bibinfo {author} {\bibfnamefont {J.}~\bibnamefont {Zhou}}, \bibinfo
  {author} {\bibfnamefont {X.~W.~D.}\ \bibnamefont {Lou}},\ and\ \bibinfo
  {author} {\bibfnamefont {Y.}~\bibnamefont {Xie}},\ }\bibfield  {title}
  {\bibinfo {title} {{Defect‐Rich MoS2 Ultrathin Nanosheets with Additional
  Active Edge Sites for Enhanced Electrocatalytic Hydrogen Evolution}},\ }\href
  {https://doi.org/10.1002/adma.201302685} {\bibfield  {journal} {\bibinfo
  {journal} {Advanced Materials}\ }\textbf {\bibinfo {volume} {25}},\ \bibinfo
  {pages} {5807} (\bibinfo {year} {2013})}\BibitemShut {NoStop}%
\bibitem [{\citenamefont {Komsa}\ and\ \citenamefont
  {Krasheninnikov}(2015)}]{Krasheninnikov.Komsa.2015}%
  \BibitemOpen
  \bibfield  {author} {\bibinfo {author} {\bibfnamefont {H.-P.}\ \bibnamefont
  {Komsa}}\ and\ \bibinfo {author} {\bibfnamefont {A.~V.}\ \bibnamefont
  {Krasheninnikov}},\ }\bibfield  {title} {\bibinfo {title} {{Native defects in
  bulk and monolayer MoS2 from first principles}},\ }\href
  {https://doi.org/10.1103/physrevb.91.125304} {\bibfield  {journal} {\bibinfo
  {journal} {Physical Review B}\ }\textbf {\bibinfo {volume} {91}},\ \bibinfo
  {pages} {125304} (\bibinfo {year} {2015})}\BibitemShut {NoStop}%
\bibitem [{\citenamefont {Lang}(1974)}]{Lang.Lang.1974}%
  \BibitemOpen
  \bibfield  {author} {\bibinfo {author} {\bibfnamefont {D.~V.}\ \bibnamefont
  {Lang}},\ }\bibfield  {title} {\bibinfo {title} {{Deep-level transient
  spectroscopy: A new method to characterize traps in semiconductors}},\ }\href
  {https://doi.org/10.1063/1.1663719} {\bibfield  {journal} {\bibinfo
  {journal} {Journal of Applied Physics}\ }\textbf {\bibinfo {volume} {45}},\
  \bibinfo {pages} {3023} (\bibinfo {year} {1974})}\BibitemShut {NoStop}%
\bibitem [{\citenamefont {Elcoro}\ \emph {et~al.}(2017)\citenamefont {Elcoro},
  \citenamefont {Bradlyn}, \citenamefont {Wang}, \citenamefont {Vergniory},
  \citenamefont {Cano}, \citenamefont {Felser}, \citenamefont {Bernevig},
  \citenamefont {Orobengoa}, \citenamefont {Flor},\ and\ \citenamefont
  {Aroyo}}]{Aroyo.Elcoro.2017}%
  \BibitemOpen
  \bibfield  {author} {\bibinfo {author} {\bibfnamefont {L.}~\bibnamefont
  {Elcoro}}, \bibinfo {author} {\bibfnamefont {B.}~\bibnamefont {Bradlyn}},
  \bibinfo {author} {\bibfnamefont {Z.}~\bibnamefont {Wang}}, \bibinfo {author}
  {\bibfnamefont {M.~G.}\ \bibnamefont {Vergniory}}, \bibinfo {author}
  {\bibfnamefont {J.}~\bibnamefont {Cano}}, \bibinfo {author} {\bibfnamefont
  {C.}~\bibnamefont {Felser}}, \bibinfo {author} {\bibfnamefont {B.~A.}\
  \bibnamefont {Bernevig}}, \bibinfo {author} {\bibfnamefont {D.}~\bibnamefont
  {Orobengoa}}, \bibinfo {author} {\bibfnamefont {G.~d.~l.}\ \bibnamefont
  {Flor}},\ and\ \bibinfo {author} {\bibfnamefont {M.~I.}\ \bibnamefont
  {Aroyo}},\ }\bibfield  {title} {\bibinfo {title} {{Double crystallographic
  groups and their representations on the Bilbao Crystallographic Server}},\
  }\href {https://doi.org/10.1107/S1600576717011712} {\bibfield  {journal}
  {\bibinfo  {journal} {Journal of Applied Crystallography}\ }\textbf {\bibinfo
  {volume} {50}},\ \bibinfo {pages} {1457} (\bibinfo {year} {2017})},\ \Eprint
  {https://arxiv.org/abs/1706.09272} {1706.09272} \BibitemShut {NoStop}%
\bibitem [{\citenamefont {Aroyo}\ \emph {et~al.}(2006)\citenamefont {Aroyo},
  \citenamefont {Kirov}, \citenamefont {Capillas}, \citenamefont {Perez-Mato},\
  and\ \citenamefont {Wondratschek}}]{Wondratschek.Aroyo.2006}%
  \BibitemOpen
  \bibfield  {author} {\bibinfo {author} {\bibfnamefont {M.}~\bibnamefont
  {Aroyo}}, \bibinfo {author} {\bibfnamefont {A.}~\bibnamefont {Kirov}},
  \bibinfo {author} {\bibfnamefont {C.}~\bibnamefont {Capillas}}, \bibinfo
  {author} {\bibfnamefont {J.}~\bibnamefont {Perez-Mato}},\ and\ \bibinfo
  {author} {\bibfnamefont {H.}~\bibnamefont {Wondratschek}},\ }\bibfield
  {title} {\bibinfo {title} {{Bilbao Crystallographic Server. II.
  Representations of crystallographic point groups and space groups}},\ }\href
  {https://doi.org/10.1107/s0108767305040286} {\bibfield  {journal} {\bibinfo
  {journal} {Acta Crystallographica Section A: Foundations of Crystallography}\
  }\textbf {\bibinfo {volume} {62}},\ \bibinfo {pages} {115} (\bibinfo {year}
  {2006})}\BibitemShut {NoStop}%
\bibitem [{\citenamefont {Zak}(1980)}]{Zak.Zak.1980}%
  \BibitemOpen
  \bibfield  {author} {\bibinfo {author} {\bibfnamefont {J.}~\bibnamefont
  {Zak}},\ }\bibfield  {title} {\bibinfo {title} {{Symmetry Specification of
  Bands in Solids}},\ }\href {https://doi.org/10.1103/physrevlett.45.1025}
  {\bibfield  {journal} {\bibinfo  {journal} {Physical Review Letters}\
  }\textbf {\bibinfo {volume} {45}},\ \bibinfo {pages} {1025} (\bibinfo {year}
  {1980})}\BibitemShut {NoStop}%
\bibitem [{\citenamefont {Zak}(1981)}]{Zak.Zak.1981}%
  \BibitemOpen
  \bibfield  {author} {\bibinfo {author} {\bibfnamefont {J.}~\bibnamefont
  {Zak}},\ }\bibfield  {title} {\bibinfo {title} {{Band representations and
  symmetry types of bands in solids}},\ }\href
  {https://doi.org/10.1103/physrevb.23.2824} {\bibfield  {journal} {\bibinfo
  {journal} {Physical Review B}\ }\textbf {\bibinfo {volume} {23}},\ \bibinfo
  {pages} {2824} (\bibinfo {year} {1981})}\BibitemShut {NoStop}%
\bibitem [{\citenamefont {Thonhauser}\ and\ \citenamefont
  {Vanderbilt}(2006)}]{Vanderbilt.Thonhauser.2006}%
  \BibitemOpen
  \bibfield  {author} {\bibinfo {author} {\bibfnamefont {T.}~\bibnamefont
  {Thonhauser}}\ and\ \bibinfo {author} {\bibfnamefont {D.}~\bibnamefont
  {Vanderbilt}},\ }\bibfield  {title} {\bibinfo {title}
  {{Insulator/Chern-insulator transition in the Haldane model}},\ }\href
  {https://doi.org/10.1103/physrevb.74.235111} {\bibfield  {journal} {\bibinfo
  {journal} {Physical Review B}\ }\textbf {\bibinfo {volume} {74}},\ \bibinfo
  {pages} {235111} (\bibinfo {year} {2006})},\ \Eprint
  {https://arxiv.org/abs/cond-mat/0608527} {cond-mat/0608527} \BibitemShut
  {NoStop}%
\bibitem [{\citenamefont {Soluyanov}\ and\ \citenamefont
  {Vanderbilt}(2012)}]{Vanderbilt.Soluyanov.2012yvd}%
  \BibitemOpen
  \bibfield  {author} {\bibinfo {author} {\bibfnamefont {A.~A.}\ \bibnamefont
  {Soluyanov}}\ and\ \bibinfo {author} {\bibfnamefont {D.}~\bibnamefont
  {Vanderbilt}},\ }\bibfield  {title} {\bibinfo {title} {{Smooth gauge for
  topological insulators}},\ }\href
  {https://doi.org/10.1103/physrevb.85.115415} {\bibfield  {journal} {\bibinfo
  {journal} {Physical Review B}\ }\textbf {\bibinfo {volume} {85}},\ \bibinfo
  {pages} {115415} (\bibinfo {year} {2012})},\ \Eprint
  {https://arxiv.org/abs/1201.5356} {1201.5356} \BibitemShut {NoStop}%
\bibitem [{\citenamefont {Tao}(2012)}]{Tao.Tao.2012}%
  \BibitemOpen
  \bibfield  {author} {\bibinfo {author} {\bibfnamefont {T.}~\bibnamefont
  {Tao}},\ }\bibfield  {title} {\bibinfo {title} {{Topics in Random Matrix
  Theory}},\ }\bibfield  {journal} {\bibinfo  {journal} {Graduate Studies in
  Mathematics}\ }\href {https://doi.org/10.1090/gsm/132} {10.1090/gsm/132}
  (\bibinfo {year} {2012})\BibitemShut {NoStop}%
\bibitem [{\citenamefont {Bradlyn}\ \emph {et~al.}(2019)\citenamefont
  {Bradlyn}, \citenamefont {Wang}, \citenamefont {Cano},\ and\ \citenamefont
  {Bernevig}}]{Bernevig.Bradlyn.2019nh}%
  \BibitemOpen
  \bibfield  {author} {\bibinfo {author} {\bibfnamefont {B.}~\bibnamefont
  {Bradlyn}}, \bibinfo {author} {\bibfnamefont {Z.}~\bibnamefont {Wang}},
  \bibinfo {author} {\bibfnamefont {J.}~\bibnamefont {Cano}},\ and\ \bibinfo
  {author} {\bibfnamefont {B.~A.}\ \bibnamefont {Bernevig}},\ }\bibfield
  {title} {\bibinfo {title} {{Disconnected elementary band representations,
  fragile topology, and Wilson loops as topological indices: An example on the
  triangular lattice}},\ }\href {https://doi.org/10.1103/physrevb.99.045140}
  {\bibfield  {journal} {\bibinfo  {journal} {Physical Review B}\ }\textbf
  {\bibinfo {volume} {99}},\ \bibinfo {pages} {045140} (\bibinfo {year}
  {2019})},\ \Eprint {https://arxiv.org/abs/1807.09729} {1807.09729}
  \BibitemShut {NoStop}%
\bibitem [{\citenamefont {Ahn}\ \emph {et~al.}(2019)\citenamefont {Ahn},
  \citenamefont {Park},\ and\ \citenamefont {Yang}}]{Yang.Ahn.2019pt}%
  \BibitemOpen
  \bibfield  {author} {\bibinfo {author} {\bibfnamefont {J.}~\bibnamefont
  {Ahn}}, \bibinfo {author} {\bibfnamefont {S.}~\bibnamefont {Park}},\ and\
  \bibinfo {author} {\bibfnamefont {B.-J.}\ \bibnamefont {Yang}},\ }\bibfield
  {title} {\bibinfo {title} {{Failure of Nielsen-Ninomiya Theorem and Fragile
  Topology in Two-Dimensional Systems with Space-Time Inversion Symmetry:
  Application to Twisted Bilayer Graphene at Magic Angle}},\ }\href
  {https://doi.org/10.1103/physrevx.9.021013} {\bibfield  {journal} {\bibinfo
  {journal} {Physical Review X}\ }\textbf {\bibinfo {volume} {9}},\ \bibinfo
  {pages} {021013} (\bibinfo {year} {2019})},\ \Eprint
  {https://arxiv.org/abs/1808.05375} {1808.05375} \BibitemShut {NoStop}%
\bibitem [{\citenamefont {Bouhon}\ \emph {et~al.}(2020)\citenamefont {Bouhon},
  \citenamefont {Wu}, \citenamefont {Slager}, \citenamefont {Weng},
  \citenamefont {Yazyev},\ and\ \citenamefont
  {Bzdušek}}]{Bouhon.Slager.2020np}%
  \BibitemOpen
  \bibfield  {author} {\bibinfo {author} {\bibfnamefont {A.}~\bibnamefont
  {Bouhon}}, \bibinfo {author} {\bibfnamefont {Q.}~\bibnamefont {Wu}}, \bibinfo
  {author} {\bibfnamefont {R.-J.}\ \bibnamefont {Slager}}, \bibinfo {author}
  {\bibfnamefont {H.}~\bibnamefont {Weng}}, \bibinfo {author} {\bibfnamefont
  {O.~V.}\ \bibnamefont {Yazyev}},\ and\ \bibinfo {author} {\bibfnamefont
  {T.}~\bibnamefont {Bzdušek}},\ }\bibfield  {title} {\bibinfo {title}
  {{Non-Abelian reciprocal braiding of Weyl points and its manifestation in
  ZrTe}},\ }\href {https://doi.org/10.1038/s41567-020-0967-9} {\bibfield
  {journal} {\bibinfo  {journal} {Nature Physics}\ }\textbf {\bibinfo {volume}
  {16}},\ \bibinfo {pages} {1137} (\bibinfo {year} {2020})}\BibitemShut
  {NoStop}%
\bibitem [{\citenamefont {Friedel}(1958)}]{Friedel.Friedel.1958}%
  \BibitemOpen
  \bibfield  {author} {\bibinfo {author} {\bibfnamefont {J.}~\bibnamefont
  {Friedel}},\ }\bibfield  {title} {\bibinfo {title} {{Metallic alloys}},\
  }\href {https://doi.org/10.1007/BF02751483} {\bibfield  {journal} {\bibinfo
  {journal} {Il Nuovo Cimento}\ }\textbf {\bibinfo {volume} {7}},\ \bibinfo
  {pages} {287} (\bibinfo {year} {1958})},\ \bibinfo {note} {supplement
  2}\BibitemShut {NoStop}%
\bibitem [{\citenamefont {Mahan}(2000)}]{Mahan.Mahan.2000}%
  \BibitemOpen
  \bibfield  {author} {\bibinfo {author} {\bibfnamefont {G.~D.}\ \bibnamefont
  {Mahan}},\ }\href@noop {} {\emph {\bibinfo {title} {{Many-Particle
  Physics}}}},\ \bibinfo {edition} {3rd}\ ed.\ (\bibinfo  {publisher} {Kluwer
  Academic / Plenum Publishers},\ \bibinfo {address} {New York},\ \bibinfo
  {year} {2000})\BibitemShut {NoStop}%
\bibitem [{\citenamefont {Economou}(2006)}]{Economou.Economou}%
  \BibitemOpen
  \bibfield  {author} {\bibinfo {author} {\bibfnamefont {E.~N.}\ \bibnamefont
  {Economou}},\ }\href@noop {} {\emph {\bibinfo {title} {{Green's functions in
  quantum physics}}}},\ Vol.~\bibinfo {volume} {7}\ (\bibinfo  {publisher}
  {Springer Science \& Business Media},\ \bibinfo {year} {2006})\BibitemShut
  {NoStop}%
\bibitem [{\citenamefont {Hewson}(1993)}]{Hewson.Hewson.1993}%
  \BibitemOpen
  \bibfield  {author} {\bibinfo {author} {\bibfnamefont {A.~C.}\ \bibnamefont
  {Hewson}},\ }\href {https://doi.org/10.1017/cbo9780511470752.025} {\emph
  {\bibinfo {title} {{The Kondo Problem to Heavy Fermions}}}}\ (\bibinfo
  {publisher} {Cambridge University Press},\ \bibinfo {year} {1993})\ pp.\
  \bibinfo {pages} {411--418}\BibitemShut {NoStop}%
\bibitem [{\citenamefont {Qi}\ \emph {et~al.}(2006)\citenamefont {Qi},
  \citenamefont {Wu},\ and\ \citenamefont {Zhang}}]{Zhang.Qi.2006}%
  \BibitemOpen
  \bibfield  {author} {\bibinfo {author} {\bibfnamefont {X.-L.}\ \bibnamefont
  {Qi}}, \bibinfo {author} {\bibfnamefont {Y.-S.}\ \bibnamefont {Wu}},\ and\
  \bibinfo {author} {\bibfnamefont {S.-C.}\ \bibnamefont {Zhang}},\ }\bibfield
  {title} {\bibinfo {title} {{Topological quantization of the spin Hall effect
  in two-dimensional paramagnetic semiconductors}},\ }\href
  {https://doi.org/10.1103/PhysRevB.74.085308} {\bibfield  {journal} {\bibinfo
  {journal} {Physical Review B}\ }\textbf {\bibinfo {volume} {74}},\ \bibinfo
  {pages} {085308} (\bibinfo {year} {2006})}\BibitemShut {NoStop}%
\bibitem [{\citenamefont {Bernevig}\ \emph {et~al.}(2006)\citenamefont
  {Bernevig}, \citenamefont {Hughes},\ and\ \citenamefont
  {Zhang}}]{Zhang.Bernevig.2006q49}%
  \BibitemOpen
  \bibfield  {author} {\bibinfo {author} {\bibfnamefont {B.~A.}\ \bibnamefont
  {Bernevig}}, \bibinfo {author} {\bibfnamefont {T.~L.}\ \bibnamefont
  {Hughes}},\ and\ \bibinfo {author} {\bibfnamefont {S.-C.}\ \bibnamefont
  {Zhang}},\ }\bibfield  {title} {\bibinfo {title} {{Quantum Spin Hall Effect
  and Topological Phase Transition in HgTe Quantum Wells}},\ }\href
  {https://doi.org/10.1126/science.1133734} {\bibfield  {journal} {\bibinfo
  {journal} {Science}\ }\textbf {\bibinfo {volume} {314}},\ \bibinfo {pages}
  {1757} (\bibinfo {year} {2006})},\ \Eprint
  {https://arxiv.org/abs/cond-mat/0611399} {cond-mat/0611399} \BibitemShut
  {NoStop}%
\bibitem [{\citenamefont {Dzyaloshinskii}(2003)}]{Dzyaloshinskii.2003}%
  \BibitemOpen
  \bibfield  {author} {\bibinfo {author} {\bibfnamefont {I.}~\bibnamefont
  {Dzyaloshinskii}},\ }\bibfield  {title} {\bibinfo {title} {{Some consequences
  of the Luttinger theorem: The Luttinger surfaces in non-Fermi liquids and
  Mott insulators}},\ }\href {https://doi.org/10.1103/PhysRevB.68.085113}
  {\bibfield  {journal} {\bibinfo  {journal} {Physical Review B}\ }\textbf
  {\bibinfo {volume} {68}},\ \bibinfo {pages} {085113} (\bibinfo {year}
  {2003})}\BibitemShut {NoStop}%
\bibitem [{\citenamefont {Gurarie}(2011)}]{Gurarie.2011}%
  \BibitemOpen
  \bibfield  {author} {\bibinfo {author} {\bibfnamefont {V.}~\bibnamefont
  {Gurarie}},\ }\bibfield  {title} {\bibinfo {title} {{Single-particle Green's
  functions and interacting topological insulators}},\ }\href
  {https://doi.org/10.1103/PhysRevB.83.085426} {\bibfield  {journal} {\bibinfo
  {journal} {Physical Review B}\ }\textbf {\bibinfo {volume} {83}},\ \bibinfo
  {pages} {085426} (\bibinfo {year} {2011})}\BibitemShut {NoStop}%
\bibitem [{\citenamefont {Blason}\ and\ \citenamefont
  {Fabrizio}(2023)}]{Fabrizio.Blason.2023}%
  \BibitemOpen
  \bibfield  {author} {\bibinfo {author} {\bibfnamefont {A.}~\bibnamefont
  {Blason}}\ and\ \bibinfo {author} {\bibfnamefont {M.}~\bibnamefont
  {Fabrizio}},\ }\bibfield  {title} {\bibinfo {title} {{Unified role of Green's
  function poles and zeros in correlated topological insulators}},\ }\href
  {https://doi.org/10.1103/PhysRevB.108.125115} {\bibfield  {journal} {\bibinfo
   {journal} {Physical Review B}\ }\textbf {\bibinfo {volume} {108}},\ \bibinfo
  {pages} {125115} (\bibinfo {year} {2023})},\ \Eprint
  {https://arxiv.org/abs/2304.08349} {2304.08349} \BibitemShut {NoStop}%
\bibitem [{\citenamefont {Wagner}\ \emph {et~al.}(2023)\citenamefont {Wagner},
  \citenamefont {Crippa}, \citenamefont {Amaricci}, \citenamefont {Hansmann},
  \citenamefont {Klett}, \citenamefont {K{\"o}nig}, \citenamefont
  {Sch{\"a}fer}, \citenamefont {Di~Sante}, \citenamefont {Cano}, \citenamefont
  {Millis}, \citenamefont {Georges},\ and\ \citenamefont
  {Sangiovanni}}]{Sangiovanni.Wagner.2023}%
  \BibitemOpen
  \bibfield  {author} {\bibinfo {author} {\bibfnamefont {N.}~\bibnamefont
  {Wagner}}, \bibinfo {author} {\bibfnamefont {L.}~\bibnamefont {Crippa}},
  \bibinfo {author} {\bibfnamefont {A.}~\bibnamefont {Amaricci}}, \bibinfo
  {author} {\bibfnamefont {P.}~\bibnamefont {Hansmann}}, \bibinfo {author}
  {\bibfnamefont {M.}~\bibnamefont {Klett}}, \bibinfo {author} {\bibfnamefont
  {E.~J.}\ \bibnamefont {K{\"o}nig}}, \bibinfo {author} {\bibfnamefont
  {T.}~\bibnamefont {Sch{\"a}fer}}, \bibinfo {author} {\bibfnamefont
  {D.}~\bibnamefont {Di~Sante}}, \bibinfo {author} {\bibfnamefont
  {J.}~\bibnamefont {Cano}}, \bibinfo {author} {\bibfnamefont {A.~J.}\
  \bibnamefont {Millis}}, \bibinfo {author} {\bibfnamefont {A.}~\bibnamefont
  {Georges}},\ and\ \bibinfo {author} {\bibfnamefont {G.}~\bibnamefont
  {Sangiovanni}},\ }\bibfield  {title} {\bibinfo {title} {{Mott insulators with
  boundary zeros}},\ }\href {https://doi.org/10.1038/s41467-023-42773-7}
  {\bibfield  {journal} {\bibinfo  {journal} {Nature Communications}\ }\textbf
  {\bibinfo {volume} {14}},\ \bibinfo {pages} {7531} (\bibinfo {year}
  {2023})},\ \Eprint {https://arxiv.org/abs/2301.05588} {2301.05588}
  \BibitemShut {NoStop}%
\bibitem [{\citenamefont {Horn}\ and\ \citenamefont
  {Johnson}(1985)}]{Johnson.Horn.1985}%
  \BibitemOpen
  \bibfield  {author} {\bibinfo {author} {\bibfnamefont {R.~A.}\ \bibnamefont
  {Horn}}\ and\ \bibinfo {author} {\bibfnamefont {C.~R.}\ \bibnamefont
  {Johnson}},\ }\href {https://doi.org/10.1017/cbo9780511810817} {\emph
  {\bibinfo {title} {{Matrix Analysis}}}}\ (\bibinfo  {publisher} {Cambridge
  University Press},\ \bibinfo {year} {1985})\BibitemShut {NoStop}%
\bibitem [{\citenamefont {Cano}\ \emph {et~al.}(2018)\citenamefont {Cano},
  \citenamefont {Bradlyn}, \citenamefont {Wang}, \citenamefont {Elcoro},
  \citenamefont {Vergniory}, \citenamefont {Felser}, \citenamefont {Aroyo},\
  and\ \citenamefont {Bernevig}}]{Bernevig.Cano.2018}%
  \BibitemOpen
  \bibfield  {author} {\bibinfo {author} {\bibfnamefont {J.}~\bibnamefont
  {Cano}}, \bibinfo {author} {\bibfnamefont {B.}~\bibnamefont {Bradlyn}},
  \bibinfo {author} {\bibfnamefont {Z.}~\bibnamefont {Wang}}, \bibinfo {author}
  {\bibfnamefont {L.}~\bibnamefont {Elcoro}}, \bibinfo {author} {\bibfnamefont
  {M.~G.}\ \bibnamefont {Vergniory}}, \bibinfo {author} {\bibfnamefont
  {C.}~\bibnamefont {Felser}}, \bibinfo {author} {\bibfnamefont {M.~I.}\
  \bibnamefont {Aroyo}},\ and\ \bibinfo {author} {\bibfnamefont {B.~A.}\
  \bibnamefont {Bernevig}},\ }\bibfield  {title} {\bibinfo {title} {{Topology
  of Disconnected Elementary Band Representations}},\ }\href
  {https://doi.org/10.1103/physrevlett.120.266401} {\bibfield  {journal}
  {\bibinfo  {journal} {Physical Review Letters}\ }\textbf {\bibinfo {volume}
  {120}},\ \bibinfo {pages} {266401} (\bibinfo {year} {2018})},\ \Eprint
  {https://arxiv.org/abs/1711.11045} {1711.11045} \BibitemShut {NoStop}%
\bibitem [{\citenamefont {Lieb}(1989)}]{Lieb.Lieb.1989}%
  \BibitemOpen
  \bibfield  {author} {\bibinfo {author} {\bibfnamefont {E.~H.}\ \bibnamefont
  {Lieb}},\ }\bibfield  {title} {\bibinfo {title} {{Two theorems on the Hubbard
  model}},\ }\href {https://doi.org/10.1103/physrevlett.62.1201} {\bibfield
  {journal} {\bibinfo  {journal} {Physical Review Letters}\ }\textbf {\bibinfo
  {volume} {62}},\ \bibinfo {pages} {1201} (\bibinfo {year}
  {1989})}\BibitemShut {NoStop}%
\bibitem [{\citenamefont {Mera}\ and\ \citenamefont
  {Ozawa}(2021)}]{Mera.Ozawa.2021}%
  \BibitemOpen
  \bibfield  {author} {\bibinfo {author} {\bibfnamefont {B.}~\bibnamefont
  {Mera}}\ and\ \bibinfo {author} {\bibfnamefont {T.}~\bibnamefont {Ozawa}},\
  }\bibfield  {title} {\bibinfo {title} {{Kähler geometry and Chern
  insulators: Relations between topology and the quantum metric}},\ }\href
  {https://doi.org/10.1103/PhysRevB.104.045104} {\bibfield  {journal} {\bibinfo
   {journal} {Physical Review B}\ }\textbf {\bibinfo {volume} {104}},\ \bibinfo
  {pages} {045104} (\bibinfo {year} {2021})}\BibitemShut {NoStop}%
\bibitem [{\citenamefont {Carlson}\ and\ \citenamefont
  {Keller}(1957)}]{Keller.Carlson.1957}%
  \BibitemOpen
  \bibfield  {author} {\bibinfo {author} {\bibfnamefont {B.~C.}\ \bibnamefont
  {Carlson}}\ and\ \bibinfo {author} {\bibfnamefont {J.~M.}\ \bibnamefont
  {Keller}},\ }\bibfield  {title} {\bibinfo {title} {{Orthogonalization
  Procedures and the Localization of Wannier Functions}},\ }\href
  {https://doi.org/10.1103/physrev.105.102} {\bibfield  {journal} {\bibinfo
  {journal} {Physical Review}\ }\textbf {\bibinfo {volume} {105}},\ \bibinfo
  {pages} {102} (\bibinfo {year} {1957})}\BibitemShut {NoStop}%
\bibitem [{\citenamefont {Haldane}(1988)}]{Haldane.Haldane.1988}%
  \BibitemOpen
  \bibfield  {author} {\bibinfo {author} {\bibfnamefont {F.~D.~M.}\
  \bibnamefont {Haldane}},\ }\bibfield  {title} {\bibinfo {title} {{Model for a
  Quantum Hall Effect without Landau Levels: Condensed-Matter Realization of
  the "Parity Anomaly"}},\ }\href {https://doi.org/10.1103/PhysRevLett.61.2015}
  {\bibfield  {journal} {\bibinfo  {journal} {Physical Review Letters}\
  }\textbf {\bibinfo {volume} {61}},\ \bibinfo {pages} {2015} (\bibinfo {year}
  {1988})}\BibitemShut {NoStop}%
\bibitem [{\citenamefont {Jha}\ \emph {et~al.}(2017)\citenamefont {Jha},
  \citenamefont {Rani},\ and\ \citenamefont {Ganesh}}]{Ganesh.Jha.2017}%
  \BibitemOpen
  \bibfield  {author} {\bibinfo {author} {\bibfnamefont {V.~B.}\ \bibnamefont
  {Jha}}, \bibinfo {author} {\bibfnamefont {G.}~\bibnamefont {Rani}},\ and\
  \bibinfo {author} {\bibfnamefont {R.}~\bibnamefont {Ganesh}},\ }\bibfield
  {title} {\bibinfo {title} {{Impurity-induced current in a Chern insulator}},\
  }\href {https://doi.org/10.1103/PhysRevB.95.115434} {\bibfield  {journal}
  {\bibinfo  {journal} {Physical Review B}\ }\textbf {\bibinfo {volume} {95}},\
  \bibinfo {pages} {115434} (\bibinfo {year} {2017})},\ \Eprint
  {https://arxiv.org/abs/1610.04642} {1610.04642} \BibitemShut {NoStop}%
\bibitem [{\citenamefont {Xiao}\ \emph {et~al.}(2010)\citenamefont {Xiao},
  \citenamefont {Chang},\ and\ \citenamefont {Niu}}]{Xiao.Niu.2010}%
  \BibitemOpen
  \bibfield  {author} {\bibinfo {author} {\bibfnamefont {D.}~\bibnamefont
  {Xiao}}, \bibinfo {author} {\bibfnamefont {M.-C.}\ \bibnamefont {Chang}},\
  and\ \bibinfo {author} {\bibfnamefont {Q.}~\bibnamefont {Niu}},\ }\bibfield
  {title} {\bibinfo {title} {{Berry phase effects on electronic properties}},\
  }\href {https://doi.org/10.1103/RevModPhys.82.1959} {\bibfield  {journal}
  {\bibinfo  {journal} {Reviews of Modern Physics}\ }\textbf {\bibinfo {volume}
  {82}},\ \bibinfo {pages} {1959} (\bibinfo {year} {2010})}\BibitemShut
  {NoStop}%
\bibitem [{\citenamefont {Culcer}\ \emph {et~al.}(2017)\citenamefont {Culcer},
  \citenamefont {Sekine},\ and\ \citenamefont
  {MacDonald}}]{Culcer.Sinova.2017}%
  \BibitemOpen
  \bibfield  {author} {\bibinfo {author} {\bibfnamefont {D.}~\bibnamefont
  {Culcer}}, \bibinfo {author} {\bibfnamefont {A.}~\bibnamefont {Sekine}},\
  and\ \bibinfo {author} {\bibfnamefont {A.~H.}\ \bibnamefont {MacDonald}},\
  }\bibfield  {title} {\bibinfo {title} {{Interband coherence response to
  electric fields in crystals: Berry-phase contributions and disorder
  effects}},\ }\href {https://doi.org/10.1103/PhysRevB.96.035106} {\bibfield
  {journal} {\bibinfo  {journal} {Physical Review B}\ }\textbf {\bibinfo
  {volume} {96}},\ \bibinfo {pages} {035106} (\bibinfo {year}
  {2017})}\BibitemShut {NoStop}%
\end{thebibliography}%
\clearpage
\onecolumngrid
\setcounter{section}{0}
\setcounter{equation}{0}
\setcounter{figure}{0}
\renewcommand{\thesection}{S\arabic{section}}
\renewcommand{\thesubsection}{\thesection.\arabic{subsection}}
\renewcommand{\thesubsubsection}{\thesubsection.\arabic{subsubsection}}
\renewcommand{\theequation}{\arabic{equation}}
\renewcommand{\thefigure}{S\arabic{figure}}
\renewcommand{\theHsection}{supp.\arabic{section}}
\renewcommand{\theHequation}{supp.\arabic{equation}}
\renewcommand{\theHfigure}{supp.\arabic{figure}}
\begin{center}
\large\bfseries Supplementary information for ``Ring states in topological materials''
\end{center}

\section{Bound state from the Green's function}\label{app:boundstate}
In this note, we review the form of bound states from a Green's function perspective, which can be found in Refs.~\cite{Economou.Economou,Hewson.Hewson.1993}. Let us take the perturbed Hamiltonian,
\begin{align}\tilde\H=\H+\V,\end{align}
with $\V$ a general perturbation of rank smaller than $\H$, and with both operators defined in the same Hilbert space.
We can define the unperturbed resolvent Green's function as
\begin{align}\G(\omega)=(\omega-\H)\inv,\end{align}
and equivalently, the perturbed Green's function 
\begin{align}\tilde\G(\omega)=(\omega-\tilde\H)\inv,\end{align}
which, when expressed in terms of the unperturbed Green's function, becomes
\begin{align}
   \tilde\G(\omega)=[1-\G(\omega)\V]\inv\G(\omega).
\end{align}
Expanding the prefactor in a power series, we obtain the Dyson expansion
\begin{align}
    \tilde\G(\omega)=\G(\omega)+\G(\omega) \V \tilde\G(\omega)=\G(\omega)+\G(\omega)\V\G(\omega)+\G(\omega)\V\G(\omega)\V\G(\omega)+...\equiv\G(\omega)+\G(\omega) \T(\omega) \G(\omega),\label{eqapp:dysonexp}
\end{align}
where the T-matrix is defined. It can be rewritten in the convenient form
\begin{align}
    \T(\omega)=\V[1-\G(\omega)\V]\inv.\label{eqapp:tmat}
\end{align}
The spectrum of the perturbed Hamiltonian corresponds to poles in the perturbed Green's function, which can occur perturbatively close to the poles of $\G(\omega)$ in the band continuum, or, alternatively, as poles originating from the infinite sum in \eqref{eqapp:dysonexp} which appear in $\T(\omega)$. Considering the latter case, we look at eigenstates of $\tilde\H$  
\begin{align}(E-\H)\ket{\varphi_E}=\V\ket{\varphi_E}\end{align}
acting with $\G(\omega)$ on both sides, we find the eigenstates must satisfy  
\begin{align}\ket{\varphi_E}=\G(E)\V\ket{\varphi_E}\quad\text{or}\quad[1-\G(E)\V]\ket{\varphi_E}=0.\end{align}
for $E$ explicitly corresponding to the poles of $\T(\omega)$ in \eqref{eqapp:tmat}. For energies $E$ outside the band continuum, the states $\ket{\varphi_E}$ are bound states, while inside the band continuum, the states $\ket{\varphi_E}$ are scattering states with a large overlap with the unperturbed states from the band. Namely, in the band continuum the solutions $\ket{\varphi_E}$ satisfy the Lippmann-Schwinger equation
\begin{align}
    \ket{\varphi_E}=\ket{\psi_E}+\G(E+i0^+)\V\ket{\varphi_E}
\end{align}
with $\ket{\psi_E}$ the unperturbed eigenstates of $\H$.

Note that we consider impurities that do not bring new quantum degrees of freedom into the Hilbert space, but rather impart a local change in the potential affecting the states in the original Hilbert space. Therefore, the bound states created by the impurity will be reflected in the total density of states of the band, which will amount to a reduced number of states in the bands in favor of states outside the band. 
The change in the density of states at frequency $\omega$ can be obtained from the imaginary part of the retarded Green's function,
\begin{align}\delta\nu(\omega)\equiv\tilde\nu(\omega)-\nu(\omega)=-{1\over \pi}\im\tr[\tilde\G(\omega+i0^+)-\G(\omega+i0^+)],\end{align}
where the trace is taken over the entire Hilbert space.
Using the convenient relation\begin{align}\tr \G(\omega)=-\partial_\omega\log\det\G(\omega),\end{align}
we find that
\begin{align}\delta\nu(\omega)={1\over\pi}\im\partial_\omega\log\det [1-\G(\omega+i0^+)\V]\inv,\end{align}
and using the standard scattering-phase definition~\cite{Friedel.Friedel.1958,Mahan.Mahan.2000,Economou.Economou}
\begin{align}\eta(\omega)=\arg\det\!\big[1-\G(\omega+i0^+)\V\big],\end{align}
the change in density of states gets the simple form
\begin{align}\delta\nu(\omega)=-{1\over\pi}\partial_\omega\eta(\omega).\end{align}
Note that to conserve the total number of states, the integrated density of states removed from a band must correspond to the number of bound states outside the band continuum. Therefore, the number of impurity-bound states is given by
\begin{align}N=-\int_{\rm bands} \delta\nu(\omega)d\omega.\end{align}
which is the sum of states removed from the individual bands,
\begin{align}N^\alpha=-\int_{\varepsilon_\alpha^{\rm min}}^{\varepsilon_\alpha^{\rm max}} \delta\nu(\omega)d\omega={1\over\pi}[\eta(\varepsilon_\alpha^{\rm max})-\eta(\varepsilon_\alpha^{\rm min})]\label{eqapp:removedstates}\end{align}
where we consider that the bands are well separated in energy by energy gaps and delimited by the band-edge energies $\varepsilon_\alpha^{\rm min}$ and $\varepsilon_\alpha^{\rm max}$.

\begin{figure}[t]
    \centering
    \hspace{.06\linewidth}(a)\hspace{.45\linewidth}(b)
    
  \includegraphics[width=.45\columnwidth]{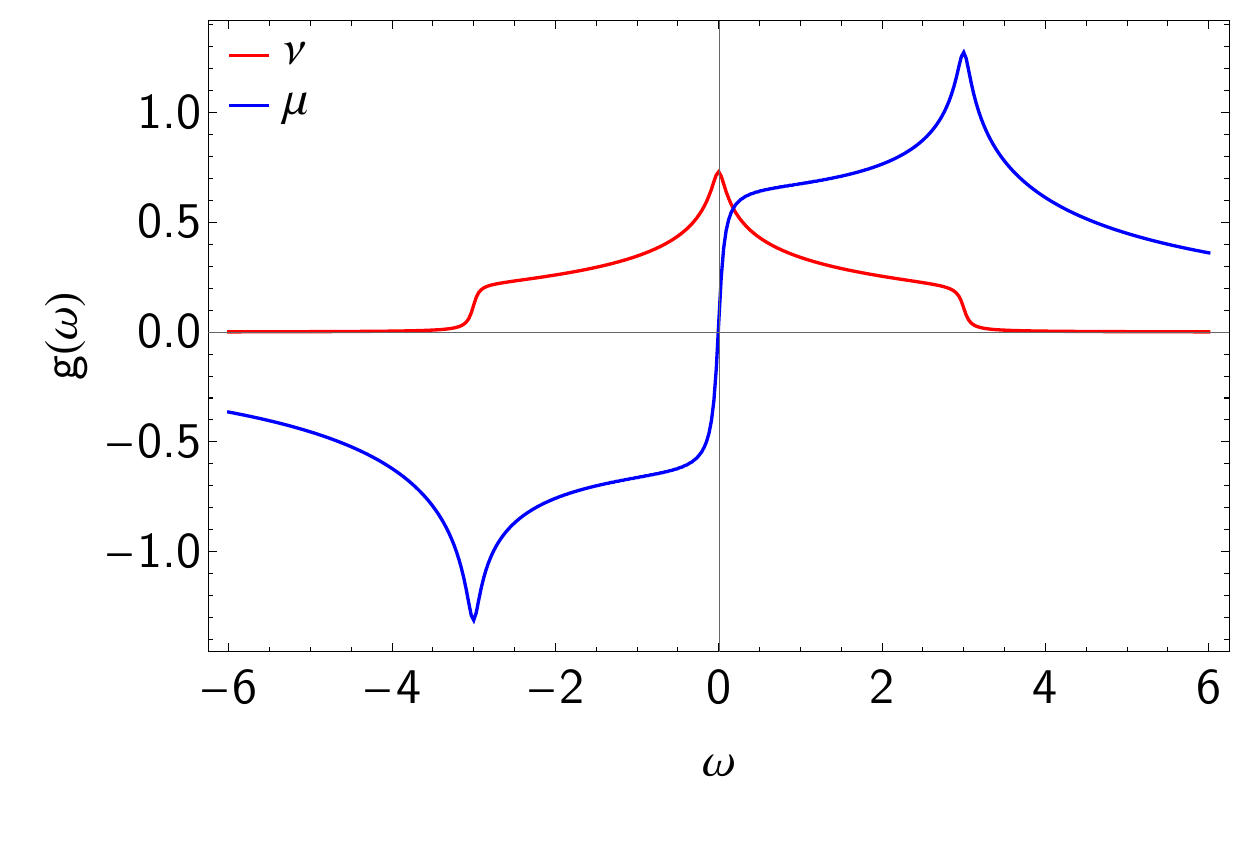}
   \includegraphics[width=.45\columnwidth]{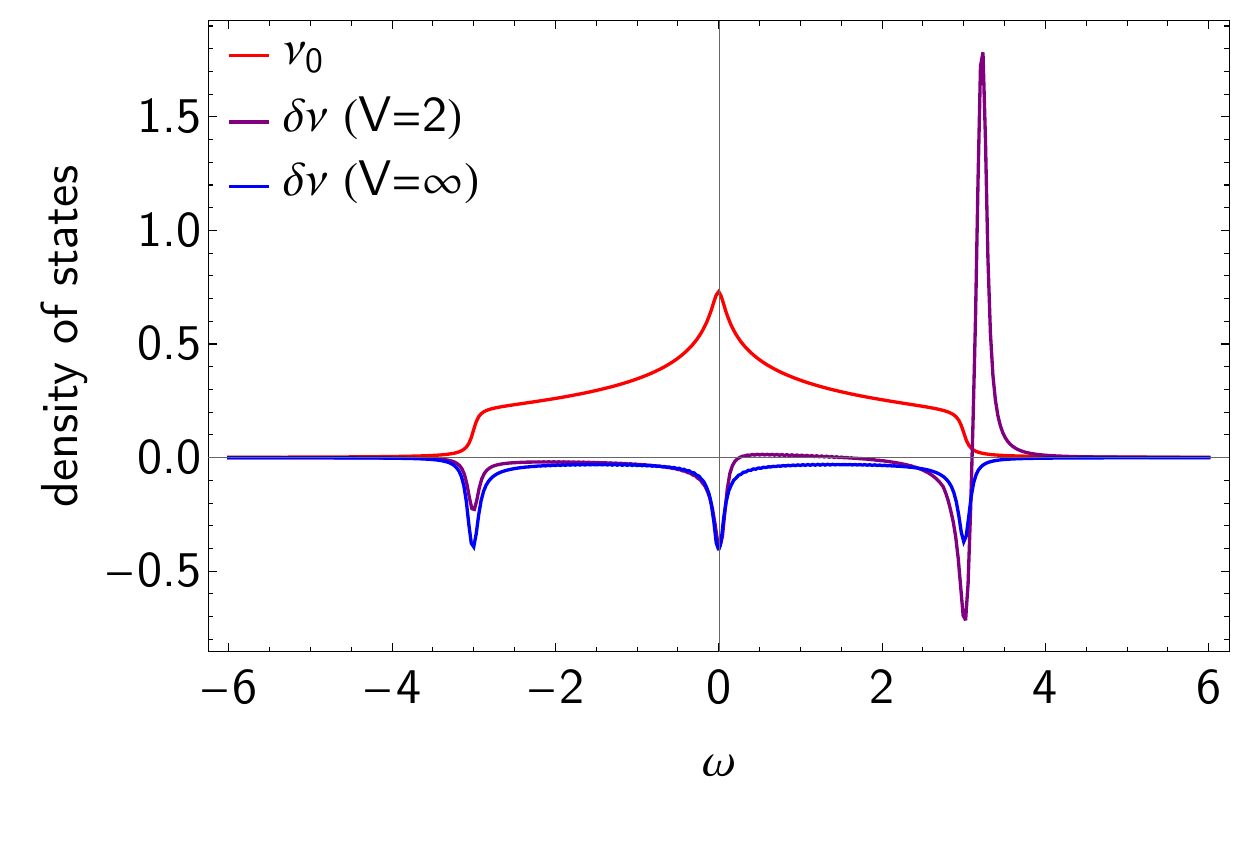}
    \caption{Local Green's function and density of states for a single atomic band, defined in \eqref{eqapp:singleband} with $t=1$.}
    \label{figapp:singleband}
\end{figure}

Now let us consider an impurity of a specific form, with a single local eigenstate $\ket{\phi^\sigma_0}$. 
The perturbation is explicitly given by \begin{align}\V=v\proj{\phi^\sigma_0},\end{align}  which amounts to adding a local energy to the orbital of symmetry character $\sigma$ in the position $\bq_0$. Choosing a basis that contains the state $\ket{\sigma}$ the $T$-matrix has a single nonzero element given by
\begin{align}t^\sigma(\omega)=\bra{\phi^\sigma_0}\T(\omega)\ket{\phi^\sigma_0}=v+v\bra{\phi^\sigma_0}\G(\omega)\ket{\phi^\sigma_0}v+...={v\over 1-g^\sigma(\omega)v}\end{align}
where
\begin{align}g^\sigma(\omega)\equiv\bra{\phi^\sigma_0}\G(\omega)\ket{\phi^\sigma_0}\end{align} 
is the impurity projected Green's function. It follows that the $T$-matrix will only have poles at the solutions of $g^\sigma(\omega)=1/v$. In particular, in the strong impurity limit $v\to\pm\infty$, its poles correspond to $g^\sigma(\omega)=0$. These solutions can be captured by computing the real and imaginary parts of the retarded impurity projected Green's function 
\begin{align}g^\sigma(\omega+i0^+)=\mu^\sigma(\omega)-i\pi\nu^\sigma(\omega).\end{align}
with $\nu^\sigma(\omega)$ the orbital-projected density of states. The density of states vanishes outside the band continuum, and therefore the bound state solutions are determined by the zeros of $\mu^\sigma(\omega)$. For $v\to\pm\infty$, we can expand $t^\sigma$ as $t^\sigma\sim -1/g^\sigma(1+1/g^\sigma v)$, and therefore, we see the phase shift is determined by the phase of $g^\sigma$. We can write
\begin{align}\delta\nu^\sigma(\omega)\sim-{1\over\pi}\partial_\omega\arg g^\sigma(\omega+i0^+),\end{align}
which can be understood as follows: 
In the gap, when $\partial_\omega\mu^\sigma(\omega)|_{\omega=E}<0$ for an energy $E$ at which there is a pole of the $t^\sigma(\omega)$, there is a positive change in density of states, that is, an added state. In the band, the change of density of states is spread throughout the band. A positive $\partial_\omega\mu^\sigma(\omega)>0$ implies that charge density is removed throughout the band, mostly from energies at which the phase of $g^\sigma(\omega)$ changes substantially, i.e., the zero of $\mu(\omega)$  and the band edges, see Fig.~\ref{figapp:singleband}. An accumulated $\pi$ phase shift across the band implies that a full state is removed from the band. Note that if $\mu^\sigma(\omega)$ does not cross zero with a positive slope through the band, there cannot be a total $\pi$ phase shift.

We should note, by making use of a Kramers-Kronig relation
\begin{align}\mu^\sigma(\omega)=P\!\int d \omega'\,{\nu^\sigma(\omega')\over \omega-\omega'}\end{align}
with $P$ the Cauchy principal value, that the information about the zeros in $\mu^\sigma(\omega)$ and consequently the information about the bound states created by a strong local potential $\V$, is completely contained in the orbital resolved density of states $\nu^\sigma(\omega)$ across the energy axis.

For concreteness, let us look at the example of a trivial two-dimensional band, obtained by a tight binding model on a square lattice, described by the Hamiltonian 
\begin{align}\H=\sum_{\ev{ij}} -tc^\dag_i c_j,\label{eqapp:singleband}\end{align}
where $c^\dag_i$ creates a particle in the state $\ket{\phi_i}$ at the Bravais lattice vector $\bR_i$. The sum is taken over the nearest neighbors.
The local Green's function, projected into the single orbital located at $\bR_0=0$ can be obtained by diagonalizing the Hamiltonian in reciprocal space
\begin{align}\varepsilon_\bk=-2t(\cos{k_x}+\cos k_y),\end{align}
and the local projected Green's function obtained from 
$g(\omega)=\bra{\phi_0}\G(\omega)\ket{\phi_0}=\int {d\bk\over(2\pi)^2} (\omega-\varepsilon_\bk )\inv$,
which is shown in Fig.~\ref{figapp:singleband}. A local perturbation of the form $\V=vc^\dag_0c_0$, leads to a bound state at an energy given by $g(\omega)=1/v$.
For small $v$ it corresponds to a state at the band edge, see Fig.~\ref{figapp:singleband}(b), which is taken to large energies as $v\to\infty$, 
leaving the band with one less state. Note that the change in the band's density of states is distributed across all energies in the band. Most of the spectral weight is removed from the band edges and the van Hove singularity where $\partial_\omega \arg g(\omega)$ is the largest. There is a total phase shift $\eta(\omega)$ of $\pi$ across the band, corresponding to one bound state following \eqref{eqapp:removedstates}.

\section{Ring state wavefunctions in the flat band limit}\label{app:flatbandexact}

In this note, we solve for the ring state energies and wavefunctions in the large $v$ limit in a simple two-band flat Hamiltonian given by
\begin{align}\H=\sum_{\alpha=1,2}\varepsilon_\alpha\P^\alpha\end{align}
with $\P^\alpha=\sum_\bk\proj{\psi^\alpha_\bk}$ the projector into the $\alpha$ band, and $\varepsilon_\alpha$ its energy, set without loss of generality to $\varepsilon_1=0$ and $\varepsilon_2=\Delta$, with $\Delta$ the spectral gap.
We consider that the Hamiltonian is perturbed by a rank-1 operator $\V$ with a single eigenstate of energy $v$, given by
\begin{align}\V=v\proj{\sigma}.\end{align}
As described in the main text, the perturbed Hamiltonian $\H+\V$ contains bound states $\ket{\varphi_E}$ at energy $E$ that satisfy
\begin{align}\ket{\varphi_E}=\G(E)\V\ket{\varphi_E}=\sum_\alpha{v\lambda_E\over E-\varepsilon_\alpha}\ket{\Upsilon^\alpha},\quad\text{with}~~ \lambda_E={\langle\sigma|\varphi_E\rangle}\label{eqapp:boundstate}\end{align}
where $\ket{\Upsilon^\alpha}\equiv \P^\alpha\ket{\sigma}$ are non-normalized band projections of the $\ket{\sigma}$ state. They satisfy $\bra{\Upsilon^\alpha}\Upsilon^\beta\rangle=s_\alpha\delta_{\alpha\beta}$, where $s_\alpha$ is the integrated band overlap with the $\sigma$ orbital over the band Bloch states, defined in Secs.~1.2 and 1.6 of the main text (see also Notes~\ref{app:proof2} and~\ref{app:proof3}).
Therefore, the state $\ket{\sigma}$ can be written as 
\begin{align}\ket{\sigma}=\sum_{\alpha=1,2}\P^\alpha\ket{\sigma}=\ket{\Upsilon^1}+\ket{\Upsilon^2}.\end{align}
From the normalization of $\ket{\sigma}$ the projected norms satisfy $s_1+s_2=1$. Let us rewrite $s_1=(1+\delta s)/2$, $s_2=(1-\delta s)/2$. An obstructed band can be distinguished from a trivial band by the requirement that $0<s_\alpha<1$ and consequently $|\delta s|<1$ for \emph{any} choice of local state $\ket\sigma$ in the Hilbert space of $\H$ that preserves the local site symmetry. 

\begin{figure}
    \centering
    \includegraphics[width=.2\columnwidth]{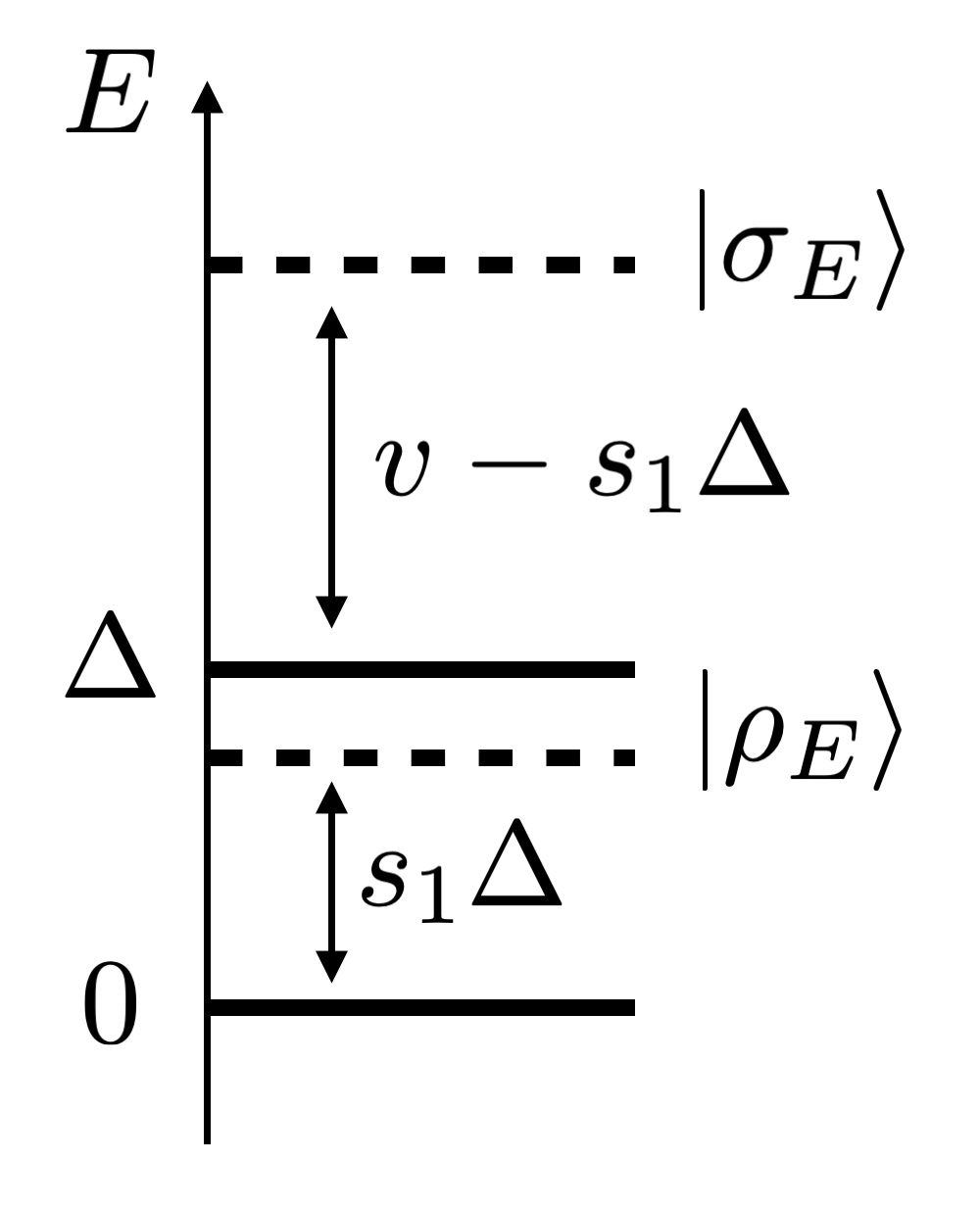}
    \caption{Energy spectrum of $\H+\V$ in the large $v$ limit for the model of two perfectly flat bands, up to order $O(1/v)$ for $v>0$.}
    \label{figapp:flatbandenergydiag}
\end{figure}

In the large $|v|$ limit, there are two possible solutions of Eq.~\eqref{eqapp:boundstate}, with energies outside the spectral continuum which are qualitatively distinct. One solution has an energy of order $O(v)$. It corresponds to a solution approaching the eigenstate of $\V$, $\langle\sigma|\varphi_E\rangle\to1$. Let us call this solution $\ket{\sigma_E}$ and write it as
\begin{align}\ket{\sigma_E}=(1+a)\ket{\Upsilon^1}+(1+b)\ket{\Upsilon^2}=\ket{\sigma}+O(1/v),\end{align}
with $a$ and $b$ small parameters of order $O(1/v)$. The parameters $a$ and $b$ depend on the energy of the bound state as a function of $v$, as it will be derived below.
The second state must be orthogonal to $\ket{\sigma_E}$, with $\langle\sigma|\varphi_E\rangle\to0$ as $v\to\infty$. The  energy of the $\ket{\rho_E}$ state remains of order $O(1)$, between the minimum and maximum eigenvalues of the unperturbed Hamiltonian $\H$. The finite energy state, $\ket{\rho_E}$ is orthogonal to $\ket{\sigma_E}$ by avoiding the impurity site $\bq_0$. Orthogonality with $\ket{\sigma}$ at $v=\pm\infty$ fixes the asymptotic form of this state to be
\begin{align}\ket{\rho}=\sqrt{s_2\over s_1}\ket{\Upsilon^1}-\sqrt{s_1\over s_2}\ket{\Upsilon^2}\label{eqapp:rho},\end{align}
satisfying $\V\ket{\rho}=0$. Note that $\ket\rho$ is not itself an eigenstate of $\H+\V$. The exact eigenstate of the perturbed Hamiltonian $\H+\V$ at a finite energy $E$, $\ket{\rho_E}$ is close to $\ket{\rho}$, deviating by the same small parameters $a$ and $b$ of order $O(1/v)$,
\begin{align}\ket{\rho_E}=(1+b){\sqrt{1-\delta s}\over\sqrt{1+\delta s}}\ket{\Upsilon^1}-(1+a){\sqrt{1+\delta s}\over\sqrt{1-\delta s}}\ket{\Upsilon^2}=\ket\rho+O(1/v),\label{eqapp:rhoe}\end{align} 
The forms above guarantee orthogonality $\bra{\sigma_E}\rho_E\rangle=0$, and both states are normalized provided
\begin{align}\frac{1}{2} (a+1)^2 (1+\delta s)+\frac{1}{2} (b+1)^2 (1-\delta s)=1.\end{align}
The large-$v$ level structure described by these two states is summarized schematically in Fig.~\ref{figapp:flatbandenergydiag}.

\begin{figure}[h!]
    \centering
    \includegraphics[width=.35\columnwidth]{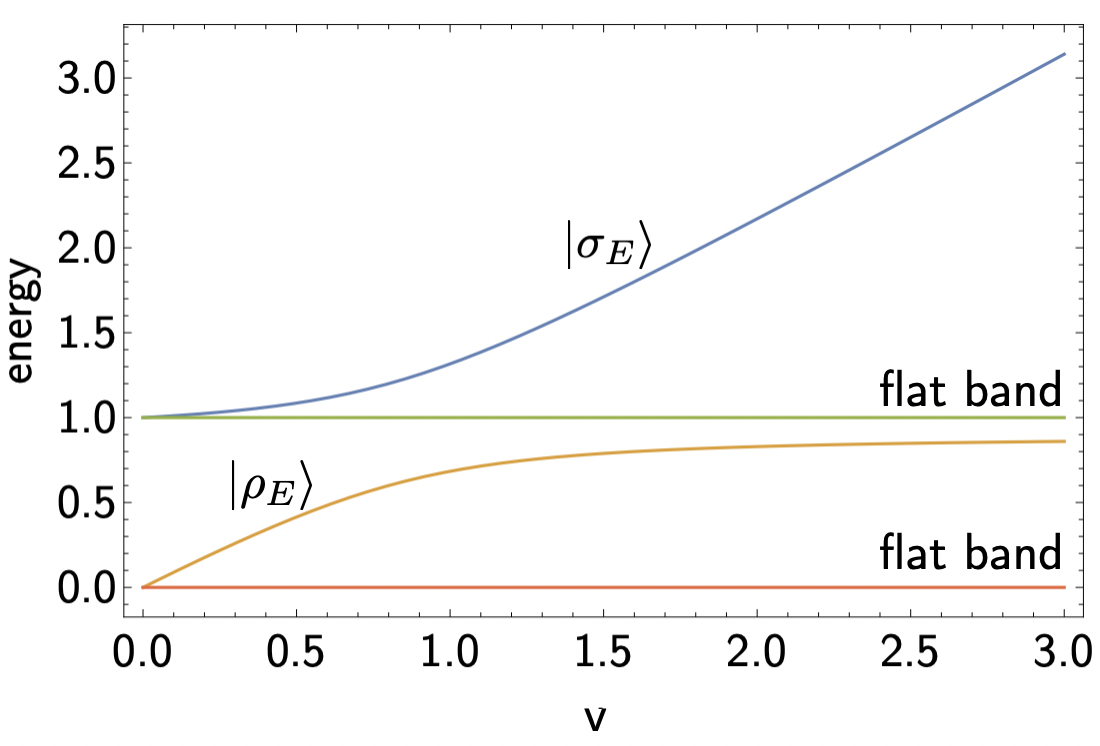}
    \caption{Evolution of the bound state energies in the flat band model for $\Delta=1$, $\delta s=0.9$.}
    \label{figapp:evolutionflatband}
\end{figure}

We now fix  $a$ and $b$  by taking both $\ket{\sigma_E}$ and $\ket{\rho_E}$ states to be eigenstates of $\H+\V$. First, we can calculate their energies from the inner product of $\ket{\sigma}$ and Eq.~\eqref{eqapp:boundstate}, which gives a quadratic equation for the energy $E$,
\begin{align}{1\over v}={s_1\over E-\varepsilon_1}+{s_2\over E-\varepsilon_2},\end{align}
with two solutions
\begin{equation}
    E_\pm=\half({\Delta+v\pm\sqrt{\Delta^2- 2v\delta s\Delta+v^2}}).\label{eq:flatbandsols}
\end{equation}
Assuming $v\delta s>0$, we find \begin{align}E_+\sim v+\Delta(1-\delta s)/2=s_2\Delta+v=\Delta+v-s_1\Delta,\end{align} and \begin{align}E_-\sim \Delta(1+\delta s)/2=s_1\Delta,\end{align} approximate to first order in $1/v$. The energies $E_\pm$ deviate from the original band energies as long as $|\delta s|\neq 1$.
Expanded to order $O(1/v)$, we can associate the largest energy with the $\ket{\sigma_E}$ state, and the lowest energy to the ring state $\ket{\rho_E}$,
\begin{align}E_\pm=\half[v(1\pm 1)+\Delta(1\mp\delta s)]\pm(1-\delta s^2){\Delta^2\over 4v}+O(1/v^2).\end{align} 
It is important to make two remarks here. First, in a topological phase, \emph{both} energies $E_\pm$ will correspond to bound states outside the original unperturbed bands. The ingap state $\ket{\rho_E}$ is separated from the lower band by $\Delta s_1$ and from the upper band by $\Delta s_2$, both of order $O(1)$ and independent of $v$, for all possible choices of $\V$. This evolution as a function of impurity strength is shown explicitly in Fig.~\ref{figapp:evolutionflatband}.
In a trivial band, there is always a choice of $\V$ such that the ring state solution does not exist. Second, the geometric overlap with the band states determines the finite energy separation of the ring state from the band edge. As we see below, it can be understood as the result of the level repulsion between the ring states and the band eigenstates. The energy of the ring state depends only on the energy gap and the overlap $s_\alpha$, becoming independent of $v$ for large enough  strength $v\gg \Delta$. 

Using the expressions for $E_\pm$, we take
\begin{align}\bra{\sigma_E}(\H+\V)\ket{\sigma_E}=E_+,\quad\bra{\rho_E}(\H+\V)\ket{\rho_E}=E_-, \end{align}
to find the solutions for the small parameters $a$ and $b$,
\begin{align}a= -{\Delta \over 2v} (1-\delta s)=-{s_2\Delta\over v}+O(1/v^2),\quad b= {\Delta\over 2v} (1+\delta s)={s_1\Delta\over v}+O(1/v^2)\end{align} 
which are both non-vanishing provided $|\delta s|<1$, as is necessarily the case for topological bands.

As a sanity check, we look at the topologically trivial flat band case with $\delta s=-1$, $s_1=0$, $s_2=1$, and $v>0$. In this case, $E_+=v+\varepsilon_2=v+\Delta$ and $E_-=\varepsilon_1=0$. Since only the $\alpha=2$ has nonzero matrix elements with $\V$, the orbital $\ket{\sigma}$ is removed from this band without obstruction or creating a ring state. Here, these formulas give $b=0$ -- so $\ket{\sigma_E}=\ket{\Upsilon^2}=\ket{\sigma}$ exactly with no finite-$v$ correction -- while $a=-\Delta/v$ is irrelevant because $\ket{\Upsilon^1}=\hat P^1\ket{\sigma}=0$ at $s_1=0$. The ring state solution does not exist, and both forms \eqref{eqapp:rho} and \eqref{eqapp:rhoe} are not well defined since they involve $\sqrt{s_2/s_1}$ with $s_1=0$. 

Finally, let us note that since any symmetry $\hat{g}$ operator that commutes with the Hamiltonian $\hat g \H=\H\hat{g}$ also commutes with the band projectors $\hat{g}\P^\alpha=\P^\alpha\hat{g}$ it follows that both $\ket{\sigma_E}$ and $\ket{\rho_E}$ transform under the same symmetry representation. Their orthogonality comes from spatially avoiding each other.

\section{Number of impurity states}\label{app:weyls}

The number of bound states expected from a local perturbation $\H+\V$ is highly constrained by the spectrum of $\H$ and $\V$ separately. We use Weyl's inequalities~\cite{Johnson.Horn.1985}, which state that the perturbed eigenvalues $E_i$, for $i=1,...,N$ 
organized in descending order (with $N$ the total number of eigenstates) satisfy
\begin{align}\varepsilon_j + v_k \le E_i \le \varepsilon_r + v_s,\quad j+k-N \ge i \ge r+s-1.\end{align}
where $\varepsilon_i, v_i$ are the eigenvalues of the unperturbed Hamiltonian and the perturbation, respectively. The unperturbed energies reach from $\varepsilon_1=\varepsilon_{\rm max}$ to $\varepsilon_N=\varepsilon_{\rm min}$. In the simple case where $\V$ is a rank 1 operator with a single nonzero eigenvalue, $v_1=v$ and $v_{k\neq1}=0$, and $v\gg\varepsilon_1$, the above inequality tells us that there is a single energy above $\varepsilon_1$. Namely, taking $i=1$ we find
\begin{align}\varepsilon_1\le E_1\le\varepsilon_1+v,\end{align}
and 
\begin{align}\varepsilon_2\le E_2\le\varepsilon_1.\end{align}
which is the tightest bound on $E_2$ given a large $v$. These inequalities imply that there can exist a single perturbed energy above the spectrum of $\H$. Analogously, if $v\ll \varepsilon_N$ there can only exist a single bound state below the unperturbed spectrum. In these limits, the previous inequalities amount to Cauchy's interlacing theorem~\cite{Johnson.Horn.1985}, which effectively removes one column and one row from the perturbed Hamiltonian. In particular, the interlacing theorem tells us that the perturbed system has energy states $E_{i}$ satisfying
\begin{align}\varepsilon_{i}\le E_{i}\le\varepsilon_{i-1}.\end{align}
and therefore, within a gap, a rank 1 perturbation $\V$ can create, at most, a single bound state.  There can only exist more than one state in the gap if there are more nonvanishing eigenstates of $\V$, as is the case for a multi-orbital or multi-site vacancy.

\section{Level repulsion}\label{app:repulsion}

\subsection{Discrete levels}

The exact energy at which the eigenvalues of a perturbed Hamiltonian $E_i$ are found depends on the structure of their wavefunctions and can be obtained by the Hadamard variation formulas~\cite{Tao.Tao.2012}. Let us interpret the strength of the perturbation $\V$ as a time $t$, and let $t\to\infty$ in the following way
\begin{align}\H(t)=\H+t\hat\V\end{align}
where $\hat\V=\proj{\sigma}$ is the normalized perturbation. For an instantaneous eigenstate $\ket{\psi_n(t)}$, we write
\begin{align}
\lambda_n(t)\equiv\langle\sigma|\psi_n(t)\rangle.
\end{align}
Thus the weight of the instantaneous level on the impurity orbital is $|\lambda_n|^2$. The eigenvalues of the (finite dimension) Hamiltonian $\H(t)$ are deformed adiabatically as a function of $t$, satisfying the differential equations
\begin{align}\dot \varepsilon_n=\bra{\psi_n}\hat\V\ket{\psi_n}=|\lambda_n|^2,\label{eqapp:velocity}\end{align}
known as the first Hadamard variation formula, and 
\begin{align}\ddot \varepsilon_n=2\sum_{m\neq n}{|\bra{\psi_m}\hat\V\ket{\psi_n}|^2\over \varepsilon_n-\varepsilon_m},\label{eqapp:energies}\end{align} 
known as the second Hadamard variation formula. 
\begin{figure}[t]
    \centering
    \includegraphics[width=.45\columnwidth]{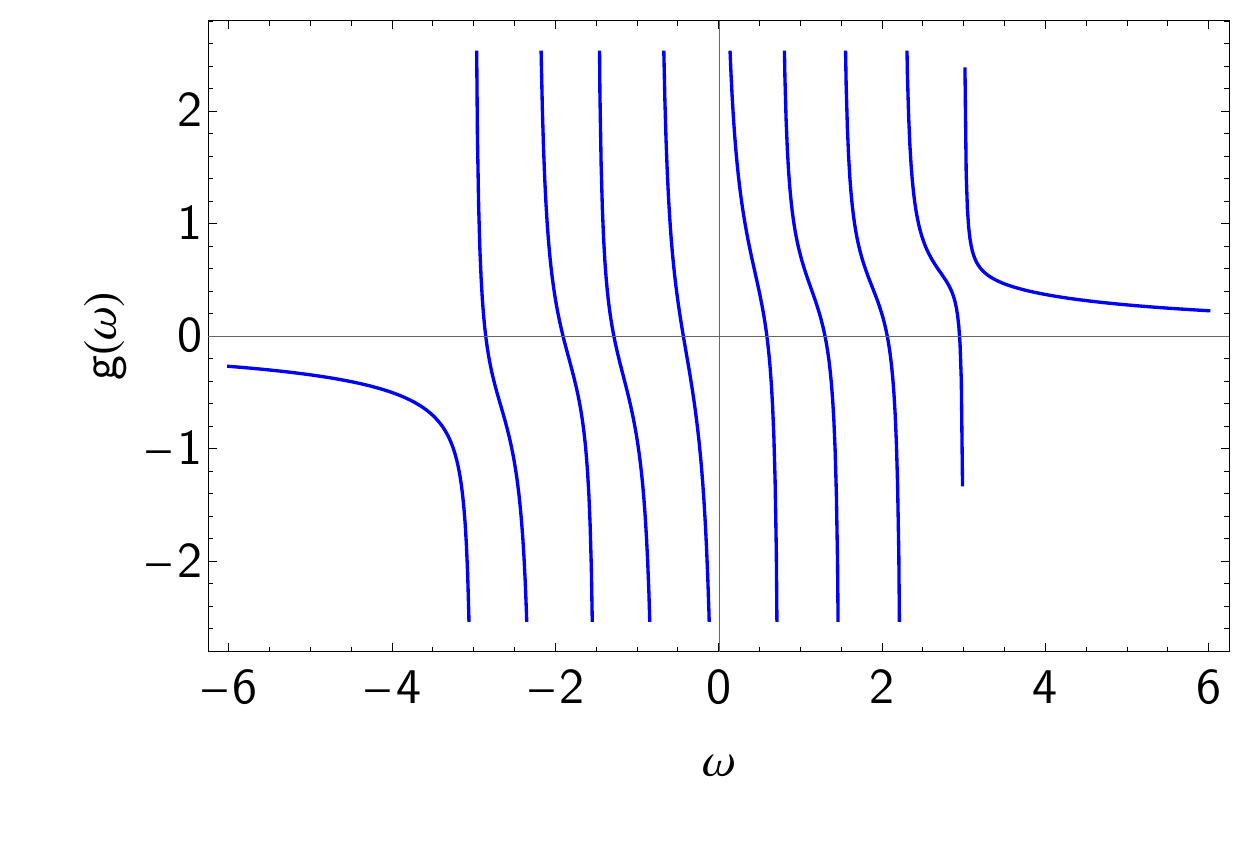}
    \caption{Poles of $g(\omega)$ in a single orbital tight binding Hamiltonian~\eqref{eqapp:singleband} with periodic boundaries and a finite size of $6\times6$. The local Green's function diverges at every energy eigenvalue and decays monotonically otherwise~\cite{Slater.Koster.1954}. There are 9 distinct poles, and a zero in between each pole corresponds to an attractive fixed point of the perturbed spectrum. Outside the band, there is one zero at $\omega=\pm\infty$ which allows for the removal of a single state from the band by a local perturbation, whose energy $E\to v$ as $v\to\pm\infty$.}
    \label{figapp:disc1band}
\end{figure}
The second formula can be interpreted as follows. The eigenvalues at a time $t$ feel a level repulsion from states close by in energy, provided the matrix elements of $\hat\V$ that couple them to $\varepsilon_n$ do not vanish. The level repulsion is expressed as an effective force that acts on the eigenenergies $\varepsilon_n$, with the opposite sign for states below or above $\varepsilon_n$.
It is particularly insightful to rewrite the previous expression with an explicit dependence on the resolvent Green's function of the instantaneous Hamiltonian,
\begin{align}\G_t(z)=\left(z-\H(t)\right)^{-1}
=\sum_m{\proj{\psi_m(t)}\over z-\varepsilon_m(t)}. \end{align}
At the pole of the state $n$, we define the principal part by removing that pole before taking the real energy limit. More precisely, $\mathcal P_n$ denotes the finite, pole-subtracted part of the resolvent at $\varepsilon_n$; in the thermodynamic notation below it becomes the usual Cauchy principal value of the retarded Green's function,
\begin{align}
\mathcal P_n\G_t(\varepsilon_n)
\equiv
\sum_{m\neq n}{\proj{\psi_m(t)}\over \varepsilon_n-\varepsilon_m}
=
\lim_{\eta\to0^+}\mathrm{Re}\left[
\G_t(\varepsilon_n+i\eta)-{\proj{\psi_n(t)}\over i\eta}
\right].
\label{eqapp:principalG}
\end{align}
The impurity-projected principal part is therefore
\begin{align}
\mathcal P_n g_t^\sigma(\varepsilon_n)
\equiv
\bra{\sigma}\mathcal P_n\G_t(\varepsilon_n)\ket{\sigma}
=
\sum_{m\neq n}{|\langle\sigma|\psi_m(t)\rangle|^2\over \varepsilon_n-\varepsilon_m}.
\label{eqapp:principalg}
\end{align}
For $\V=\proj{\sigma}$, Eq.~\eqref{eqapp:energies} is then
\begin{align}
    \ddot\varepsilon_n=2|\lambda_n|^2\,\mathcal P_n g_t^\sigma(\varepsilon_n).
\label{eqapp:forceprincipal}
\end{align}
Thus the force is the real principal part of the projected Green's function, with the pole of the state whose motion is being computed removed. Note that if $\lambda_n=0$ the impurity does not couple with the $\ket{\psi_n}$ eigenstate and therefore, its energy remains unchanged. Furthermore, the energy $\varepsilon_n$ is not repelled from states that do not couple to the impurity. 

In a finite size system, typical extended levels with $\varepsilon_n$ within the band, experience an opposing force from the energy levels $\varepsilon_m$ above and below $\varepsilon_n$, proportional to $|\lambda_n|^2|\lambda_m|^2(\varepsilon_n-\varepsilon_m)\inv$, which vanishes at a finite energy in between the unperturbed poles since the total force diverges with opposite signs when approaching consecutive poles, see Fig.~\ref{figapp:disc1band}. The level repulsion is of the order $O(1/N)$, as expected in a band continuum. This implies that given a local perturbation with a single eigenstate, all eigenstates are weakly perturbed to find a new energy that sits in between the original unperturbed energies.
However, the states at band edges behave differently. At infinitesimal time $t$, a state at the band edge with a nonvanishing overlap $\lambda_n$ will be pushed out of the band since it does not experience a counteracting force. 
At $t\to\infty$ this eigenstate energy will converge to an attractive fixed point of Eq.~\eqref{eqapp:energies}, where both sides of the equation vanish outside the band continuum.

\begin{figure}
    \centering
    \includegraphics[width=.45\columnwidth]{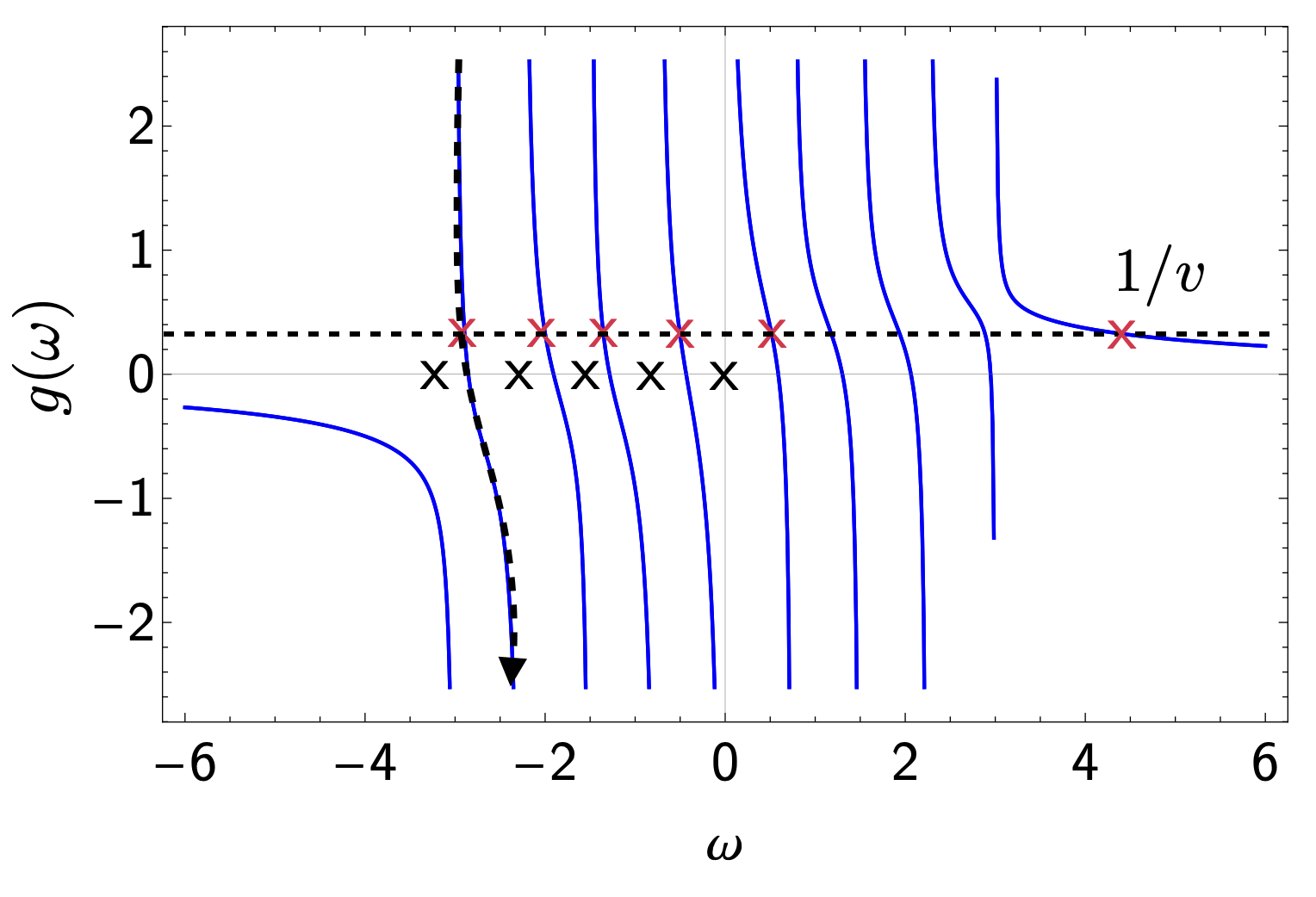}
    \caption{Evolution of the local spectrum of a single band given a local perturbation $\V=v\proj{\sigma}$. The unperturbed poles are marked with a black cross, and the perturbed solutions are marked in red. A single solution lies outside the band, which corresponds to the bound state, whose strong coupling fixed point is at $v=\pm\infty$. This is consistent with the fact that only one single eigenenergy of the perturbed system exceeds the unperturbed bandwidth, as discussed in Note~\ref{app:weyls}.}
    \label{figapp:evolutionstates}
\end{figure}

As the state $\varepsilon_n$ is removed from the band continuum, it becomes localized around the impurity. One would expect the level repulsion from each delocalized state to be small. Therefore in a trivial multiband system, the localized state can cross high-energy bands and find its final energy at $E=\pm\infty$. However, as we have argued in the main text, $g(\omega)$ can have finite energy zeros within energy gaps. These fixed points, which we denote \emph{ingap spectral attractors} indicate that the bands above and below the final energy of the localized state exert an equal force on this state such that it cannot be removed from the gap by varying $v$.
The discrete spectral flow associated with this rank-one removal is illustrated in Fig.~\ref{figapp:evolutionstates}.

\subsection{Thermodynamic limit}

In the thermodynamic limit, the vast majority of states are weakly affected by $\V$, remaining extended through the entire sample. Therefore, we are only interested in the effect of $\V$ on states outside the band continuum, which are bound at the impurity site. There are two closely related, but distinct, Green's functions entering this discussion. First, the exact Hadamard force for an isolated instantaneous eigenstate of $\H(v)=\H+v\proj{\sigma}$ is obtained from the real principal part of the pole-subtracted resolvent of the \emph{perturbed} Hamiltonian. Let
\begin{align}
g_v^\sigma(z)\equiv \bra{\sigma}(z-\H(v))^{-1}\ket{\sigma}.
\end{align}
At an isolated eigenvalue $E_n(v)$, this projected resolvent contains the pole $|\lambda_n|^2/(z-E_n)$. We therefore define
\begin{align}
\widetilde\mu^\sigma_v(E_n)
\equiv
\mathrm{Re}\,\mathcal P_n g_v^\sigma(E_n+i0^+)
=
\sum_{m\neq n}{|\langle\sigma|m_v\rangle|^2\over E_n-E_m},
\label{eqapp:mutildedef}
\end{align}
where $\mathcal P_n$ removes the pole of $\ket{n_v}$. In a continuum notation this is precisely the Cauchy principal value: if
\begin{align}
g_v^\sigma(E+i0^+)=
P\!\int dE'\,{\nu_v^\sigma(E')\over E-E'}-i\pi\nu_v^\sigma(E),
\end{align}
where $P\!\int$ denotes the Cauchy principal value. Thus $\mathrm{Re}\,g_v^\sigma(E+i0^+)=P\!\int dE'\,\nu_v^\sigma(E')/(E-E')$, with the isolated pole removed when $E=E_n$. The Hadamard formula then gives
\begin{align}
\partial_v^2E_n
=2|\lambda_n|^2\,\widetilde\mu^\sigma_v(E_n),
\label{eqapp:instantforce}
\end{align}
where $\lambda_n=\langle\sigma|n_v\rangle$. This is the literal Newton equation: it is the net level repulsion exerted by the real principal part of the other instantaneous eigenstates.

Second, the impurity bound-state branch outside the clean spectrum is determined by the \emph{unperturbed} projected Green's function,
\begin{align}
g^\sigma(E(v))=\bra{\sigma}(E(v)-\H)^{-1}\ket{\sigma}={1\over v}.
\label{eqapp:cleanpole}
\end{align}
Inside a gap, $g^\sigma=\mu^\sigma$ is real. Differentiating \eqref{eqapp:cleanpole} with respect to $v$ gives
\begin{align}
\partial_E\mu^\sigma(E)\,\partial_vE=-{1\over v^2}.
\label{eqapp:firstclean}
\end{align}
Using Hellmann's theorem, $\partial_vE=|\langle\sigma|\varphi_E\rangle|^2\equiv\lambda_E^2$, this becomes
\begin{align}
\lambda_E^2=-{1\over v^2\partial_E\mu^\sigma(E)}.
\label{eqapp:lambdaclean}
\end{align}
Differentiating once more yields the exact clean-branch curvature
\begin{align}
\partial_v^2E
=
{2\over v^3\partial_E\mu^\sigma(E)}
-
{\partial_E^2\mu^\sigma(E)\over v^4[\partial_E\mu^\sigma(E)]^3}.
\label{eqapp:cleanbranch1}
\end{align}
Since the branch obeys $\mu^\sigma(E)=1/v$, Eq.~\eqref{eqapp:cleanbranch1} can be written as
\begin{align}
\partial_v^2E
=
-2\lambda_E^2\mu^\sigma(E)
-
{\partial_E^2\mu^\sigma(E)\over\partial_E\mu^\sigma(E)}\lambda_E^4.
\label{eqapp:energyg}
\end{align}
Equation~\eqref{eqapp:energyg} shows explicitly what is meant by the linearized force used in the main text. Near a simple vacancy-limit zero $E_\infty$, we may write
\begin{align}
\mu^\sigma(E)=\partial_E\mu^\sigma(E_\infty)(E-E_\infty)+O((E-E_\infty)^2).
\end{align}
Keeping this linearized $\mu^\sigma(E)$ removes the curvature term $\partial_E^2\mu^\sigma$ from Eq.~\eqref{eqapp:energyg} and gives the Newton equation
\begin{align}
\partial_v^2E=-2\lambda_E^2\mu^\sigma(E).
\label{eqapp:energygapprox}
\end{align}
For the full nonlinear $\mu^\sigma(E)$, the first correction is the second term of Eq.~\eqref{eqapp:energyg}, which is $O(\lambda_E^4)$ provided $\partial_E^2\mu^\sigma/\partial_E\mu^\sigma$ remains finite. The minus sign in Eq.~\eqref{eqapp:energygapprox} is a consequence of using $v$ rather than $1/v$ as the flow variable. It does not change the attractor criterion: a zero with $\partial_E\mu^\sigma<0$ is stable under the strong-coupling flow $g^\sigma(E)=1/v$.

\begin{figure}
    \centering
    \includegraphics[width=\linewidth]{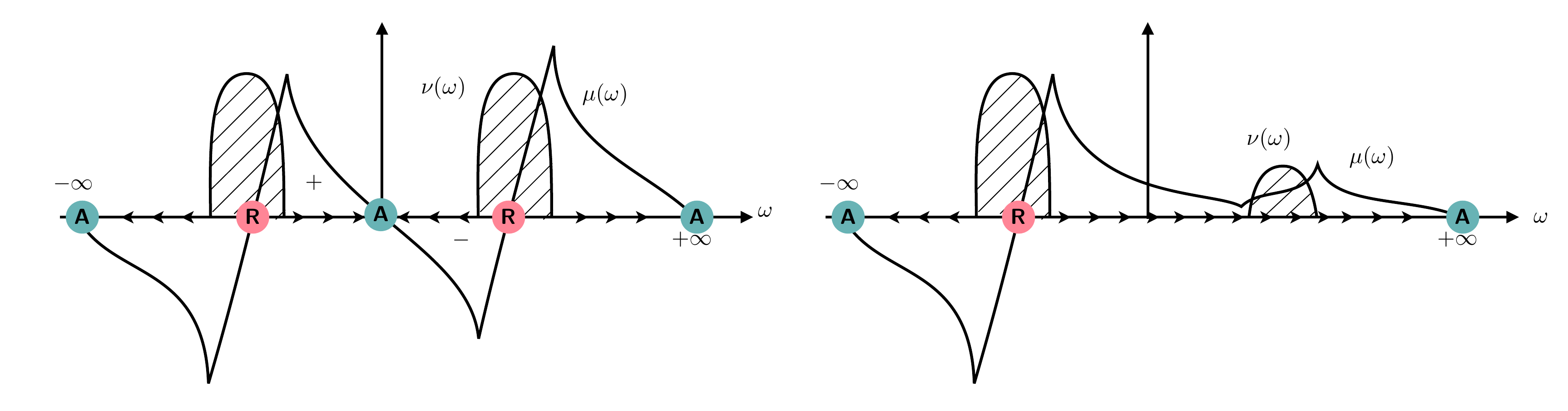}
    \caption{Scheme of the general form of the impurity projected Green's function in dispersive bands. Zeros of $\mu(\omega)$ correspond to spectral attractors (A) or repellers (R) according to their slope, fixed points of Eq.~\eqref{eqapp:energies}. The arrows indicate the flow of the bound state created by the impurity to the strong impurity limit. The existence of an ingap attractor is required for topological bands, independently of their dispersion; while it is absent or can be removed symmetrically and adiabatically in unobstructed bands.}
    \label{fig:schemetopovstriv}
\end{figure}

Again, also in the thermodynamic limit there is always a fixed point at $E=\pm\infty$, but important are instances in which zero occurs in the spectral gap of $\H$. These correspond to points at which the bands above and below the bound state repel it in equal strength. In these cases, the energy at which the bound state exists is a fixed point that depends only on the eigenstate of the impurity and only weakly on its strength. Figure~\ref{fig:schemetopovstriv} summarizes this force-flow picture for trivial and topological dispersive bands.

Let us return to the exactly solvable flat band limit of Sec.~\ref{app:flatbandexact}. A zero in $\mu^\sigma(\omega)$ exists provided both $s_\alpha$ do not vanish. Let us rewrite the degenerate band energies in a symmetric way $\varepsilon_1=-\Delta/2$ and $\varepsilon_2=+\Delta/2$. The effective force on the state at energy $E$ for an attractive potential is proportional to
\begin{align}\mu^\sigma(E)={s_1\over E+\Delta/2}+{s_2\over E-\Delta/2}={E-\Delta\delta s/2\over E^2-(\Delta/2)^2}\end{align}
There is a spectral attractor at $E=\Delta\delta s/2$ provided $|\delta s|<1$, as is the case for topological bands for every possible $\sigma$. A finite bandwidth of the topological bands will quantitatively change the effective force experienced by the ingap state and, consequently, change the energy of the fixed point.

\section{Examples of $\mu^\sigma(\omega)$ in topological band insulators}\label{secapp:examples}

In this section, we look at the evolution of the real part of the local Green's function in distinct models and parameters. We have seen that $\mu^\sigma(\omega)$ can be interpreted as the effective force experienced by the band eigenstates in the presence of a local perturbation of rank 1 with nonzero eigenstate $\ket\sigma$. Its zeros correspond to fixed points in the strong coupling limit to the local impurity.
As a guiding principle, we choose the state $\sigma$ that maximizes the overlap with the filled bands of energy below the chemical potential. The chemical potential is without loss of generality set to zero in all models.

For the trivial band presented in Fig.~\ref{figapp:singleband}, there is a single zero in the real part $\mu(\omega)=\re g(\omega+i0^+)$ in the middle of the band, corresponding to an unstable fixed point for the eigenvalues in the presence of a local scalar potential. That is, the band tends to spread its density of states above and below the band.

\subsection{Chern insulator from a band inversion}

\begin{figure}[h!]
    \centering
    \includegraphics[width=.45\columnwidth]{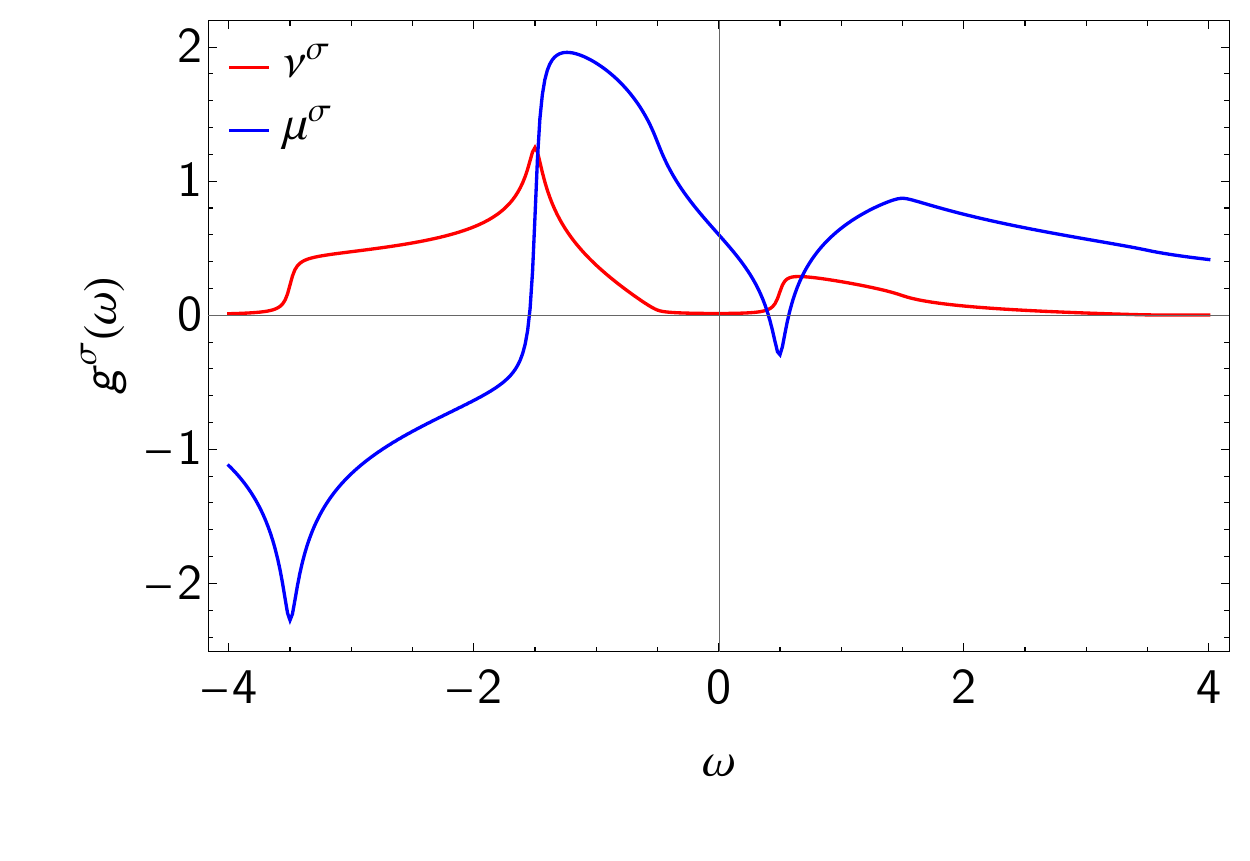}
    \includegraphics[width=.45\columnwidth]{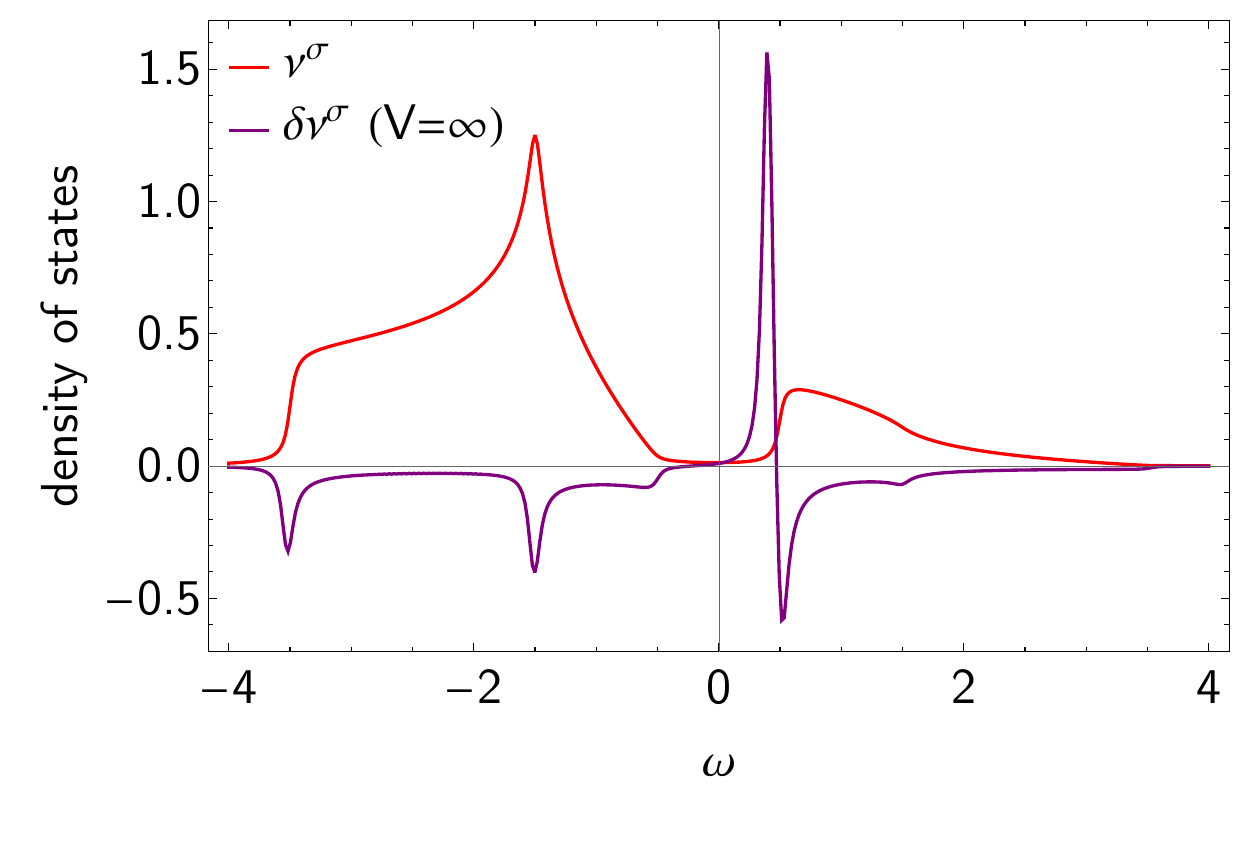}

    \includegraphics[width=.45\columnwidth]{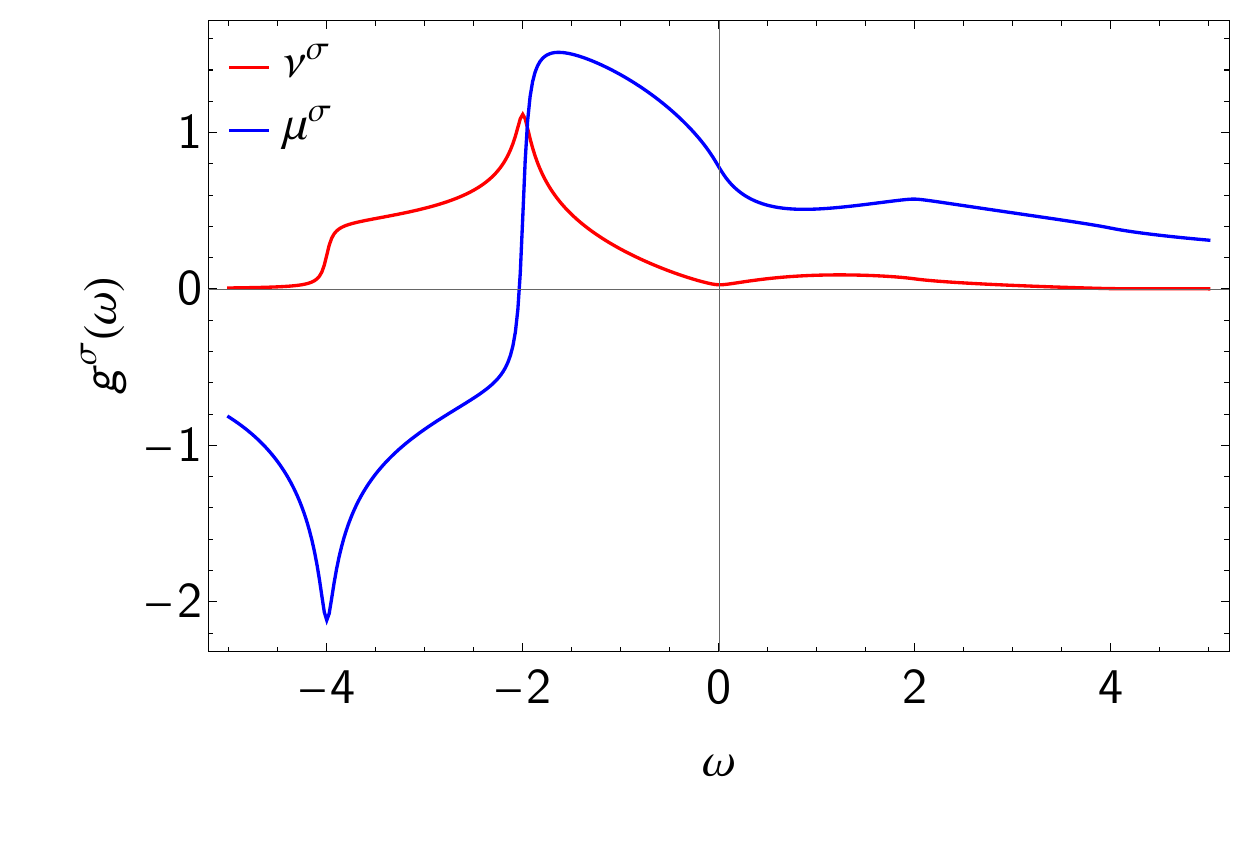}
    \includegraphics[width=.45\columnwidth]{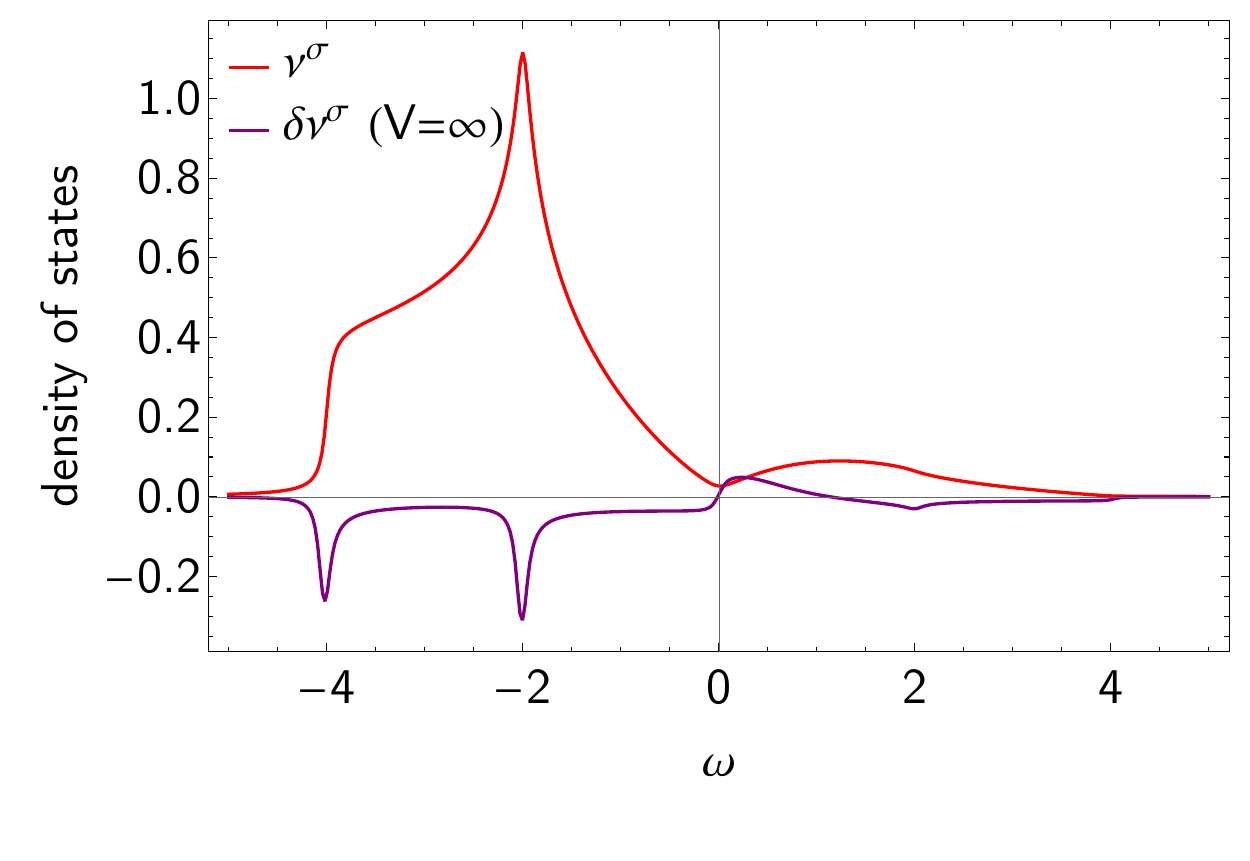}
    
    \includegraphics[width=.45\columnwidth]{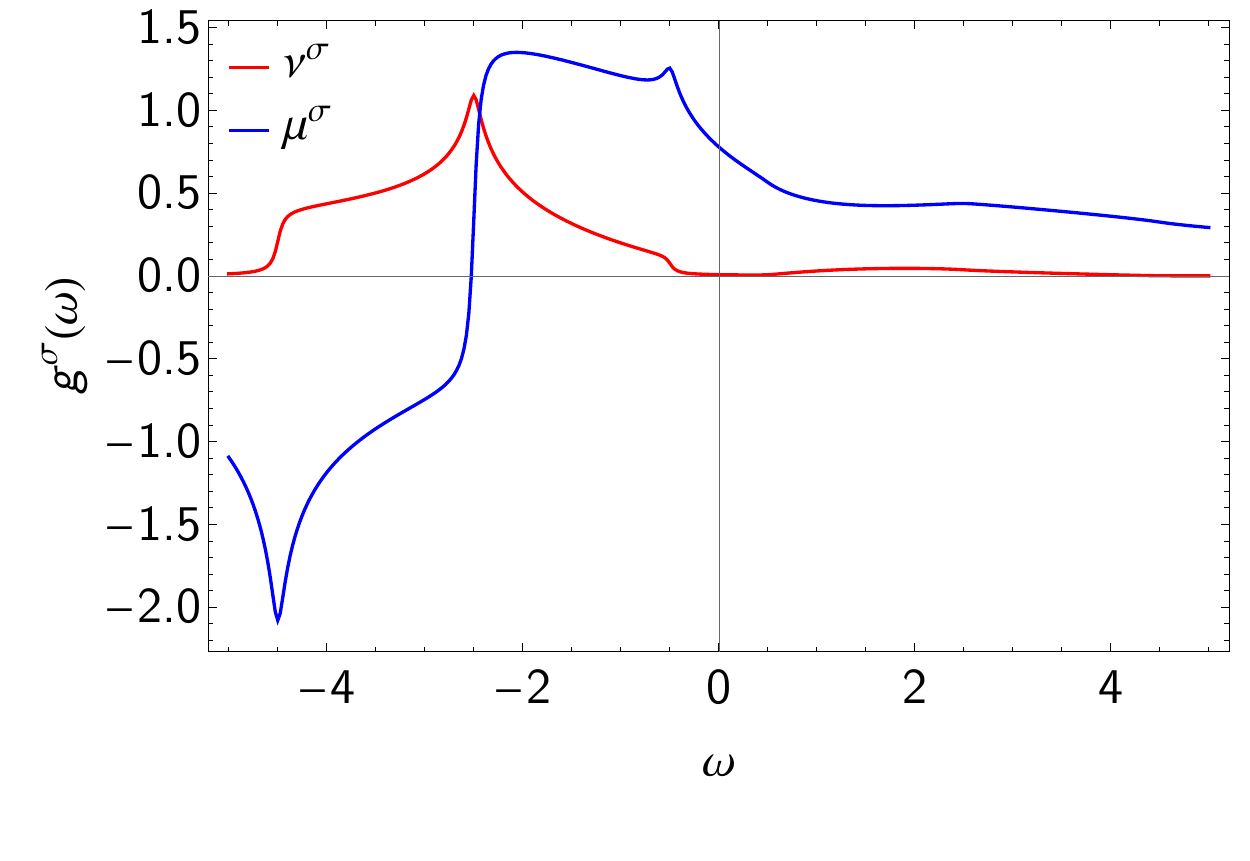}
    \includegraphics[width=.45\columnwidth]{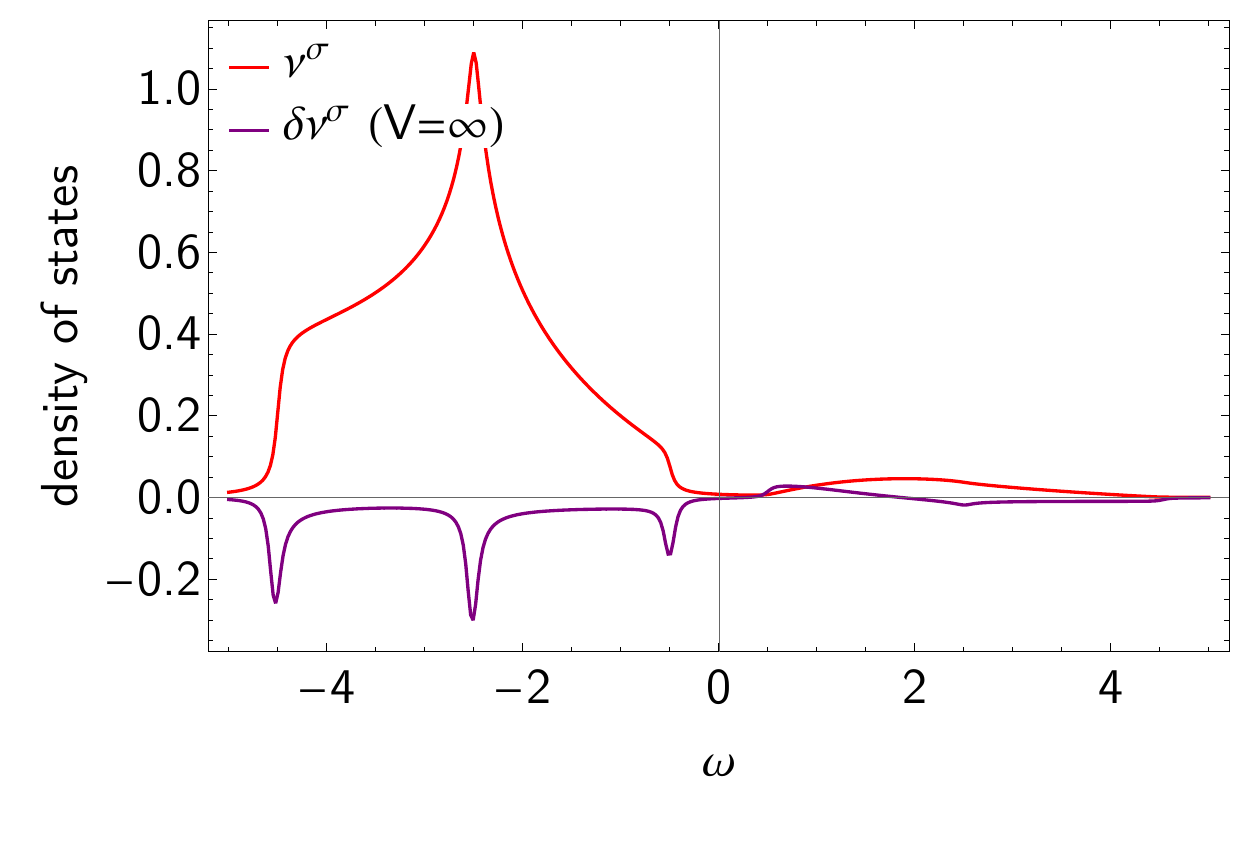}
    \caption{Left: Evolution of the local projected Green's function $g^\sigma(\omega)=\mu^\sigma(\omega)-i\pi\nu^\sigma(\omega)$ as a function of the mass parameter $m$ across a topological phase transition. Right: The density of states, and the change in the density of states in the strong impurity limit. Upper row: $M=-0.5$, topological phase; Middle row: $M=0$, topological phase transition; lower row: $M=+0.5$, trivial phase. Across the topological phase transition, the zero of $\mu^\sigma(\omega)$ is removed, and with it the ingap bound state at $v\to\infty$, present in the topological case as a sharp peak of positive change in density of states, $\delta\nu^\sigma>0$. Only exactly at the topological phase transition $\mu^\sigma(\omega)$ loses its divergence and is smooth across the Dirac point. Note that we assume that the ring state does not overlap trivially with a band which it shares no matrix elements with.
    }
    \label{figapp:cherng}
\end{figure}

\begin{figure}[h!]
    \centering
    \includegraphics[width=.45\columnwidth]{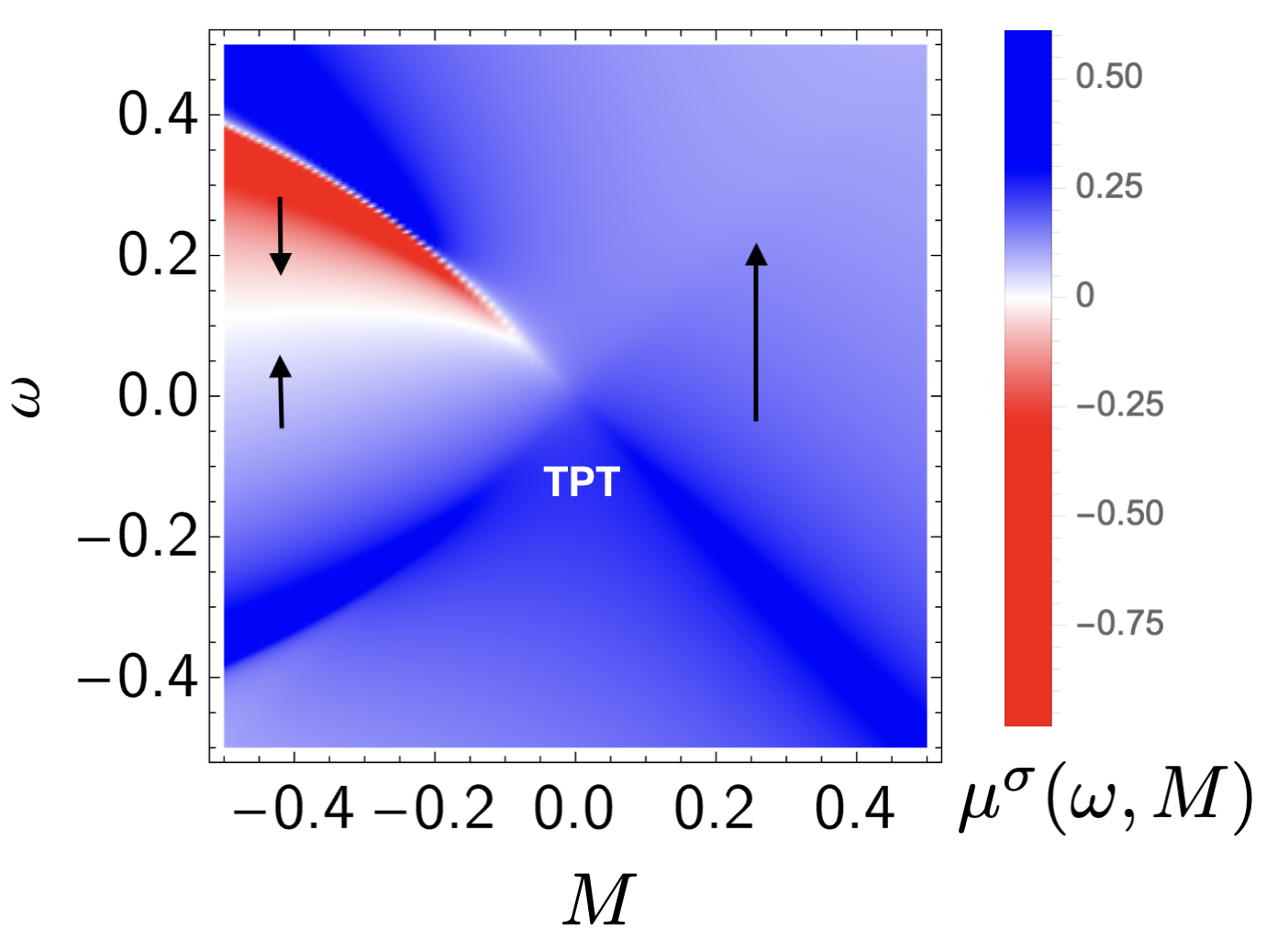}
    \caption{Phase diagram of $\mu^\sigma(\omega)$ for $\sigma=s$ in the (continuum) Chern insulator model for small mass. The topological phase transition (TPT) is characterized by the appearance of a zero of negative slope (attractor), and one of positive slope (repeller) together a region $\mu^\sigma(\omega)<0$, which corresponds to an effective negative force on ingap bound states that traps the ring state in the gap, due to the ingap stable fixed point.}
    \label{figapp:phasediagmu}
\end{figure}

Let us consider a two-band model on a square lattice~\cite{Zhang.Qi.2006}, given by
\begin{align}\H=\sum_{ij}h_{ij}^{\sigma\tau}c^\dag_{i\sigma}c_{j\tau}\end{align}
with $c^\dag_{i\sigma}$ creating an electron in the orbital state $\ket{{\phi_i^\sigma}}$ with $i,j$ spatial indices and $\sigma,\tau$ orbital indices labelling the eigenstates of $\sigma_z$, $\sigma_z\ket{s}=+\ket{s}$ and $\sigma_z\ket p=-\ket p$. In reciprocal space, 
\begin{align}\H=\sum_\bk  c^\dag_\bk H(\bk) c_\bk,\end{align} with $c^\dag_\bk=(c^\dag_{\bk,s},c^\dag_{\bk,p})$ and the momentum space  Hamiltonian taking the simple two-band form
\begin{align}H(\bk)=(m+4b-2b\cos k_x-2b\cos k_y)\sigma_z-\upsilon\sin k_x \sigma_x-\upsilon\sin k_y\sigma_y\label{eqapp:qwzmodel}\end{align}
Taking $b,\upsilon>0$, the model admits a Chern insulator phase for $-8b<m<0$, where the Chern number changes sign at $m=-4b$. Around $m=0$, the Hamiltonian can be expanded around $k=0$, becoming
\begin{align}H(\bk)=(m+bk^2)\sigma_z-\upsilon\bk\cdot\bsigma+O(k^3)\label{eqapp:cherncont}\end{align}
which is the same model used in Eq.~\eqref{eq:cherncont} in the main text. At $m<0$, the filled band has a majority of $s$ character. A topological phase transition occurs by closing the spectral gap at $k=0$, at which point the filled Bloch eigenstate at $k=0$ switches orbital character. This structure can be appreciated by looking at the $s-$orbital projected local Green's function, which can be seen in Fig.~\ref{figapp:cherng}. The real $\mu^\sigma(\omega)$ for $\sigma=s$ acquires a zero of negative slope in the band gap at the topological phase, corresponding to a spectral attractor at a finite energy, a scenario unique to multiband Hamiltonians. This can be seen explicitly in Fig.~\ref{figapp:phasediagmu} by the appearance of a negative region bounded by zeros of the opposite slope. In the equations of motion defined in the previous section, this zero corresponds to a stable fixed point at a finite energy at which the bound state converges to in the strongly coupled impurity limit.

\begin{figure}
    \centering
    \includegraphics[width=.7\linewidth]{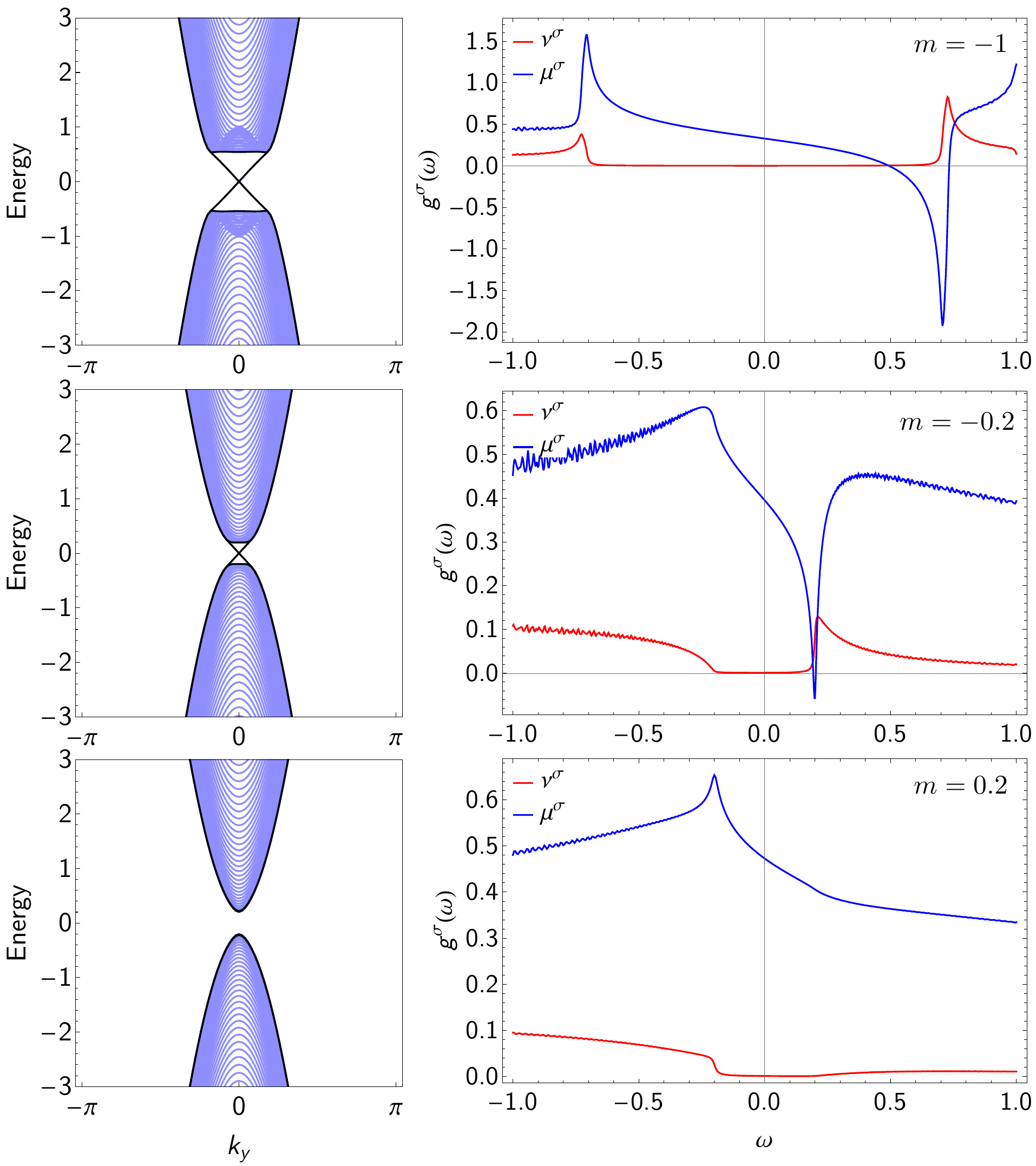}
    \caption{Energy spectrum of Hamiltonian \eqref{eqapp:qwzmodel} with open boundaries and impurity projected Green's function, with impurity state defined by $\sigma_z\ket{\sigma}=-\ket{\sigma}$. In all three plots we choose $\upsilon=1$ and $b=3$. Note that the shape of the band edge in the topological phase does not influence the existence of the zero in $\mu^\sigma(\omega)$, which is created in the gap closing transition.}
    \label{figapp:shape}
\end{figure}

In Fig.~\ref{figapp:cherng} we take a closer look at the evolution of the density of states $\nu^\sigma(\omega)$ and $\mu^\sigma(\omega)$ across the band inversion, where the band edges with opposite orbital character are interchanged. Since we look at a two-dimensional model, the projected density of states is expected to diverge at the band edge associated with the $\sigma$ band. At the topological phase transition, the top band edge of the $s$ band is interchanged with the bottom band edge of the $p$ band, leading to a shift in the divergence in $\mu^\sigma(\omega)$ from the filled to the empty band. This process comes together with the creation of the ingap zero and the protected ring state in the presence of a strong impurity.
In Fig.~\ref{figapp:shape} we show that the shape of the band edge does not affect the existence of a zero in $\mu^\sigma(\omega)$. 

\subsection{Topological insulator from a split band representation}\label{secapp:kanemele}
Another example worth considering is the case of graphene, with a very small energy gap determined by spin-orbit coupling. In this case, the topological bands are not formed by a band inversion around high symmetry momenta, rather, their topology results from the partial filling of a split elementary band representation~\cite{Bernevig.Cano.2018}. The model is given by Kane-Mele's model~\cite{Mele.Kane.2005} written in terms of the nearest-neighbor and next-nearest neighbor hopping terms
\begin{align}\H=\H_0+\H_R+\H_{KM}\end{align}
with 
\begin{align}\H_0=-t\sum_{\ev{ij}}{c^\dag_{i}}{c_{j}}\end{align}
for $c_i^\dag=(c^\dag_{i\up},c^\dag_{i\downarrow})$. Note there are two sublattices, and a total of $4N$ Bloch states. The Rashba Hamiltonian is given by
\begin{align}\H_R={\alpha\over 2i}\sum_{\ev{ij}}{c^\dag_{i}}(\bsigma\times \vec d_{ij})_z{c_{j}}\end{align}
for $\vec d_{ij}=\bq_i-\bq_j$ 
and a staggered hopping that gaps the Dirac points at $K$ and $K'$ momenta
\begin{align}\H_{KM}={\beta\over 2i}\sum_{\ev{\ev{ij}}}{c^\dag_{i}}(\nu_{ij}\sigma_z){c_{j}}\end{align}
for $\nu_{ij}=\pm1$ alternating clockwise for the six next nearest neighbors to a given site. 
\begin{figure}[t]
    \centering
    \includegraphics[width=.45\columnwidth]{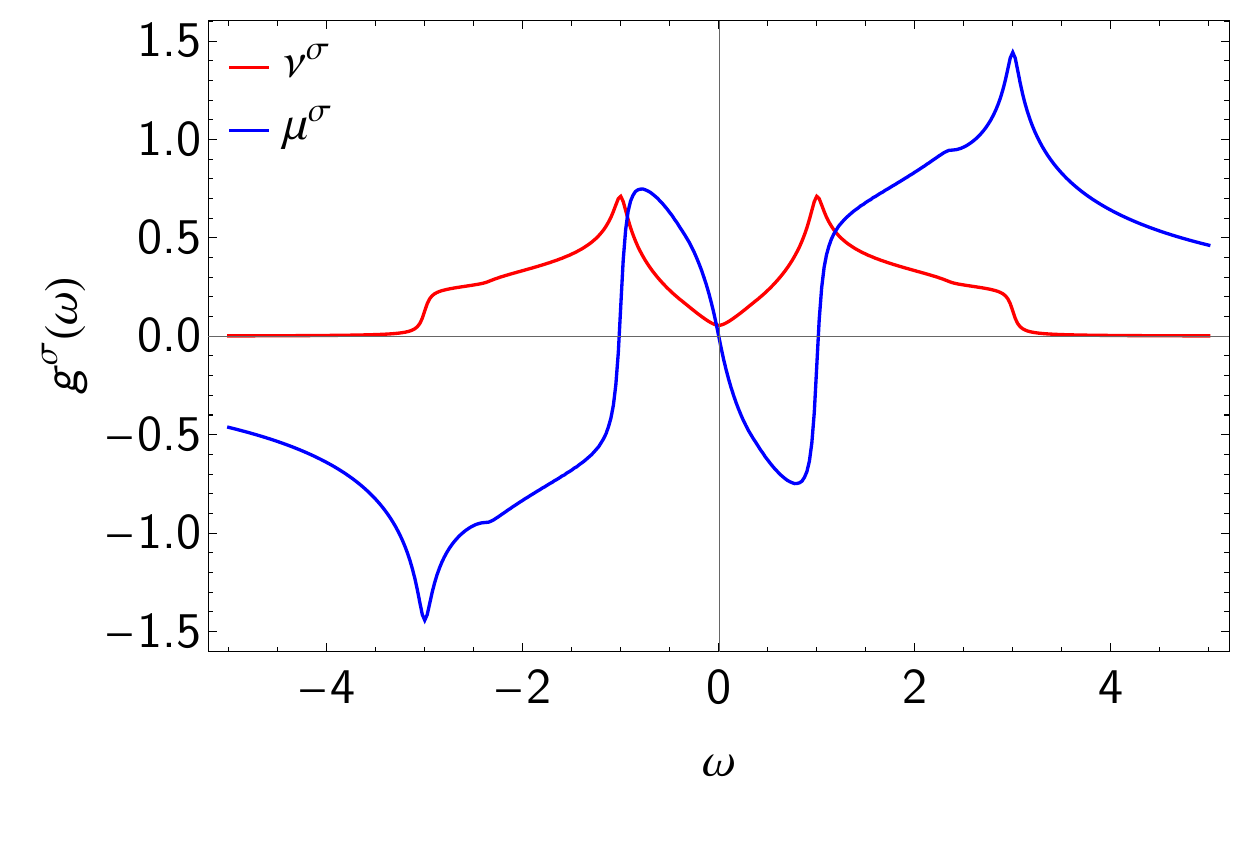}
    \includegraphics[width=.45\columnwidth]{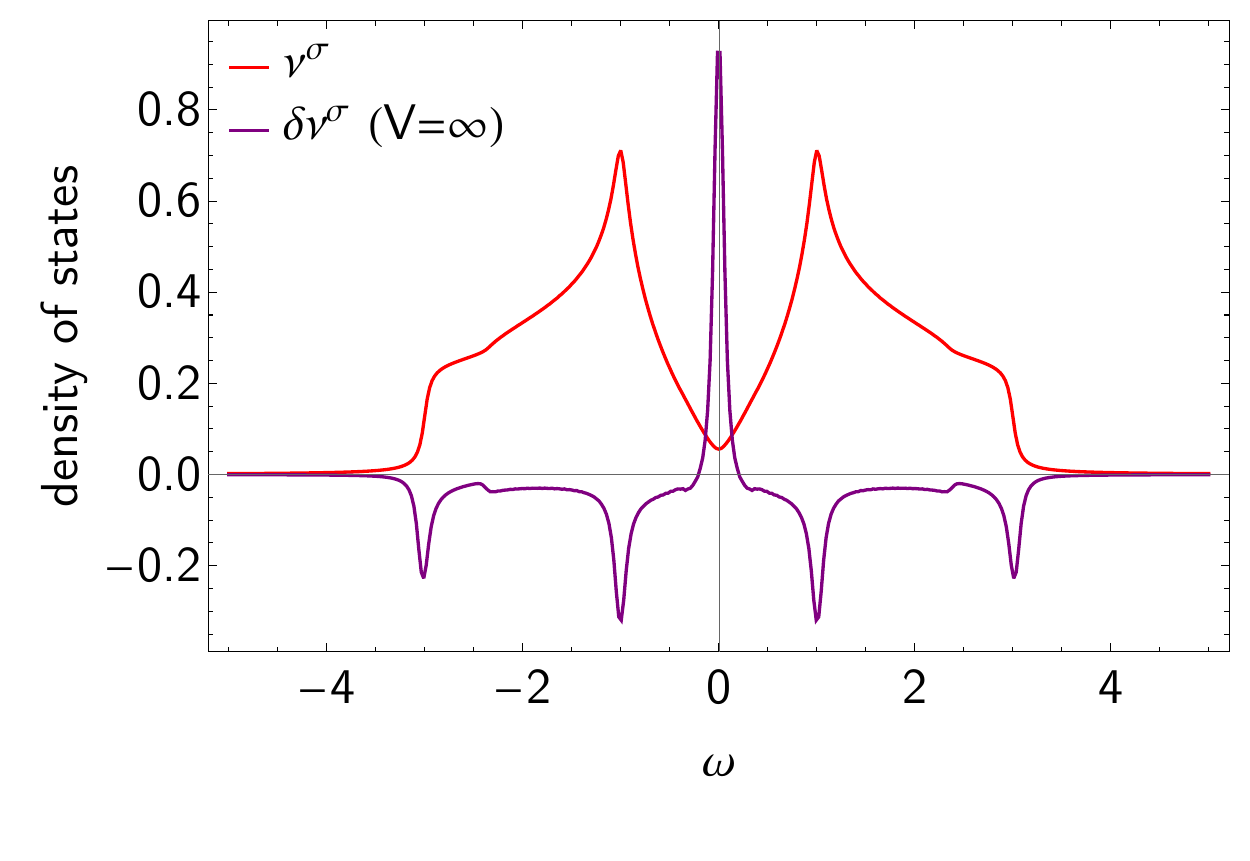}

    \includegraphics[width=.45\columnwidth]{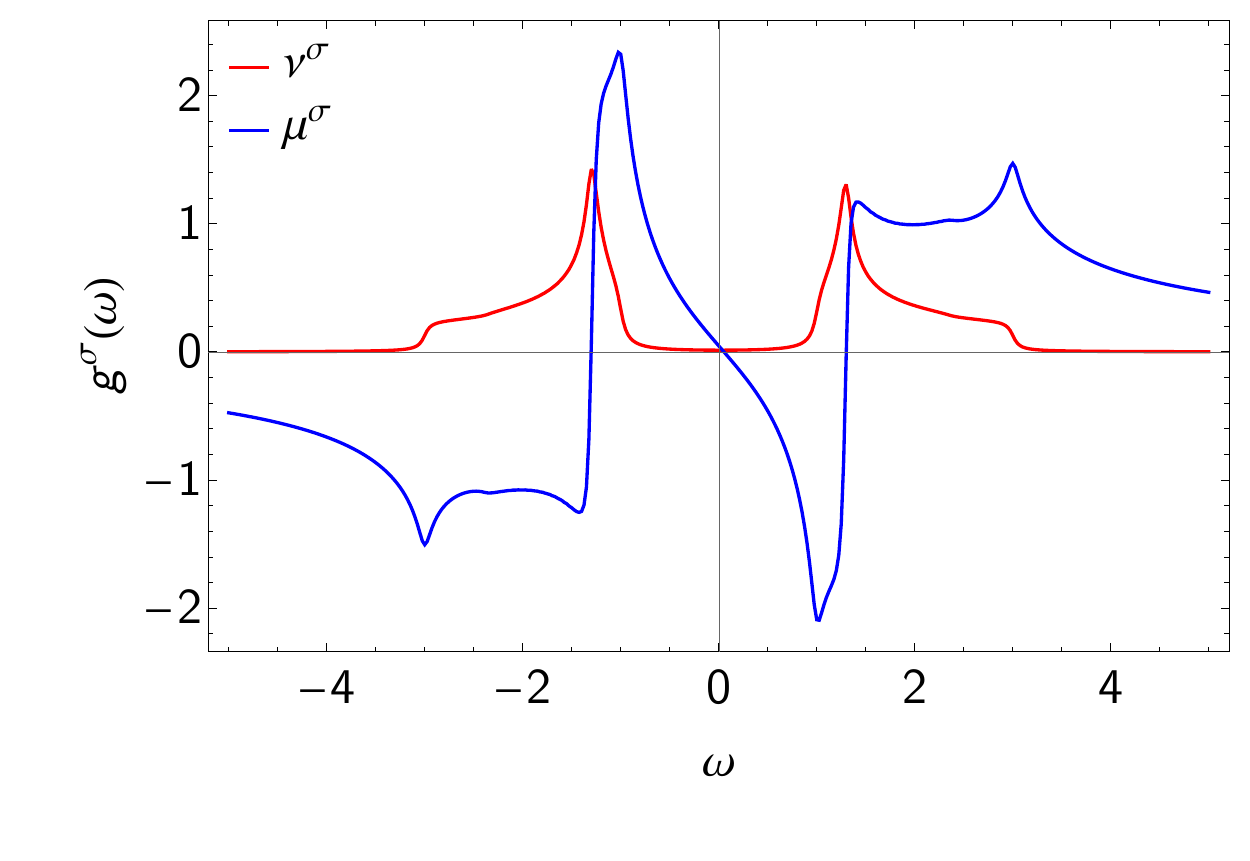}
    \includegraphics[width=.45\columnwidth]{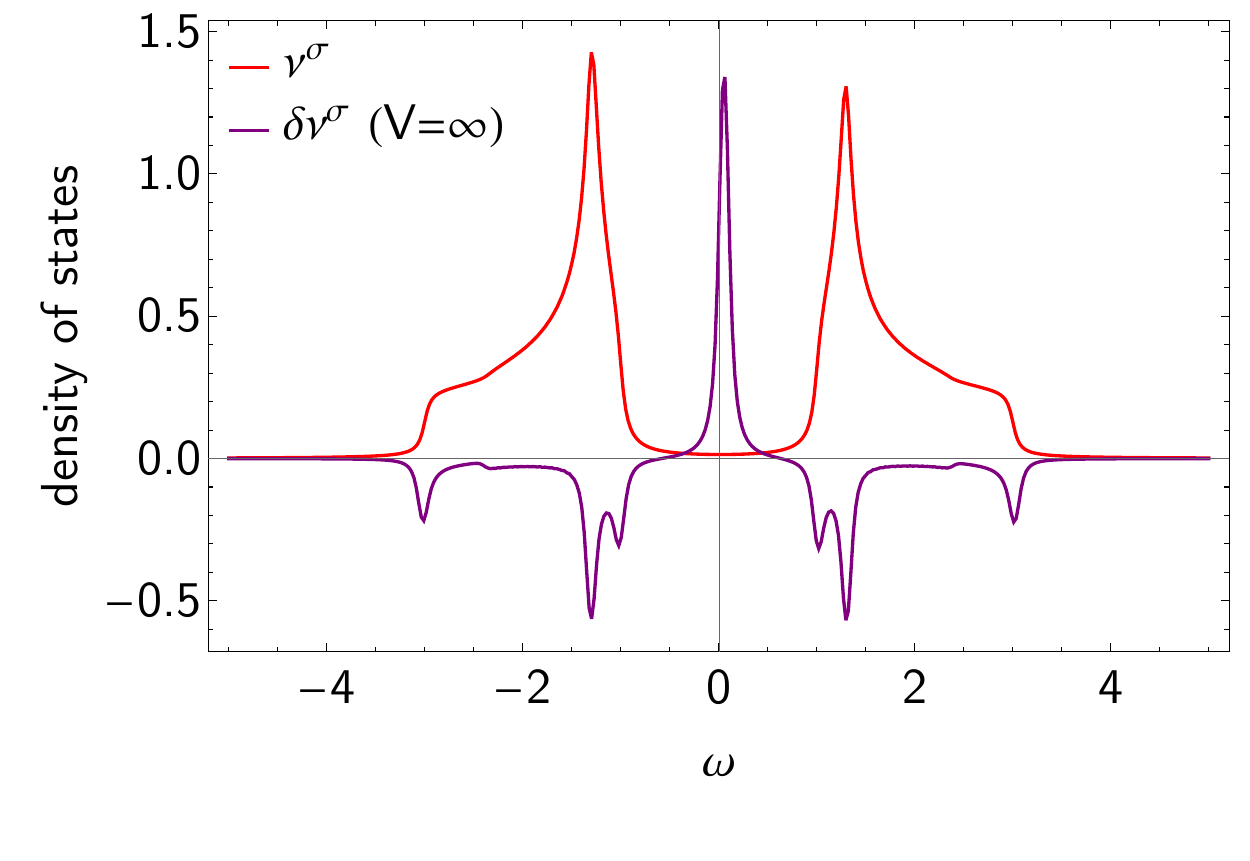}
    
    \caption{Local projected Green's function for the Kane-Mele model, coinciding with graphene for $\beta=0$ (top row) and with a $\Z_2$ topological insulator with $\beta=1$ (bottom row). The ingap ring state in the gapped phase of the Kane-Mele model has the same origin as the Dirac point resonant state known to exist for vacancies in graphene.}
    \label{figapp:kanemelechern}
\end{figure}
We can consider a vacancy in a single site $\bq_0$ located in one of the sublattices, which affects a Kramers pair, \begin{align}\V=v\left(\proj{\phi_0^\up}+\proj{\phi_0^\downarrow}\right)\end{align} which will lead to the appearance of a Kramers pair of ring states at $E=0$ in the strongly coupled impurity limit. An index theorem~\cite{Lieb.Lieb.1989} guarantees the symmetric spectrum, and the zero energy state can be understood by the fact that a state of definite sublattice is equally spread over the filled and empty bands. Therefore the level repulsion from the band into the ring state of definite sublattice is identical from both bands. In Fig.~\ref{figapp:kanemelechern} we plot the impurity projected local Green's function
for two values of $\beta$, which determine the gap. Interestingly, we find that the zero energy state is robust independently of the magnitude of the spectral gap. In the Dirac semimetal limit, the ring state becomes the well-known zero-energy resonance expected in graphene~\cite{Balatsky.Wehling.2014}.

We can appreciate from the present analysis that the topological protection of the ring states originates from the Wannier obstruction of the filled bands to construct symmetric states at a given sublattice and that the spectral fixed point generated by a local impurity is insensitive to the magnitude of the spectral gap; it is removed only when the gap closes or the protecting symmetry is broken, and is otherwise protected by the entire bandwidth of the bands. In the next appendices, we discuss how the zeros in $\mu^\sigma(\omega)$ emerge in topologically obstructed bands.

\section{Wannier obstruction and orbital projected states}\label{app:essential}

Let us consider the flattened Bloch Hamiltonian defined by the band projector as 
\begin{align}\H=\sum_\alpha\varepsilon_\alpha\P^\alpha, \quad \P^\alpha=\sum_\bk\proj{\psi_{\alpha\bk}}.\end{align} 
A topological obstruction implies that it cannot be adiabatically connected to any flat band atomic limit, whose flat band Hamiltonian is defined by
\begin{align}\H_{\rm AL}=\sum_\sigma\varepsilon_\sigma\P^\sigma,\quad\P^\sigma=\sum_i\proj{\phi^\sigma_i}\end{align}
for $\sigma$ a local orbital that preserves the local site symmetry group of the site $\bq_i$~\cite{Vanderbilt.Marzari.2012}. Both Hamiltonians preserve the same spatial and internal symmetries. The possible $\H_{\rm AL}$ are exhaustively classified~\cite{Bernevig.Bradlyn.2017}, and each band $\P^\sigma$ transforms under a band representation of the space group~\cite{Zak.Zak.1980,Zak.Zak.1981}.

Focusing on a specific band $\P^\alpha$ and a specific atomic limit $\P^\sigma$ of the same dimension, if there is an adiabatic deformation between the two, then there is a smooth and periodic unitary transformation $U$ such that
\begin{align}\P^\alpha=U\P^\sigma U^\dag\end{align}
whose (diagonal) matrix elements in the momentum representation $U(\bk)=\bra{\bk}U\ket{\bk}$ are smooth and periodic. A simple diagnosis of whether such unitary transformation exists was proposed in~\cite{Vanderbilt.Thonhauser.2006} by considering the projected states 
\begin{align}\ket{\Upsilon_i^{\sigma\alpha}}\equiv \P^\alpha\ket{\phi^\sigma_i}=\P^\alpha\P^\sigma\ket{\bq_i}\label{app:projstates}\end{align}
where $\ket{\bq_i}$ is a position eigenstate. Ref.~\cite{Vanderbilt.Thonhauser.2006} proposes to check their linear (in)dependence. The logic is the contrapositive: an \emph{irremovable} singularity of the Gram matrix $S_{ij}=\bra{\Upsilon_i^{\sigma\alpha}}\Upsilon_j^{\sigma\alpha}\rangle$ (equivalently, a protected zero of $s_{\alpha\bk}^\sigma$ at some $\bk$ that cannot be lifted by any symmetry-preserving change of the trial orbital $\ket{\phi^\sigma}$) rules out any smooth periodic $U$ with $\P^\alpha=U\P^\sigma U^\dag$, and therefore rules out a symmetry-preserving adiabatic path between $\P^\alpha$ and $\P^\sigma$. The converse is weaker than the obstruction direction: a trivial bundle can exhibit accidental zeros of $s_{\alpha\bk}^\sigma$ for a poorly chosen trial orbital, but those zeros are always removable by deforming $\ket{\phi^\sigma}$ within the same symmetry-compatible class; only the protected, deformation-invariant zeros encode the topological obstruction. Namely, if $U$ does not exist, the projected states $\ket{\Upsilon_i^{\sigma\alpha}}$ necessarily span a space \emph{smaller} than the original band, which means that the overlap matrix
\begin{align}S_{ij}=\bra{\Upsilon_i^\alpha}\Upsilon_j^\alpha\rangle\label{app:overlapmat}\end{align}
is singular with $\det S=0$, with irremovable zero eigenstates. The diagonal entries of $S$ correspond to the norm of the projected states, $s_\alpha=\bra{\Upsilon_i^\alpha}\Upsilon_i^\alpha\rangle$. Since $\bq_i$ form a lattice of symmetry-related sites, $\bq_i=g_{ij}\bq_j$ for $g_{ij}$ an element of the space group, it follows that the norm $s_i$ is independent of the lattice site.
Inserting \eqref{app:projstates} into \eqref{app:overlapmat} we can write the overlap matrix in the following form
\begin{align}S_{ij}=\bra{\bq_i}\S\ket{\bq_j}\end{align}
where $\S$ is the positive semidefinite overlap operator, which we define in a basis independent way as follows
\begin{align}\S\equiv(\P^\alpha\P^\sigma)^\dagger(\P^\alpha\P^\sigma)=\P^\sigma\P^\alpha\P^\sigma.\end{align}
Its singular nature implies that 
\begin{align}\rk \S<\rk\P^\alpha=\rk\P^\sigma\end{align}
Since both $\P^\alpha$ and $\P^\sigma$ are diagonalized in the momentum basis, so is $\S$; rank deficiency $\det\S=0$ therefore forces the diagonal element to vanish at some $\bk_0$, i.e.\ $s_{\alpha\bk_0}^\sigma=0$. Explicitly,
\begin{align}s_{\alpha\bk}^\sigma=\bra{\bk}\S\ket{\bk}=\sum_{ij}\exp\{i\bk\cdot(\bq_i-\bq_j)\}S_{ij}=|\!\bra{\sigma}\psi_{\alpha\bk}\rangle|^2\end{align}
which necessarily vanishes for some value of $\bk$. While it is possible that $s_{\alpha\bk}^\sigma$ has zeros that are not topologically protected, the topological statement is that at least one of these zeros is irremovable, for any choice of symmetric, periodic and smooth unitary rotation $U$ we may do either in $\P^\sigma$ or $\P^\alpha$. For the specific case of a band with Chern number $C\neq 0$, Ref.~\cite{Mera.Ozawa.2021} shows that the total vorticity of the projected frame around the zeros of $s_{\alpha\bk}^\sigma$ equals $C$, so the zeros must exist with vorticities summing to $C$ and cannot be removed by any smooth periodic gauge.

The lack of an exponential localized Wannier basis of a given orbital character can be seen by trying to construct it explicitly from the projected states. These are not orthogonal. The orthogonalization process~\cite{Keller.Carlson.1957} gives a set of orthogonal states
\begin{align}\ket{w_i}=\sum_{j}[S^{-1/2}]_{ij}\ket{\Upsilon_j}\end{align} 
which can only be constructed as long as $\det S\neq0$. Rotating the atomic band by any smooth $U$ will not solve this problem. Therefore, the Wannier functions defined by
\begin{align}
    \ket{w_{i}}=\sum_\bk ~e^{i\bk\cdot\bq_i}e^{i\chi(\bk)}\ket{\psi_{\alpha\bk}}
\end{align}
for an arbitrary smooth and periodic phase $\chi(\bk)$ cannot be described as exponentially localized orbitals and necessarily decay algebraically away from their center at $\bq_i$~\cite{Vanderbilt.Marzari.2012}.
The lack of exponential localization of any choice of symmetric $\ket{w_i}$ is reflected in the impossibility of orthogonalizing the projected states $\ket{\Upsilon_i^{\sigma\alpha}}=\P^\alpha\P^\sigma\ket{\bq_i}$ for any symmetry allowed choice of $\P^\sigma$.

A second necessary ingredient for our discussion is that 
\begin{align}\sum_\alpha\P^\alpha=\sum_\sigma\P^\sigma=\id\end{align}
which implies in particular that for every point in the Brillouin zone,
\begin{align}\sum_{\alpha} s_{\alpha\bk}^\sigma=1\end{align}
which can be read in the following way: if the projected weight $s^\sigma_{\alpha\bk}$ vanishes at some $\bk$ in one Bloch band, then by completeness some other band must carry weight at that $\bk$. This is not a pointwise statement that the complementary bands have a second Gram-matrix zero at the same momentum; rather, the missing orbital weight is redistributed among the other projectors in $\sum_\alpha\P^\alpha=\id$. To conclude that this weight lies in bands flanking the topological gap, we restrict to the band representation $\rho_{\bq,\sigma}$ generated by $\ket{\sigma}$; an obstructed band representation has its constituent bands split by topological gaps, and the orbital then has nonzero overlap with the bands on either side of each such gap. In the rank-one two-band setting this reduces to $s_1+s_2=1$ with both $s_\alpha>0$. The spectral consequence follows only after this geometric statement is combined with the energy ordering of the bands: the same impurity projected spectral measure must carry weight below and above the gap. Note that in our notation
\begin{align}s_\alpha={1\over N}\sum_\bk s_{\alpha\bk}^\sigma\end{align}
such that $s_\alpha$ are always valued between 0 and 1.

\section{Topological stability of ingap ring states}\label{app:proof}

In this section, we prove that a Wannier obstruction in the band representation generated by the impurity orbital leads to the existence of finite energy zeros of the real part of the impurity projected Green's function $\mu^\sigma(\omega)$, corresponding to spectral attractors in the presence of strongly coupled perturbations.
Our proof comes in two steps. First, we consider the flat band limit and show that a zero of $\mu^\sigma(\omega)$ must exist in the topological gap. Second, we show that ingap zeros cannot be removed adiabatically.

\subsection{Necessity of zeros in $\mu^\sigma(\omega)$ in a topological gap}\label{app:proof1}

Let us consider a topological flat band Hamiltonian defined by the projector
\begin{align}\H=\sum_\alpha\varepsilon_\alpha\P^\alpha, \quad \P^\alpha=\sum_\bk\proj{\psi_{\alpha\bk}}.\end{align} 
The resolvent Green's function is given by
\begin{align}
    \G(\omega)=\sum_\alpha{\P^\alpha\over \omega-\varepsilon_\alpha}
\end{align}
where all the poles collapse into a single energy $\varepsilon_\alpha$. For a rank 1 impurity of the form $\V=v\proj{\phi_0^\sigma}$, the impurity projected Green's function is given by
\begin{align}
    g^\sigma(\omega)=\sum_\alpha{s_\alpha\over\omega-\varepsilon_\alpha}
\end{align}
giving a residue of $s_\alpha=N\inv\tr \P^\alpha\P^\sigma$ with $N$ the number of states in the band. The pole weights satisfy $0\le s_\alpha\le1$ and $\sum_\alpha s_\alpha=1$, which assumes, without loss of generality, that there is a single $\sigma$ orbital per site in the system. 
This formula is the energy-space counterpart of the projected-frame statement in Note~\ref{app:essential}. For two flat bands that flank a gap, $g^\sigma(\omega)=s_1/(\omega-\varepsilon_1)+s_2/(\omega-\varepsilon_2)$. If $s_1,s_2>0$, the two terms have opposite signs for $\varepsilon_1<\omega<\varepsilon_2$ and produce the unique zero $E_\infty=(s_1\varepsilon_2+s_2\varepsilon_1)/(s_1+s_2)$. Thus the Gram-matrix obstruction supplies the nonzero residues $s_\alpha$, while the zero of $g^\sigma$ follows from the ordering of the corresponding poles in energy.

\begin{figure}[H]
    \centering
    \includegraphics[width=.45\linewidth]{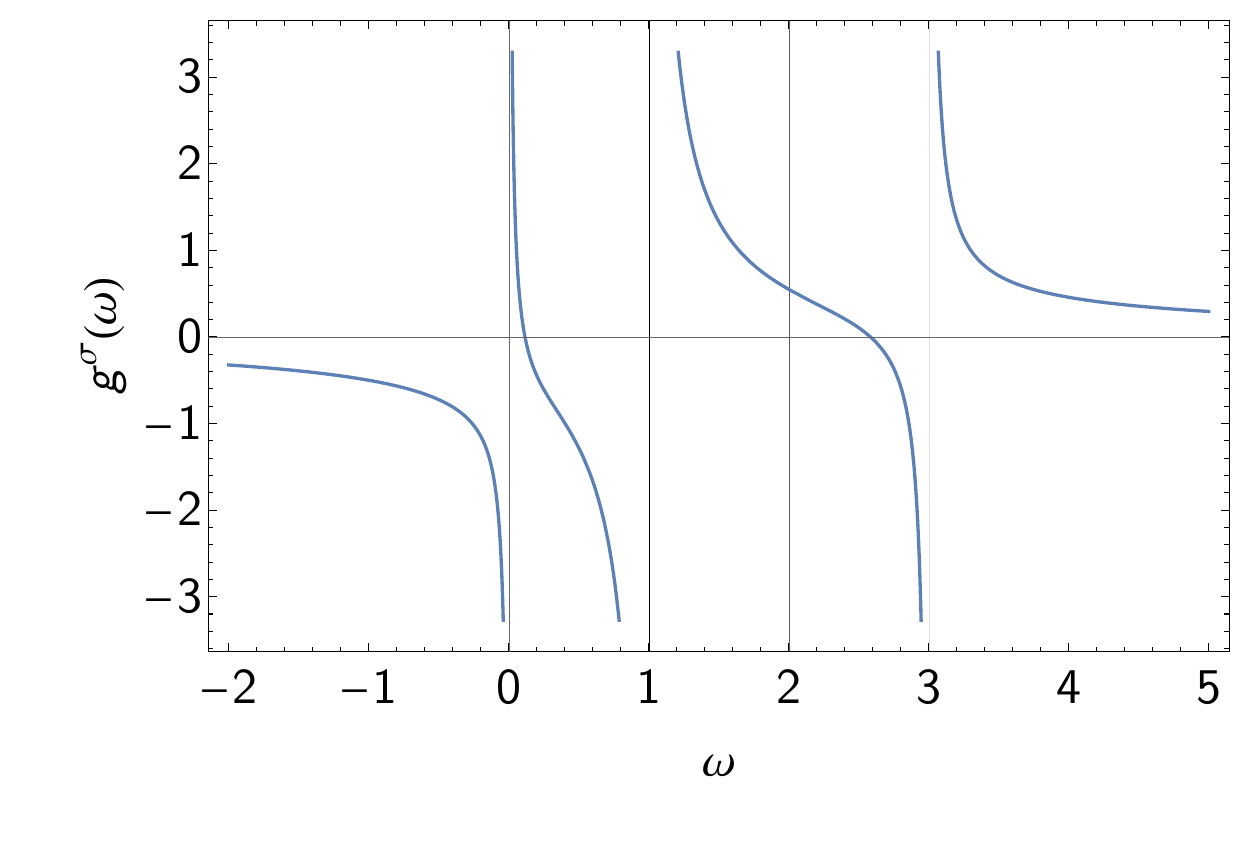}
    \caption{Example of the impurity projected Green's function for a four-band flat band Hamiltonian characterized by $s_{\alpha}=(0.1,0.7,0,0.2)$ at energies $\varepsilon_\alpha=(0,1,2,3)$. Only the bands at energies $0, 1$ and $3$ are visible to the impurity $\V$. Here $\ket{\sigma}$ generates a band representation supported on the essential set $\{0,1,3\}$, with band $2$ a bystander; the topological gap of interest must lie within the essential set for ingap zeros of $\mu^\sigma(\omega)$ to be enforced by obstruction.}
    \label{figapp:flatbandgengsig}
\end{figure}

If there is a topological obstruction, then $s_\alpha<1$ for all $\alpha$ and $s_\alpha>0$ for at least two bands separated by a topological gap (in the rank-one two-band setting; the multi-band generalization is given in Note~\ref{app:essential} via band representations). Keeping only bands for which $s_\alpha\neq0$, we get,

\begin{align}
    g^\sigma(\omega)=\sum_{\alpha:s_\alpha\neq0}{s_\alpha\over\omega-\varepsilon_\alpha}
\end{align}
Since $s_\alpha>0$ on at least one band on each side of the gap, $g^\sigma(\omega)$ has a positive-weight pole both below and above the gap. Cauchy pole-interlacing for the positive-residue meromorphic function above then guarantees exactly one zero between any two consecutive positive-weight poles, with slope $\partial_\omega g^\sigma(\omega)=-\sum_\alpha s_\alpha/(\omega-\varepsilon_\alpha)^2<0$. The two positive-weight poles flanking the gap bracket no other positive-weight pole, so the zero lies in the gap and is an attractor. Equivalently, $g^\sigma(\omega)$ has an opposite sign below and above the band, corresponding to a $\pi$ shift across $\varepsilon_\alpha$. In between essential bands, there are topologically robust zeros of $\mu^\sigma(\omega)=\re g^\sigma(\omega+i0^+)$ at energies $E_a$ confined to the gap between the bands. These zeros satisfy $\partial_\omega\mu^\sigma(\omega)|_{E_a}<0$ which characterizes a spectral \emph{attractor} in the sense discussed above. As $v\to\pm\infty$ the spectrum of the perturbed Hamiltonian $\H+\V$ will get a state pinned at these attractors in the form of a ring state. Figure~\ref{figapp:flatbandgengsig} illustrates this finite-band interlacing structure in a four-band flat-band example, including the case where one band is invisible to the impurity because its residue vanishes.

\subsection{Stability of zeros in $\mu^\sigma(\omega)$ in dispersive bands}\label{app:proof2}

Throughout this note we assume that the topological obstruction enforces $\nu^\sigma\not\equiv 0$ on each gap-adjacent band, so that the band-edge sign structure of Eq.~\eqref{eq:edgesign} is maintained under any symmetric, gap-preserving deformation. The failure of this assumption requires the asymptotic decoupling $\nu^\sigma\to 0$ identically on one band, which is available only in trivial bands; this is the second route to removal discussed in Sec.~\ref{sec:protection}.

\paragraph{Stability theorem.}
The flat-band argument above gives a representative of the phase in which the zero of $g^\sigma$ is explicit. The dispersive question is whether this zero can be removed while the spectral gap remains open. Let $\H(t)$ be a continuous, symmetry-preserving deformation with a gap $(E_v(t),E_c(t))$, and define the impurity projected Green's function by the positive spectral measure of $\ket{\sigma}$,
\begin{align}
g_t^\sigma(z)=\bra{\sigma}(z-\H(t))^{-1}\ket{\sigma}
=\int {d\nu_t^\sigma(E)\over z-E}.
\end{align}
Here $z$ is the complex spectral parameter. On the real axis we take the retarded limit $z=\omega+i0^+$ and write $g_t^\sigma(\omega+i0^+)=\mu_t^\sigma(\omega)-i\pi\nu_t^\sigma(\omega)$. Inside the gap, $\nu_t^\sigma(\omega)=0$ and $g_t^\sigma(\omega)=\mu_t^\sigma(\omega)$ is real. Moreover,
\begin{align}
\partial_\omega\mu_t^\sigma(\omega)
=-\int {d\nu_t^\sigma(E)\over(\omega-E)^2}<0
\label{eq:mustability}
\end{align}
on the active impurity channel. Therefore $\mu_t^\sigma(\omega)$ is strictly decreasing in the gap. If the projected spectral measure has the topological edge-sign structure
\begin{align}
\mu_t^\sigma(E_v^+)>0,\qquad \mu_t^\sigma(E_c^-)<0,
\label{eq:edgesign}
\end{align}
then the intermediate value theorem gives a unique ingap zero $E_a(t)$, and Eq.~\eqref{eq:mustability} shows that it is an attractor, $\partial_\omega\mu_t^\sigma(E_a)<0$. Such a zero cannot annihilate with another zero inside the gap, because monotonicity forbids zero pairs there. It can disappear only by reaching a band edge, which is precisely where the edge signs in Eq.~\eqref{eq:edgesign} can change. Thus the flat-band zero persists under dispersion as long as the gap remains open and the local spectral weight forced by the obstructed band representation remains on both sides of the gap. Removing it requires closing the relevant topological gap, breaking the protecting symmetry, or changing the local representation coupled to the impurity.

We now show that, under the obstruction assumption above, the attractor cannot annihilate with the adjacent-band repeller through a local saddle-node at a regular band edge; the attractive fixed point is therefore lost only when the bulk gap closes, as realized in the explicit topological transitions studied below.
First, let us assume the zero exists at an ingap energy $E_a$, crossing zero with a negative slope $(\mu^\sigma)'(E_a)<0$ where the prime indicates derivative with respect to the energy, reaching the conduction band edge, which we set without loss of generality to be $\varepsilon^{\rm min}=0$, with a negative value $\mu^\sigma(0)<0$. Since we know that above the conduction band $\mu^\sigma(+\infty)>0$, we know there is a zero of positive slope at an energy $E_r>0$ inside the conduction band, $(\mu^\sigma)'(E_r)>0$. Here the subscripts indicate ``attractor" or ``repeller" according to the sign of the slope of $\mu^\sigma(\omega)$. We may ask, can the zero of the band and the zero of the gap come together and annihilate each other, effectively removing the ingap spectral attractor without a topological, or band gap closing, phase transition? The answer is \emph{no}. To see this, we note that if such a point would exist, the two zeros, the attractor, and the repeller would meet at the band edge (or slightly above) with zero slope, satisfying at the annihilation point
\begin{align}
\mu^\sigma(E_a)=\mu^\sigma(E_r)=0,\quad(\mu^\sigma)'(E_a)=(\mu^\sigma)'(E_r)=0\label{eqapp:anihilate}
\end{align}
To see if this is possible, let us assume a power-law form for the density of states near the conduction band edge. Namely, we say that $\nu^\sigma(\omega)\sim\omega^\gamma$ with $\gamma\geq0$, for $\omega$ defined from $0$ up to a maximum value of $W$, the conduction bandwidth, such that the total number of states that match the impurity character is $n=W^{1+\gamma}/{(1+\gamma)}$. 

We can write $\mu^\sigma(\omega)$ as a function of the density of states of the valence and conduction bands separately,  
\begin{align}\mu^\sigma(\omega)=P\!\int d\omega'\,{\nu^\sigma(\omega')\over \omega-\omega'}=\mu^\sigma_v(\omega)+P\!\int_0^W d\omega'\,{\nu^\sigma(\omega')\over \omega-\omega'}\end{align}
where the first term is finite, $\mu^\sigma_v(\omega)\sim 1/(\omega-\varepsilon_v)$ for $\varepsilon_v\ll0$ a representative energy of the valence band, and therefore at $\omega$ close to the conduction band $\omega\sim0$ is just a finite positive number. Similarly, $(\mu^\sigma_v)'(\omega)\sim -1/(\omega-\varepsilon_v)^2$, which is a finite negative number at the band edge. The second term may be divergent at the band edge, and therefore we focus on it. Looking at the slope of $\mu(\omega)$ as we approach the band edge from below $\omega\to0^-$, we find that
\begin{align}(\mu^\sigma)'(\omega)=(\mu^\sigma_v)'(\omega)+\int_0^W d\omega' {(\nu^\sigma)'(\omega')\over \omega-\omega'}.\end{align}

For $\gamma=0$, the density of states is constant. Treating the band edge as an idealized step of height $\nu_0$ supported on $0<\omega'<W$ (and zero otherwise), direct integration gives $\mu^\sigma(\omega)=\nu_0\ln|\omega/(\omega-W)|$ and $(\mu^\sigma)'(\omega)=\nu_0/\omega-\nu_0/(\omega-W)$, so $(\mu^\sigma)'(\omega)\sim\nu_0/\omega\to-\infty$ as $\omega\to0^-$. The slope is therefore negative approaching the band edge from below, and \eqref{eqapp:anihilate} cannot be satisfied. 

Since we look at the impurity projected density of states, a step function is not necessarily the correct choice. Let us say that it grows linearly, $\nu(\omega)\sim\omega$. In this case, $\mu'(\omega)\sim -\log(1-W/\omega)$, with a logarithmic divergence at the band edge, and therefore $\mu'(0^-)<0$ and the condition \eqref{eqapp:anihilate} cannot be satisfied. For larger powers $\gamma>1$, the density of states grows very slowly, and we can expect that the condition \eqref{eqapp:anihilate} may be satisfied. However, the limit is well defined and can be computed in a closed form
\begin{align}\lim_{\omega\to0^-}(\mu^\sigma)'(\omega)\sim-{1\over \gamma-1}\, W^{\gamma-1}.\end{align}
That is, at a band edge, ${\mu^\sigma}'(\omega)$ is \emph{strictly negative} for any algebraic growth of the density of states $\nu^\sigma\sim\omega^\gamma$. Therefore, it is impossible to satisfy \eqref{eqapp:anihilate} at a band edge.

The power-law form $\nu^\sigma\sim\omega^\gamma$ is chosen for concreteness. The same Hilbert-transform analysis shows that $(\mu^\sigma)'(0^-)<0$ for any positive-measure $d\nu^\sigma$ on the band, including band edges where $\nu^\sigma$ approaches zero faster than any power (for example, exponentially): the band-interior weight makes $(\mu^\sigma)'$ strictly negative and finite, and the near-edge suppression only strengthens this conclusion. The condition for the protection to fail is therefore not ``fast enough'' band-edge decay but the strict atomic-limit case in which $\nu^\sigma$ vanishes identically on the band, which the Wannier obstruction blocks at any finite parameter.

Under the obstruction assumption above, the additional zeros created at the topological phase transition are therefore lost with gap closing in the explicit transitions illustrated below.
We can see the topological phase transition in Figs.~\ref{figapp:phasediagmu} and \ref{figapp:cherng}. At that point two concurrent things happen: the repeller annihilates the attractor, and the band gap closes. Importantly we note that the repeller of the lower band remains unaffected across the transition.

In conclusion, the number of finite energy spectral attractors is stable against dispersion deformations, and cannot be changed adiabatically, unless $\nu^\sigma$ is made to fully vanish within the band. This is only possible if a topological obstruction does not prevent it.

\subsection{Topologically enforced adiabatic discontinuity between unperturbed and perturbed bands}\label{app:proof3}

We now show that the ring state, even if it enters the band continuum, turns into a resonance which cannot be adiabatically removed. That is, the finite energy attractor in $\mu^\sigma(\omega)$ necessarily remains even if the bulk density of states is finite $\nu^\sigma(\omega)$ at its energy.

Let us first consider the case where the attractor does not overlap with the band, such that the perturbed band is composed of extended scattering eigenstates $\ket{\tilde\psi_n^\alpha}$, which assume the general form~\cite{Economou.Economou} \begin{align}\ket{\tilde\psi^\alpha_\bk}=\ket{\psi^\alpha_\bk}+\G(E+i0^+)\V\ket{\tilde\psi^\alpha_\bk},\label{eqapp:scatteringstates}\end{align} 
with $\ket{\psi^\alpha_\bk}$ the unperturbed Bloch states. Independently of spatial dimension, in a finite size lattice with $N$ sites if a rank-1 impurity $\V$ is strongly coupled to the band, that is $\G(E+i0^+)\V$ has an eigenstate of order $O(1)$ when $v\to\infty$, it leads to a $\pi$ shift across the band from the removal of a single state from the band continuum. There are only $N-1$ solutions of the type in \eqref{eqapp:scatteringstates}, stemming from the zeros of $\G(\omega)$ in between the unperturbed poles. On the other hand, if the perturbation caused by $\V$ on the band is weak, there is no bound state removed from the band and the sign of $g(\omega)$ is the same on both edges of the band, a zero overall phase shift. In this case, Eq.~\eqref{eqapp:scatteringstates} has $N$ solutions corresponding to weak perturbations from the original Bloch states.
For $v\to\pm\infty$, let us write formally the band projector operator at $v\to\infty$,
\begin{align}\Tilde{\P}^\alpha=\sum_{n=1}^{N-1}\proj{\tilde{\psi}^\alpha_\bk}\end{align}
The projector into the perturbed topological band is related to the unperturbed ones by
\begin{align}\sum_\alpha\Tilde{\P}^\alpha+\proj{\sigma_E}+\proj{\rho_E}=\sum_\alpha\P^\alpha=\id.\label{appeq:completeness}\end{align}
Bracketing both sides of the completeness equality \eqref{appeq:completeness} with the impurity eigenstate $\ket{\sigma}$, we find in the limit $v\to\pm\infty$
\begin{align}\sum_\alpha\bra{\sigma}\Tilde{\P}^\alpha\ket{\sigma}+1=\sum_\alpha s_\alpha=1\label{eqapp:sumrule}\end{align}
where we used $|\bra{\sigma}\sigma_E\rangle|^2\to 1$ and $|\bra{\sigma}\rho_E\rangle|^2\to 0$.
Eq.~\eqref{eqapp:sumrule} implies that $\bra{\sigma}\Tilde{\P}^\alpha\ket{\sigma}=0$ for every band $\alpha$ of the perturbed system. This is simply a restatement that $\ket{\sigma_{E=\infty}}=\ket{\sigma}$ is an eigenstate of the perturbed Hamiltonian $\tilde\H$ with an energy far beyond the band energies.

Let us choose without loss of generality to bring the ring state $\ket{\rho_E}$ into the (empty) band $\alpha=2$. If it is possible to remove the zeros of $\mu^\sigma(\omega)$ by annihilating one attractor and one repeller at the empty band, the perturbation $\V$ did not remove a state from the band, and all the perturbed states in the band can be adiabatically deformed back into clean Bloch states.
That is, $\tilde\P^{\alpha=2}+\proj{\rho_E}$ must span the same space as $\P^{\alpha=2}$. On the other hand, from completeness, also the perturbed filled band $\tilde\P^{\alpha=1}$ together with the state $\ket{\sigma_E}$ must span the same space as the unperturbed filled band $\P^{\alpha=1}$. That is, we have shown that in the limit $v\to\infty$, if the ring state can be absorbed by one of the topological bands, the local state $\ket{\sigma_{E=\infty}}=\ket{\sigma}$ must be contained in the unperturbed band subspace $\P^{\alpha=1}$ of the other topological band, which is a contradiction.

We have proven in the previous paragraph that the attractor of $\mu^\sigma(\omega)$ cannot be removed simply by absorbing it into one band. The ring state is robust to any spectral deformation that will not re-invert the bands by removing the additional zeros of $\mu^\sigma(\omega)$ created at the phase transition. Note that if the attractor is buried into a band, it will result in a resonance, which means that when the ring state overlaps with the band it gains a finite lifetime~\cite{Economou.Economou} given by
\begin{align}\Gamma(E)={2\pi\nu^\sigma(E)\over\left|(\mu^\sigma)'(E)\right|}.\end{align}

Let us consider for example graphene~\cite{Balatsky.Wehling.2014}, which we have seen in Fig.~\ref{figapp:kanemelechern} to retain its attractor at the Dirac point even if the gap is fully closed. The ring state in this case gains a polynomial decay and has infinite lifetime only at $v=\pm\infty$ since at the Dirac point $\nu^\sigma(\omega)$ vanishes. For any smaller $v$ the resonance is slightly shifted above or below the Dirac point and gains a finite lifetime. Alternatively, when the density of states grows slower 
than a linear, $\gamma>1$, it is possible that the attractor moves into the band. This is expected for three dimensional Weyl semimetals where $\nu(\omega)\propto\omega^2$ at energies close to the Weyl point. The resonance expected in this case is also topologically protected.

\section{Exact solutions for ring state wavefunctions in the two-dimensional Chern insulator}\label{app:chernGF}

We derive the exact spatial structure of the ring state wavefunction by computing the real-space Green's function of the continuum Chern insulator. This calculation reveals the origin of the $e^{i\varphi}$ phase winding and provides the radial functions $g(r)$ and $f(r)$ that determine the ring state properties, including the local currents derived in the following section.

\subsection{The continuum model and its length scales}

We work with the two-band continuum Hamiltonian
\begin{align}
    h(\bk)=(m+bk^2)\sigma_z - \upsilon\bk\cdot\bsigma,
\end{align}
i.e.\ Eq.~\eqref{eq:cherncont} of the main text, where the $\sigma_z$ eigenstates $\ket{s} = (1,0)^T$ and $\ket{p} = (0,1)^T$ represent two orbitals with opposite parity. The parameter $m$ is the mass (gap), $b$ controls the quadratic dispersion, and $\upsilon$ is the $s$-$p$ coupling strength. The model is topological with Chern number $C=(\sgn b - \sgn m)/2$, equal to $-\sgn(m)$ when $b>0$, so the topological phase has $mb<0$.

The real-space Green's function is
\begin{equation}
\G(\omega;\br,0)=\bra{\br}\G(\omega)\ket{0}=\int {d{\vec k}\over (2\pi)^2}\frac{\omega\sigma_0+({m}+b k^2)\sigma_z - \upsilon{\vec k}\cdot{\vec \sigma}}{\omega^2-(m+bk^2)^2-\upsilon^2k^2}~e^{i{\vec k}\cdot{\vec r}}.
\end{equation}
The denominator factors exactly as $-b^2(k^2+\kappa_1^2)(k^2+\kappa_2^2)$, where
\begin{align}
\kappa_{1,2}^2 = \frac{(\upsilon^2+2mb)\mp\sqrt{(\upsilon^2+2mb)^2-4b^2(m^2-\omega^2)}}{2b^2}\label{eq:kappaexact}
\end{align}
are the roots of $\omega^2-(m+bk^2)^2-\upsilon^2k^2=0$ in $k^2$. Their symmetric polynomials take the compact form
\begin{align}
\kappa_1^2 + \kappa_2^2 = \frac{\upsilon^2+2mb}{b^2}, \qquad \kappa_1^2\,\kappa_2^2 = \frac{m^2-\omega^2}{b^2},\label{eq:kappavieta}
\end{align}
so the product is gap-controlled (vanishing at the band edge $\omega \to \pm m$), while the sum mixes the band-curvature scale $\upsilon^2/b^2$ with the band-inversion scale $2m/b$. In particular, at $\omega=0$ the product reduces to $\kappa_1\,\kappa_2\big|_{\omega=0} = |m|/|b| = k_{\rm inv}^2$, where $k_{\rm inv} = \sqrt{|m/b|}$ is the band-inversion momentum (the zero of $m+bk^2$ in the topological phase, where the orbital character is fully mixed; see below). The band-inversion momentum is thus the geometric mean of the two pole scales. We carry out the entire calculation below in terms of $\kappa_1$ and $\kappa_2$; all closed-form Bessel-function expressions are exact in these variables. In the scale-separated regime $\upsilon^2 \gg b\sqrt{m^2-\omega^2}$, the two roots split into a small gap scale and a large UV scale,
\begin{align}
\kappa_1 \simeq \frac{\sqrt{m^2-\omega^2}}{\upsilon}, \qquad \kappa_2 \simeq \frac{\upsilon}{b},\label{eq:kdef}
\end{align}
recovering the long-distance and short-distance momentum scales $k_1, k_2$ used in the main text. These correspond to two length scales with distinct physical origins:

\begin{itemize}
\item $1/\kappa_1$: the long-distance scale, set by the gap. This controls the exponential decay of the ring state.
\item $1/\kappa_2$: the short-distance scale where the curvature term dominates; the true UV cutoff $1/\Lambda$ is parametrically shorter.
\end{itemize}
The band inversion at $k_{\rm inv}$ (which requires $m$ and $b$ to have opposite signs) is the origin of the nontrivial topology.

\subsection{Diagonal Green's function}

Let us first calculate the local diagonal Green's function, which is pathological without a lattice regularization. To do so, we truncate the integral at large momenta by a momentum cutoff $\Lambda$. At $\br=0$ the off-diagonal angular integral vanishes, so this local object is
\begin{equation}
\G_\Lambda(\omega;0,0)=-{1\over 2\pi b^2}\int_0^\Lambda d{k}k\frac{ \omega\sigma_0+({m}+b k^2)\sigma_z}{(k^2+\kappa_1^2)(k^2+\kappa_2^2)}.
\end{equation}
We split it into the same two diagonal building blocks used below,
\begin{equation}
\G_\Lambda(\omega;0,0)=-{1\over 2\pi b^2}\left[A_0(\omega)+bB_0(\omega)\right],
\end{equation}
with
\begin{align}A_0(\omega)={\omega\sigma_0+m\sigma_z\over \kappa_1^2 - \kappa_2^2}\log(\kappa_1/\kappa_2),\end{align}
and
\begin{align}B_0(\omega)=\left({-\kappa_1^2\log \kappa_1+\kappa_2^2\log \kappa_2\over \kappa_1^2 - \kappa_2^2}+{\log\Lambda}\right){\sigma_z}.\end{align}
The subscript $0$ denotes the local, cutoff-regularized value at the impurity site. The logarithmic divergence of $\G_\Lambda(\omega;0,0)$ is not present for the lattice problem, and we can fix the cutoff to fit the numerical result.

\begin{figure}
    \centering
    \includegraphics[width=.4\columnwidth]{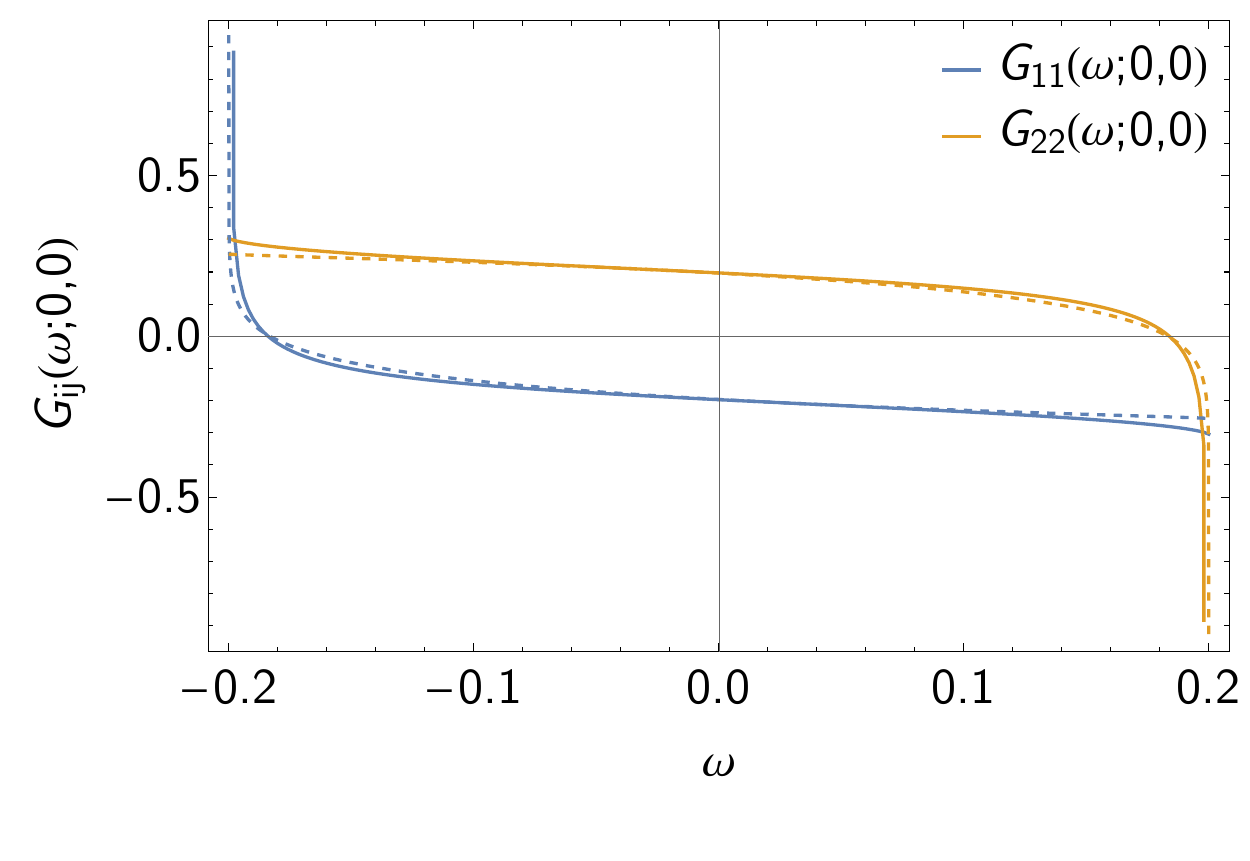}
    \includegraphics[width=.4\columnwidth]{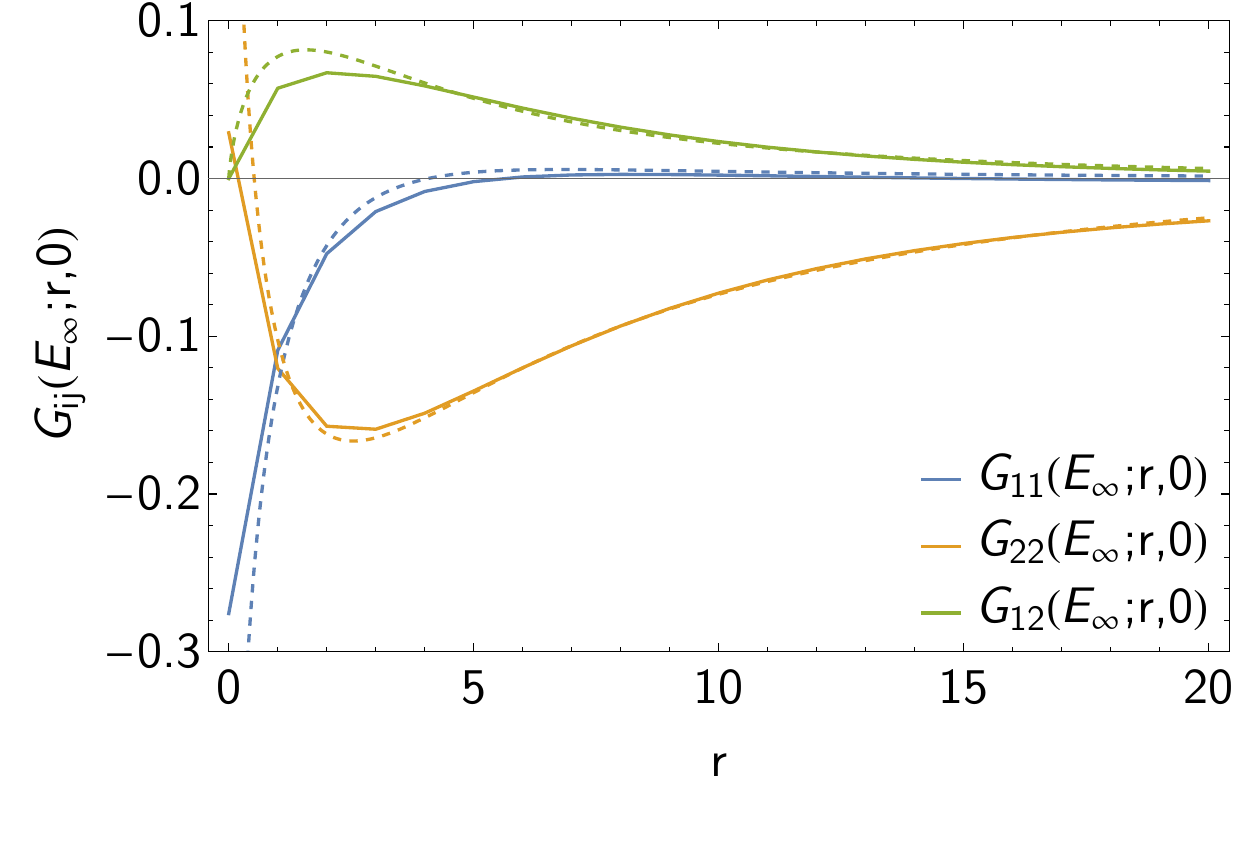}
    \caption{Comparison between the numerical calculation of $\G_{ij}(\omega;r,0)$ from the tight-binding model in a mesh of $200\times200$ sites in the Brillouin zone, and analytical expressions of the ingap Green's function in the continuum approximation. We use a cutoff value of $\Lambda=5$ at $r=0$ and no cutoff in the right figure, which means we can see the nonphysical logarithmic divergence as $r\to0$. Both figures are in the Chern phase with $m=-0.2$, $b=1.5$, and $\upsilon=1$. The continuum cutoff for this figure is $\Lambda=5$ (matched to the lattice ring-state energy); the resulting zero energy is $E_\infty(\Lambda=5)\approx -0.192$. Throughout Note~\ref{app:chernGF} we instead use $\Lambda=20$, which suppresses Gibbs ringing in the radial profile and gives $E_\infty(\Lambda=20)\approx -0.199$; the lattice value is $E_{\rm TB}\approx -0.193$ ($\Lambda$-independent), reproduced to within the logarithmic $\Lambda$-dependence of the continuum cutoff theory by $\Lambda=5$.}
    \label{fig:numericalandanal}
\end{figure}

The analytical expressions can be compared to the numerical calculation of Green's function in Fig.~\ref{fig:numericalandanal}.

Let us now look at the radial dependence by decomposing the operator as $\G(\omega;\br,0)=G(\omega;r)+F(\omega;\br)$, where $G$ is diagonal in the orbital basis and $F$ is off diagonal. For $r>0$ the diagonal continuum integral can be evaluated in the $\Lambda\to\infty$ limit,
\begin{align}
G(\omega;r)=-{1\over 2\pi b^2}\int_0^\infty dk ~ k J_0(kr)\frac{\omega\sigma_0+({m}+b k^2)\sigma_z}{(k^2+\kappa_1^2)(k^2+\kappa_2^2)}
\end{align}
where $J_0(x)$ is the Bessel function of the first kind.
The block $A(\omega;r)$ is obtained from the partial-fraction identity
\begin{align} 
{1\over(k^2+\kappa_1^2)(k^2+\kappa_2^2)}={1\over(\kappa_2^2-\kappa_1^2)}\left[\frac{1}{k^2+\kappa_1^2}-\frac{1}{k^2+\kappa_2^2}\right]
\end{align}
Then, the integral is tabulated, and it can be explicitly written as $G(\omega;r)=-[A(\omega;r)+bB(\omega;r)]/(2\pi b^2)$ with
\begin{align}
A(\omega;r)=-\frac{\omega\sigma_0+{m}\sigma_z}{\kappa_1^2-\kappa_2^2}
\left [K_0(\kappa_1 r)-K_0(\kappa_2 r) \right ].
\end{align}
Note that as the energy reaches the gap $\omega\to \pm m$, $A(\omega;r)$ vanishes for one of the diagonal entries.
The curvature block $B(\omega;r)$ follows by treating the $k^2$ in the numerator as $-\nabla^2=-(\partial^2_r+\frac{1}{r}\partial_r)$. Then, using $\nabla^2 K_0(kr)=k^2 K_0(kr)$,
\begin{align}B(\omega;r)=\frac{\kappa_1^2 K_0(\kappa_1r)-\kappa_2^2 K_0(\kappa_2r)}{(\kappa_1^2-\kappa_2^2)}\sigma_z.\end{align}
The local quantities above are the short-distance versions of the same two blocks: $A_0(\omega)=\lim_{r\to0}A(\omega;r)$, while $B_0(\omega)$ is the cutoff-regularized local value of $B(\omega;r)$. Using $K_0(x)=-\log(x/2)-\gamma_{\rm E}+O(x^2\log x)$,
\begin{align}B(\omega;r)=B_0(\omega)-\left[\log(\Lambda r/2)+\gamma_{\rm E}\right]\sigma_z+O(r^2\log r),\end{align}
so $B_0$ is not a new contribution; it is the same curvature block with its logarithmic UV divergence absorbed into the cutoff. At the origin, the diagonal entries define the local projected Green's functions $g^s(\omega)\equiv\bra{s}\G_\Lambda(\omega;0,0)\ket{s}$ and $g^p(\omega)\equiv\bra{p}\G_\Lambda(\omega;0,0)\ket{p}$. Away from the origin, we write the $s$-channel radial extension as $g(\omega;r)\equiv\bra{s}G(\omega;r)\ket{s}$, with the cutoff understood when this radial function is evaluated at $r=0$.
The diagonal elements still diverge logarithmically as $r\to0$. This is an ultraviolet divergence, which could be regularized in one of two ways: either cut it off by making the $K_0(k r)\sim K_0(k/\kappa_2)$ for $r<1/\kappa_2$ (the short distance cutoff), or add a quartic $(k^4)$ term to the dispersion. This divergence is not present in the tight-binding model, and it is not important. After regularizing the ultraviolet divergence, the diagonal elements have three regimes: Around $1/\kappa_2$ they are constant (short scale). Between $1/\kappa_2$ and $1/\kappa_1$ (intermediate scale) they decrease logarithmically with $r$, and after that (long scale) they go down exponentially. The short scale is dominated by the $\kappa_2^2 K_0(\kappa_2r)$ term while the long scale term is dominated by the $\kappa_1^2 K_0(\kappa_1r)$ term, so these elements of the Green's function change sign as $r$ grows.

The continuum theory itself has only one internal momentum scale, $\kappa_2\sim \upsilon/b$, at which the Dirac kinetic energy balances the curvature kinetic. The strong-impurity (vacancy) ring state, however, is a renormalization phenomenon: it requires UV weight beyond $\kappa_2$ to be supported, and so the cutoff $\Lambda$ is a genuine free parameter of the continuum theory. The divergence in $\G(\omega;0,0)$ is only logarithmic, so $E_\infty$ depends on $\Lambda$ as $\log\Lambda$ while the radial profile for $r\gtrsim 1/\kappa_2$ is only logarithmically sensitive. Throughout this note we fix $\Lambda=20$, chosen by a smoothness criterion (the radial profile of $g$ becomes monotone to its peak only for $\Lambda\gtrsim 3$, and Gibbs ringing from the hard cutoff decays as $1/\Lambda^2$, rendered invisible at $\Lambda=20$). The lattice ring-state energy in Fig.~\ref{fig:numericalandanal} is reproduced to within the logarithmic $\Lambda$-dependence: $E_\infty(\Lambda=5)\approx-0.192$ and $E_\infty(\Lambda=20)\approx-0.199$ are both in-gap, and the physical statements below (dominance of hybridization current, radial profile for $r\gtrsim 1/\kappa_2$) are $\Lambda$-independent.

\begin{figure}[h]
    \centering
    \includegraphics[width=\columnwidth]{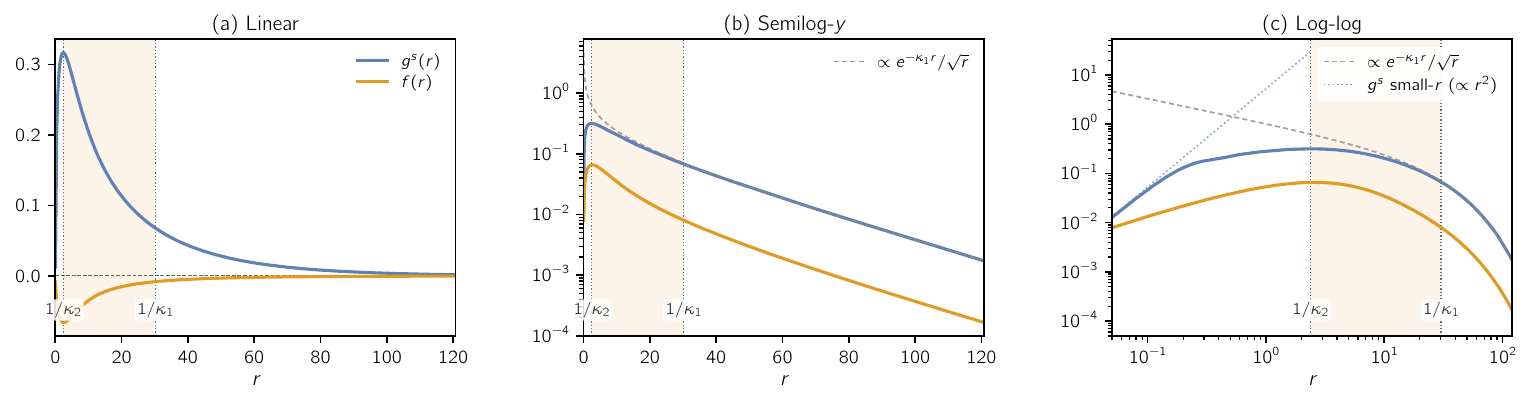}
    \caption{Radial decay regimes of the continuum Chern Green's functions at $m=-0.2$, $b=1.5$, $\upsilon=1$, $E=E_\infty\approx-0.199$ ($\Lambda=20$). (a)~Semilog-$y$: $|g(r)|$ and $|f(r)|$ versus $r$, with reference Bessel tail $\sim e^{-\kappa_1 r}/\sqrt{r}$. (b)~Log-log on the same data. (c)~Local decay rate $-d\log|g|/dr$ with horizontal references at $\kappa_1$, $\kappa_2$ and the Bessel-tail prediction $\kappa_1+1/(2r)$. Shaded band: the inversion window $1/\kappa_2 < r < 1/\kappa_1$.}
    \label{fig:logscale}
\end{figure}

The three decay regimes are illustrated in Fig.~\ref{fig:logscale}, which plots $|g(r)|$ and $|f(r)|$ on semilog and log-log axes. The shaded region marks the inversion window $1/\kappa_2 < r < 1/\kappa_1$. Outside this window the diagonal radial amplitude is exponentially suppressed; the Bessel asymptote $\sim e^{-\kappa_1 r}/\sqrt{r}$ (dashed) is reached for $r\gtrsim 2/\kappa_1$. Inside the window the closed-form $K_0$ representation of $g$ has a sign change inherited from the cancellation of the two $K_0$ terms; the renormalized $g_\Lambda$, obtained from the same $\Lambda$-truncated integral that fixes $E_\infty$, is smooth on $[0,\infty)$ and pins $g_\Lambda(E_\infty;0)=g^s_\Lambda(E_\infty)=0$. The local decay rate $-d\log|g|/dr$ (panel c) approaches the long-distance asymptote $\kappa_1+1/(2r)$ from below; for these parameters there is no clean intermediate regime where the rate sits at $\kappa_2$, so the inversion window is best understood as a crossover between the UV cutoff and the gap-controlled tail, not as a separate decay rate.

\subsection{Off-diagonal Green's function: the origin of $e^{i\varphi}$}\label{app:offdiag}

The off-diagonal terms of the Green's function encode the $s$-$p$ mixing that gives rise to the $e^{i\varphi}$ phase winding in the ring state. This phase factor is not an intrinsic property of atomic $p$-orbitals; it is inherited from the momentum-space structure of the $\bk\cdot\bsigma$ coupling.

The off-diagonal part $F(\omega;\br)$ involves the matrix integral
\begin{align}F(\omega;\br)=-{1\over 2\pi b^2}\int d\bk\frac{\upsilon(\sigma_x\partial_x+\sigma_y\partial_y)e^{i\bk\cdot\br}}{(k^2+\kappa_1^2)(k^2+\kappa_2^2)}.\end{align}
To evaluate this, note that the coupling $\bk\cdot\bsigma = k_x\sigma_x + k_y\sigma_y$ can be written in polar momentum coordinates as $k(e^{i\theta_k}\sigma_- + e^{-i\theta_k}\sigma_+)/2$, where $k_x + ik_y = ke^{i\theta_k}$. When Fourier transformed, the angular integral over $\theta_k$ picks out the $e^{\pm i\varphi}$ dependence:
\begin{align}F(\omega;\br) = \frac{i\,f(\omega;r)}{2}\left[e^{i\varphi}\sigma_- + e^{-i\varphi}\sigma_+\right],\label{eq:FApp}\end{align}
where $\br=re^{i\varphi}$ and $\sigma_\pm\equiv\sigma_x\pm i\sigma_y$. The $e^{\pm i\varphi}$ pairing with $\sigma_\mp$ is inherited directly from the polar decomposition of $\bk\cdot\bsigma$: the angular integral over $\theta_k$ pairs $e^{i\theta_k}\sigma_-$ with $e^{i\varphi}$ and $e^{-i\theta_k}\sigma_+$ with $e^{-i\varphi}$. Both $n=\pm 1$ Bessel identities $\int d\theta_k\, e^{\pm i\theta_k}\,e^{ikr\cos(\theta_k-\varphi)}=2\pi i\, e^{\pm i\varphi} J_1(kr)$ supply the same factor of $i$ (the $i^{-1}$ from $n=-1$ is cancelled by $J_{-1}=-J_1$), so the two terms combine with a relative plus sign; the overall plus combination is also dictated by hermiticity of the underlying matrix integrand. Acting on $\ket{s}=(1,0)^T$ then yields the $p$-component $ie^{i\varphi}f(\omega;r)\ket{p}$ with winding $e^{i\varphi}$, consistent with Eq.~\eqref{eq:ringfull} and Fig.~\ref{fig:ring}. The key result is that the $e^{i\varphi}$ factor emerges from the angular structure of the $\bk\cdot\bsigma$ coupling in momentum space, not from any intrinsic orbital property.

These terms do not diverge at $r=0$; rather, they start as $r \log r$ at the short scale, continue as $1/r$, and then decay exponentially. At the long scale all elements of the Green's function decay exponentially.

\subsection{The ring state wavefunction}

The finite-energy bound-state wavefunction is obtained from the Green's function acting on the impurity orbital, evaluated at an energy $E(v)$ satisfying $g^s(E(v))=1/v$. For an $s$-type impurity ($\ket{s} = (1,0)^T$), the Green's-function representative of the state takes the form
\begin{align}
\rho_E(\br) = g(E;r)\ket{s} + ie^{i\varphi}f(E;r)\ket{p},
\end{align}
where $g(E;r)\equiv\bra{s}\G(E;\br,0)\ket{s}$ is the radial extension of the local projected Green's function and $f(E;r)$ is the off-diagonal radial amplitude of the $p$-orbital. The physical wavefunction is obtained by normalizing this Green's-function representative, $\ket{\rho_E^{\rm norm}}=\ket{\rho_E}/\sqrt{\langle\rho_E|\rho_E\rangle}$; below, $\rho_E(\br)$ denotes the representative before this final normalization unless explicitly stated. The overall normalization is fixed after constructing this Green's-function spinor and does not affect the node or the angular winding. For $r>0$, the diagonal block $G(E;r)$ above gives the exact radial function
\begin{align}
g(E;r)=
{(E+m-b\kappa_1^2)K_0(\kappa_1 r)
-(E+m-b\kappa_2^2)K_0(\kappa_2 r)
\over 2\pi b^2(\kappa_1^2-\kappa_2^2)},\label{eq:gApp}
\end{align}
where $\kappa_{1,2}$ are evaluated at the same energy $E$ through Eq.~\eqref{eq:kappaexact}. At the impurity site the unregularized $K_0$ form is replaced by the same cutoff-regularized local quantity used in the bound-state condition, $g_\Lambda(E;0)=g^s(E)$. From the calculation above, the exact off-diagonal radial function is
\begin{align}
f(E;r) = \frac{\upsilon[\kappa_1 K_1(\kappa_1 r) - \kappa_2 K_1(\kappa_2 r)]}{2\pi b^2(\kappa_2^2 - \kappa_1^2)}.\label{eq:fApp}
\end{align}
On the ring branch, the solution of $g^s(E(v))=1/v$ approaches the finite zero $E_\infty$ of $\mu^s$ as $v\to\infty$, rather than scaling with $v$. The zeroth order in the $1/v$ expansion is therefore the vacancy-limit wavefunction
\begin{align}
\rho_\infty(\br)\equiv\rho_{E_\infty}(\br)
=g(E_\infty;r)\ket{s}+ie^{i\varphi}f(E_\infty;r)\ket{p}.
\end{align}
For any fixed $r>0$, the two functions in this limiting wavefunction are finite and independent of $v$ in the strong-impurity limit. At the origin, orthogonality to the removed $s$ orbital fixes the cutoff-regularized core condition $g_\Lambda(E_\infty;0)=g^s(E_\infty)=0$, while the regular vanishing of $f(E_\infty;r)$ at $r=0$ follows from the cancellation of the leading $1/r$ singularities of the two $K_1$ terms. The factor $i$ is a phase convention from the Fourier transform of the off-diagonal Green's function, and the $e^{i\varphi}$ factor is inherited from the off-diagonal Green's function (Sec.~\ref{app:offdiag}). Since $f(E_\infty;r)$ remains $O(v^0)$, this winding survives in $\rho_\infty(\br)$ itself; it is not a subleading correction in $1/v$. For large but finite $v$, assuming the zero of $\mu^s$ at $E_\infty$ is simple, we use $g^s=\mu^s$ inside the gap and define $\mu'(E_\infty)\equiv\left.\partial_E\mu^s(E)\right|_{E=E_\infty}=\partial_Eg^s(E_\infty)$. Then
\begin{align}
E(v)=E_\infty+{1\over v\,\mu'(E_\infty)}+O(v^{-2}),
\end{align}
so the Green's-function representative has the expansion
\begin{align}
\rho_E(\br)=\rho_\infty(\br)
+{1\over v\,\mu'(E_\infty)}
\partial_E\left[g(E;r)\ket{s}+ie^{i\varphi}f(E;r)\ket{p}\right]_{E=E_\infty}
+O(v^{-2}).
\end{align}
This correction is proportional to $1/v$ and to the inverse local slope $1/\mu'(E_\infty)$. For the parameters used in Fig.~\ref{fig:ring}, $\mu'(E_\infty)\simeq -69.5$, so the steep zero of $\mu^s$ makes the finite-$v$ wavefunction converge rapidly to the vacancy ring-state shape.

\subsection{Weak-impurity band-edge states}

It is useful to distinguish the strong-vacancy limit from the weak-impurity part of the same finite-energy branch. For a local orbital channel $\sigma=s,p$, a bound state outside the continuum satisfies
\begin{align}
g^\sigma(E(v))=\bra{\sigma}\G_\Lambda(E(v);0,0)\ket{\sigma}={1\over v},
\end{align}
and its wavefunction is, up to normalization,
\begin{align}
\rho_E^\sigma(\br)=\G_\Lambda(E;\br,0)\ket{\sigma}.
\end{align}
For $\sigma=s$ this reduces to the form in Eq.~\eqref{eq:ringfull}. For $\sigma=p$ the same Green's-function blocks appear, with the off-diagonal winding complex conjugated by the interchange of source orbital. The weak-impurity limit is useful mainly for identifying the branches of Eq.~\eqref{eq:ringfull}. In the $s$ channel, the low-energy solution lies in the gap and approaches the lower band edge as $v\to0^+$. It is then an ordinary shallow band-edge impurity state: one pole scale $\kappa_1$ becomes small, the decay length is long, and the profile is dominated by its band-edge tail. As $v$ is increased, this same branch moves away from the edge and approaches the vacancy-limit spectral attractor $E_\infty$, where $g^s(E_\infty;0)=0$ and the profile develops the ring-shaped core suppression. For the parameters used in Fig.~\ref{fig:ring}, $m=-0.2$, $b=1.5$, $\upsilon=1$, and $\Lambda=20$, we find $E_\infty=-0.1989048$. On the gap branch, $v=0.1$ gives $E_{\rm gap}-m\simeq 2.8\times10^{-58}$, while $v=100$ gives $E_{\rm gap}\simeq -0.1990394$, already close to $E_\infty$. The same scalar equation also has an upper solution above the band, or above the cutoff edge in the continuum regularization. That branch is not the ring state: at weak impurity it is cutoff/band-edge dominated, while at strong impurity it becomes the high-energy local impurity branch described next.

\begin{figure}[h]
    \centering
    \includegraphics[width=\columnwidth]{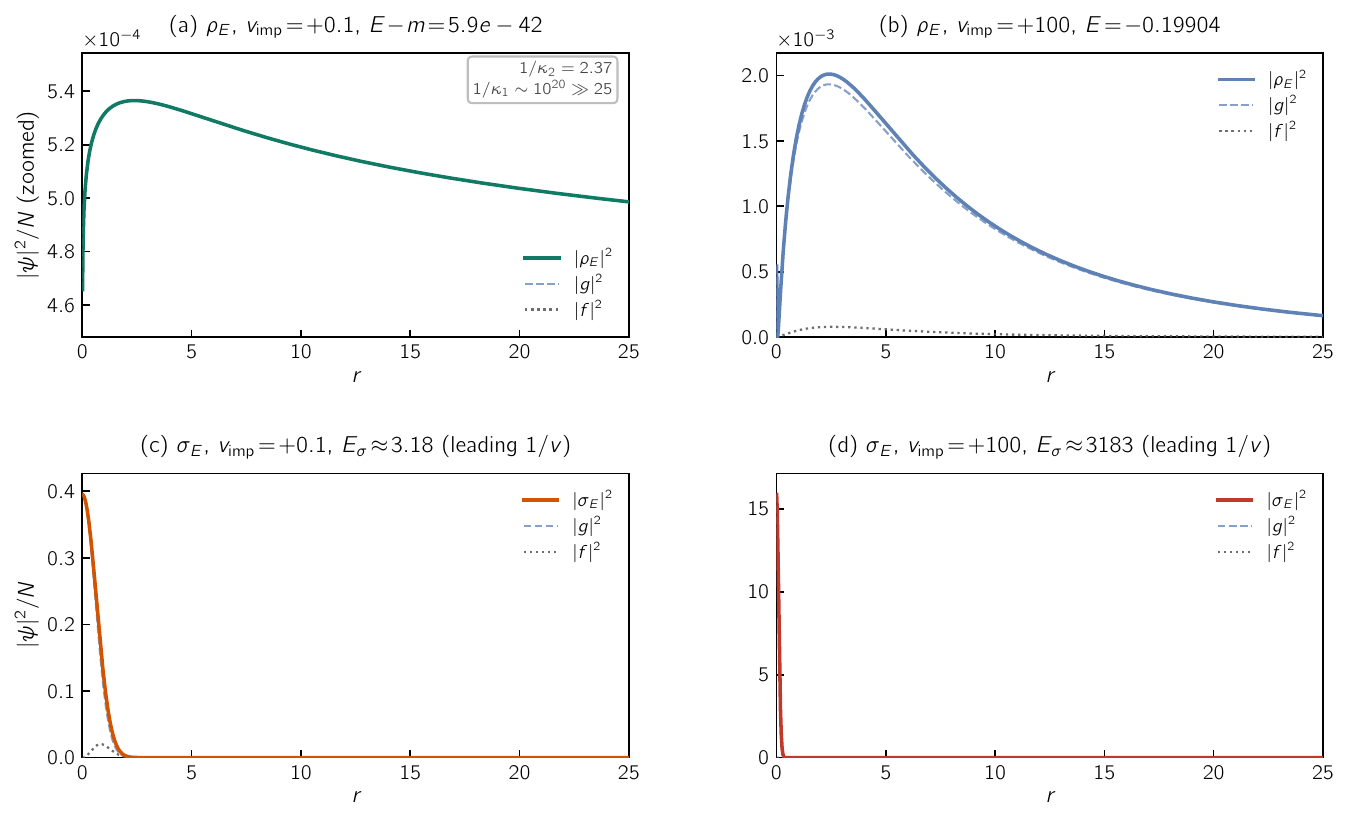}
    \caption{Normalized radial probability density $|\rho|^2/\mathcal{N}$ and $|\sigma|^2/\mathcal{N}$ of $s$-orbital impurity states at $\Lambda=20$, with $|\rho|^2 = g(E;r)^2 + f(E;r)^2$ and $\mathcal{N}=\int_0^\infty 2\pi r\,|\psi|^2\,dr$. Each panel uses its own $y$-axis scale. Top row (gap branch, $g(E)=1/v$): (a) $v=+0.1$, weak-impurity band-edge state with $E\!-\!m\!\simeq\!2.8\!\times\!10^{-58}$; broad envelope on the IR-log background with a $K_0(\kappa_2 r)$ bump at $r\!\sim\!1/\kappa_2\!\approx\!2.4$, with $\mathcal{N}$ astronomically large because the state extends to $1/\kappa_1\!\gg\!1$. (b) $v=+100$, vacancy-like state with $E\!\simeq\!-0.1990394\!\approx\!E_\infty$; the ring profile, with peak at the ring scale $r\!\approx\!2.4$ on $\ket{s}$. Bottom row (upper-solution branch of the same scalar equation, described in the next subsection): mollified leading-$1/v$ asymptote $\sigma_{1/v}(\br)\!\approx\!\delta^{(2)}(\br)\ket{s}$. (c) $v=+0.1$: formal $E\!\approx\!\Lambda^2 v/(4\pi)=3.2$ inside the conduction continuum at this $\Lambda$ -- the panel is the asymptotic schematic, not a true above-band bound state. (d) $v=+100$: formal $E\!\approx\!3\!\times\!10^{3}$ above $E_c^{\rm top}\!\simeq\!600$; at this $\Lambda$ one pole scale $\kappa_2$ turns imaginary so the closed-form $K_0$ decay mixes with an oscillatory component, and the mollified delta is the cleanest visual of the localized-on-$\ket{s}$ character.}
    \label{fig:analyticalwavefuncs}
\end{figure}

\subsection{High-energy $1/v$ branch}

Let us now describe the ordinary high-energy solution of the same scalar equation for an $s$-orbital vacancy. The impurity is $\V=v\proj{s,0}$, and the bound-state condition is
\begin{align}
g^s(E)=\bra{s}\G(E;0,0)\ket{s}={1\over v},
\end{align}
Besides the finite-energy solution discussed above, this equation has a high-energy solution. Let $\omega=E_{1/v}$ denote its energy. Since $g^s(E)\sim1/E$ at large energy, the bound-state condition gives
\begin{align}
{1\over \omega}\simeq {1\over v},\qquad \omega\simeq v.
\end{align}
The corresponding Green's-function spinor has the same formal structure as Eq.~\eqref{eq:ringfull}, but evaluated at this high-energy solution,
\begin{align}
\G(\omega;\br,0)\ket{s}
=g(\omega;r)\ket{s}+ie^{i\varphi}f(\omega;r)\ket{p}.
\end{align}
Because $\omega$ is much larger than the Hamiltonian scale on this branch, the Green's function is dominated by the scalar part $1/\omega$. Here $\H$ denotes the real-space single-particle Hamiltonian operator whose momentum-space matrix is $h(\bk)$. Expanding the resolvent gives
\begin{align}
\G(\omega)
=\left(\omega-\H\right)^{-1}
={1\over \omega}\left[1+{\H\over \omega}+\cdots\right]
\simeq {1\over v}\left[1+{\H\over v}+O(v^{-2})\right].
\end{align}
In real space, the identity term in Eq.~\eqref{eq:cherncont} has kernel $\delta^{(2)}(\br)$. The first correction is the kernel of $\H$, obtained by replacing $\bk\to-i\nabla$ in $h(\bk)$ and acting on the same delta function. For Eq.~\eqref{eq:cherncont},
\begin{align}
h(-i\nabla)=\left(m-b\nabla^2\right)\sigma_z
+i\upsilon\left(\partial_x\sigma_x+\partial_y\sigma_y\right).
\end{align}
Acting on the $s$ orbital then gives the diagonal and off-diagonal pieces below:
\begin{align}
g(\omega;r)=
{1\over v}\delta^{(2)}(\br)
+{1\over v^2}\left(m-b\nabla^2\right)\delta^{(2)}(\br)
+O(v^{-3}),
\end{align}
while the off-diagonal component gives
\begin{align}
ie^{i\varphi}f(\omega;r)
= {i\upsilon\over v^2}
\left(\partial_x+i\partial_y\right)\delta^{(2)}(\br)
+O(v^{-3}).
\end{align}
Thus the Green's-function representative before normalizing the state is
\begin{align}
\G(\omega;\br,0)\ket{s}
= {1\over v}\delta^{(2)}(\br)\ket{s}+O(v^{-2}).
\end{align}
After normalizing the state, the leading factor $1/v$ is removed. In Hilbert-space notation the high-energy branch therefore approaches the local orbital, whose continuum representative is
\begin{align}
\sigma_{1/v}(\br)=\delta^{(2)}(\br)\ket{s}+O(v^{-1}),
\end{align}
These equations only organize the high-energy solution in powers of $1/v$. The formal winding structure appears in the off-diagonal component, but on this branch it is parametrically small: it is $O(v^{-2})$ in the unnormalized Green's-function representative and $O(v^{-1})$ after normalization. This is the key distinction from the finite-energy ring state in Fig.~\ref{fig:analyticalwavefuncs}(a): there the energy approaches $E_\infty$, the functions $g(E_\infty;r)$ and $f(E_\infty;r)$ are independent of $v$ in the vacancy limit, and the off-diagonal winding survives instead of being pushed into a subleading high-energy correction.

\section{Local currents around a vacancy}\label{app:haldane}

We derive the local current carried by the ring state in the continuum Chern insulator~\eqref{eq:cherncont} and reproduce it for a vacancy in the Haldane model~\cite{Haldane.Haldane.1988}. The occupied ring state carries a circulating current around the defect together with a $2\pi$ winding of the relative phase between its orbital components, both fixed by the off-diagonal impurity Green's function and both tracking $\mathrm{sign}(C)$. These are local signatures of the bulk Chern texture, read directly from the orbital winding of the Green's function instead of from linear response, and they survive when the topological domain shrinks to a single vacancy. They are component-resolved quantities in the physical $s,p$ basis, not a new gauge-invariant charge of the impurity state. Jha, Rani, and Ganesh~\cite{Ganesh.Jha.2017} computed impurity currents in the perturbative regime; the strong-vacancy limit here is the complementary case, where the bound state is pinned to the spectral attractor. The current by itself does not separate a topological host from an inhomogeneous trivial one~\cite{Rachel.Diop.2020}; what makes it diagnostic is its locking to the attractor zero of $\mu^\sigma$, which the trivial host does not have.

\subsection{Continuum model}

The velocity operator is obtained from the Hamiltonian~\eqref{eq:cherncont} as $\vec{v} = \partial h/\partial \bk$:
\begin{align}
\vec{v} = 2b\bk\sigma_z - \upsilon\bsigma.\label{eq:velocity}
\end{align}
In real space, the current $\vec{j} = \re[\psi^\dagger \vec{v} \psi]$ becomes~\cite{Xiao.Niu.2010,Culcer.Sinova.2017}
\begin{align}
\vec{j}(\br) = {2b}\im\left[\psi^\dagger \sigma_z \nabla\psi\right] - {\upsilon}\psi^\dagger \bsigma \psi.\label{eq:currentfull}
\end{align}
The first term is the phase gradient contribution: it arises from the spatial variation of the wavefunction phase and is diagonal in the orbital basis ($\sigma_z$). The second term is the hybridization contribution: it arises from the pseudospin texture of the mixed $s$-$p$ state and requires coherence between both orbital components.

\begin{figure}[h]
    \centering
    \includegraphics[width=\columnwidth]{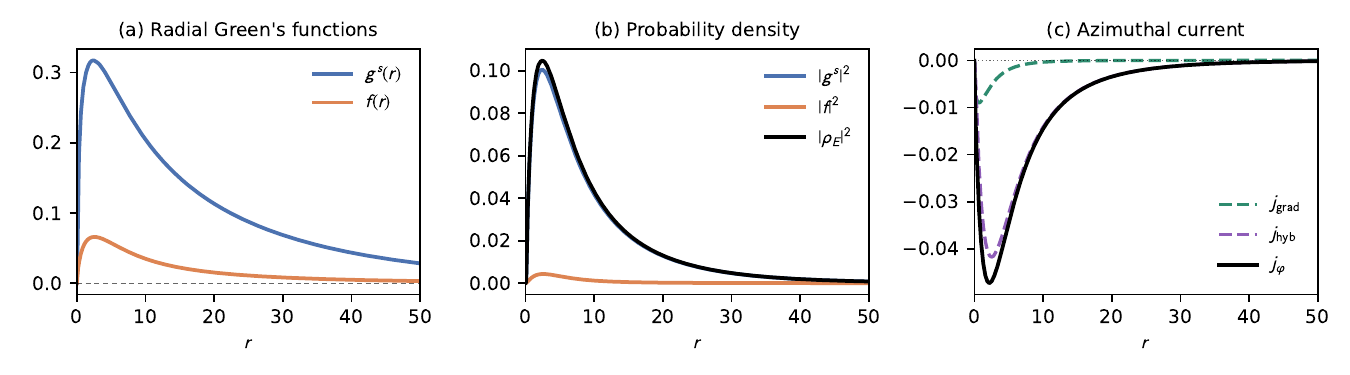}
    \caption{Radial structure of the ring state in the continuum Chern insulator ($m = -0.2$, $\upsilon = 1$, $b = 1.5$, $\Lambda=20$, $E_\infty\approx -0.199$). The two relevant length scales are $1/\kappa_2\approx 2.4$ (inner ring radius, set by the curvature/Dirac crossover $\kappa_2\sim \upsilon/b$) and $1/\kappa_1\approx 30$ (decay length of the long-distance Bessel tail $\sim e^{-\kappa_1 r}/\sqrt{r}$). (a)~Green's functions $g(r)$ and $f(r)$. (b)~Probability density showing dominant $s$-character. (c)~Azimuthal current decomposition: the phase gradient term $j_{\rm grad}$ (from $e^{i\varphi}$ winding) and hybridization term $j_{\rm hyb}$ (from $s$-$p$ interference) have the same sign ($j_{\rm grad}<0$, $j_{\rm hyb}<0$), with the hybridization term dominating in magnitude; the net circulation is clockwise.}
    \label{fig:radial}
\end{figure}

For the ring state wavefunction Eq.~\eqref{eq:ringfull}, $\rho = g(r)\ket{s} + i e^{i\varphi} f(r)\ket{p}$ with real radial functions, the azimuthal current receives contributions from both terms:
\begin{align}
j_\varphi(r) = \underbrace{-\frac{2b|f|^2}{r}}_{j_{\rm grad}} \underbrace{- 2\upsilon\, g f}_{j_{\rm hyb}}.\label{eq:jphi}
\end{align}
Both terms are controlled by the same radial functions $g$ and $f$ defined in Note~\ref{app:chernGF}. The phase-gradient term $j_{\rm grad}$ follows from the $e^{i\varphi}$ phase of the $p$ component, while the hybridization term $j_{\rm hyb}$ comes from interference between the $s$ and $p$ amplitudes. For the parameters used in Fig.~\ref{fig:radial},  the two terms have the same sign in the ring window ($f,g>0$, so $j_{\rm hyb}=-2\upsilon gf<0$) and the hybridization contribution dominates in magnitude, producing a net clockwise current. For the two-band continuum model \eqref{eq:cherncont}, the local current reverses under the time-reversed partner obtained by the joint reversal $(m,b)\to(-m,-b)$.

\subsection{Off-diagonal component vortex}

Consider the wavefunction structure from Eq.~\eqref{eq:ringfull}: $\rho(\br) = g(r)\ket{s} + ie^{i\varphi}f(r)\ket{p}$. The scalar amplitude $g(E;r)$ is the radial extension of the diagonal Green's function in the removed orbital channel, and the strong-impurity state is selected by the zero $g(E_\infty;0)=g^s(E_\infty)=0$. The vortex is not in this scalar piece. It appears in the complementary, off-diagonal response $Q_s\G(E;\br,0)\ket{s}$, where $Q_s=1-\proj{s}$; in the two-band model this is precisely the $p$ component $ie^{i\varphi}f(r)\ket{p}$. Writing $\rho_p(\br)=ie^{i\varphi}f(r)$, projecting onto this component gives
\begin{align}
n_\varphi = \frac{1}{2\pi}\oint d\varphi\, \partial_\varphi \arg(\rho_p) = 1
\end{align}
provided the loop is taken at a radius where $f(r)\neq 0$, so that $\arg(\rho_p)$ is well defined, because the off-diagonal Green's function inherits the angular dependence of the $k_x+i k_y$ hybridization (Sec.~\ref{app:offdiag}). In the rotational continuum model, the off-diagonal Green's function gives a real-space image of the projected-frame singularity discussed in Note~\ref{app:essential}. The winding is a property of the $p$ component in the physical $s,p$ basis, not of the full spinor, which is single valued after a $2\pi$ rotation.

The local current in Fig.~\ref{fig:radial} is the corresponding response of the mixed $s$-$p$ spinor. In the ring window, the hybridization term in Eq.~\eqref{eq:jphi} gives clockwise circulation, and the circulation reverses for the time-reversed continuum model.

\subsection{Higher-winding continuum example}

As an illustration of the same mechanism, we consider a quadratic two-band continuum model whose off-diagonal hybridization winds twice around the origin,
\begin{align}
h_2(\bk)=(m+bk^2)\sigma_z-\upsilon(k_x^2-k_y^2)\sigma_x-2\upsilon k_xk_y\sigma_y,
\label{eq:cherncontC2}
\end{align}
or equivalently $h_{sp}=-\upsilon k_-^2$ and $h_{ps}=-\upsilon k_+^2$, with $k_\pm=k_x\pm i k_y$. Unlike the linear $C=1$ Dirac model of Sec.~\ref{sec:chern}, here $d_\perp(\bk)\sim k^2$ and $d_z(\bk)\sim bk^2$ at large $k$, so $\hat d(\bk)$ retains angular dependence at the UV cutoff: this finite-cutoff continuum model is not compactified to a $C=2$ invariant on its own, and its integer label is fixed by a lattice or UV completion. We use it to illustrate the orbital-winding mechanism at higher winding. Its Green's function denominator is
\begin{align}
D_2(E,k)=E^2-(m+bk^2)^2-\upsilon^2k^4.
\end{align}
For an $s$-orbital vacancy, the bound state has the form
\begin{align}
\rho_{C=2}(\br)=g_2(r)\ket{s}+e^{2i\varphi}f_2(r)\ket{p},
\end{align}
where, with a finite momentum cutoff $\Lambda$,
\begin{align}
g_2(r)&={1\over 2\pi}\int_0^\Lambda \dd k\,k\,{E+m+bk^2\over D_2(E,k)}J_0(kr),\\
f_2(r)&={\upsilon\over 2\pi}\int_0^\Lambda \dd k\,k^3\,{J_2(kr)\over D_2(E,k)}.
\end{align}
Thus the $p$-component winds twice in this orbital basis, while its radial profile differs from that of the linear $C=1$ model.

The current must also be recomputed from the velocity operator of Eq.~\eqref{eq:cherncontC2}. For real radial functions $g_2$ and $f_2$, the azimuthal current is
\begin{align}
j_\varphi^{(2)}(r)=
-{4b\,f_2(r)^2\over r}
+2\upsilon\left[
g_2(r)\left(f_2'(r)+{2f_2(r)\over r}\right)-f_2(r)g_2'(r)
\right].
\label{eq:jphiC2}
\end{align}
The first term is the phase-gradient current for a double winding. The second term is the hybridization current and contains radial derivatives because the $C=2$ Hamiltonian is quadratic in momentum. Here primes denote derivatives with respect to the radial coordinate, $f_2'(r)\equiv\partial_r f_2(r)$ and $g_2'(r)\equiv\partial_r g_2(r)$.

\begin{figure}[h]
    \centering
    \includegraphics[width=.8\columnwidth]{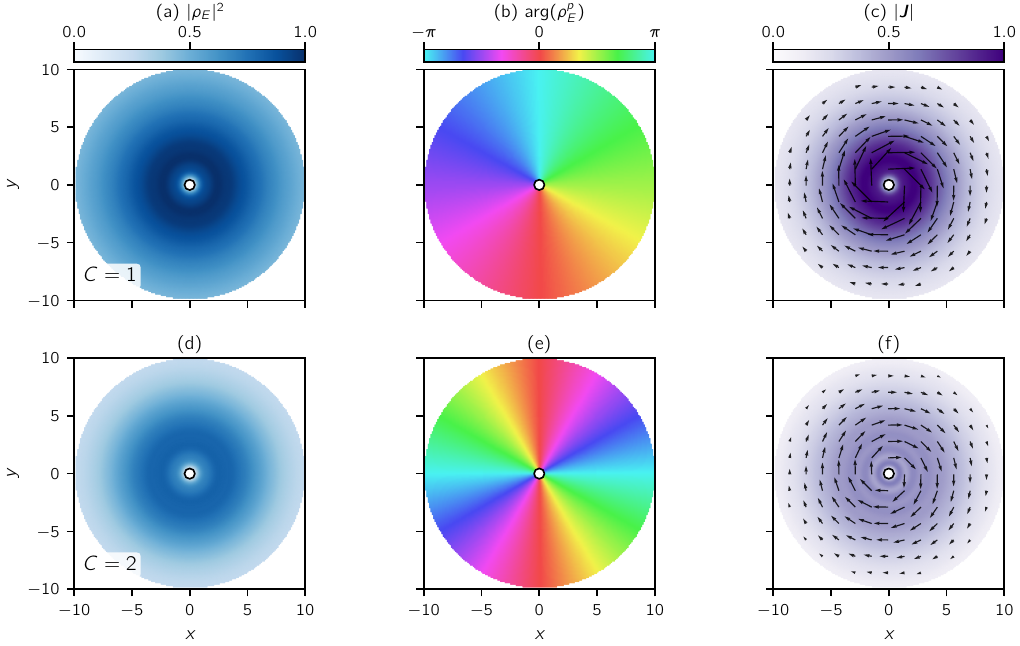}
    \caption{Comparison of ring states for the linear $C = 1$ continuum model (top row, a--c) and the quadratic $C = 2$ continuum model in Eq.~\eqref{eq:cherncontC2} (bottom row, d--f). (a,d) Probability density $|\rho_E|^2$. The $C=2$ density is recalculated from $D_2(E,k)$ and is not assumed to be identical to the $C=1$ density. (b,e) Phase of the $p$-component arg$(\rho_E^p)$, showing single winding ($n_\varphi = 1$) for $C = 1$ and double winding ($n_\varphi = 2$) for $C = 2$. (c,f) Current magnitude $|\mathbf{J}|$ with arrows indicating the circulation obtained from the appropriate current operator. Parameters: $m=-0.2$, $b=1.5$, $\upsilon=1$, $\Lambda=5$ (kept here for comparability with the lattice ring-state energy; cf.~Fig.~\ref{fig:radial} which uses $\Lambda=20$ to suppress Gibbs ringing), with impurity energies $E_{C=1}\approx -0.192$ and $E_{C=2}\approx -0.076$. The component winding is fixed by the angular structure of the hybridization, while the radial profiles and energies retain the logarithmic cutoff dependence discussed in Note~\ref{app:chernGF}.}
    \label{fig:chern2}
\end{figure}

Figure~\ref{fig:chern2} compares the linear $C=1$ model and the quadratic $C=2$ model, recalculating the density and current in both cases using the same finite cutoff. The lower row shows the double component winding expected from the quadratic hybridization, while the current pattern is obtained from Eq.~\eqref{eq:jphiC2} rather than from the $C=1$ current formula.

\subsection{Haldane model numerics}

To compare the continuum picture with a lattice model, we present numerical calculations for the Haldane model~\cite{Haldane.Haldane.1988} on a honeycomb lattice with a single vacancy. The Haldane Hamiltonian is
\begin{align}
H = -t\sum_{\langle ij\rangle} c_i^\dagger c_j + t_2\sum_{\langle\langle ij\rangle\rangle} e^{i\phi_{ij}} c_i^\dagger c_j + M\sum_i \xi_i c_i^\dagger c_i,
\end{align}
where $t$ is the nearest-neighbor hopping, $t_2$ is the next-nearest-neighbor hopping with phase $\phi_{ij} = \pm\phi$ (sign determined by chirality), and $M$ is the sublattice mass with $\xi_i = \pm 1$ for A/B sublattices. The model is topological with Chern number $C = \pm 1$ when $|M| < 3\sqrt{3}t_2\sin\phi$.

We consider a finite system of $N\times N$ unit cells with open boundary conditions and introduce a vacancy by adding a large on-site potential $V_{\rm imp} = 100t$ to a single A-site at the center. The system is diagonalized exactly. The local DOS at site $i$ is $\nu_i = \bra{i} \P \ket{i}$, where $\P = \sum_{n\in\rm occ} \proj{\psi_n}$ is the projector onto occupied states; to highlight the ring state contribution, we plot the deviation from the bulk average $\delta\nu_i = \nu_i - \bar{\nu}_{\rm bulk}$. Here $\nu_i$ is an occupied-projector equilibrium response at fixed chemical potential, summed over all states below the Fermi level, rather than an energy-resolved STM LDOS or the single-eigenstate density of the isolated continuum ring state of Sec.~\ref{sec:chern}; the two diagnostics agree on the ring geometry in the topological phase but differ in spectral content. The bond current between sites $i$ and $j$ is
\begin{align}
J_{ij} = 2\,\mathrm{Im}\!\left[H_{ij}\P_{ji}\right],
\end{align}
up to orientation convention, with $\P_{ij}\equiv\bra{i}\P\ket{j}$ the single-particle density matrix. The local current vector at site $i$ is $\vec{J}_i = \sum_j (\br_j - \br_i) J_{ij}$, summed over neighboring sites.

\begin{figure}[h]
    \centering
    \includegraphics[width=0.6\columnwidth]{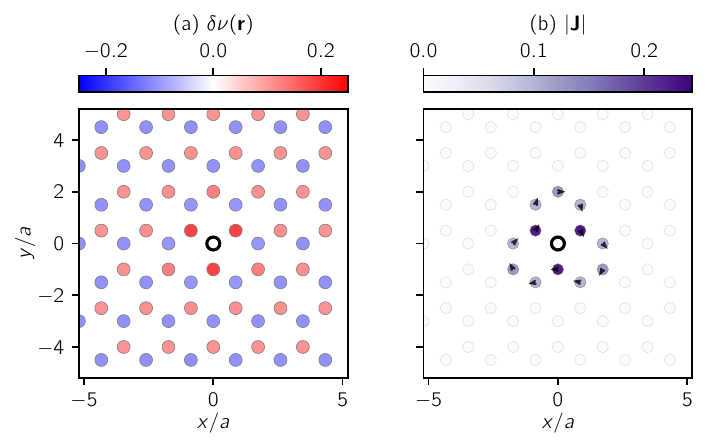}
    \includegraphics[width=0.6\columnwidth]{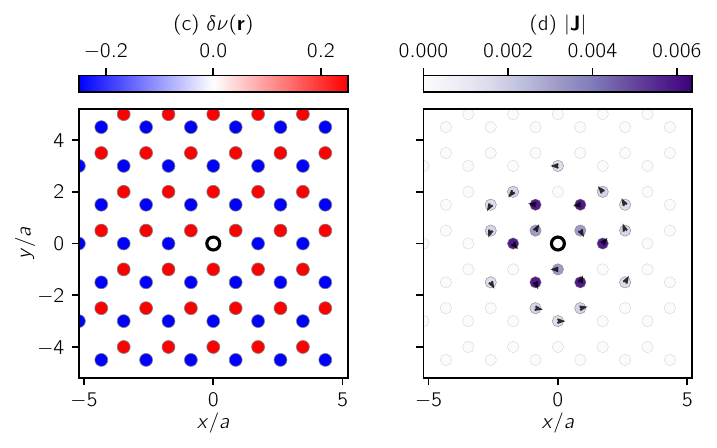}
    \caption{Comparison of vacancy states in the Haldane model ($N=21$). Left: topological phase ($M=0.5t$, $t_2=0.3t$, $C=1$) showing (a) coherent, azimuthally uniform ring of DOS enhancement $\delta\nu$ and (b) circulating current. Right: trivial phase ($M=2.0t$, $t_2=0.1t$, $C=0$) showing (c) scattered Friedel-like DOS oscillations without coherent ring structure and (d) absence of comparable circulating currents. The coherent ring pattern combined with persistent current is used here as a local signature of the topological phase.}
    \label{fig:haldane_ring}
\end{figure}

Figure~\ref{fig:haldane_ring} compares the topological and trivial phases. In the topological phase ($M = 0.5t$, $t_2 = 0.3t$, $\phi = \pi/2$, giving $C = 1$), the vacancy binds a ring state with two local signatures: (i) a \emph{coherent}, azimuthally uniform ring of enhanced DOS surrounding the vacancy, and (ii) a persistent circulating current, with chirality consistent with the continuum Chern texture. In the trivial phase ($M = 2.0t$, $t_2 = 0.1t$, $C = 0$), the vacancy induces Friedel-like oscillations that can have larger amplitude but lack the coherent ring structure: the DOS fluctuations are scattered and anisotropic rather than forming a uniform ring. The trivial phase also shows no comparable circulating current. The coherent ring together with the persistent circulation therefore distinguishes the topological phase from the trivial one.

\section{Dimension independence of ring states}\label{app:dimension}

In this note, we establish that the spectral attractor mechanism applies in any spatial dimension $d$ in which the band-edge sign conditions below are satisfied, arising from the Wannier obstruction encoded in band inversion rather than from the specific form of band-edge singularities. We further show that band inversion creates zeros of the projected Green's function not only inside the gap, where the attractor resides, but also inside each band, where these zeros act as repellers. This structure provides a sharp distinction between topological and trivial insulators that holds in any dimension.

\subsection{General formulation}

To make the role of dimensionality explicit, we write the $d$-dimensional continuum Hamiltonian using the Clifford algebra,
\begin{align}\label{eq:hdim}
h(\bk) = \upsilon\sum_{i=1}^d k_i \Gamma_i + (m + bk^2)\Gamma_0,
\end{align}
where $\Gamma_0, \Gamma_1, \ldots, \Gamma_d$ satisfy $\{\Gamma_\mu, \Gamma_\nu\} = 2\delta_{\mu\nu}$. The virtue of this formulation is that no explicit matrix representation is needed: the algebra alone determines all physical quantities, and the dimension $d$ enters only as a parameter. The spectrum is $\varepsilon_\pm(\bk) = \pm\sqrt{\upsilon^2 k^2 + (m+bk^2)^2}$, yielding a gap of $2|m|$. Band inversion occurs when $mb < 0$, which we take to define the topological phase.

For an impurity that couples to the $\Gamma_0 = +1$ orbital, which we call the $s$-orbital, the projected Green's function takes the form
\begin{align}
g^s(\omega) &= \int \frac{d^dk}{(2\pi)^d} \frac{\omega + m + bk^2}{\omega^2 - \upsilon^2k^2 - (m+bk^2)^2} \nonumber\\
&= \Omega_d \int_0^\infty dk\, k^{d-1} \, \mathcal{I}(k, \omega),\label{eq:gsdim}
\end{align}
where $\Omega_d = S_{d-1}/(2\pi)^d$ with $S_{d-1} = 2\pi^{d/2}/\Gamma(d/2)$ the surface area of the unit $(d{-}1)$-sphere. The key observation is that the integrand
\begin{align}
\mathcal{I}(k, \omega) = \frac{\omega + m + bk^2}{\omega^2 - \upsilon^2k^2 - (m+bk^2)^2}
\end{align}
is entirely independent of dimension. All dependence on $d$ enters through the radial measure $k^{d-1}dk$ and the geometric prefactor $\Omega_d$. This separation will allow us to identify which features of the spectral attractor are universal and which depend on dimensionality.

\subsection{Band-edge behavior and UV regularization}

The behavior of $g^s(\omega)$ near the band edges depends on dimension through the density of states. Near a band edge $E_{\rm edge}$, setting $\Delta\omega = |E_{\rm edge} - \omega| \to 0$, the projected Green's function scales as $g^s(\omega) \sim (\Delta\omega)^{(d-2)/2}$. For $d < 2$ this represents a divergence, for $d = 2$ it becomes logarithmic, and for $d > 2$ the function remains finite at the band edge. In low dimensions, arbitrarily weak impurities can bind states; for $d > 2$ a threshold impurity strength is required. However, this distinction affects only the threshold behavior, not the existence of the spectral attractor.

To obtain finite values at the band edges for $d > 2$, we regularize the continuum model with an ultraviolet cutoff $k_\Lambda$ that mimics the finite Brillouin zone of a lattice. The integral in Eq.~\eqref{eq:gsdim} then runs from $0$ to $k_\Lambda$. Defining the momentum scale
\begin{align}
\alpha^2 = \frac{\upsilon^2 + 2mb}{b^2},
\end{align}
the projected Green's function at the band edges evaluates to
\begin{align}
\mu^s(E_v) &= -\frac{\Omega_d}{b^2} \int_0^{k_\Lambda} dk\, k^{d-3} \frac{2m + bk^2}{\alpha^2 + k^2}, \\
\mu^s(E_c) &= -\frac{\Omega_d}{b} \int_0^{k_\Lambda} dk\, \frac{k^{d-1}}{\alpha^2 + k^2}.
\end{align}
For $d = 3$, these integrals give
\begin{align}
\mu^s(E_v) &= -\frac{\Omega_3}{b^2}\left[\left(\frac{2m}{\alpha} - b\alpha\right)\arctan\frac{k_\Lambda}{\alpha} + bk_\Lambda\right], \\
\mu^s(E_c) &= -\frac{\Omega_3}{b}\left[k_\Lambda - \alpha\arctan\frac{k_\Lambda}{\alpha}\right].
\end{align}
For our model parameters $m = -1$, $b = 0.5$, $\upsilon = 1.5$, and cutoff $k_\Lambda = 2$, we obtain $\alpha \approx 2.24$ and the numerical values $\mu^s(E_v) \approx +0.095$ and $\mu^s(E_c) \approx -0.037$. The opposite signs confirm the existence of a zero inside the gap. For the parameter range and physical cutoff used here ($k_\Lambda$ of order the inversion scale $\sim\sqrt{|m/b|}$), varying $k_\Lambda$ changes the magnitudes while preserving the sign structure that guarantees the spectral attractor; pushing $k_\Lambda$ far beyond the inversion scale would let the $bk^2$ piece in the valence-edge integrand dominate and can in principle flip the sign of $\mu^s(E_v)$, so the relevant statement is that the lattice (or a physically motivated regularization at the inversion scale) preserves the topological sign assignment.

\subsection{Sign structure and the gap attractor}

Inside the gap, where there are no band states, the integrand in Eq.~\eqref{eq:gsdim} has no poles on the real $k$-axis. Consequently, $g^s(\omega)$ is purely real throughout the gap: $g^s = \mu^s$ with $\nu^s = 0$. The existence of a zero in $\mu^s(\omega)$ follows from a sign argument that depends only on the orbital content at the band edges, not on any special symmetry such as particle-hole symmetry.

As $\omega$ approaches the valence band edge $E_v$ from inside the gap, the integral is dominated by small $k$ where the band eigenstates have definite orbital character. In the topological phase with $m < 0$, the valence band at small $k$ has predominantly $s$-character. Near $\omega = E_v$, the numerator evaluates to $\omega + m + bk^2 \approx 2m < 0$ at small $k$, while the denominator $\omega^2 - \upsilon^2k^2 - (m+bk^2)^2$ is negative throughout the gap. A negative numerator divided by a negative denominator yields a positive integrand, and therefore $\mu^s(E_v^+) > 0$.

The situation reverses at the conduction band edge. As $\omega \to E_c^-$, the numerator at $k = 0$ becomes $\omega + m \to 0^-$, while for any $k > 0$ the numerator $\omega + m + bk^2$ is positive. Since the denominator remains negative, the integrand is negative for all $k > 0$, giving $\mu^s(E_c^-) < 0$.

\begin{figure}[h]
    \centering
    \includegraphics[width=.8\columnwidth]{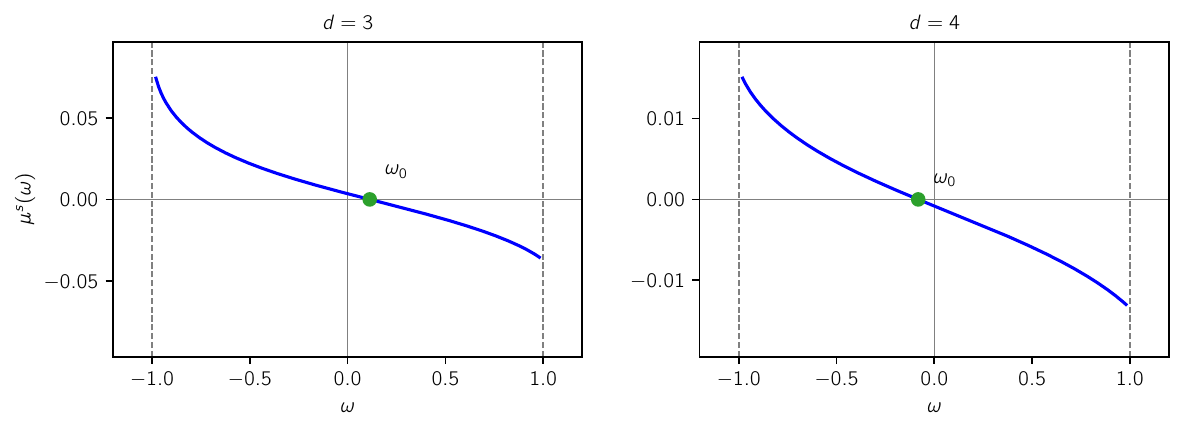}
    \caption{Real part of the projected Green's function $\mu^s(\omega)$ inside the gap for the $d$-dimensional Hamiltonian~\eqref{eq:hdim} with $m = -1$, $b = 0.5$, $\upsilon = 1.5$, and UV cutoff $k_\Lambda = 2$. Dashed lines mark the band edges $E_v = -1$ and $E_c = +1$. The finite values $\mu^s(E_v) > 0$ and $\mu^s(E_c) < 0$ guarantee a zero $\omega_0$ inside the gap, marking the spectral attractor. (a)~$d = 3$: $\omega_0 \approx 0.11$. (b)~$d = 4$: $\omega_0 \approx -0.08$.}
    \label{figapp:mudimension}
\end{figure}

The opposite signs at the two band edges guarantee, by continuity, a zero $\mu^s(\omega_0) = 0$ at some energy $\omega_0$ inside the gap. This is the spectral attractor (denoted $E_a$ in Note~\ref{app:proof}; its flat-band limit is the $E_\infty$ of Note~\ref{app:proof1}). The position of $\omega_0$ is not fixed at mid-gap by any symmetry; it is determined by the competition between valence and conduction contributions and generically lies away from the gap center. The zero exists because band inversion redistributes orbital weight between the bands. Figure~\ref{figapp:mudimension} illustrates this behavior for $d = 3$ and $d = 4$.

\subsection{Zeros inside the bands: repellers}

The sign analysis extends beyond the gap. Inside each band, $\mu^s(\omega)$ also exhibits zeros that arise from the same topological origin.

The spectral sum rule provides universal constraints on the asymptotic behavior: $\mu^s(\omega) \to 1/\omega$ as $|\omega| \to \infty$, so that $\mu^s \to 0^+$ as $\omega \to +\infty$ and $\mu^s \to 0^-$ as $\omega \to -\infty$. These limits hold in any dimension and for any band structure.

For $d > 2$, where $\mu^s$ remains finite at the band edges, the function is continuous across the gap-to-band boundaries (the discontinuity appears only in the imaginary part $\nu^s$, which turns on inside the bands). The signs established at the band edges from inside the gap therefore persist just inside each band: $\mu^s(E_v^-) > 0$ when approaching the valence edge from below, and $\mu^s(E_c^+) < 0$ when approaching the conduction edge from above.

Combining these observations with the asymptotic behavior reveals additional sign changes. In the valence band, $\mu^s$ evolves from $0^-$ at $\omega \to -\infty$ to a positive value at $\omega \to E_v^-$. By continuity, there must exist an energy $\omega_v < E_v$ where $\mu^s(\omega_v) = 0$. Similarly, in the conduction band, $\mu^s$ evolves from a negative value at $\omega \to E_c^+$ to $0^+$ at $\omega \to +\infty$, requiring a zero at some $\omega_c > E_c$.

\begin{figure}[h]
    \centering
    \includegraphics[width=.55\columnwidth]{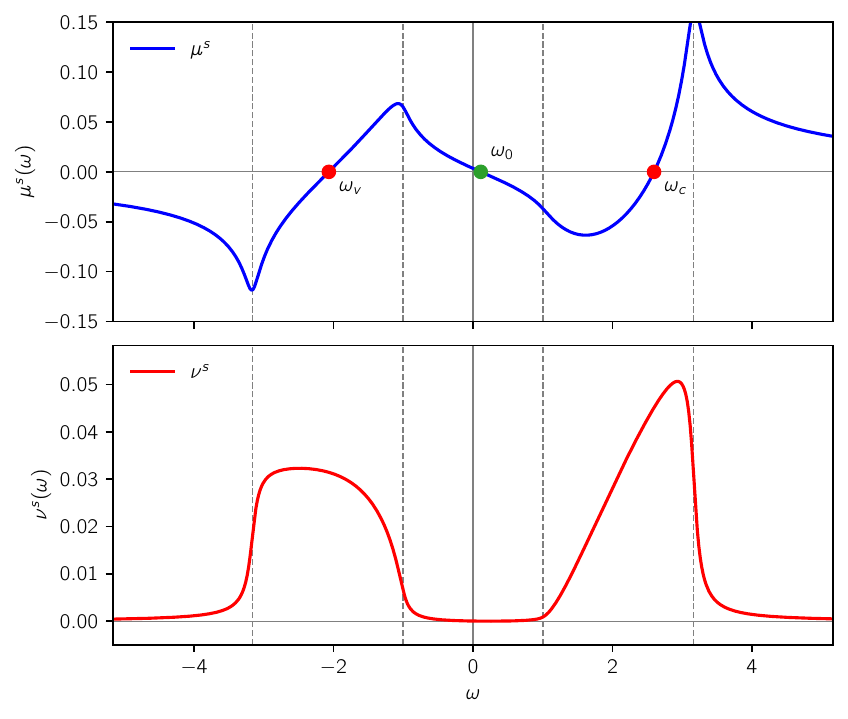}
    \caption{(a)~Real part $\mu^s(\omega)$ of the projected Green's function for the $d = 3$ Hamiltonian~\eqref{eq:hdim} with UV cutoff $k_\Lambda = 2$. Dashed lines mark the band edges: inner edges $E_v$, $E_c$ and outer bandwidth edges $\pm E_\Lambda$. Three zeros are visible: valence repeller $\omega_v$ (red), gap attractor $\omega_0$ (green), and conduction repeller $\omega_c$ (red). (b)~Spectral function $\nu^s(\omega) \geq 0$, nonzero only inside the bands. Outside the bandwidth, $\mu^s \to 1/\omega$ as required by the spectral sum rule.}
    \label{figapp:murepellers}
\end{figure}

These band zeros act as repellers for impurity states. As the impurity strength varies, resonances within the bands flow toward these zeros but cannot cross them, effectively bounding the accessible energy range for impurity-induced spectral features. Figure~\ref{figapp:murepellers} illustrates this structure for $d = 3$, showing the three zeros of $\mu^s(\omega)$ together with the spectral function $\nu^s(\omega)$.

\subsection{Topological versus trivial phases}

The number of zeros of $\mu^s(\omega)$ provides a sharp distinction between topological and trivial insulators. In the topological phase, where band inversion distributes $s$-orbital weight across both bands, sign changes occur in all three spectral regions, yielding three zeros: one in the valence band at $\omega_v < E_v$, one in the gap at $\omega_0$, and one in the conduction band at $\omega_c > E_c$.

The trivial phase presents a stark contrast. When $mb > 0$ the band edges are not inverted: at $k=0$ the $s$-orbital sits entirely on one side of the gap (in the conduction band for $m>0$, in the valence band for $m<0$ with $b<0$), and at the corresponding band edge the numerator $\omega + m + bk^2$ keeps the same sign approached from inside the gap. The $s$-orbital still acquires weight in both bands away from $k=0$ through the $\upsilon k_i\Gamma_i$ coupling, but the band-edge sign structure differs from the topological case: the numerator no longer changes sign between $E_v$ and $E_c$, so $\mu^s$ does not cross zero inside the gap. The same band-edge sign argument shows that the single zero is forced inside the band hosting the $k=0$ $s$-weight (the conduction band for $m>0$, $b>0$), while in the opposite band $\mu^s$ retains a definite sign. The trivial phase therefore has a single zero, compared to three zeros in the topological phase: the distinction is in the band-edge orbital-content signs set by inversion, not in any complete absence of $s$-weight from one band.

\subsection{Dimension independence}

The arguments above rely on three ingredients: (i) the orbital content at band edges, set by band inversion; (ii) the asymptotic behavior $\mu^s \to 1/\omega$ from the spectral sum rule; and (iii) the continuity of $\mu^s$ across band edges for $d > 2$. None depends on spatial dimension. The measure $k^{d-1}dk$ affects magnitudes and threshold behavior, but not the signs that determine zeros.

One subtlety: the $1/\omega$ asymptotics requires finite bandwidth. In continuum models the bands extend to infinite energy, so the UV cutoff $k_\Lambda$ plays a physical role: it mimics the finite Brillouin zone of a lattice, restoring the correct asymptotic structure. Without regularization, the $k^{d-1}$ measure weights large momenta so heavily that high-energy contributions mask the sign structure near band edges. This is an artifact of the continuum approximation; in any real material with finite bandwidth, the sign argument applies.

We conclude that within this two-orbital Dirac family, band inversion creates an attractor inside the gap and, under the band-edge sign conditions assumed above, two additional repellers inside the bands, while the trivial phase has only the single high-energy zero implied by the $1/\omega$ asymptotics. The argument is dimension-independent in the sense that the measure $k^{d-1}dk$ does not change the relevant signs, provided the UV regularization is set at the band-inversion scale, as occurs automatically on a lattice. We do not claim this counting extends to more general topological band structures without checking the corresponding band-edge sign assignments.

\section{Edge states as Green's function zeros}\label{app:edgezeros}

In Sec.~\ref{sec:beyondrank1} we argued that clustered vacancies hybridize ring states into dispersive chiral modes, and that a complete wall of vacancies reproduces the chiral edge mode of the Chern insulator. Here we demonstrate this explicitly: the edge-state dispersion coincides with the zeros of the impurity-projected Green's function~\cite{Balents.Slager.2015}, now parameterized by the conserved momentum along the wall.

\subsection{Wall defect and factorization by momentum}

Consider the lattice Chern insulator~\eqref{eqapp:qwzmodel} on a square lattice with sites $\bq=(x,y)$ and orbitals $\sigma\in\{s,p\}$. We introduce a wall of vacancies along the $x$-axis at position $y=y_0$, removing all orbitals on that row:
\begin{align}
\V = v \sum_{x,\sigma} \proj{\phi^\sigma_{x,y_0}},\qquad v\to\infty.
\label{eq:walldefect}
\end{align}
Since $\V$ is translationally invariant along $x$, the crystal momentum $k_x$ remains a good quantum number. Introducing the Fourier-transformed basis states
\begin{align}
\ket{\phi^\sigma_{k_x,y_0}} = \frac{1}{\sqrt{N_x}} \sum_x e^{ik_x x}\ket{\phi^\sigma_{x,y_0}},
\end{align}
the perturbation factorizes as
\begin{align}
\V = v \sum_{k_x,\sigma}\proj{\phi^\sigma_{k_x,y_0}}.
\end{align}
Since also the clean Hamiltonian $\H$ is diagonal in $k_x$, the $T$-matrix and bound-state conditions factorize by $k_x$. At each $k_x$, the problem reduces to the rank-2 impurity problem studied in Sec.~\ref{sec:beyondrank1}, with the impurity-projected Green's function
\begin{align}
g^{\sigma\tau}(\omega;k_x) = \bra{\phi^\sigma_{k_x,y_0}}(\omega-\H)^{-1}\ket{\phi^\tau_{k_x,y_0}}.
\label{eq:gsurf}
\end{align}
The $g^{\sigma\tau}$ are the elements of a $2\times 2$ matrix $g(\omega;k_x)$ in the orbital indices $\sigma,\tau\in\{s,p\}$, depending parametrically on $k_x$.

\subsection{Bound state condition and edge mode dispersion}

In the vacancy limit $v\to\infty$, the bound-state energies at each $k_x$ are determined by~\cite{Balents.Slager.2015}
\begin{align}
\det\, g(E;k_x) = 0.
\label{eq:edgezerocond}
\end{align}
The solutions $E(k_x)$ trace out the edge-state dispersion. To understand its structure, define the eigenvalues $\mu_\pm(\omega;k_x)$ of the real symmetric matrix $\re\, g(\omega;k_x)$. In the spectral gap, where $ g$ is real, these eigenvalues decrease monotonically with $\omega$ (Sec.~\ref{sec:beyondrank1}). Their zeros,
\begin{align}
\mu_\pm(E;k_x) = 0,
\label{eq:muzero_edge}
\end{align}
generalize the single-impurity condition $\mu^\sigma(E_\infty)=0$ from Sec.~\ref{sec:levelrepulsion}, now yielding a dispersion rather than a pinning energy.

Each zero of $\mu_\pm$ is either an \emph{attractor} or a \emph{repeller}, depending on the sign of $\partial_\omega\mu_\pm$ at the crossing. An attractor has $\mu_\pm>0$ below and $\mu_\pm<0$ above ($\partial_\omega\mu_\pm<0$): the vacancy wall binds a state at this energy, and small perturbations push it back. A repeller has the opposite sign order, $\mu_\pm<0$ below and $\mu_\pm>0$ above ($\partial_\omega\mu_\pm>0$): the zero is unstable and does not pin a bound state. These two types of zeros come in pairs. At a critical momentum $k_x^*$ near the gap edge, the attractor and repeller branches merge and annihilate; beyond $k_x^*$, no in-gap zero exists and the vacancy wall does not bind a state at that momentum.

The total scattering phase shift $\arg(\det\,g)$ jumps by $\pi$ at each zero of $\mu_\pm$, whether attractor or repeller. The net phase accumulated across the gap counts the difference between states removed from and returned to the continuum, connecting the edge-state dispersion to the Friedel sum rule of Sec.~\ref{sec:dispersive}.

$ g(\omega;k_x)$ depends on the \emph{bulk} band structure alone. The edge mode dispersion is entirely encoded in the bulk Green's function; introducing the boundary merely selects which of its zeros become physical bound states.

\subsection{Evaluation by momentum integration}

The projected Green's function~\eqref{eq:gsurf} is evaluated by integrating the bulk resolvent over the transverse momentum $k_y$ at fixed $k_x$:
\begin{align}
g^{\sigma\tau}(\omega;k_x) = \int_{-\pi}^{\pi}\frac{dk_y}{2\pi}\, \left[\omega+i0^+-h(k_x,k_y)\right]^{-1}_{\sigma\tau},
\label{eq:gsurf_kint}
\end{align}
where $h(k_x,k_y)$ is the $2\times2$ Bloch Hamiltonian~\eqref{eqapp:qwzmodel}. This is a one-dimensional integral of a $2\times2$ matrix, computable from the bulk band structure alone with no reference to boundaries or ribbons.

\subsection{Numerical demonstration}\label{app:edgezeros:numerics}

We evaluate~\eqref{eq:gsurf_kint} numerically for the lattice Chern insulator~\eqref{eqapp:qwzmodel} with parameters $m=-0.5$, $\upsilon=1$, $b=1.5$ (the same parameters used in Fig.~\ref{fig:impurityhyb}), and broadening $\eta=0.03$.

\begin{figure}[h]
    \centering
    \includegraphics[width=\columnwidth]{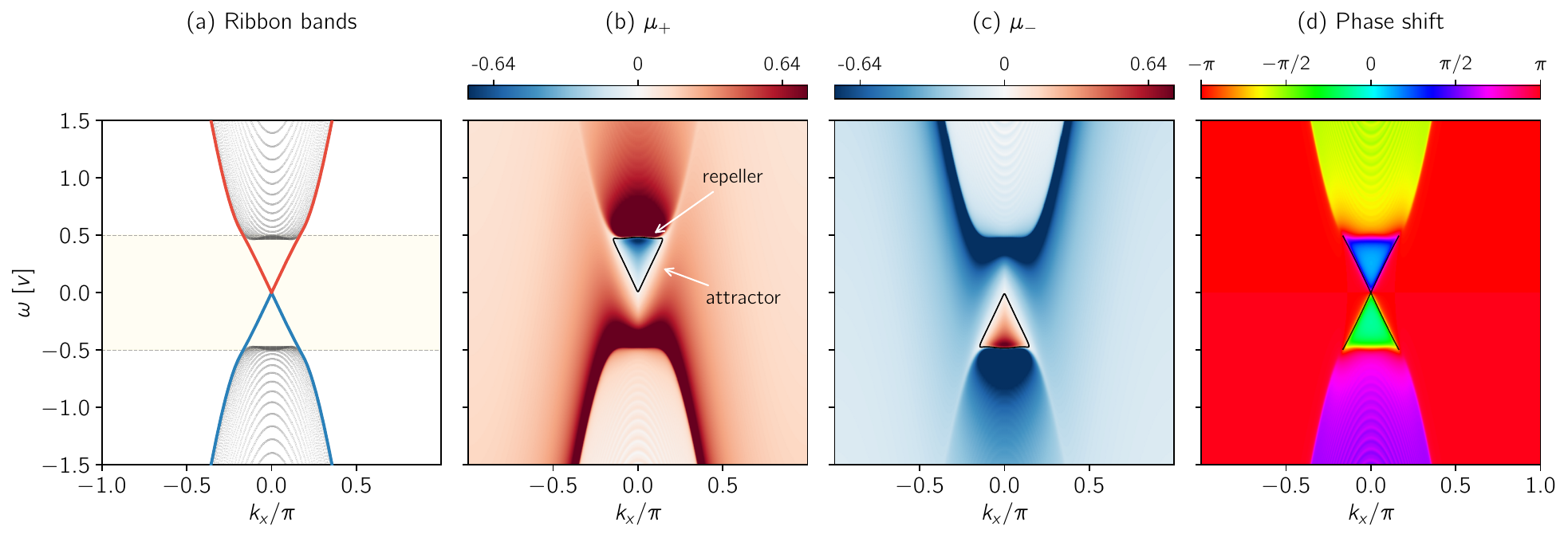}
    \caption{Edge states as zeros of the bulk-projected Green's function for the lattice Chern insulator~\eqref{eqapp:qwzmodel} with $m=-0.5$, $\upsilon=1$, $b=1.5$. (a)~Ribbon band structure ($L_y=60$, open boundaries in $y$). Gray: bulk sub-bands; red/blue: chiral edge modes. (b,c)~Eigenvalues $\mu_+$ and $\mu_-$ of $\re\,g^{\sigma\tau}(\omega;k_x)$ computed from the $k_y$-integrated bulk resolvent~\eqref{eq:gsurf_kint}. A zero with $\mu>0$ below and $\mu<0$ above (red$\to$blue) is an attractor: the energy at which the vacancy wall binds a state. A zero with the opposite sign order (blue$\to$red) is a repeller. At a critical momentum $k_x^*$, the attractor and repeller branches merge and annihilate at the band edge; beyond $k_x^*$, no in-gap state exists. (d)~Total phase shift $\arg(\det\,g^{\sigma\tau})$: the $\pi$ jumps at each zero count the states removed from the continuum by the vacancy wall, generalizing the Friedel sum rule to the dispersive boundary. Broadening $\eta=0.03$.}
    \label{fig:edgezeros}
\end{figure}

Figure~\ref{fig:edgezeros}(a) shows the ribbon band structure ($L_y=60$, open boundaries in $y$) for reference: the chiral edge mode crosses the bulk gap near $k_x=0$, consistent with $C=-\mathrm{sign}(m)=+1$ (Note~\ref{app:chernGF}). Panels~(b) and~(c) show the eigenvalues $\mu_+(\omega;k_x)$ and $\mu_-(\omega;k_x)$ of the real symmetric matrix $\re\,g^{\sigma\tau}(\omega;k_x)$, computed from the bulk integral~\eqref{eq:gsurf_kint} with no reference to boundaries. Each eigenvalue changes sign along a line that traces one branch of the edge-mode dispersion; these are the spectral attractors where the projected Green's function crosses zero. The zero contours coincide with the edge-state spectrum from panel~(a).

Panel~(d) shows the total phase shift $\arg(\det\,g^{\sigma\tau})$, which jumps by $\pi$ at each attractor line. These jumps count the states removed from the bulk continuum by the vacancy wall, connecting the edge-state dispersion to the Friedel sum rule of Sec.~\ref{sec:dispersive}.

This is the multi-vacancy generalization of the ring-state mechanism. At a single vacancy, the bound-state energy is pinned at $\mu^\sigma(E_\infty)=0$; for a wall of vacancies, the same condition~\eqref{eq:edgezerocond} yields a dispersion $E(k_x)$ parameterized by the conserved momentum along the wall. Ring states at isolated vacancies (Fig.~\ref{fig:impurityhyb}a,d) are the $k_x$-independent limit of this dispersion; as vacancies coalesce into a wall, their pinning energies fan out into the dispersing edge mode (Fig.~\ref{fig:impurityhyb}c,f).
\end{document}